\documentclass[11pt,a4paper]{article}

\usepackage{amsmath}
\usepackage{amsfonts}
\usepackage{amssymb}
\usepackage{amsthm}
\usepackage{graphicx}
\usepackage{mathtools}
\usepackage{dsfont}
\usepackage{algorithm}
\usepackage{tabularx, booktabs}
\usepackage{caption}
\usepackage{algorithmic}
\usepackage{enumitem}
\usepackage{threeparttable} 
\usepackage{lmodern}
\usepackage{multirow}
\usepackage{array}
\usepackage{makecell}
\usepackage{microtype}
\usepackage[left=2cm,right=2cm,top=2cm,bottom=2cm]{geometry}
\usepackage[hyperfootnotes=false]{hyperref}
\hypersetup{hidelinks,
  pdftitle={Distribution-Free Symmetry Detection in Multi-Reference Alignment at Low SNR},
  pdfauthor={Nir Sharon and Sheng Xu}}
\graphicspath{{./Figs/}}
\usepackage{placeins}
\usepackage{subcaption}
\usepackage{xcolor}
\usepackage{comment}
\usepackage{titling}
\usepackage{multicol}
\newtheorem{theorem}{Theorem}

\newtheorem{lemma}[theorem]{Lemma}
\newtheorem{corollary}[theorem]{Corollary}
\newtheorem{proposition}[theorem]{Proposition}

\newtheorem{definition}{Definition}
\newtheorem{remark}{Remark}

\newcommand{\0}{\mathbf{0}}
\newcommand{\s}{\mathbf{s}}
\newcommand{\x}{\mathbf{x}}
\newcommand{\y}{\mathbf{y}}
\newcommand{\z}{\mathbf{z}}
\newcommand{\e}{\mathbf{e}}
\renewcommand{\u}{\mathbf{u}}
\newcommand{\vv}{\mathbf{v}}
\newcommand{\w}{\mathbf{w}}
\newcommand{\p}{\mathbf{p}}
\newcommand{\q}{\mathbf{q}}
\newcommand{\g}{\mathbf{g}}
\newcommand{\T}{\mathsf{T}}
\newcommand{\F}{\mathcal{F}}
\newcommand{\Fb}{\mathbf{F}}
\newcommand{\Ub}{\mathbf{U}}
\newcommand{\Wb}{\mathbf{W}}
\newcommand{\Cb}{\mathbf{C}}
\newcommand{\Db}{\mathbf{D}}
\newcommand{\Ib}{\mathbf{I}}
\newcommand{\Ab}{\mathbf{A}}
\newcommand{\Eb}{\mathbf{E}}
\newcommand{\E}{\mathbb{E}}
\renewcommand{\P}{\mathbb{P}}
\newcommand{\ri}{\mathrm{i}}
\newcommand{\R}{\mathbb{R}}
\newcommand{\Z}{\mathbb{Z}}

\newcommand{\C}{\mathbb{C}}
\newcommand{\sH}{\mathsf{H}}
\newcommand{\cC}{\mathcal{C}}
\newcommand{\cX}{\mathcal{X}}
\newcommand{\cI}{\mathcal{I}}
\newcommand{\cJ}{\mathcal{J}}
\newcommand{\dd}{\mathrm{d}}
\newcommand{\1}{\mathds{1}}

\newcommand{\bOmega}{\mathbf{\Omega}}
\newcommand{\bpsi}{\boldsymbol{\psi}}
\newcommand{\bmu}{\boldsymbol{\mu}}
\newcommand{\beps}{\boldsymbol{\varepsilon}}
\newcommand{\bbeta}{\boldsymbol{\eta}}
\newcommand{\Var}{\mathrm{Var}}

\newcommand{\SO}[1]{\mathcal{SO}({#1})}

\newcommand{\Cd}[1]{\mathsf{C}_{#1}}
\newcommand{\Dm}[1]{\mathsf{D}_{#1}}
\newcommand{\Dmj}[2]{\mathsf{D}_{#1,#2}}
\newcommand{\Sym}{\mathrm{Sym}}
\newcommand{\Fix}[1]{V^{#1}}

\begin{document}

\title{\textbf{Distribution-Free Symmetry Detection in Multi-Reference Alignment at Low SNR}}
\author{Nir Sharon \and Sheng Xu\thanks{Corresponding author. Email: \texttt{sxu@utdallas.edu}}}
\date{}

\maketitle
\begin{abstract}
Symmetry can reveal the structure of a signal, but detecting it becomes difficult when observations are noisy and randomly transformed. We study how to infer symmetry directly from multi-reference alignment data with additive isotropic Gaussian noise of known variance, without aligning the observations or knowing how the transformations were sampled.

For cyclic symmetry, we identify a simple statistical signature that yields tests with exact finite-sample error guarantees. We demonstrate their performance even at very low signal-to-noise ratios. Dihedral symmetry presents a deeper challenge: we show that, for some signals, invariant moments below a sharp order threshold cannot distinguish reflectional from purely rotational symmetry. For generic signals, however, this distinction emerges much earlier. We use these insights to develop tests for dihedral symmetry with asymptotic error control under explicit nondegeneracy conditions. Together, the results clarify what we can learn about symmetry from noisy, unaligned data and suggest a path toward symmetry inference in cryo-EM.
\end{abstract}

\section{Introduction}

Detecting symmetry means determining which transformations of a signal leave it unchanged. Across the physical and biological sciences, symmetry is a key structural prior: it governs how molecules fold, how crystals repeat, and how objects can be efficiently represented from incomplete observations. In cryo-electron microscopy (cryo-EM), for instance, a macromolecule's rotational symmetry is often described by a finite point group, which dramatically reduces the adequate search space in 3-D reconstruction. Using the correct symmetry prior improves the signal-to-noise ratio (SNR) through averaging and prevents systematic reconstruction artifacts. In practice, however, the assumed symmetry is rarely inferred from the imaging data itself; instead, practitioners typically rely on \emph{a priori} biological knowledge, such as homology to previously studied complexes or known symmetries within a protein family, to select a symmetry group. This indirect process can lead to misspecified symmetry assumptions and motivates the need for statistical procedures that can \emph{infer symmetry directly from data}. To study this question in a controlled and analytically tractable setting, we turn to the \emph{multi-reference alignment (MRA)} model~\cite{abbe2018multireference, bandeira2023estimation, bendory2017bispectrum, perry2019sample}, a canonical simplified model of cryo-EM where one observes many noisy, randomly rotated (and possibly reflected) copies of an unknown 1-D signal. This model has been studied extensively as a testbed for estimation under group actions, including its sample complexity in the low-SNR regime~\cite{perry2019sample} and method-of-moments recovery from invariants~\cite{abbe2018multireference, bandeira2023estimation, bendory2017bispectrum}. Our goal is to identify the underlying symmetry group of the signal from such observations without requiring alignment (due to high noise levels) or prior knowledge of the sampling distribution.

Cyclic and dihedral rotation groups are prominent among the symmetries recorded for biological assemblies in the Protein Data Bank~\cite{rcsbPDB}. We study their one-dimensional MRA analogues as a first step toward statistical symmetry inference. When it comes to detecting these symmetries in cryo-EM, most computational techniques rely on \emph{multivariate statistical analysis} (MSA), developed initially to extract structural variability from noisy projection images of randomly oriented macromolecules~\cite{van1984multivariate}. Subsequent extensions of MSA enabled symmetry identification from side- and tilted-view particles, as well as from stoichiometrically heterogeneous samples~\cite{costa2008novel}. More recently, \cite{reboul2020point} introduced a complementary approach that infers point-group symmetry directly from reconstructed charge-density maps, after particle classification and 3-D reconstruction, by applying candidate symmetry operations and quantifying the resulting map correspondence. Representation-theoretic characterizations of point symmetries in cryo-EM~\cite{gurevich2009representation, chang2013} have further examined when and how symmetry can be recovered. The present work develops a statistical testing framework for cyclic and dihedral MRA, motivated by the symmetry-assignment problem in cryo-EM. We formulate the MRA invariance hypotheses as a multiple-testing problem and obtain error guarantees that hold uniformly over the unknown transformation distribution under the stated Gaussian noise model. The tests operate directly on noisy MRA observations. Extending these guarantees to projected cryo-EM data requires additional analysis, outlined in Section~\ref{sec:conclusion}. Beyond cryo-EM, symmetry detection is a long-studied problem in signal processing, computer vision, and geometry processing~\cite{chen2026archsym, liu2010computational, loy2006detecting, mitra2013symmetry}.

We divide the problem of symmetry detection into three layers. (i) \textbf{Group level}. We first identify the relevant transformation groups. For one-dimensional signals, we focus on the cyclic group $\Cd{d}$ (consisting of rotations) and the dihedral group $\Dm{d}$ (rotations plus reflections). (ii) \textbf{Signal level}. For a fixed unknown signal $\x$, its symmetry subgroup is the maximal subgroup of either $\Cd{d}$ or $\Dm{d}$ that leaves $\x$ unchanged. (iii) \textbf{Statistical level}. Given an observation model and noise, we ask which signal symmetries are identifiable from the distribution of the data. In the high-noise regime, where alignment is unreliable, we pursue identifiability via low-order moments that are invariant to the unknown group elements. This leads to concrete, testable patterns in the Fourier domain.

Our first main result shows that \textbf{second-order statistics} already suffice for distribution-free symmetry detection in the \textbf{cyclic case}. In particular, the diagonal of the Fourier-domain second moment equals the power spectrum of $\x$ and reveals a zero pattern indexed by the subgroup: if $\x$ has $\Cd{m}$ symmetry, then $\F\x[\ell]=0$ for all nonmultiples of $m$, independently of the (unknown) distribution of rotations. Leveraging this fact, we construct sequential hypothesis tests that traverse the subgroup lattice as a directed acyclic graph (DAG), achieving strong, finite-sample family-wise error rate (FWER) control via sequential rejection. The tests have $\chi^2$ null distributions with known scaling and exhibit exponential power under fixed alternatives.

For the \textbf{dihedral case}, which extends MRA to incorporate reflections~\cite{bendory2022dihedral}, we first characterize the detectable equivalence classes and their implication relations. Then, we encode the class in a DAG to be ready for multiple testing. For the symmetry detection in the dihedral case, we prove a fundamental limitation: on the sparsest $\Cd{m}$-invariant spectra, no moments of order below $d/m$can separate dihedral from cyclic symmetry. Therefore, no moment order that is bounded independently of $d/m$ separates the symmetry labels on all signals (see Theorem~\ref{thm:third_moment_failure}). The bound is sharp: moments of order at most $d/m$ determine the dihedral label on every $\Cd{m}$-invariant signal (Proposition~\ref{prop:order_dm_sufficient}), so $d/m$ is the smallest moment order that separates the symmetry labels on the entire class. This obstruction is non-generic: on generic signals, third-order reflection-consistent phase invariants already separate $\Dm{m}$ from $\Cd{m}$, and we can extract them without knowledge of the group distribution (see Proposition~\ref{prop:third_order_generic}). We build our tests on a \emph{fourth-order} form of these distribution-free phase invariants, which treats the collapsed and the parity-split regimes within a single framework and yields practical tests distinguishing $\Dm{m}$ from $\Cd{m}$ (and, when applicable, the two dihedral parity classes) without assuming a known shift or reflection distribution. These tests are asymptotic Wald tests, and we state precisely the nondegeneracy conditions under which their combination along the subgroup lattice controls the family-wise error rate (Proposition~\ref{prop:FWER_dihedral}). Table~\ref{tab:moment_order} provides an overview of our analysis and outlines the different study cases and the moment order needed for symmetry detection.

The paper's \textbf{main contributions} are as follows:
\begin{itemize}[itemsep=1pt, topsep=1pt]
\item \emph{Moment-based invariants for symmetry.} We identify simple Fourier-domain patterns of the second, third, and fourth moments that fully determine cyclic symmetry, split dihedral equivalence classes, and remain usable at very low SNR without alignment.
\item \emph{DAG-structured multiple testing with FWER control.} Exploiting subgroup lattices, we adapt sequential rejection procedures~\cite{goeman2010sequential,meijer2015multiple} for hierarchically structured hypotheses~\cite{meinshausen2008hierarchical} to test families of symmetry hypotheses efficiently and rigorously. The cyclic procedure controls the family-wise error rate strongly and in finite samples. The dihedral procedure combines asymptotic Wald tests; we prove asymptotic family-wise error control under explicit nondegeneracy conditions on the true hypotheses that the procedure can reach, and we identify the nested-null configurations in which these conditions fail (Section~\ref{sec:FWER_dihedral}).
\item \emph{Finite-sample guarantees and empirical validation.} The cyclic tests admit closed-form $\chi^{2}$ thresholds under Gaussian noise and achieve exponential power under alternatives; the dihedral Wald tests are asymptotically calibrated and consistent on nondegenerate signals. Simulations illustrate the exact cyclic tests at very low SNR and the stabilized dihedral procedure across noise levels.
\item \emph{Path to cryo-EM.} We outline the additional steps needed to extend the MRA invariants and testing logic to symmetry inference in cryo-EM, where symmetry priors can materially improve reconstructions.
\end{itemize}

\begin{table}[htbp]
\centering
\setlength{\extrarowheight}{2pt}
\begin{tabularx}{\textwidth}{@{}l l c c >{\raggedright\arraybackslash}X@{}}
\toprule
 & & \multicolumn{2}{c}{Moment order} & \\
\cmidrule(lr){3-4}
Setting & Regime & Generic & All signals & Comments\\
\midrule
Cyclic MRA & \textemdash & $2$ & $2$
  & Detection of $\Cd{m}$-invariance from the power spectrum; \emph{exact} and \emph{distribution-free} in finite samples (Thm.~\ref{thm:test_size_power}).\\
\addlinespace
\multirow{2}{*}{Dihedral MRA}
 & \makecell[l]{collapsed\\($d/m$ odd)} & $3$ & $\tfrac{d}{m}$
  & Separating $\Dm{m}$ from $\Cd{m}$. On generic signals the reflection-consistent phases are visible at third order, and the separating quantities are distribution-free (Prop.~\ref{prop:third_order_generic}). Order $d/m$ separates the classes on every $\Cd{m}$-invariant signal (Prop.~\ref{prop:order_dm_sufficient}), and no smaller order does so on spectra supported on a single conjugate pair (Thm.~\ref{thm:third_moment_failure}). Our tests use a fourth-order form of the phase invariants, asymptotically valid under unknown $\rho$ (Thm.~\ref{thm:DMRA_Dm_testing}; see Remark~\ref{rmk:third_vs_fourth}).\\
\addlinespace
 & \makecell[l]{parity-split\\($d/m$ even)} & \makecell[c]{$3$\ \text{\scriptsize($d/m\ge 6$)}\\$4$\ \text{\scriptsize($d/m=4$)}} & $\tfrac{d}{m}$
  & Also separates $\Dm{m}^{0}$ from $\Dm{m}^{1}$. Our tests use a signed fourth-order parity witness (Thm.~\ref{thm:DMRA_Dm_testing_parity_split}); a third-order witness exists when $d/m\ge 6$ (Remark~\ref{rmk:third_order_parity}). Order $d/m$ resolves both the dihedral label and the parity on every signal with exact rotational symmetry $\Cd{m}$ (Prop.~\ref{prop:order_dm_sufficient}); on a single conjugate pair all lower-order parity witnesses vanish (Remark~\ref{rmk:worst_case_d_m}).\\
\bottomrule
\end{tabularx}
\caption{Moment order needed for symmetry detection in MRA, and the guarantees attached to each case. ``Generic'' is the lowest order at which the invariant moments separate the relevant classes on generic signals (explicit nonvanishing conditions, Definition~\ref{def:generic_signals}). ``All signals'' is the smallest order $R$ such that invariant moments of order at most $R$ determine the label on \emph{every} signal in the class: order $d/m$ always suffices (Proposition~\ref{prop:order_dm_sufficient}), and it is required by spectra supported on a single conjugate pair, for which no smaller order separates the classes (Theorem~\ref{thm:third_moment_failure}, via Theorem~\ref{thm:general_invariant_moment_criterion}). In every case the separating quantities can be estimated without knowledge of the group distribution $\rho$.}
\label{tab:moment_order}
\end{table}

\paragraph{Organization.}
The remainder of the paper is organized as follows. Section~\ref{sec:Problem_formulation} formalizes MRA and the symmetry-detection problem. Section~\ref{sec:general_invariant_moment_principle} develops an invariant-moment framework for symmetry inference and studies conditions for uniform label recovery. Section~\ref{sec:invariants} specializes this framework to MRA, recording moment invariants and their Fourier-tensor representations. Section~\ref{sec:cyclic} develops distribution-free tests for cyclic MRA. Section~\ref{sec:dihedral} treats dihedral MRA, covering detectable symmetry classes, Fourier phase patterns, the sharp population moment order for symmetry identification, and fourth-order testing procedures. The testing procedures in Sections~\ref{sec:Detection_PMRA} and \ref{sec:Detection_DMRA} address a complementary statistical objective to exact label recovery: simultaneously assessing a family of invariance hypotheses under family-wise error control. Their guarantees hold in finite samples for cyclic MRA and asymptotically under stated nondegeneracy conditions for dihedral MRA. Section~\ref{sec:conclusion} sketches a roadmap from these results to cryo-EM. Proofs and multiple-testing details appear in the appendices.

\paragraph{Notations.}
The symbol $\triangleq$ denotes ``is defined as''. Denote $\R_{\geq 0}$ and $\Z_{\geq 0}$ as the sets of non-negative real numbers and integers, respectively. For integers $a$, $b$, and $c$, write $a \equiv b \pmod{c}$ to mean that $c$ divides $a - b$. For any integers $a$ and $b$, we write $a \mid b$ if $a$ divides $b$, and $a \nmid b$ otherwise. Throughout the paper, all divisors are understood to be positive unless otherwise specified.

Denote $\Delta_d \triangleq \{ \vv \in \R_{\geq 0}^d : \sum_{i=0}^{d-1} \vv[i] = 1 \}$ as the $(d-1)$-dimensional probability simplex.  
Let $\odot$ denote the entrywise (Hadamard) product between vectors or matrices of the same shape, and $\ast$ denote the (circular) convolution operator. For two groups $G_1$ and $G_2$, we write $G_2 \leq G_1$ if $G_2$ is a subgroup of $G_1$, and $G_2 < G_1$ if $G_2$ is a proper subgroup of $G_1$ (that is, $G_2 \leq G_1$ but $G_2 \neq G_1$). We use $\xrightarrow{p}$ to denote convergence in probability,  $\xrightarrow{d}$ to denote convergence in distribution, and
$\stackrel{d}{=}$ to denote equality in distribution.

For $d\geq 1$, denote $\cI_d\triangleq\{0,1,\ldots,d-1\}$. For any vector $\vv\in\R^d$ (or $\C^d$), denote $\vv[k]$ as its $(k+1)$-th entry for $k\in\cI_d$. That is, $\vv=(\vv[0],\vv[1],\ldots,\vv[d-1])^\T$. Similarly, for any matrix $\mathbf{M}\in\R^{d_1\times d_2}$ (or $\C^{d_1\times d_2}$), denote $\mathbf{M}[k_1,k_2]$ as its $(k_1+1,k_2+1)$-th entry for $k_1\in\cI_{d_1}$ and $k_2\in\cI_{d_2}$. Denote the transpose and conjugate transpose of a complex matrix $\mathbf{M}$ by $\mathbf{M}^{\T}$ and $\mathbf{M}^{\sH}$, respectively. For any tensor $\mathbf{T}\in\R^{d_1\times\ldots\times d_r}$ (or $\C^{d_1\times\ldots\times d_r}$) of order $r$, denote $\mathbf{T}[k_1,\ldots,k_r]$ as its $(k_1,\ldots,k_r)$-th entry for $k_j\in\cI_{d_j}$, $j=1,\ldots,r$. We adopt the convention that all indices are interpreted modulo the corresponding dimension. That is, for a vector $\vv\in\R^d$ (or $\C^d$), $\vv[k] = \vv[k \bmod d]$ for any integer $k$, and similarly for matrices and higher-order tensors. We denote the discrete Fourier transform of any such vector, matrix, or tensor by $\mathcal{F}$, and its inverse Fourier transform by $\mathcal{F}^{-1}$, with the explicit definitions given in Section~\ref{sec:Fourier_convention}.

Denote by $\Cb_d\in\R^{d\times d}$ the cyclic permutation matrix, i.e., $\Cb_d[i,j]=\1\{i-j\equiv 1 \pmod{d}\}$. We may omit the subscript $d$ in the sequel if there is no ambiguity. For any vector $\vv \in \mathbb{R}^d$ (or $\mathbb{C}^d$), define $\Cb_{\vv}\in\R^{d\times d}$ (or $\C^{d\times d}$) as the \emph{circulant matrix} generated by $\vv$, whose $(i,j)$-th entry is given by
\begin{align*}
\Cb_{\vv}[i,j]=\vv[(i-j) \bmod d], \quad \text{for } i,j \in \cI_d.
\end{align*}
Define $\Db_{\vv}\in\R^{d\times d}$ (or $\C^{d\times d}$) as the diagonal matrix generated by $\vv$, whose $(i,i)$-th diagonal entry is given by $\Db_{\vv}[i,i]=\vv[i]$ for $i\in \cI_d$ and zeros elsewhere. For any $r\geq 1$, define $\mathbf{T}_{\vv}^{(r)} \in (\R^d)^{\otimes r}$ (or $(\C^d)^{\otimes r}$) as the order-$r$ diagonal tensor whose $\underbrace{(i, i, \ldots, i)}_{r\ \text{times}}$-th diagonal entry is given by $\mathbf{T}_{\vv}^{(r)}\underbrace{[i, i, \ldots, i]}_{r\ \text{times}}=\vv[i]$ for $i\in \cI_d$ and zeros elsewhere. In particular, the case $r=1$ recovers the vector $\vv$ itself and the case $r=2$ recovers the diagonal matrix: $\mathbf{T}_{\vv}^{(2)} = \Db_{\vv}$. 

\section{Problem formulation} \label{sec:Problem_formulation}

\subsection{Multireference alignment and orbit recovery problems} \label{sec:MRAandOrbitRecovery}

We study the problem of (discrete) \textbf{multireference alignment (MRA)}. Let $\x\in\R^d$ be an unknown signal of interest, representing samples of a continuous function at $d$ equally spaced points on the unit circle. Let $G$ be a known finite subgroup of the orthogonal group $\mathsf{O}(2)$ that preserves this sampling grid, thereby acting on $\x$ via index permutations induced by rotations (circular shifts) and reflections. By classical classification results (e.g., \cite{conway2003quaternions}), every finite subgroup of $\mathsf{O}(2)$ is, up to conjugacy, either a \textbf{cyclic group} $\Cd{k}$ (rotations by multiples of $2\pi/k$) or a \textbf{dihedral group} $\Dm{k}$ (which includes both such rotations and reflections), for integers $k \geq 1$. In this discrete setting, all admissible rotation groups are subgroups of $\Cd{d}$, while all admissible groups containing reflections are subgroups of $\Dm{d}$. We focus on these two full symmetry groups, taking $G=\Cd{d}$ or $G=\Dm{d}$.

We observe $N$ independent noisy measurements of $\x$ under random \textbf{unknown} group actions:
\begin{equation}\label{eq:MRA_model}
    \y_j  =  g_j \circ\x  + \beps_j,  \quad g_j \in G, \quad j=1,\ldots, N ,
\end{equation}
where $\{g_j\}_{j=1}^N$ are i.i.d. latent group elements drawn from an \textbf{unknown} probability distribution $\rho : G \to \R_{\geq 0}$. We allow $\rho$ to be \textit{non-uniform}, so that some rotations or reflections may occur more frequently than others. The noise vectors $\{\beps_j\}_{j=1}^N$ are i.i.d. isotropic Gaussian with variance $\sigma^2$ per entry. Throughout, we assume that $\sigma^2$ is known and the signal-to-noise ratio is measured by
\begin{equation}\label{eq:SNR}
\mathrm{SNR}\triangleq \frac{\|\x\|_2^2}{d\,\sigma^2},
\end{equation}
the average signal energy per entry divided by the noise variance. Throughout, ``distribution-free refers to validity over the unknown distribution $
ho$ of transformations; the noise distribution is the specified isotropic Gaussian model with known variance. 

We distinguish between the two important special cases, depending on the choice of the group $G$, as follows:

\paragraph{Cyclic MRA (C-MRA): $G = \Cd{d}$.} In this case, the group $G = \Cd{d}$ is the \textit{cyclic group} of order $d$. Namely, $G$ is a finite group generated by the plane rotation of $2\pi/d$ radians. In other words, $G\equiv \langle r\rangle = \{1,r,\ldots,r^{d-1}\}$ with a rotation $r$ of order $d$. The action of any element $r^\ell \in G$ corresponds to a rotation of $\ell (2\pi/d)$ radians. This action on the signal $\x\in\R^d$ is defined as 
\begin{equation*}
    (r^\ell \cdot \x)[k] = \x[(k-\ell)\bmod d], \quad \text{for }\ell,k\in \cI_d.
\end{equation*}
For more details on this case, see, e.g.,~\cite{abbe2018multireference}.

\paragraph{Dihedral MRA (D-MRA): $G = \Dm{d}$.} Here, $G = \Dm{d}$ is the \textit{dihedral group} of order $2d$, representing the full symmetry group of a regular $d$-gon in $\mathbb{R}^2$. This group is generated by a rotation $r$ of order $d$ and a reflection $s$ of order $2$. (In abstract algebra, $\Dm{2d}$ refers to this same dihedral group. We use the geometric convention $\Dm{d}$ in this paper.) The rotation is defined as in the cyclic group $\Cd{d}$, corresponding to a rotation by angle $2\pi/d$. The reflection acts on $\x\in\R^d$ as 
\[ (s\cdot \x)[\ell] = \x[-\ell \bmod d], \quad\text{for } \ell\in \cI_d. \] 
Since rotation does not commute with reflection, the group $\Dm{d}$ is non-abelian. Specifically, the rotation and reflection satisfy the relation $rs = sr^{-1}$. In summary, the generators form $\Dm{d} \equiv \langle r, s \mid r^d = s^2 = 1,\, srs = r^{-1} \rangle = \{1,r,\ldots, r^{d-1}, s, rs,\ldots, r^{d-1}s\}$. For more details on this case, see, e.g.,~\cite{bendory2022dihedral}.

\paragraph{Connection to orbit recovery.}
The MRA model is a special case of the more general \emph{orbit recovery} problem. Let $\x\in\mathbb R^d$ be an unknown signal and let $G$ be a known compact group acting continuously and linearly on $\mathbb R^d$, with action denoted by $g\circ\x$. We assume that the action is orthogonal with respect to the Euclidean inner product:
\begin{equation}\label{eq:orthogonal_action}
\langle g\circ\boldsymbol{u},g\circ\boldsymbol{v}\rangle
=\langle\boldsymbol{u},\boldsymbol{v}\rangle,
\qquad g\in G,\quad\boldsymbol{u},\boldsymbol{v}\in\mathbb R^d.
\end{equation}
The observations follow the model
\begin{equation}\label{eq:orbit_model}
\y_j=A(g_j\circ\x)+\beps_j\in\mathbb R^m,
\qquad j=1,\ldots,N,
\end{equation}
where $A\in\mathbb R^{m\times d}$ is a known linear observation operator (e.g., identity, projection, or measurement map). The latent group elements $\{g_j\}_{j=1}^N$ are i.i.d. with an unknown probability measure $\rho$ on $G$, not necessarily the Haar measure. The noise vectors $\{\beps_j\}_{j=1}^N$ are i.i.d. noise vectors and independent of the group elements. The C-MRA and D-MRA models correspond to $A=\mathbf I_d$ and the permutation actions of $\Cd{d}$ and $\Dm{d}$, respectively. 

For any $h\in G$, replacing $\x$ by $h\circ\x$ and $\rho$ by
the distribution of $gh^{-1}$, where $g\sim\rho$, leaves the
observation law unchanged. Thus, the signal is identifiable at best
up to its orbit
\[
    G\circ\x\triangleq\{g\circ\x:g\in G\}.
\]
The orbit recovery problem seeks to determine this orbit from
the observations; see, e.g.,
\cite{bandeira2023estimation, fan2024maximum, xu2025misspecified}. More general observation operators $A$ may introduce additional
identifiability issues beyond the global group-action ambiguity.

A particularly important application of orbit recovery, and a key motivation for studying the MRA model, arises in \emph{cryo-Electron Microscopy (cryo-EM)}. In cryo-EM, the goal is to reconstruct a 3-D molecular structure from a large collection of 2-D projection images taken at unknown orientations. The underlying object is the molecular Coulomb potential, and each image is a tomographic line-integral projection of a rotated copy of that potential. In this setting, $\x$ represents the structure, $G=\mathsf{SO}(3)$ describes its 3-D rotational action, and $A$ maps the rotated structure to an image. Cyclic and dihedral MRA provide simplified models with circular shifts and possibly reflections, retaining the latent group action while taking $A=\mathbf I_d$. Studying MRA provides valuable and essential theoretical and algorithmic insights into more complex applications such as cryo-EM.

Section~\ref{sec:general_invariant_moment_principle} develops
an invariant-moment framework for symmetry detection for general orbit recovery \eqref{eq:orbit_model} with $A=\mathbf I_d$, while
Sections~\ref{sec:invariants}--\ref{sec:dihedral} then specialize
this framework to C-MRA and D-MRA and develop the corresponding
hypothesis testing procedures.

\subsection{The symmetry detection problem}\label{sec:symmetry_detection_problem}

The definitions below are formulated for the general orbit recovery model~\eqref{eq:orbit_model}. A subgroup of $G$ describes an invariance of $\x$ if every
element of that subgroup leaves $\x$ unchanged.
To capture all such transformations, we define the symmetry of $\x$ as follows.

\begin{definition}[Symmetry of a signal]\label{def:signal_symmetry}
    The \textbf{symmetry (group)} of $\x$ is the unique maximal subgroup $H \leq G$ that satisfies
\begin{equation} \label{eqn:sym}
 \x = h\circ \x, \quad \text{for every } h \in H.
\end{equation}
Namely, no subgroup $H'$ satisfies~\eqref {eqn:sym} with $H < H'$. Equivalently, $H$ is the \textbf{stabilizer} of $\x$,
\begin{equation}\label{eqn:stabilizer}
\Sym(\x) \coloneqq \{g\in G : g\circ\x = \x\},
\end{equation}
and we write $H = \Sym(\x)$; see, e.g., \cite[Chapter 4]{milne2021group}. In this case, we say that $\x$ has (exactly) \textbf{$H$-symmetry}.
\end{definition}

When the group describes spatial rotations and reflections fixing a common point, conventionally the origin, the stabilizer is also called the \textit{point-group symmetry} of the underlying object. For C-MRA and D-MRA, this 
interpretation is obtained
by placing the signal indices at the vertices of a regular
polygon, so that circular shifts and reflections correspond
to rotations and reflections fixing its center. For a three-dimensional molecular structure, exact $\Cd{k}$-symmetry means that there exists an axis such that rotations by integer multiples of $2\pi/k$ radians (or $360/k$ degrees) about this axis leave $\x$ invariant, and moreover $\Cd{k}$ is the \emph{maximal} subgroup of $G$ with this invariance property. Here, $\Cd{k}$ corresponds to an axial group in the classification of discrete point groups in three dimensions. 

\begin{remark}[Exact symmetry versus invariance]\label{rmk:exact_vs_invariance}
Throughout, ``$\x$ has $H$-symmetry'' always means the \emph{exact} statement $\Sym(\x)=H$, as in Definition~\ref{def:signal_symmetry}. We use a distinct term for the weaker property: we say $\x$ is \textbf{$H$-invariant} (or has \emph{at least} $H$-symmetry) if
\begin{equation}\label{eqn:fixed_space}
h\circ\x=\x \quad\text{for all } h\in H, \qquad\text{or equivalently}\quad \x\in\Fix{H}\coloneqq\{\y : h\circ\y=\y \ \text{for all } h\in H\}.
\end{equation}
For the linear actions considered here, $\Fix{H}$ is the fixed-point subspace of $H$ (the set of signals left unchanged by every element of $H$). Thus, $H$-invariance means $H\le\Sym(\x)$, whereas exact $H$-symmetry means equality. For example, in C-MRA, a signal with exact $\Cd{m}$-symmetry is $\Cd{m'}$-invariant for every $\Cd{m'}\le\Cd{m}$, but its exact symmetry group is only $\Cd{m}$.
\end{remark}

For every fixed signal $\x$, its stabilizer $\Sym(\x)$ (equivalently, its symmetry) is
uniquely determined.
The difficulty in symmetry detection concerns identifying
this subgroup from the observations.
In particular, a globally transformed signal $\x'=g\circ\x$
has its own unique stabilizer,
\[
    \Sym(\x')=g\Sym(\x)g^{-1},
\]
which may differ from $\Sym(\x)$.
Under the orbit recovery model~\eqref{eq:orbit_model},
however, the transformation of the signal can be absorbed
into the unknown distribution of the latent group elements,
leaving the observation law unchanged.
Thus, different signals with different conjugate stabilizers
can be statistically indistinguishable.
This motivates considering conjugacy classes of stabilizers.

\begin{definition}[Detectable signal symmetry classes]
\label{def:detectable_symmetry}
Let $H_1,H_2 \leq G$ be subgroups that occur as exact symmetry
groups of signals in $\R^d$ under the specified group action.
We declare $H_1 \sim H_2$ if there exists $g\in G$ such that,
for every signal $\x$ with $\Sym(\x)=H_1$, the transformed
signal $g\circ\x$ satisfies
\[
    \Sym(g\circ\x)=H_2.
\]
Since
\[
    \Sym(g\circ\x)=g\,\Sym(\x)\,g^{-1},
\]
this condition is equivalent to $H_1$ and $H_2$ being
conjugate in $G$, namely,
\[
    H_2=gH_1g^{-1}
    \qquad \text{for some } g\in G.
\]
The corresponding equivalence class
\[
    [H]\coloneqq\{gHg^{-1}:g\in G\}
\]
is called a \textbf{detectable signal symmetry class},
or simply a \textbf{symmetry class}.
The symmetry class of $\x$ is $[\Sym(\x)]$,
which is invariant under replacing $\x$ by any other
representative of its orbit under $G$.
\end{definition}

\begin{remark}[Exact symmetry groups versus symmetry classes]
\label{rmk:sgp_symm}
In C-MRA, the acting group $G$ is abelian, so
\[
    gHg^{-1}=H
    \qquad \text{for every } H \leq G \text{ and } g \in G.
\]
Consequently, every symmetry class contains a single subgroup.
Although a global cyclic shift changes the signal, it leaves
the signal's stabilizer unchanged.
The ambiguity in the signal's overall position therefore
introduces no ambiguity in its symmetry subgroup.

In D-MRA, some subgroups have distinct conjugates.
For example, rotating a signal can change its reflection axis
and hence its reflection symmetry subgroup.
The original and rotated signals each have a unique stabilizer,
but the observations cannot distinguish which of these conjugate
subgroups is the stabilizer of the underlying signal.
The ambiguity therefore concerns statistical identification,
not the uniqueness of the stabilizer for a fixed signal.
This issue arises only when the stabilizer is not normal in $G$;
purely rotational subgroups of the dihedral group remain normal
and have singleton symmetry classes.
\end{remark}

To simplify notation, whenever a symmetry class is a singleton, namely $[H]=\{H\}$, we identify the
symmetry class with its unique subgroup and write $H$ instead
of $[H]$.
In particular, brackets are unnecessary throughout C-MRA.
When non-singleton classes arise, as in D-MRA, we specify the
relevant classes explicitly and retain class notation as needed; see Section~\ref{sec:DSC_DMRA}. We therefore formulate symmetry detection at the level of
symmetry classes, with the above convention for singleton classes.

\begin{definition}[Symmetry detection problem]\label{def:symmetry_detection}
Let $H=\Sym(\x)$.
Given the samples~\eqref{eq:orbit_model}, the symmetry
detection problem is to determine the symmetry class $[H]$
among those realizable by signals in the model.
When $[H]=\{H\}$, this amounts to determining the subgroup $H$.
\end{definition}

\paragraph{Invariance hypotheses in MRA.}
For the C-MRA and D-MRA models, we consider the finite collection
of symmetry classes realizable under the corresponding group action.
The estimation target is the exact symmetry class of $\x$.
To infer this target from noisy observations, we assess a family of
invariance hypotheses, each asking whether the signal possesses
at least a specified candidate symmetry, up to a global group action.

For a candidate class $[H]$, the invariance hypothesis asserts that
some representative of the signal orbit is $H$-invariant, or equivalently,
\[
    gHg^{-1}\leq \operatorname{Sym}(\x)
    \qquad \text{for some } g\in G.
\]
For singleton classes this reduces to $H\leq \operatorname{Sym}(\x)$,
as in Remark~\ref{rmk:exact_vs_invariance}.

On these candidate classes, define
\[
    [H_1]\leq [H_2]
    \quad\Longleftrightarrow\quad
    H_1\leq gH_2g^{-1}
    \text{ for some } g\in G.
\]
For C-MRA and D-MRA, this relation is a partial order.
For singleton classes it reduces to ordinary subgroup inclusion.
If $[H_1]\leq [H_2]$, then the invariance hypothesis for $[H_2]$
implies that for $[H_1]$.
The exact symmetry class $[\operatorname{Sym}(\x)]$ is the greatest
candidate class whose invariance hypothesis is true: it is itself
a true candidate, and every other true candidate lies below it.
Thus, the collection of true invariance hypotheses determines
the exact symmetry label.

With finite noisy data, multiple testing provides a statistical
approach to symmetry inference through simultaneous assessment
of the invariance hypotheses.
Their nesting supplies the hierarchy for DAG-structured testing:
the subgroup-inclusion hierarchy in Section~\ref{sec:Detection_PMRA}
for C-MRA, and the corresponding class hierarchy in
Section~\ref{sec:Detection_DMRA} for D-MRA.
The testing procedures report the maximal non-rejected candidates as candidate estimates of the exact symmetry group or class. Several incomparable maximal candidates may remain. Under the conditions established in the corresponding sections, family-wise error control bounds the probability of rejecting any true invariance hypothesis. It does not control non-rejection of false invariance hypotheses or the error probability of selecting an exact symmetry label. In particular, a non-rejected symmetry is not certified as present; power against false invariance hypotheses is essential when using the output to choose a reconstruction prior.

We develop and analyze this framework in the challenging
\textit{high-noise regime}, with the goal of inferring the
detectable symmetry class of the underlying signal in the sense
of Definition~\ref{def:detectable_symmetry}.

\subsection{The significance of symmetry detection}

Identifying the symmetry of a signal is crucial for efficient and accurate recovery since it reveals structural constraints that reduce the solution space and help resolve ambiguities in reconstruction. Symmetry enables averaging over equivalent configurations, enhancing signal quality, and allows for more compact representations, reducing computational cost. In addition, in cryo-EM, leveraging symmetry not only improves recovery fidelity but also aids interpretation of the underlying biological structure.

To illustrate the above, we provide a few examples, beginning with graphical illustrations of the 3-D reconstruction problem in cryo-EM. In these examples, we emphasize the critical role of accurate symmetry information in the performance of the recovery algorithm. For the case of cryo-EM, we employ the commonly used RELION package~\cite{zivanov2022bayesian}, which uses an Expectation-Maximization (EM) algorithm~\cite{sigworth1998maximum, scheres2012relion} to maximize a regularized likelihood within a Bayesian framework. The algorithm iteratively aligns and averages thousands of particle images while refining their orientations and the underlying 3-D structure, resulting in a high-resolution density map. As with other leading cryo-EM software, RELION requires the user to manually specify the assumed symmetry. We present several 3-D reconstructions generated under different symmetry priors. The two cryo-EM examples shown are based on real-world datasets obtained from the Electron Microscopy Public Image Archive (EMPIAR)~\cite{iudin2016empiar}. In each case, we utilized the available online datasets, which comprised a few tens of thousands of selected particle images.

\begin{figure}[htb]
    \centering
    \subfloat[Imposing symmetry]{\includegraphics[width=0.58\textwidth]{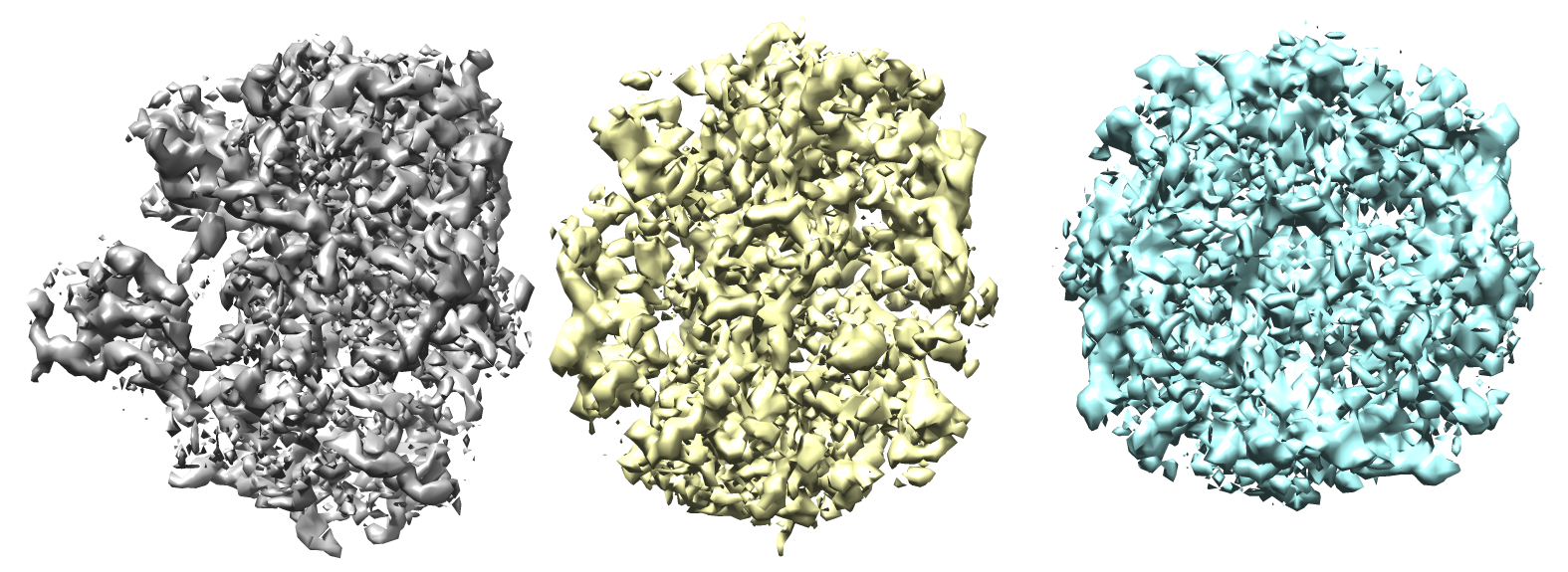} \label{fig:10028} }
    \qquad
    \subfloat[Fitting asymmetry]{\includegraphics[width=0.3\textwidth]{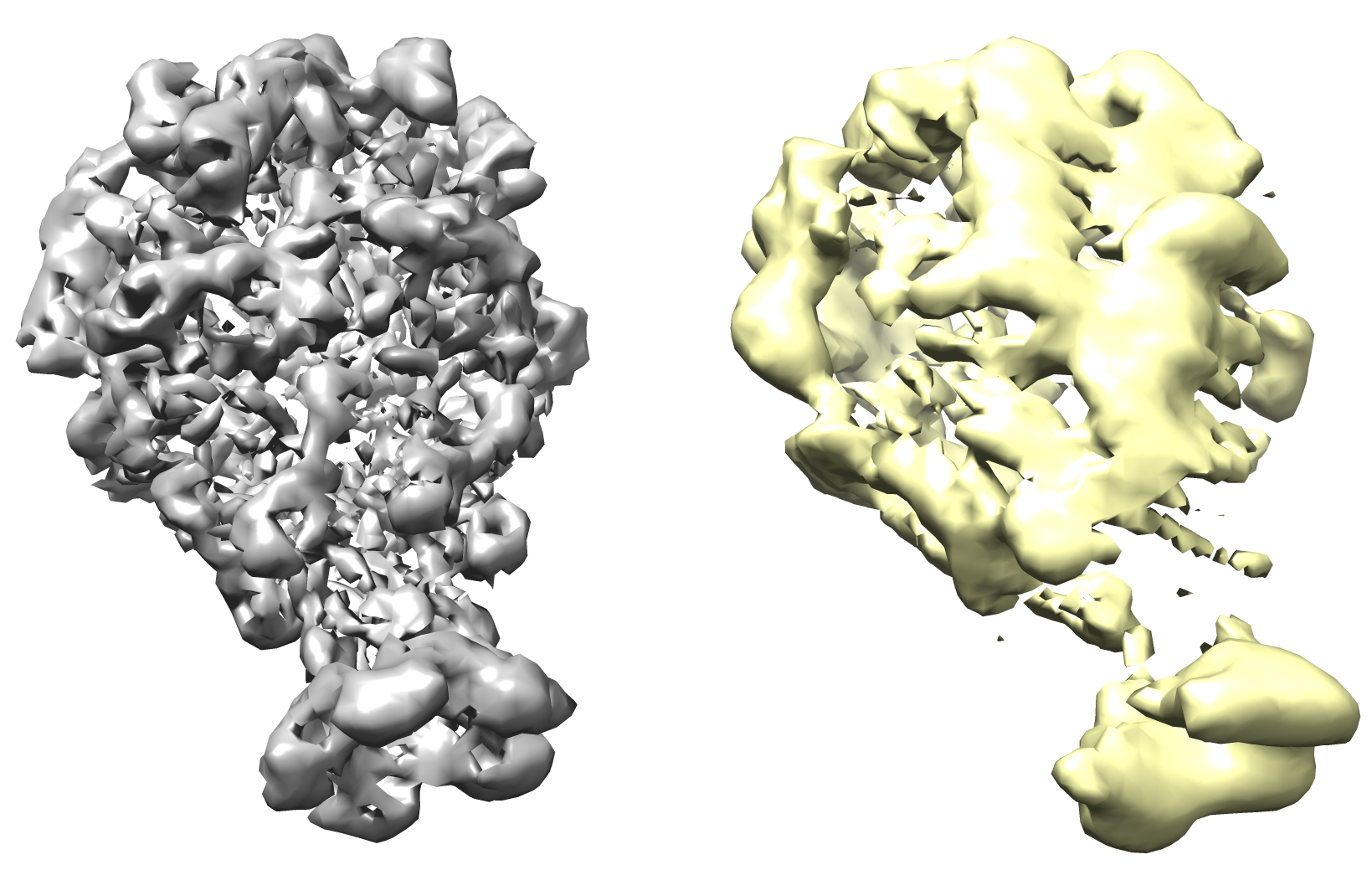} \label{fig:10089} }
    \caption{Demonstration of the effect of incorrect symmetry assignment in cryo-EM reconstruction. The original objects (gray) are shown alongside reconstructions where inappropriate symmetries are enforced: two distinct symmetries (yellow and blue, left) or no symmetry/asymmetry (yellow, right).}
    \label{fig:wrong_prior}
\end{figure}

In the first example, we use the 80S ribosome dataset from EMPIAR-10028~\cite{empiar10028}, which consists initially of about $105k$ particles. This molecule lacks point-group symmetry, as illustrated by the leftmost reconstruction in Figure~\ref{fig:10028}. Using the same dataset, we then impose two non-trivial symmetry constraints: $\Cd{2}$ and $\Cd{4}$ (twofold and fourfold rotational symmetry, respectively). The corresponding reconstructions are shown in the middle and right panels of Figure~\ref{fig:10028}. While all volumes exhibit high resolution, the differences among them result solely from the imposed symmetry, and using a spurious symmetry can introduce artifacts, highlighting the importance of choosing a suitable symmetry prior.

In the second example, we examine the protein TcdA1, available as EMPIAR-10089~\cite{empiar10089}, which is known to possess $\Cd{5}$ symmetry. The reconstruction assuming this symmetry is shown on the left in Figure~\ref{fig:10089}. For comparison, we also perform a reconstruction without enforcing any symmetry (i.e., assuming an asymmetric volume), shown on the right. This result exhibits markedly lower resolution and incomplete structural features, underscoring the importance of accurate symmetry specification.

\begin{figure}[htb]
    \centering
    \subfloat[An asymmetric signal of length $d=60$ and its two symmetric versions, having $\Cd{20}$ and $\Dm{20}$ symmetry.]{\includegraphics[width=0.45\textwidth]{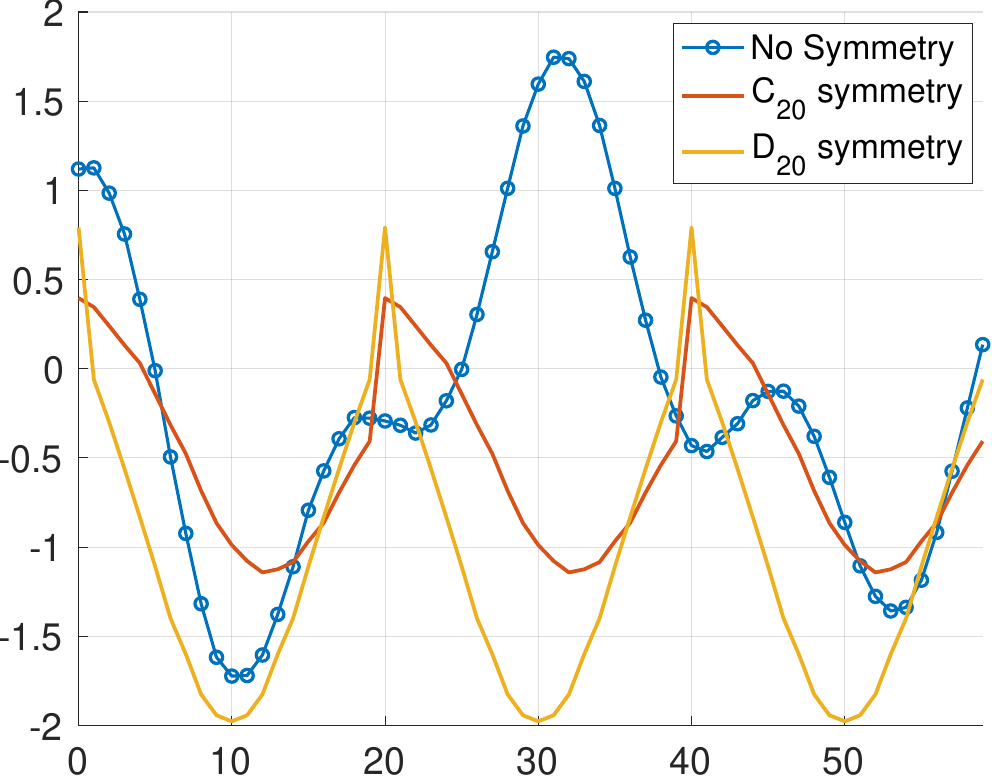} \label{fig:DMRA_signal} }
    \quad
    \subfloat[The error rates, measured in Root Mean Squared Error (RMSE), as functions of the noise levels, given in Signal-to-Noise Ratio (SNR).]{\includegraphics[width=0.48\textwidth]{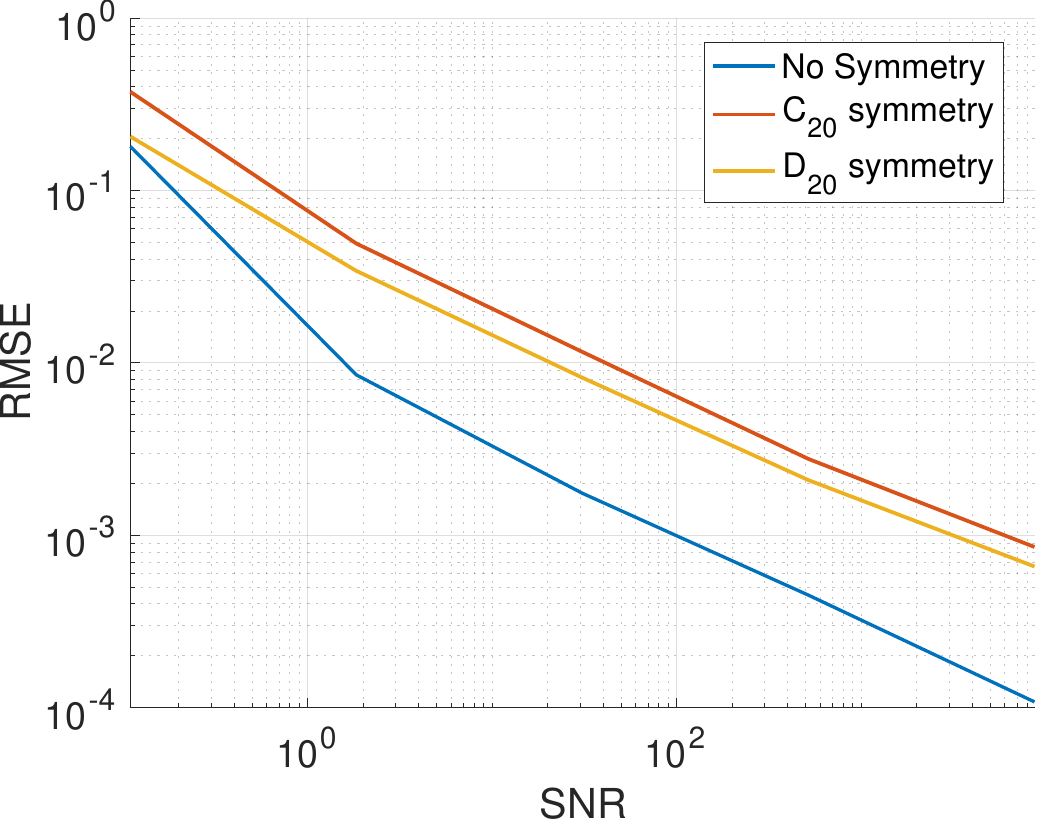} \label{fig:DMRA_errors} }
    \caption{The effect of symmetry in Dihedral MRA reconstruction without a prior. On the left, the signals. On the right, we observe how the cases of symmetry, when applied to the EM algorithm without a correct prior, lead to larger errors.}
    \label{fig:MRA_wrong_prior}
\end{figure}

We now turn our attention to the D-MRA model. In this example, we present an asymmetric target signal and its two symmetric versions, as shown in Figure~\ref{fig:DMRA_signal}. Then we recover the signal using the Expectation-Maximization (EM) method; see~\cite{bendory2022dihedral}. Figure~\ref{fig:DMRA_errors} shows the error rates, with symmetric signals having higher error than the original asymmetric signal. 

In summary, these cryo-EM and MRA examples demonstrate that accurately determining the symmetry of a 3-D structure or an MRA signal prior to reconstruction is crucial for achieving reliable results, high resolution, and structural validity. They further highlight the importance of inferring symmetry directly from the observed data, so that the symmetry assumptions imposed during reconstruction are supported by the data themselves. Although this paper focuses on MRA, the proposed approach is intended to lay the groundwork for a concise and statistically robust framework for purely data-driven symmetry inference applicable to cryo-EM and beyond.

\section{An invariant-moment framework for symmetry inference}
\label{sec:general_invariant_moment_principle}

This section develops the invariant-moment framework underlying
our approach to symmetry inference.
We study identification and recovery of symmetry labels, while
establishing the invariant features and estimation tools used
in the later sections.

\subsection{Invariant features under high noise}\label{sec:invariant_features}

To motivate our approach to symmetry detection, we first consider the usual objective of orbit recovery: estimating the signal up to a global group action. In the observation setting of \eqref{eq:orbit_model}, a standard alignment-based approach first estimates the latent transformations and then aligns and averages the observations to recover the signal~\cite{perry2019sample,shkolnisky2012viewing}. This approach can be effective at low noise levels, when accurate alignment is possible. At high noise levels, however, individual transformations can be difficult to estimate reliably, even when the underlying signal is known~\cite{singer2018mathematics}.

A related concern arises in the classical incidental-parameter problem illustrated by the Neyman--Scott example~\cite{neyman1948consistent}: when the number of nuisance parameters increases with the sample size, joint maximum likelihood estimation can yield inconsistent estimates of the parameters of interest. Such difficulties highlight the value of approaches that account for nuisance variables without estimating them individually. One possibility is to integrate them out of the likelihood~\cite{berger1999integrated}. Another is to construct estimators from sample averages of suitable invariant features, as in moment-based estimation. We focus on the latter approach.

To implement this approach, we seek features that are unchanged
by the latent group action, so that transformed copies of the
signal have the same feature values. Formally, a function $\phi:\R^d\to\R$ is called a \emph{$G$-invariant feature} if
\[
    \phi(g\circ\z)=\phi(\z)
    \qquad\text{for every }g\in G\text{ and }\z\in\R^d.
\]
If $\phi$ is also a polynomial, we call it a
\emph{$G$-invariant polynomial}. Such a feature is constant on each signal orbit. In particular, $\phi(g_j\circ\x)=\phi(\x)$, so its value on a noiseless observation does not depend on the latent transformation.

Returning to symmetry inference, we use invariant features to mitigate the influence of nuisance parameters in high-noise settings. Our goal is to infer the symmetry (class) defined in Section~2.3 directly from the observations, without reconstructing the full signal orbit. The next subsection develops this framework for compact orthogonal group actions. Section~\ref{sec:invariants} then specializes it to MRA.

\subsection{Identification and exact recovery of symmetry labels}\label{sec:recovery}

We now develop the approach of Section~\ref{sec:invariant_features} using polynomial invariant features. These features can be organized by degree through invariant moments. Our target is the exact symmetry label of the signal, namely the stabilizer conjugacy class defined in Section~\ref{sec:symmetry_detection_problem}. We first characterize when population invariant moments determine
this label and show how these moments can be estimated from noisy
observations without knowledge of the transformation distribution.
We then give sufficient conditions under which these estimates
yield uniform recovery of the label.

Throughout this subsection, we consider the full-observation model~(4), with $A=\mathbf I_d$ and an arbitrary unknown probability distribution $\rho$ on $G$. The group $G$ is compact and acts continuously, linearly, and orthogonally on $\R^d$. In particular, the group action preserves the Euclidean inner product and norm (see~\eqref{eq:orthogonal_action}).

\paragraph{Signal classes and symmetry targets.} A signal class is a set $\cX\subseteq\R^d$ specifying the admissible signals; it need not be a linear subspace.

\begin{definition}[$G$-invariant signal class]
Let $G$ act on $\R^d$. A set $\cX\subseteq\R^d$ is called a \emph{$G$-invariant signal class} if
\[
g\circ\x\in\cX
\qquad\text{for every }\x\in\cX\text{ and }g\in G.
\]
Equivalently, $\cX$ is a union of $G$-orbits.
\end{definition}

This is exactly the closure condition on the set: it does not require $g\circ\x=\x$ for individual signals. Examples include $\cX=\R^d$ and $\cX=\{\x\in\R^d:\|\x\|_2\leq B\}$, with $B>0$, since orthogonal actions preserve the norm.

The target of symmetry detection is the stabilizer conjugacy class defined in Section~\ref{sec:symmetry_detection_problem}. Let $\cC$ denote the collection of such classes realizable on $\cX$, and define
\[
\kappa:\cX\to\cC,
\qquad
\kappa(\x)=[\Sym(\x)].
\]
We refer to $\kappa(\x)$ as the \emph{symmetry label} of $\x$. Since
\[
\Sym(g\circ\x)=g\Sym(\x)g^{-1},
\]
the symmetry label is constant on each signal orbit:
\begin{align}\label{eq:invariant_label}
\kappa(g\circ\x)=\kappa(\x),\qquad \text{for every }\x\in\cX\text{ and }g\in G.
\end{align}

\paragraph{Polynomial invariant features and moments.}
A systematic way to construct polynomial features satisfying the invariance condition of Section~\ref{sec:invariant_features} is to average polynomials over the group. Let $\rho_0$ denote the Haar probability measure on $G$, namely the probability measure unchanged by left or right multiplication by a fixed group element. Thus, for every measurable $E\subseteq G$ and $h\in G$,
\[
\rho_0(hE)=\rho_0(Eh)=\rho_0(E).
\]
For finite $G$, $\rho_0$ is the uniform distribution, assigning probability $1/|G|$ to each element. For a real polynomial $p:\R^d\to\R$, define its group average, also called its \emph{Reynolds average}, by
\[
(\mathcal R_G p)(\z)\triangleq \int_G p(g\circ\z)\,\dd\rho_0(g).
\]
The average is an invariant polynomial of degree at most $\deg (p)$ and equals $p$ whenever $p$ is already invariant. All polynomial features are defined on the entire space $\R^d$, even when the admissible signals are restricted to a subset $\cX$.

The coordinate products $\z[k_1]\cdots \z[k_r]$ are degree-$r$ monomials. For an integer $r\geq1$, the tensor $\z^{\otimes r}$ is the order-$r$ tensor of these products:
\[
(\z^{\otimes r})[k_1,\ldots,k_r]
=\z[k_1]\cdots \z[k_r],\qquad k_1,\ldots,k_r\in \cI_d=\{0,\ldots,d-1\}.
\]
In particular, $\z^{\otimes1}=\z$ and $\z^{\otimes2}=\z\z^{\mathsf T}$. 

For any probability measure $\rho$ on $G$ and integer $r\geq 1$, define the order-$r$ noiseless population moment tensor by
\begin{align*}
M_r(\z,\rho) &= \E_{g\sim \rho} \Big[\underbrace{(g\circ \z)\otimes \cdots \otimes (g\circ \z)}_{(r-1)\ \text{times}}\otimes (\overline{g\circ \z})\Big] \\
 &= \E_{g\sim \rho} \left[(g\circ \z)^{\otimes r} \right]\in\R^{d\times\ldots\times d}, \quad \z\in\R^d.
\end{align*}
where the first expression is valid for general complex-valued signals, while the second holds when $\z$ (and hence $g\circ\z$) is real-valued.
Its Haar-averaged counterpart is the \emph{invariant moment tensor}
\[
M_r(\z,\rho_0)=\int_G(g\circ\z)^{\otimes r}\,\mathrm{d}\rho_0(g).
\]
For finite groups, $M_r(\z,\rho_0)=|G|^{-1}\sum_{g\in G}(g\circ\z)^{\otimes r}$.
For $R\geq1$, write
\[
M_{\leq R}(\z,\rho)=\bigl(M_1(\z,\rho),\ldots,M_R(\z,\rho)\bigr).
\]
The \emph{invariant moment vector} $M_{\leq R}(\z,\rho_0)$ is obtained by concatenating the entries of these tensors at $\rho=\rho_0$.

Each scalar coordinate of the map $\z\mapsto M_r(\z,\rho_0)$ is the group average of a degree-$r$ monomial, and hence is a $G$-invariant feature of the kind introduced in Section~\ref{sec:invariant_features}. It is a homogeneous polynomial of degree $r$, meaning that all its monomials have total degree $r$, or is identically zero. Right invariance of $\rho_0$ gives
\[
M_r(h\circ\z,\rho_0)=M_r(\z,\rho_0),\qquad h\in G,\quad\z\in\R^d.
\]

The following lemma shows that these moment coordinates contain all the information supplied by polynomial invariant features of bounded degree. The proof is deferred to Appendix~\ref{sec:proof_invariant_moments_polynomials}.

\begin{lemma}[Invariant moments and invariant polynomials]\label{lem:invariant_moments_polynomials}
For every $R\geq1$, equality $M_{\leq R}(\x,\rho_0)=M_{\leq R}(\x',\rho_0)$ holds if and only if every real $G$-invariant polynomial of degree at most $R$ takes the same value at $\x$ and $\x'$.
\end{lemma}

\paragraph{Identification from invariant moments.}
By Lemma~\ref{lem:invariant_moments_polynomials}, restricting attention to $M_{\leq R}(\cdot,\rho_0)$ retains precisely the information available from all polynomial invariant features of degree at most $R$. To determine the symmetry label, this information must distinguish signals with different symmetry classes.

\begin{definition}[Moment separation of symmetry labels]\label{def:moment_separation}
We say that invariant moments of order at most $R$ separate $\kappa$ on $\cX$ if
\[
M_{\leq R}(\x,\rho_0)= M_{\leq R}(\x',\rho_0)
\quad\Longrightarrow\quad
\kappa(\x)=\kappa(\x'),\qquad\x,\x'\in\cX.
\]
\end{definition}
Thus, signals with different symmetry labels must have different moment vectors. Signals with the same label may have different moment vectors and may belong to different orbits. This definition imposes no positive lower bound on the distance between moment vectors with different labels.

For a subset $S\subseteq\cX$, write
\[
M_{\leq R}(S,\rho_0)=\{ M_{\leq R}(\x,\rho_0):\x\in S\}
\]
for its \emph{moment image}, the set of moment vectors attainable by signals in $S$. A map from $M_{\leq R}(\cX,\rho_0)$ to $\cC$ assigns a label to each attainable noiseless moment vector.

\begin{theorem}[Invariant-moment criterion for symmetry detection]\label{thm:general_invariant_moment_criterion}
Let $G$ be a compact group acting continuously, linearly, and orthogonally on $\R^d$, and let $\rho_0$ be its Haar probability measure. Let $\cX\subseteq\R^d$ be $G$-invariant, and let $\kappa(\x)=[\Sym(\x)]$ denote the symmetry class. For $R\geq1$, there exists a map
$\tau_R:M_{\leq R}(\cX,\rho_0)\to\cC$ such that
\[
\tau_R\bigl(M_{\leq R}(\x,\rho_0)\bigr)=\kappa(\x)
\qquad\text{for every }\x\in\cX
\]
if and only if moments of order at most $R$ separate $\kappa$ on $\cX$. If separation fails, no rule based only on $M_{\leq R}(\x,\rho_0)$ recovers $\kappa$ for every $\x\in\cX$.

Moreover, there exists a finite order $R_0$, depending only on the group action, such that $M_{\leq R_0}(\x,\rho_0)$ determines the orbit $G\circ\x$ for every $\x\in\R^d$, and hence determines $[\Sym(\x)]$.
\end{theorem}

The map $\tau_R$ describes identification from exact moments. Statistical recovery from estimated moments also requires control of the effect of estimation error, addressed below. The finite-order assertion uses the existence of a finite collection of invariant polynomials whose values distinguish distinct compact orbits~\cite{bandeira2023estimation}; Appendix~\ref{sec:proof_general_invariant_moment_criterion} provides the proof and explains this invariant-theoretic property. 

\paragraph{Estimation under Gaussian noise.}
The preceding results concern exact invariant moments of the signal. To estimate these moments from observations, we use the sample-averaging approach described in Section~\ref{sec:invariant_features}, with polynomial features corrected for additive Gaussian noise.

Consider the observations of form \eqref{eq:orbit_model} with $A=\mathbf I_d$ satisfying
\begin{align}\label{eq:orbit_model_direct}
\y_j=g_j\circ\x+\beps_j,\qquad
g_j\overset{\mathrm{i.i.d.}}{\sim}\rho,\qquad
\beps_j\overset{\mathrm{i.i.d.}}{\sim}\mathcal N(0,\sigma^2\mathbf I_d),
\quad j=1,\ldots,N,
\end{align}
with independent latent transformations and noise. We assume that $\sigma^2$ is known. The symbols $\E_{\rho}$ and $\P_{\rho}$ denote expectation and probability under this model with $\x$ fixed. Their dependence on $\sigma$ is suppressed.

For a $G$-invariant polynomial $p$, we have $p(g_j\circ\z)=p(\z)$. However, additive Gaussian noise can bias polynomial features. For a polynomial $p$ of degree at most $r$, define its Gaussian correction by the finite sum
\[
\mathcal D_\sigma p
=\sum_{k=0}^{\lfloor r/2\rfloor}
\frac{(-\sigma^2/2)^k}{k!}\Delta^k p,
\qquad
\Delta p=\sum_{i=0}^{d-1}\frac{\partial^2p}{\partial \z[i]^2}.
\]
Here $\Delta$ is the Euclidean Laplacian, $\Delta^0p=p$, and $\Delta^k$ denotes $k$ repeated applications of $\Delta$, with derivatives taken with respect to the real coordinates of the argument. Each application either gives zero or lowers polynomial degree by at least two, so $\Delta^k p=0$ when $2k>r$. We also write
$\mathcal D_\sigma=\exp(-\sigma^2\Delta/2)$, with the exponential interpreted through this terminating series on polynomials.

For a $G$-invariant polynomial $p$, define the associated \emph{Gaussian-noise-corrected invariant feature} to be the function
\[
q:\R^d\to\R,\qquad q(\z)=(\mathcal D_\sigma p)(\z).
\]
The value $p(\x)$ is the noiseless feature to be estimated; $q(\y_j)$ is the corrected feature value computed from observation $\y_j$. As shown in Lemma~\ref{lem:estimation_invariant} below, $q$ remains $G$-invariant and its observed value has expectation $p(\x)$. The correction removes noise bias in expectation; an individual $q(\y_j)$ still has sampling variability. For example,
\[
p(\z)=\|\z\|_2^2
\quad\Longrightarrow\quad
q(\z)=\|\z\|_2^2-d\sigma^2.
\]
In this case, $\E\|\y_j\|_2^2=\|\x\|_2^2+d\sigma^2$, and evaluating $q$ at $\y_j$ subtracts the expected noise energy.

For a fixed finite collection of $G$-invariant polynomials $p_1,\ldots,p_K$, define the feature vectors
\[
\boldsymbol p=(p_1,\ldots,p_K)^{\mathsf T},\qquad
q_\ell=\mathcal D_\sigma p_\ell,\qquad
\boldsymbol q=(q_1,\ldots,q_K)^{\mathsf T}.
\]
Thus $\boldsymbol q(\y_j)$ is the random vector of corrected feature values from one observation. The scalar and vector sample estimators are, respectively,
\[
\widehat p_N=\frac1N\sum_{j=1}^Nq(\y_j),\qquad
\widehat{\boldsymbol p}_N=\frac1N\sum_{j=1}^N\boldsymbol q(\y_j).
\]
All these corrected functions and estimators use the known noise variance $\sigma^2$.

\begin{lemma}[Estimation of invariant polynomial features]\label{lem:estimation_invariant}
For every real $G$-invariant polynomial $p$, its corrected feature $q=\mathcal D_\sigma p$ is $G$-invariant and
\[
\E_{g\sim \rho,\beps}[q(\y_j)]=p(\x)
\]
for every probability measure $\rho$ on $G$ and the expectation is over both $g_j$ and $\beps_j$. Consequently, for every fixed $\x$ and $\rho$,
\[
\widehat p_N\xrightarrow{p}p(\x),\qquad
\widehat{\boldsymbol p}_N\xrightarrow{p}\boldsymbol p(\x)
\qquad\text{as }N\to\infty.
\]
For the finite feature vector $\boldsymbol q$ defined above,
\[
\boldsymbol q(\y_j)\stackrel{d}{=}\boldsymbol q(\x+\beps),
\qquad \beps\sim\mathcal N(0,\sigma^2\mathbf I_d).
\]
Thus, for fixed $\x$, the joint distribution of the corrected feature values, including their covariances, does not depend on $\rho$.
\end{lemma}

Applying Lemma~\ref{lem:estimation_invariant} to the coordinate
polynomials of $M_r(\,\cdot\,,\rho_0)$ gives the corrected
invariant-moment estimator
\begin{align}
\widehat M_r^{\mathrm{corr}}(\rho_0)
&=\frac1N\sum_{j=1}^N
\bigl(\mathcal D_\sigma[M_r(\,\cdot\,,\rho_0)]\bigr)(\y_j),\label{eq:corrected_invariant_moment_estimator}\\
\E_{g\sim \rho,\beps}\bigl[\widehat M_r^{\mathrm{corr}}(\rho_0)\bigr]
&=M_r(\x,\rho_0).\notag
\end{align}
Here $\mathcal D_\sigma$ acts coordinatewise on the first
argument of $M_r$. The argument $\rho_0$ records the Haar
averaging used to construct the invariant features; the
expectation identity holds for every sampling distribution
$\rho$ on $G$. Collecting orders through $R$, define
\[
\widehat M_{\leq R}^{\mathrm{corr}}(\rho_0)
=\bigl(\widehat M_1^{\mathrm{corr}}(\rho_0),\ldots,
\widehat M_R^{\mathrm{corr}}(\rho_0)\bigr).
\]
This is an unbiased, consistent estimator of
$M_{\leq R}(\x,\rho_0)$, and its construction requires
$G$ and $\sigma^2$, but not $\rho$.

The full invariant-moment vector describes the information
available from polynomial invariant features of degree at
most $R$. The MRA procedures developed below use selected
features, expressed in Fourier coordinates, and functions
of their empirical estimates. Each selected feature is a
linear combination of invariant-moment coordinates, possibly
with a constant term. Its corrected empirical average can
be computed directly, without constructing the full moment
tensors. Sections~\ref{sec:cyclic} and~\ref{sec:dihedral} determine which selected features suffice
for the relevant symmetry questions and develop tests based
on their estimates. 

\paragraph{An upper bound for symmetry-label estimation.}

To translate moment estimation accuracy into a uniform bound on label
error, we impose a positive distance between the moment images of
different label classes.

For $a\in\cC$, let
\[
\cX_a=\{\x\in\cX:\kappa(\x)=a\}
\]
be its \emph{label class}, with moment image $M_{\leq R}(\cX_a,\rho_0)$.

Distances between moment vectors use the Euclidean norm of their concatenated entries. Equivalently, for a tuple of tensors,
\[
\|(\mathbf T_1,\ldots,\mathbf T_R)\|^2
=\sum_{r=1}^R\|\mathbf T_r\|_F^2,
\]
where $\|\mathbf T_r\|_F^2$ is the sum of its squared entries. For nonempty sets $S,T$ in this space, define
\[
\mathrm{dist}(\boldsymbol u,S)=\inf_{\boldsymbol v\in S}\|\boldsymbol u-\boldsymbol v\|,
\qquad
\mathrm{dist}(S,T)=\inf_{\boldsymbol u\in S,\,\boldsymbol v\in T}
\|\boldsymbol u-\boldsymbol v\|.
\]

Theorem~\ref{thm:general_invariant_moment_criterion} concerns identification from exact invariant moments. To obtain uniform recovery from estimated moments, we strengthen
the moment separation condition by requiring a positive separation
margin. This allows moment estimation error bounds to be translated
into classification error bounds, yielding the uniform upper bound
result below in Proposition~\ref{prop:uniform_upper}.
The proof is deferred to Appendix~\ref{sec:proof_uniform_upper}.

\begin{proposition}[Uniform recovery upper bound]\label{prop:uniform_upper}
Fix an integer $R\geq1$. Suppose $\cX$ is bounded, $\cC$ is finite with at least two elements, and each label class $\cX_a$ is nonempty. Assume that the \emph{moment separation margin}
\[
\delta_R\triangleq\min_{a\ne b}
\mathrm{dist}\bigl(M_{\leq R}(\cX_a,\rho_0),M_{\leq R}(\cX_b,\rho_0)\bigr)
\]
is positive. There exists a measurable estimator $\widehat\kappa_N$ based on corrected invariant-moment estimates of orders up to $R$, whose construction does not
require $\rho$, such that for every $N\geq1$ and known noise level
$\sigma\geq0$,
\begin{equation}\label{eq:uniform_upper}
\sup_{\x\in\cX,\,\rho}
\P_\rho\{\widehat\kappa_N\ne\kappa(\x)\}
\leq\min\Big\{1,2\exp\!\Big(
-\frac{N\delta_R^2}{K_R(1+\sigma^{2R})}\Big)\Big\}.
\end{equation}
Here $K_R$ depends only on $G$, $R$, and a bound on
$\|\x\|$ over $\cX$, and is independent of $\sigma$, $N$, $\x$, and $\rho$. The supremum is over all admissible signals and all probability
measures $\rho$ on $G$.

\end{proposition}
As an immediate consequence, for each fixed $\sigma$, we have uniform consistency as
$N\to\infty$.

\begin{remark}[Positive moment margin]
A sufficient condition for a positive margin is that the finitely many label
classes are compact and moments of order at most $R$ separate their labels.
Their moment images are then disjoint compact sets with positive pairwise
distances. Compactness of $\cX$ alone does not suffice, since exact symmetry
classes need not be closed. For example, in the C-MRA and D-MRA models, the unrestricted signal space has zero moment margin for every finite $R$, since signals with different symmetry labels can have arbitrarily close moment vectors. This lack of uniform separation motivates the hypothesis-testing approach (also see the discussion in Remark~\ref{rmk:moments_to_tests}).
\end{remark}

\paragraph{Minimax estimation of symmetry labels.}
We measure the difficulty of estimating the symmetry label by the minimax
misclassification risk
\[
    \mathcal E_N(\sigma;\cX)
    \triangleq\inf_{\widetilde\kappa_N}\;
    \sup_{\x\in\cX,\,\rho}
    \P_\rho\{\widetilde\kappa_N\ne\kappa(\x)\},
\]
where the infimum is over all measurable $\cC$-valued estimators
based on the $N$ observations, with access to the group, signal
class, and noise level, but not to $\x$ or $\rho$. Proposition~\ref{prop:uniform_upper} bounds this risk
from above. For a lower bound, it suffices to consider two differently
labelled signals under Haar sampling, since $\rho_0$ is included in the
supremum over $\rho$. The following bound applies to all estimators,
including those using the full observations. The proof is given in Appendix~\ref{sec:proof_matching_lower}.

\begin{proposition}[A lower bound from matching low-order moments]\label{prop:matching_lower}
Suppose $\x,\x'\in\cX$ have different symmetry labels and, for some
integer $R\geq1$,
\[
M_r(\x,\rho_0)=M_r(\x',\rho_0),\qquad 1\leq r<R,
\]
\[
\Delta_R\triangleq\|M_R(\x,\rho_0)-M_R(\x',\rho_0)\|_F>0.
\]
Under the observation model~\eqref{eq:orbit_model_direct}, there is a $\sigma_0>0$, depending on
$\x$, $\x'$, and $R$, such that for every $\sigma\geq\sigma_0$ and $N\geq1$,
\begin{equation}\label{eq:minimax_lower}
\mathcal E_N(\sigma;\cX)
\geq\frac14\exp\left(-\frac{2N\Delta_R^2}{R!\,\sigma^{2R}}\right),
\end{equation}
For $R=1$, the lower-order moment-matching condition is vacuous.
\end{proposition}

We now combine this lower bound with
Proposition~\ref{prop:uniform_upper} to determine the minimax sample complexity
\[
N_\eta(\sigma;\cX)
\triangleq\min\{N\geq1:\mathcal E_N(\sigma;\cX)\leq\eta\}.
\]
When a positive moment margin holds at the smallest separating order $R$,
this order determines the optimal high-noise sample complexity. Minimality of $R$ supplies the moment-matching pair required by
Proposition~\ref{prop:matching_lower}. Appendix~\ref{sec:proof_sample_complexity} gives the proof.

\begin{theorem}[Minimax sample complexity at high noise]\label{thm:sample_complexity}
Fix a bounded signal class $\cX$ with a finite set of at least two
realizable symmetry labels. Let $R$ be the smallest positive integer
for which $M_{\leq R}(\,\cdot\,,\rho_0)$ separates these labels on
$\cX$, and suppose the moment separation margin $\delta_R$ is positive.
There exist constants $c_1,c_2,\sigma_0>0$, depending only on the fixed
signal class, $G$, and $R$, such that for every $\sigma\geq\sigma_0$
and $0<\eta\leq1/16$,
\begin{equation}\label{eq:minimax_sample_complexity}
c_1\sigma^{2R}\log\frac1\eta
\leq N_\eta(\sigma;\cX)
\leq c_2\sigma^{2R}\log\frac1\eta.
\end{equation}
The constants are independent of $\sigma$ and $\eta$.
\end{theorem}

\begin{remark}[From invariant moments to hypothesis tests]\label{rmk:moments_to_tests}
The preceding minimax recovery guarantees require a positive uniform moment margin, which need not follow from population identification. In C-MRA, second-order moments identify the cyclic symmetry label at the population level (Section~\ref{sec:structure_PMRA}), yet the Fourier coefficients distinguishing different labels may be arbitrarily small, bringing the corresponding moment vectors arbitrarily close. The recovery bound in Proposition~\ref{prop:uniform_upper} deteriorates as the margin decreases.

To infer symmetry without imposing this global separation, we apply the invariant-feature approach developed here to assess invariance hypotheses. With finite noisy data, the procedures in Sections~\ref{sec:Detection_PMRA} and~\ref{sec:Detection_DMRA} assess these hypotheses simultaneously under family-wise error control. The true hypotheses determine the symmetry label through the hierarchy described in Section~\ref{sec:symmetry_detection_problem}. These procedures control erroneous rejection of true invariance hypotheses without requiring a global positive moment margin: the corresponding sections establish finite-sample control for C-MRA and asymptotic control under stated nondegeneracy conditions for D-MRA. This provides an error-controlled approach to symmetry inference even when uniform label recovery over the signal class cannot be guaranteed.
\end{remark}

\section{Moments and invariants in MRA} \label{sec:invariants}

We now specialize the framework of Section~\ref{sec:general_invariant_moment_principle} to real signals $\x\in\R^d$ under circular shifts and, in D-MRA, reflections. The acting group is $\Cd{d}$ or $\Dm{d}$, and the target remains the stabilizer conjugacy class $\kappa(\x)=[\Sym(\x)]$ defined in Section~\ref{sec:symmetry_detection_problem}. We first distinguish the population moments under the sampling distribution from their invariant counterparts and sample estimators. We then express these quantities in Fourier coordinates and identify polynomial features whose values are independent of the latent transformation distribution. Sections~\ref{sec:cyclic} and~\ref{sec:dihedral} determine which features identify the relevant symmetries and develop statistical tests based on their empirical estimates.

\subsection{Population and sample moments}

In the MRA model \eqref{eq:MRA_model}, the observations $\y_1,\ldots,\y_N$ are modeled as i.i.d. realizations of a random variable~$Y=g\circ \x+\beps$ with isotropic Gaussian noise $\beps$. This random variable depends on a latent nuisance parameter $g$, which is assumed to be drawn from a probability distribution $\rho$ over the group $G$. Therefore, the $r$-th (noiseless) population moment of $Y$ can be expressed as a function of the signal $\x$ and the distribution $\rho$, according to the generative model~\eqref{eq:MRA_model},
\begin{align}\label{eq:population_moment}
 M_r(\x,\rho) &\triangleq \E_{g\sim \rho} \Big[\underbrace{(g\circ \x)\otimes \cdots \otimes (g\circ \x)}_{(r-1)\ \text{times}}\otimes (\overline{g\circ \x})\Big] \\
 &= \E_{g\sim \rho} \left[(g\circ \x)^{\otimes r} \right]\in\R^{d\times\ldots\times d}, \quad r=1,2,\ldots, \notag
\end{align}
where the first expression is valid for general complex-valued signals, while the second holds when $\x$ (and hence $g\circ\x$) is real-valued. In this case, $(g\circ \x)^{\otimes r}$ denotes the $r$-fold outer product of $g\circ \x$. Next, the observable $r$-th sample moment
\begin{equation}\label{eq:sample_moment}
\widehat{M}_r\triangleq\frac{1}{N}\sum_{i=1}^{N} \y_i^{\otimes r}\in\R^{d\times\ldots\times d}, \quad r=1,2,\ldots,
\end{equation}
can be computed from the data. If $N$ is large enough, then by the law of large numbers, the sample moments approximate the population moments under appropriate centering, even in the presence of substantial noise. 

We propose that the population moments exhibit unique patterns characteristic of each symmetry subgroup. These patterns can be interpreted as invariant features, which remain stable under the group action. Leveraging this characterization, we aim to identify the patterns in the sample moments and thereby infer the underlying symmetry of $\x$ under generic conditions on $\rho$. Importantly, since the Fourier transform is an invertible linear transformation that preserves all information, we further analyze the transformed moments in the Fourier domain, where the inherent structure often becomes more tractable and such symmetry-indicative patterns tend to be more pronounced and interpretable.

\subsection{Fourier transform and tensor multiplication}\label{sec:Fourier_convention}

Recall $\cI_d=\{0,1,\ldots,d-1\}$. Denote $w_d=e^{-\frac{2\pi\ri}{d}}$ as the primitive $d$-th root of unity in which $\ri^2=-1$. Denote by $\mathbf{W}_d\in\C^{d\times d}$ the diagonal matrix with diagonal entries $\mathbf{W}_d[\ell,\ell]=w_d^{\ell}$ and by $\w_d$ the column vector with entries $\w_d[\ell]=w_d^{\ell}$ for $\ell\in \cI_d$. Define the symmetric matrix $\Fb_d\in\C^{d\times d}$ as follows
\begin{align*}
\Fb_d[\ell_1,\ell_2]=w_d^{\ell_1 \ell_2}, \quad\text{for } \ell_1,\ell_2\in \cI_d.
\end{align*}
We then express the $d$-point discrete Fourier transform matrix as $\Fb_d$. We may omit the subscript $d$ in the sequel if there is no ambiguity. For any $\vv\in\C^d$, its discrete Fourier transform and inverse Fourier transform are given by
\begin{align}\label{eq:Fourier_transform_vector}
    \F \vv\triangleq \Fb \vv \quad \text{and} \quad \F^{-1} \vv\triangleq \frac{1}{d} \Fb^{\sH} \vv.
\end{align}
For any $\mathbf{M}\in \C^{d\times d}$, its discrete Fourier transform and inverse Fourier transform are given by
\begin{align}\label{eq:Fourier_transform_matrix}
    \F \mathbf{M}\triangleq \Fb \mathbf{M} \Fb^{\sH}, \quad \text{and} \quad \F^{-1} \mathbf{M}\triangleq \frac{1}{d^2} \Fb^{\sH} \mathbf{M} \Fb.
\end{align}
Further, denote the normalized Fourier and inverse Fourier transform matrices as $\Ub_d\triangleq \Fb_d/\sqrt{d}$ and $\Ub_d^{\sH}= \Fb_d^{\sH}/\sqrt{d}$. Of note, $\Ub_d$ is a symmetric unitary matrix. 

\begin{remark}[Conjugate symmetry property for real signals]\label{rmk:conjugate_symmetry}
If $\x \in \R^d$ is a real signal, then its Fourier transform satisfies the \emph{conjugate symmetry} property
\[
\F \x[-\ell] \;=\; \overline{\F \x[\ell]}, \quad\text{for } \ell \in \cI_d.
\]
In particular, the DC component $\F \x[0]$ is always real. Conversely, if a complex-valued signal $\x \in \C^d$ has a Fourier transform $\F \x$ that satisfies the conjugate symmetry property, then $\x$ must be real-valued.
Hence, the conjugate symmetry of the Fourier spectrum is a \emph{necessary and sufficient condition} for the signal $\x$ to be real.
\end{remark}

Building on the vector and matrix cases, we next define the discrete Fourier transform of tensors of general order $r$. We first introduce the notion of the $n$-mode product of a tensor with a matrix.
\begin{definition}\label{def:n-mode_tensor_multiplication}
Let $r\geq 1$. Let $\mathbf{T} \in \mathbb{R}^{d_1 \times \cdots \times d_r}$ (or $\mathbb{C}^{d_1 \times \cdots \times d_r}$) be a tensor of order $r$, and let $\mathbf{M} \in \mathbb{R}^{d_n' \times d_n}$ (or $\mathbb{C}^{d_n' \times d_n}$) be a matrix. The $n$-mode (matrix) product of $\mathbf{T}$ with $\mathbf{M}$, denoted by  $\mathbf{T}\times_n \mathbf{M}$, is a tensor of size $d_1\times\ldots \times d_{n-1}\times d_n'\times d_{n+1}\times \ldots\times d_r$, obtained by multiplying each mode-$n$ fiber of $\mathbf{T}$ with $\mathbf{M}$. Elementwise, we have
\begin{align*}
(\mathbf{T}\times_n \mathbf{M})[i_1,\ldots,i_{n-1},i_n',i_{n+1},\ldots,i_r]=\sum_{i_n=0}^{d_n-1} \mathbf{T}[i_1,\ldots,i_{n-1},i_n,i_{n+1},\ldots,i_r]\cdot \mathbf{M}[i_n',i_n].
\end{align*}
\end{definition}

\begin{remark}\label{rmk:tensor_multiplication_order2}
In particular, for $r=1$, the $n$-mode product reduces to a matrix-vector product. Specifically, for a vector (a tensor of order 1) $\vv\in\C^d$ and $\mathbf{M}\in\C^{d'\times d}$,
\begin{align*}
\vv\times_1 \mathbf{M}=& \mathbf{M} \vv.
\end{align*}
For $r=2$, the $n$-mode product reduces to the standard matrix-matrix product. Specifically, for a matrix (a tensor of order 2) $\mathbf{T}\in \C^{d_1\times d_2}$ and $\mathbf{M}\in\C^{d_n'\times d_n}$,
\begin{align*}
\mathbf{T}\times_1 \mathbf{M}=& \mathbf{M} \mathbf{T},\\
\mathbf{T}\times_2 \mathbf{M}=&\mathbf{T} \mathbf{M}^\T.
\end{align*}
\end{remark}

Now we are ready to introduce the discrete Fourier transform of tensors of general order $r$.
\begin{definition}\label{def:tensor_Fourier_transform}
For $r\geq 2$, define the discrete Fourier transform of a tensor $\mathbf{T}$ of order $r$ as
\begin{align*}
\F \mathbf{T} \triangleq \mathbf{T}\times_1 \Fb \ldots \times_{r-1} \Fb \times_r \overline{\Fb},
\end{align*}
and the inverse Fourier transform is given by
\begin{align*}
\F^{-1} \mathbf{T}\triangleq \mathbf{T}\times_1 \frac{1}{d}\overline{\Fb} \ldots \times_{r-1} \frac{1}{d}\overline{\Fb} \times_r \frac{1}{d}\Fb.
\end{align*}
For $r=1$, the discrete Fourier transform and its inverse of a vector $\vv$ is given by $\F \vv\triangleq \vv\times_1 \Fb$ and $\F^{-1} \vv \triangleq \vv\times_1 \frac{1}{d}\overline{\Fb}$.
\end{definition}

\begin{remark}
In particular, for $r=1,2$, the definitions above are equivalent to those in \eqref{eq:Fourier_transform_vector} and \eqref{eq:Fourier_transform_matrix}, as noted in Remark~\ref{rmk:tensor_multiplication_order2}. Moreover, the use of $\overline{\Fb}$ in the $r$-th mode (and similarly in the inverse transform) is a matter of convention. Since the discrete Fourier transform is an invertible linear transformation, different choices--e.g., using $\Fb$ or $\overline{\Fb}$ in different modes--are information-equivalent, as long as the forward and inverse transforms are defined consistently.
\end{remark}

\subsection{Moment invariants and their Fourier representations}\label{sec:moment_invariant}

Since the group $G$ is finite in the MRA model, any probability distribution $\rho$ over $G$ can be represented unambiguously as a vector indexed by the group elements. For notational simplicity, we continue to denote this vector representation by $\rho$. We distinguish the two cases of interest: the \emph{cyclic group} $\Cd{d}$ and the \emph{dihedral group} $\Dm{d}$.

For the cyclic group $\Cd{d} = \langle r \rangle$, which has order $|\Cd{d}| = d$, the distribution $\rho$ can be viewed as a length-$d$ vector, i.e. $\rho\in \Delta_d$, with components
\[
\rho[\ell] = \rho(r^\ell), \quad\text{for } \ell \in\cI_d.
\]
For the dihedral group $\Dm{d} = \langle r, s \mid r^d = s^2 = e,\, srs = r^{-1} \rangle$, which has order $|\Dm{d}| = 2d$, the distribution $\rho$ is represented as a length-$2d$ vector, i.e. $\rho\in \Delta_{2d}$. We decompose this vector into two length-$d$ parts:
\[
\rho = \begin{pmatrix} \p \\ \q \end{pmatrix}, \quad \text{where } \p[\ell] = \rho(r^\ell), \quad \q[\ell] = \rho(r^\ell s), \quad\text{for } \ell \in\cI_d.
\]

We are now ready to express the $r$-th population moment defined in \eqref{eq:population_moment} using the $n$-mode tensor product and, correspondingly, its Fourier transform as in Definition~\ref{def:tensor_Fourier_transform}. The proof is provided in Appendix \ref{sec:proof_moment_expression}. 

\begin{proposition}\label{prop:moment_expression}
Let $\x \in \C^d$ be a general complex signal and $r \geq 2$.

\begin{itemize}
    \item[(a)] (\textbf{C-MRA})  
    Suppose $\rho \in \R^d$ is a probability distribution over the cyclic group $\Cd{d}$. Then the $r$-th population moment $M_r(\x, \rho)$ is given by
    \begin{align}\label{eq:population_moment_tensor_PMRA}
    M_r(\x, \rho) = \mathbf{T}^{(r)}_{\rho} \times_1 \Cb_{\x} \cdots\times_{r-1} \Cb_{\x} \times_r \Cb_{\overline{\x}},
    \end{align}
    where $\Cb_{\x}$ is the circulant matrix generated by $\x$, and $\mathbf{T}^{(r)}_{\rho}$ is the order-$r$ diagonal tensor with $\rho[i]$ on the $\underbrace{(i, i, \ldots, i)}_{r\ \text{times}}$-th diagonal, and zeros elsewhere. Its Fourier transform is given by
    \begin{align}\label{eq:population_moment_tensor_Fourier_PMRA}
    \F M_r(\x, \rho) = \F \mathbf{T}^{(r)}_{\rho} \times_1 \Db_{\F\x} \ldots \times_{r-1} \Db_{\F\x} \times_r \Db_{\overline{\F \x}}.
    \end{align}
    In particular, elementwise,
    \begin{align*}
    (\F \mathbf{T}^{(r)}_{\rho})[i_1,\ldots,i_r]=(\F\rho) [i_1+\ldots+i_{r-1}-i_r].
    \end{align*}

    \item[(b)] (\textbf{D-MRA})  
    Suppose $\rho \in \R^{2d}$ is a probability distribution over the dihedral group $\Dm{d}$, and decompose it as $\rho = (\p^\T, \q^\T)^\T$, where $\p[\ell] = \rho(r^\ell)$ and $\q[\ell] = \rho(r^\ell s)$ for $\ell \in \cI_d$. Then the $r$-th population moment $M_r(\x, \rho)$ is given by
    \begin{align}\label{eq:population_moment_tensor_DMRA}
    M_r(\x, \rho) = \mathbf{T}^{(r)}_{\p} \times_1 \Cb_{\x}  \cdots \times_{r-1} \Cb_{\x} \times_r \Cb_{\overline{\x}}
    + \mathbf{T}^{(r)}_{\q} \times_1 \Cb_{s \x} \times_2 \Cb_{s  \x} \cdots \times_r \Cb_{\overline{s \x}},
    \end{align}
    where $s \x$ is the reflection of $\x$. Its Fourier transform is given by
    \begin{align}\label{eq:population_moment_tensor_Fourier_DMRA}
    \F M_r(\x, \rho) =& \F\mathbf{T}^{(r)}_{\p} \times_1 \Db_{\F\x} \ldots \times_{r-1} \Db_{\F\x} \times_r \Db_{\overline{\F \x}}\notag\\
    &+ \F\mathbf{T}^{(r)}_{\q} \times_1 \Db_{\overline{\F \x}} \ldots \times_{r-1} \Db_{\overline{\F \x}} \times_r \Db_{\F \x}.
    \end{align}
\end{itemize}
\end{proposition}
Although Proposition~\ref{prop:moment_expression} assumes $r \geq 2$, the case $r = 1$ is a natural special case. For clarity, we state it separately below.
\begin{remark}
For $r=1$, the first population moment and its Fourier transform reduce to simpler forms. Specifically,
\begin{itemize}
    \item[(a)] (\textbf{C-MRA}) We have
    \begin{align*}
M_1(\x,\rho)=&\mathbf{T}^{(1)}_\rho \times_1 \Cb_\x= \Cb_\x\rho,\\
\F M_1(\x,\rho)=&\F\mathbf{T}^{(1)}_\rho \times_1 \Db_{\F\x}=\F\rho \times_1 \Db_{\F\x}= \F \x  \odot \F \rho.
\end{align*}
    \item[(b)] (\textbf{D-MRA}) We have
    \begin{align*}
M_1(\x,\rho)=&\mathbf{T}^{(1)}_{\p} \times_1 \Cb_\x+\mathbf{T}^{(1)}_{\q} \times_1 \Cb_{s\x}= \Cb_\x \p+\Cb_{s\x} \q,\\
\F M_1(\x,\rho)=&\F\mathbf{T}^{(1)}_{\p} \times_1 \Db_{\F\x}+\F\mathbf{T}^{(1)}_{\q} \times_1 \Db_{\overline{\F\x}}\\
=&\F\p \times_1 \Db_{\F\x}+\F\q \times_1 \Db_{\overline{\F\x}}= \F \x  \odot \F \p +\overline{\F\x} \odot \F\q.
    \end{align*}
\end{itemize}
\end{remark}

Now we present a corollary showing that, under the C-MRA model with uniform distribution over $\Cd{d}$, the first three population moments and their Fourier transforms recover the commonly used moment invariants: the Direct Current (DC) component, power spectrum, and bispectrum, see, e.g. \cite[Section II]{bendory2017bispectrum}. Here, we still consider general complex signal $\x\in\C^d$. The proof of this result is provided in Appendix \ref{sec:proof_population_moment_PMRA_uniform}.

\begin{corollary}\label{cor:population_moment_PMRA_uniform}
Under the C-MRA model with uniform $\rho=\rho_0$, the first three population moment $M_1(\x,\rho_0)\in\C^d$, $M_2(\x,\rho_0)\in\C^{d\times d}$, and $M_3(\x,\rho_0)\in\C^{d\times d\times d}$ satisfy, for any $i\in \cI_d$,
\begin{align*}
M_1(\x,\rho_0)[i]=& \frac{1}{d}\sum_{\ell=0}^{d-1} \x[\ell], \\
M_2(\x,\rho_0)[i,i-k]=& \frac{1}{d}\sum_{\ell=0}^{d-1} \x[\ell]\overline{\x[\ell-k]}, \quad \text{for }k \in \cI_d,\\
M_3(\x,\rho_0)[i,i+k_1,i-k_2]=& \frac{1}{d}\sum_{\ell=0}^{d-1} \x[\ell]\x[\ell+k_1]\overline{\x[\ell-k_2]}, \quad\text{for } k_1,k_2 \in \cI_d.
\end{align*}
Furthermore, their Fourier transforms give nonzero entries:
\begin{align*}
(\F M_1(\x,\rho_0))[0]=& (\F\x)[0]=\sum_{\ell=0}^{d-1}\x[\ell], \\
(\F M_2(\x,\rho_0))[k,k]=&|(\F\x)[k]|^2, \quad \text{for } k\in \cI_d,\\
(\F M_3(\x,\rho_0))[k_1,k_2,k_1+k_2]=& (\F\x)[k_1]\cdot (\F\x)[k_2] \cdot \overline{(\F\x)[k_1+k_2]},\quad \text{for } k_1,k_2\in \cI_d,
\end{align*}
which correspond to the DC component, power spectrum, and bispectrum of the signal $\x$.
\end{corollary}

\section{Symmetry detection in Cyclic MRA} \label{sec:cyclic}

\subsection{Moment structure under cyclic subgroup symmetries}\label{sec:structure_PMRA}
As the first instance of model~\eqref{eq:MRA_model}, we consider the C-MRA case. Here $\x\in\R^d$ is indexed by the vertices of a regular $d$-gon, and circular shifts represent rotations of this sampling grid. The group $G=\Cd{d}$, which is the so-called \textit{cyclic group} of order $d$, is a finite group generated by the plane rotation of $2\pi/d$ radians, that is, $G \cong \mathbb{Z}_d$. In other words, $G\equiv \langle r\rangle = \{1,r,\ldots,r^{d-1}\}$ with a rotation $r$ of order $d$. The action of any element from $r^\ell \in G$, corresponding to a rotation of $\ell (2\pi/d)$ radians, on $\x$ is defined as $(r^\ell \cdot \x)[k] = \x[(k-\ell)\bmod d]$ for $k,\ell\in \cI_d$.

The subgroup $H$ of $G$ takes the form of $\Cd{m} = \langle r^{\frac{d}{m}} \rangle$ where $m$ is some divisor of $d$. In other words, a signal that is $\Cd{m}$-invariant is equivalent to having a period of $d/m$. Notably, $\Cd{1}$ corresponds to the case of no symmetry. Hence, distinct subgroups $\Cd{m}$ give rise to distinct fixed-point subspaces $\Fix{\Cd{m}}$, so no two subgroups of $\Cd{d}$ leave the same set of signals invariant. Consequently, exact symmetry and invariance are unambiguously linked here: the exact point group $\Sym(\x)$ is the largest $\Cd{m}$ for which $\x\in\Fix{\Cd{m}}$. According to Definition~\ref{def:detectable_symmetry},
each subgroup $\Cd{m}$ is only equivalent to itself. Hence, the set of 
detectable signal symmetries is simply
\begin{align*}
\{\Cd{m} : m \mid d\}.
\end{align*}

Moreover, this allows us to establish \textit{implication relationships} between the invariance properties. Namely, we have
\begin{align*}
\text{$\Cd{m_2}$-invariance} \Rightarrow \text{(``implies'')} \ \text{$\Cd{m_1}$-invariance}, \quad \text{when } m_1 \mid m_2,
\end{align*}
because $m_1 \mid m_2$ implies $\Cd{m_1}\le\Cd{m_2}$, hence $\Fix{\Cd{m_2}}\subseteq\Fix{\Cd{m_1}}$. In other words, any signal that is invariant under $\Cd{m_2}$ 
is automatically invariant under $\Cd{m_1}$. (Note that no maximality is asserted here: this is a statement about invariance, not about exact symmetry.) This implication structure plays a crucial role in our subsequent hypothesis testing framework 
(see Section~\ref{sec:DAG_construction_cyclic}).

Recall that in Fourier space, the action of $r^{\frac{d}{m}}$ gives
\[   \F (r^{\frac{d}{m}} \cdot \x)[\ell] = \exp(-2 \pi i \ell /m  ) \F \x[\ell]. \]
Thus, if the signal $\x$ is invariant under the action $r^{\frac{d}{m}}$, then its Fourier transform satisfies
\begin{align*}
\F \x[\ell] = \exp(-2 \pi i \ell/m  ) \F \x[\ell], ~~~\text{for}~\ell\in \cI_d.
\end{align*}
This equality can hold only if either $\F \x[\ell]=0$ or $\exp(-2 \pi i \ell/m  )=1$, i.e., when $\ell/m$ is an integer. This condition similarly applies to any other element in $\Cd{m}$, where the rotations $r^{(td)/m}$ for integer $t$ impose the same requirement. Hence, for any divisor $m$ of $d$, we define the set of all indices that are multiples of $m$:
\begin{align}\label{eq:nonzeros_set}
\cI_{m,d}\triangleq\{0,m,\ldots,d-m\}.
\end{align}
We also define its complement with respect to $\cI_d$ as
\begin{align}\label{eq:zeros_set}
\cI_{m,d}^c\triangleq  \cI_d\setminus \cI_{m,d},
\end{align}
which consists of all indices that are not multiples of $m$. 
We conclude with the following proposition. The detailed proof is given in Section \ref{sec:proof_zeros_in_Fourier_MRA}.

\begin{proposition}\label{prop:zeros_in_Fourier_MRA}
Let $m$ be any divisor of $d$, and let $\x\in\R^d$.
Then $\x$ is $\Cd{m}$-invariant if and only if
\begin{equation}\label{eqn:zeros_in_Fourier_MRA}
\F\x[\ell]=0
\quad\text{for every }\ell\in\cI_{m,d}^c.
\end{equation}
Equivalently, $\x$ is not $\Cd{m}$-invariant if and only if
$\F\x[\ell]\neq0$ for some $\ell\in\cI_{m,d}^c$.

In particular, in the C-MRA setting, if $\x$ has exact
$\Cd{m'}$-symmetry, meaning $\Sym(\x)=\Cd{m'}$ with $m'\mid d$,
then $\x$ is $\Cd{m}$-invariant if and only if $m\mid m'$.
Consequently, if $m\nmid m'$, there exists
$\ell\in\cI_{m,d}^c$ such that $\F\x[\ell]\neq0$.
\end{proposition}

Following Section~\ref{sec:moment_invariant}, let $\rho \in \Delta_d$ denote the underlying probability distribution over the group $G = \Cd{d}$, where $\rho[\ell]$ is the probability assigned to the group element $r^\ell$ for $\ell \in \cI_d$. The $r$-th population moment $M_r(\x,\rho)$ is defined as the expectation $\E_{g\sim \rho}[(g\circ \x)^{\otimes r}]$ of the $r$-th order tensor product of $(g\circ \x)$, where the expectation is taken over the latent group element drawn from $\rho$, see \eqref{eq:population_moment}.

It is shown in~\cite{abbe2018multireference} that the first two moments are sufficient to ensure uniqueness under the C-MRA model when the group elements are drawn from a generic distribution. Motivated by this, we analyze the structure of the first and second population moments and their Fourier transforms. The following lemma specializes the general expression in Proposition~\ref{prop:moment_expression} to the first and second moments in the C-MRA setting.

\begin{lemma}\label{lem:population_moments_MRA}
Under the C-MRA setting, the first and second population moments $M_1(\x,\rho), M_2(\x,\rho)$ have the expressions:
\begin{equation}\label{eqn:moments}
M_1(\x,\rho) = \x \ast \rho = \Cb_{\x} \rho, \quad M_2(\x,\rho) = \Cb_{\x} \Db_{\rho} \Cb_{\x}^{\sH}= \Cb_{\x} \Db_{\rho} \Cb_{\x}^\T.
\end{equation}
Their Fourier transforms have the expressions:
\begin{equation} \label{eqn:moments_fourier}
\F M_1(\x,\rho) = \F \x  \odot \F \rho  , \quad \F M_2(\x,\rho) = \Db_{\F \x}  \Cb_{\F\rho} \Db_{\overline{\F \x}}.
\end{equation}
\end{lemma}

Combining Proposition~\ref{prop:zeros_in_Fourier_MRA} and Lemma~\ref{lem:population_moments_MRA}, we immediately obtain the following corollary. While Lemma~\ref{lem:population_moments_MRA} shows that the first Fourier moment $\F M_1(\x, \rho)$ depends on the distribution~$\rho$, Corollary~\ref{cor:zeros_in_moment} reveals that the diagonal entries of the second Fourier moment are independent of~$\rho$. This observation motivates the development of distribution-free methods for symmetry detection over the cyclic group as will be discussed in detail in Section~\ref{sec:Detection_PMRA}.

\begin{corollary}\label{cor:zeros_in_moment}
Let $m$ be any divisor of $d$. Under the C-MRA setting, if a signal $\x$ is $\Cd{m}$-invariant, then the zero patterns in the Fourier transforms of the first and second population moments take the form:
\begin{align}
\F M_1(\x,\rho) [\ell]  =&  0, \quad \text{for} \quad \ell \in \cI_{m,d}^c,\label{eqn:zeros_in_first_moment}\\
\F M_2(\x,\rho)[\ell_1, \ell_2]   =&  0, \quad \text{for} \quad \ell_1,\ell_2 \in \cI_{m,d}^c.\label{eqn:zeros_in_second_moment}
\end{align}
In particular, the diagonal entries of the second moment take the form
\begin{align}\label{eqn:diag_in_second_moment}
\F M_2(\x,\rho)[\ell, \ell]= |\F \x [\ell]|^2,\quad \text{for }\ell\in \cI_d,
\end{align}
which vanish for all $\ell \in \cI_{m,d}^c$.
\end{corollary}

\begin{proof}
By Lemma \ref{lem:population_moments_MRA}, for $\ell,\ell_1,\ell_2 \in \cI_d$, we obtain
\begin{align*}
\F M_1(\x,\rho)[\ell] =&  \F \x[\ell]  \F \rho[\ell],\\
\F M_2(\x,\rho)[\ell_1, \ell_2] =& \F \x[ \ell_1  ] \overline{\F \x[\ell_2]}  \, \F\rho[\ell_1-\ell_2].
\end{align*}
Combining with \eqref{eqn:zeros_in_Fourier_MRA} proves the first two results. For the last one, we prove it by noting that $\F \rho[0] = 1$, as long as $\rho$ is a probability distribution, i.e., $\sum_{\ell=0}^{d-1} \rho[\ell] = 1$.
\end{proof}

\subsection{Distribution-free symmetry detection for Cyclic MRA}\label{sec:Detection_PMRA}

We present a multiple testing procedure for symmetry detection in the C-MRA problem over the cyclic group, which \textit{strongly} controls the familywise error rate (FWER). We mainly utilize the multiple testing method for hypotheses structured in a directed acyclic graph (DAG), as introduced by \cite{meijer2015multiple} and detailed in Section \ref{sec:SRA}. Under this setting, the symmetry detection involves simultaneously testing a collection of hypotheses, each stating that the signal is $\Cd{m}$-invariant,
\begin{equation}\label{eqn:cyclic_hypothesis}
H_m : \ \x\in\Fix{\Cd{m}} \quad\bigl(\text{equivalently } \Cd{m}\le\Sym(\x)\bigr), \qquad m\mid d,~ m\geq 2,
\end{equation}
that is, the ``at least $\Cd{m}$-symmetry'' property of Remark~\ref{rmk:exact_vs_invariance}. We test invariance rather than exact symmetry precisely because, by the implication structure of Section~\ref{sec:structure_PMRA}, the invariance hypotheses are nested along the subgroup lattice ($m_1\mid m_2 \Rightarrow H_{m_2}\subseteq H_{m_1}$), which is what renders the DAG-structured procedure well defined; the maximal non-rejected groups are reported afterward as candidate estimates of $\Sym(\x)$, as explained in Section~\ref{sec:symmetry_detection_problem}. If every tested hypothesis is rejected, the candidate is the trivial group $\Cd{1}$. Our method contains three main components: (i) Generate hypothesis DAG for the cyclic group, (ii) Compute the raw p-value for each hypothesis by evaluating the observed test statistic against its chi-squared null distribution, and (iii) Perform sequential rejection. In (i), we construct a DAG with a proper topological ordering for all candidate 
symmetry hypotheses
\[
\{\Cd{m} : m \mid d,~ m\geq 2\},
\]
that is, the set of detectable signal symmetries with the trivial symmetry 
$\Cd{1}$ excluded (since $\Cd{1}$ corresponds to no symmetry and therefore does not need to be detected). Combined with the raw p-values obtained in step (ii), the sequential rejection algorithm (Algorithm \ref{alg:SR}) is applied to determine the rejection set, yielding the maximal non-rejected candidate symmetries of the signal $\x$. The implementation details of steps (i) and (ii) can be found in Sections~\ref{sec:DAG_construction_cyclic} and~\ref{sec:p-value_calculation_CMRA}, respectively, while step (iii) is deferred to Appendix~\ref{sec:SRA}.

\subsubsection{Generating hypothesis DAG}\label{sec:DAG_construction_cyclic}
All of the hypotheses naturally form a DAG, with parent-child relationships in the graph determined by the implication relationships as discussed in Section~\ref{sec:structure_PMRA} among these invariance hypotheses. In this DAG, each vertex represents a single invariance hypothesis $H_m$ (labeled by $\Cd{m}$), and the directed edges represent the implication relationships. Recall the implication relationships before: 
the falsehood of $\Cd{m_1}$-invariance implies the falsehood of $\Cd{m_2}$-invariance whenever $m_1 \mid m_2$. We orient the edges so that $\Cd{m_1}$ is always placed as a descendant 
of $\Cd{m_2}$. To achieve this, we proceed in the following two main steps: (i) Find all the prime factors of $d$, and (ii) Build the DAG. The resulting DAG is actually equivalent to the Hasse diagram of the subgroup lattice of $\Cd{d}$ with the leaf vertex $\Cd{1}$ removed (see Algorithm \ref{alg:DAG_MRA_cyclic} for details). 
 
{\bf Step (i):} Given the dimension $d$ of our target signal, set $\texttt{current}=d$. We start with the smallest prime number, which is $2$. While $d$ is divisible by $2$, divide $\texttt{current}$ by $2$ and record $2$ as a prime factor. Continue this process until $\texttt{current}$ is no longer divisible by $2$. Then we move to check the next prime number. For each prime $p$, while $\texttt{current}$ is divisible by $p$, divide $\texttt{current}$ by $p$ and record $p$ as a prime factor. Continue this process until $p^2>\texttt{current}$. If $\texttt{current}$ is still greater than $1$ after this step, then $\texttt{current}$ itself is a prime factor. 

{\bf Step (ii):} We begin with the root vertex $\Cd{d}$. For each prime factor $p\not= d$ of $d$, we create a new child vertex $\Cd{d/p}$ and add a directed edge from $\Cd{d}$ to $\Cd{m/p}$. Next, we move to each newly created vertex $\Cd{m}$, which has not yet been assigned children and is not a leaf vertex, i.e., $\Cd{p}$ for a prime $p$, in the DAG. For each prime factor $p\not= d$ of $d$:
\begin{itemize}[itemsep=2pt, topsep=2pt]
    \item If $p$ divides $m$ and $\Cd{m/p}$ is not already in the DAG, we create a new child vertex $\Cd{m/p}$ and add a directed edge from $\Cd{m}$ to $\Cd{m/p}$;
    \item If $p$ divides $m$ but $\Cd{m/p}$ is already in the DAG, we add a directed edge from $\Cd{m}$ to $\Cd{m/p}$.
\end{itemize}
The process terminates once all the vertices of candidate symmetries, except for the leaf vertices $\Cd{p}$ (with $p$ prime), have been assigned children.

The pseudo-code containing the two main steps above is shown in Algorithm \ref{alg:DAG_MRA_cyclic}. 

\begin{algorithm}[htb!]
    \caption{Generating Hypothesis DAG for Cyclic Group}\label{alg:DAG_MRA_cyclic}
\begin{algorithmic}[1]
\REQUIRE{Integer $d$.}

\STATE \textbf{Step (i): Find all prime factors of $d$}
\STATE \textbf{Function} \texttt{findPrimeFactors(d)}
    \STATE Initialize an empty list \texttt{primeFactors}.
    \STATE Set $\texttt{current}=d$.
    \STATE Set $p=2$.
    \WHILE{\( p^2 \leq \texttt{current} \)}
        \IF{$p$ divides $\texttt{current}$}
            \STATE Add \texttt{p} to \texttt{primeFactors}.
            \WHILE{$p$ divides $\texttt{current}$}
                \STATE $\texttt{current}=\texttt{current}/p$.
            \ENDWHILE
        \ELSE
            \STATE $p=p+1$.
        \ENDIF
    \ENDWHILE
    \IF{$\texttt{current}> 1$}
        \STATE Add \texttt{current} to \texttt{primeFactors}.
    \ENDIF
\STATE \textbf{return} \texttt{primeFactors}.
\STATE \textbf{Step (ii): Build the DAG}
    \STATE \textbf{Function} \texttt{DAGBuilderCyclic(d)}
    \STATE Set \texttt{primeFactors} to \texttt{findPrimeFactors(d)}.
    \STATE Initialize an empty dictionary $\texttt{DAGCyclic}$ with key $d$ and value an empty list.
    \STATE Initialize a $\texttt{queue}$ with the root node $d$.
    \WHILE{$\texttt{queue}\not=\emptyset$}
        \STATE $\texttt{current} = \texttt{queue.pop(0)}$
        \FOR{$p$ in $\texttt{primeFactors}$}
            \IF{$p$ divides $\texttt{current}$ and $\texttt{current}\not=p$}
                \STATE $\texttt{child} = \texttt{current} / p$;
                \STATE Add $\texttt{child}$ to $\texttt{DAGCyclic[current]}$.
                \IF{$\texttt{child}$ is not already in $\texttt{DAGCyclic}$}
                    \STATE Add $\texttt{child}$ to $\texttt{DAGCyclic}$ with value an empty list.
                    \STATE Append $\texttt{child}$ to $\texttt{queue}$.
                \ENDIF
            \ENDIF
            
        \ENDFOR
    \ENDWHILE
    \STATE \textbf{return} $\texttt{DAGCyclic}$.
\end{algorithmic}
\end{algorithm}

\begin{figure}[htb!]
\centering
\includegraphics[width=0.5\textwidth]{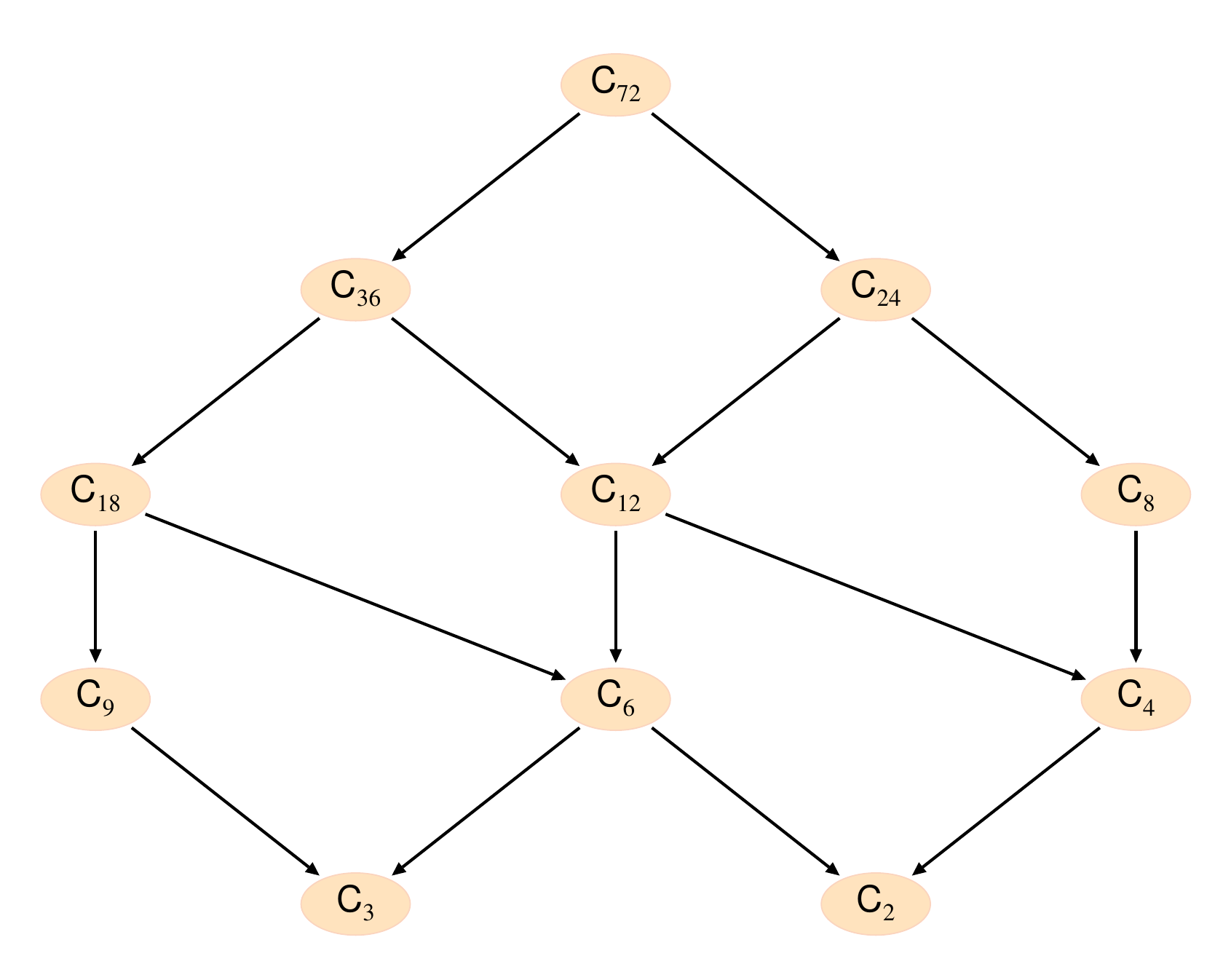}
\caption{Example of the hypothesis DAG for the cyclic group for $d=72$.}\label{fig:Cyclic72}
\end{figure}

An example of the hypothesis DAG for the cyclic group for $d=72$ is shown in Figure \ref{fig:Cyclic72}. A valid topological ordering for this DAG is as follows: 
\begin{align*}
\Cd{72},\Cd{36},\Cd{24},\Cd{18},\Cd{12},\Cd{8},\Cd{9},\Cd{6},\Cd{4},\Cd{3},\Cd{2}.
\end{align*}

\subsubsection{Chi-squared test and raw p-value calculation for single cyclic hypothesis}\label{sec:p-value_calculation_CMRA}
Recall $\cI_{m,d}^c=\{0,1,\ldots,d-1\}\setminus \{0,m,\ldots,d-m\}$, which includes all the indices except those multiples of $m$. For each vertex corresponding to the $\Cd{m}$-symmetry in the hypothesis DAG, consider the natural test statistic $\widehat{B}_m$, which equals the sum of all the diagonal entries indexed by $\cI_{m,d}^c$ in the Fourier transform of the second-order sample moment, defined by
\begin{align}\label{eq:test_stat}
\widehat{B}_m = \sum_{\ell\in \cI_{m,d}^c}\F\widehat{M}_2[\ell,\ell].
\end{align}
By definition, the diagonal elements in $\F\widehat{M}_2$ are real-valued, ensuring that $\widehat{B}_m$ is also real-valued. For an arbitrary distribution $\rho$ of the nuisance parameters $\{g_j\}_{j=1}^N$ over the cyclic group, under the null hypothesis that the signal $\x$ exhibits $\Cd{m}$ symmetry, it can be shown that $\widehat{B}_m$ follows a scaled chi-squared distribution. Specifically, for a given significance level $\alpha$ (e.g., $0.05$), let $t_k(\alpha)$ be the corresponding critical value of a standard chi-squared test of degree $k$, i.e.,
\begin{align*}
\mathbb{P}\Big(\chi_k^2\geq t_k(\alpha)\Big)=\alpha.
\end{align*}
Define the corresponding test for hypothesis $\Cd{m}$-symmetry as 
\begin{align*}
\phi_m = \1\Big\{\widehat{B}_m\geq \gamma\cdot t_{Nd(1-1/m)}(\alpha) \Big\}~~~\text{with}~~~\gamma=\frac{\sigma^2 d}{N}.
\end{align*}
Then we have the following theorem, which ensures valid size control and favorable asymptotic power for the single testing procedure $\phi_m$. The proof is deferred to Appendix \ref{sec:proof_test_size_power}.

\begin{theorem}\label{thm:test_size_power}
For any divisor $m\geq 2$ of $d$ and under the C-MRA setting, let $\phi_m$ be the test given above, and let
$E_m(\x)\triangleq\sum_{\ell\in \cI_{m,d}^c}|\F\x[\ell]|^2$
denote the energy of $\x$ in the frequencies probed by the test. Let $\E_{\Cd{m}}$ be the expectation under the hypothesis that $\x$ is $\Cd{m}$-invariant (equivalently $E_m(\x)=0$). Then, for any distribution $\rho$ of the group actions, the test has exact size
\begin{align}
\E_{\Cd{m}} [\phi_m] = \alpha.
\end{align}
Moreover, for any fixed signal $\x$ that is not $\Cd{m}$-invariant, i.e. $E_m(\x)>0$ (in particular, any $\x$ with exact $\Cd{m'}$-symmetry,
where $m'\mid d$ and $m\nmid m'$), the power satisfies
\begin{align}
\E [\phi_m]\geq 1- \exp(-CN),
\end{align}
where $C>0$ may depend on $d,m,\sigma$ and the pattern energy $E_m(\x)$.
\end{theorem}

Accordingly, the raw p-value for testing the hypothesis $\Cd{m}$-invariant can be computed as
\begin{align}\label{eq:raw_pvalue}
p_m = \mathbb{P}(\gamma\cdot\chi^2_{Nd(1-1/m)}> \widehat{B}_m).
\end{align}

\begin{remark}
The statistic $\widehat{B}_m$ is chosen because, under the C-MRA model, it admits an exact finite-sample distribution, yielding a scaled chi-squared test without requiring covariance estimation. For large sample sizes, one may alternatively invoke the central limit theorem and apply a Wald-type test for asymptotic inference. See, for example, Section \ref{sec:p-value_calculation_DMRA_collapsed} and \ref{sec:p-value_calculation_DMRA_parity_split}.
\end{remark}

\subsection{Numerical demonstration for the finite sample case}

To illustrate our detection approach, we provide a numerical example. In this example, we fix a symmetric signal and simulate data with various noise levels. At each level, we inspect the output of our method and conclude its success rate in finding the correct symmetry. The code for this example, as well as our complete method implementation and figure-generating scripts, is available at \url{https://github.com/nirsharon/symmetry_detection_MRA.git}.

The example details are as follows. We randomly fix a signal of length $d=84$, as each entry is uniformly picked in $[0,1]$. Then, we use $m=21$ and symmetrize the signal by averaging it after applying each element in $\Cd{m}$. Then, we set 25 noise levels, and for each, we simulate $10,000$ observations to estimate the second moment. The SNR level~\eqref{eq:SNR}, $\|\x\|_2^2/(d\sigma^2)$, ranges (roughly) between $8$ and $0.003$. We repeat each test case for a given noise level $20$ times to ensure robust results. 

\begin{figure}[htb!]
\centering
\includegraphics[width=0.9\textwidth]{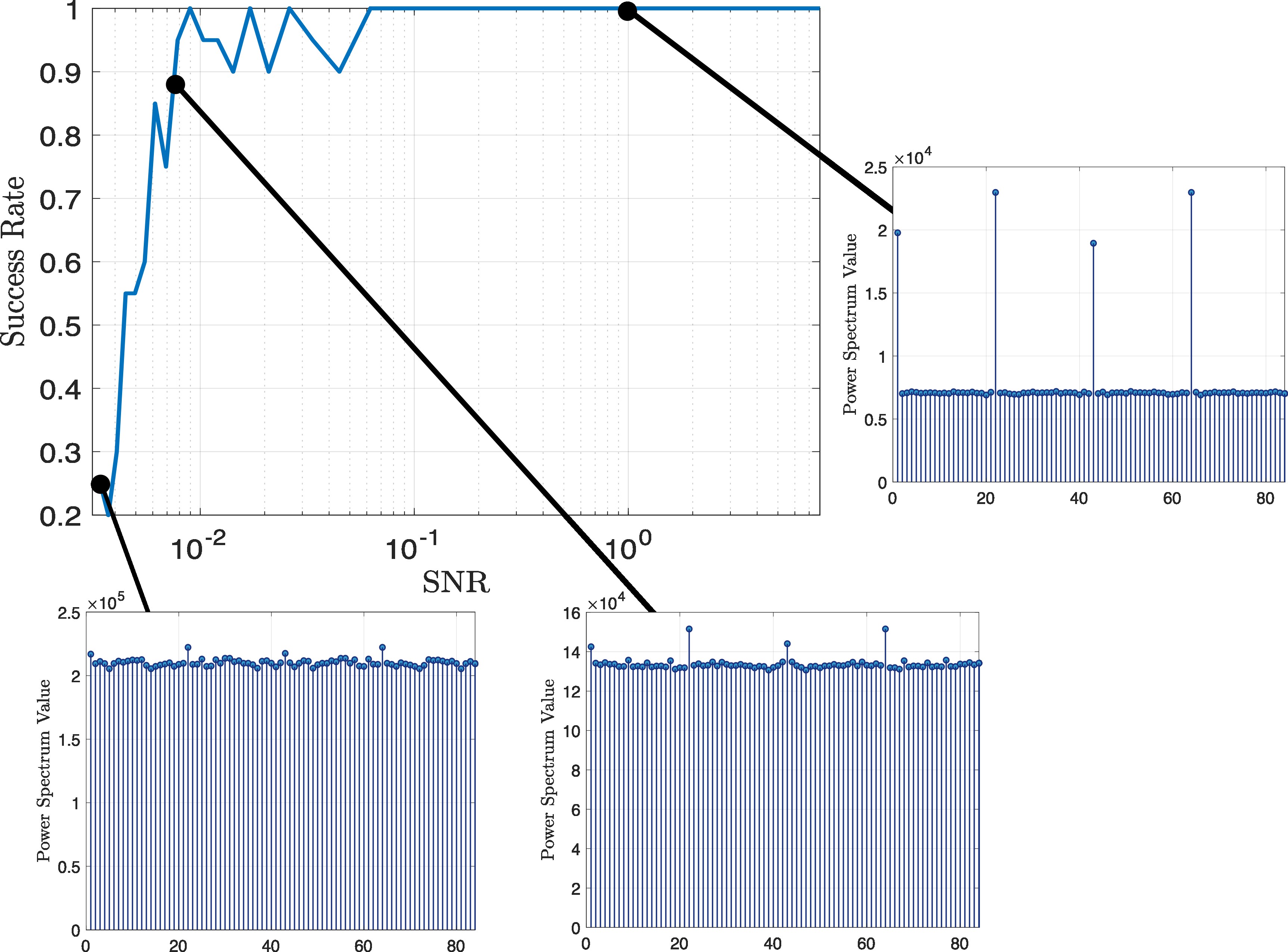}
\caption{Performance of the C-MRA symmetry detection method across varying noise levels. The top-left panel depicts the success rate as a function of SNR. Representative power spectra at selected SNR levels are shown around the main plot.} \label{fig:succ_rate}
\end{figure}

In Figure~\ref{fig:succ_rate}, we depict the success rate as a function of the SNR. The graph shows that sequential rejection ensures near-perfect results up to SNR levels of $1/100$. Then, the success deteriorates quickly. In three places, we also include a snapshot of the noisy power spectrum, which helps explain this decline, as high noise swamps the nonzero pattern. 

\section{Symmetry detection in Dihedral MRA} \label{sec:dihedral}

Passing from rotations to the full dihedral group introduces two new phenomena: equivalence classes of detectable symmetries and phase constraints beyond the cyclic zero pattern. We begin by characterizing which dihedral symmetries are distinguishable from data in D-MRA and how they imply one another along the subgroup lattice (Section~\ref{sec:DSC_DMRA}). We then describe the Fourier-domain signature of dihedral subgroup $\Dmj{m}{j}$: the same cyclic zeros together with reflection-consistent phases on the multiples of $m$. We prove that on the sparsest spectra no moments of order below $d/m$separate dihedral from cyclic symmetry, whereas moments of order $d/m$ do so on every $\Cd{m}$-invariant signal, and on generic signals third-order phase invariants already suffice, without knowledge of the group distribution (Section~\ref{sec:Limitation_DMRA}). We then construct distribution-free fourth-order phase statistics, which treat the collapsed and the parity-split regimes uniformly, use them for practical, DAG-based testing over dihedral and cyclic classes, and state precisely the family-wise error guarantee that the resulting asymptotic tests support (Section~\ref{sec:Detection_DMRA}).

\subsection{Detectable symmetry classes and implication relationships}\label{sec:DSC_DMRA}

Recall the D-MRA model from Section~\ref{sec:MRAandOrbitRecovery}. The group $G=\Dm{d}$, known as the \textit{dihedral group} of order \(2d\), is the group of symmetries of a regular $d$-gon in $\R^2$, including both rotations and reflections. It is a finite group generated by a rotation $r$ of order $d$ and a reflection $s$, and is presented as $\Dm{d} \equiv \langle r, s \mid r^d = s^2 = 1,\, srs = r^{-1} \rangle = \{1,r,\ldots, r^{d-1}, s, rs,\ldots, r^{d-1}s\}$. As candidate symmetry groups $H$ of the signal, we focus on the two families of subgroups of $G$ that together exhaust all possible cases relevant to our setting:
\begin{itemize}
    \item[(a)] \textbf{Cyclic subgroups:} $\Cd{m} = \langle r^{\frac{d}{m}} \rangle$, for any divisor $m$ of $d$;
    \item[(b)] \textbf{Dihedral subgroups:} $\Dmj{m}{j}$, for any divisor $m$ of $d$ and $j=0,\ldots,d/m-1$, defined as
    \begin{equation*}
        \Dmj{m}{j} = \langle r^{\frac{d}{m}}, r^js \rangle = \{r^{\frac{d}{m}},r^{\frac{2d}{m}},\ldots,1, r^js, r^{j+\frac{d}{m}}s, r^{j+\frac{2d}{m}}s,\dots , r^{j+d-\frac{d}{m}}s \}.
    \end{equation*}
\end{itemize}
Notably, $\Cd{m}$ is a proper subgroup of $\Dmj{m}{j}$ for any $j$, i.e., $\Cd{m}<\Dmj{m}{j}$. The trivial subgroup $\Cd{1}$, which corresponds to the case of no symmetry, is included in the cyclic case. The full group $\Dm{d}$, which corresponds to the constant signal, is included in the dihedral case by noting that $\Dm{d}\equiv \Dm{d,0}$. 

Having described all possible subgroups of $\Dm{d}$, we now turn to the question of distinct signal symmetries in the D-MRA setting (see also Remark~\ref{rmk:sgp_symm}). In contrast to the cyclic case, several special cases arise in the dihedral setting that require careful attention. In C-MRA, each distinct subgroup $\Cd{m}$ corresponds to a unique set of signals exhibiting that symmetry (see Definition~\ref{def:signal_symmetry}). For the dihedral case, this remains true in general: the set of signals invariant under $\Cd{m}$ is different from that for $\Dmj{m}{j}$. However, this distinction may break down in certain special cases, where a weaker symmetry already enforces a stronger one. We summarize them in the following proposition.

\begin{proposition}[Signal symmetry degeneracy]\label{prop:signal_symmetry_equivalence} 
Let $\x \in \R^d$ be any real signal. The cyclic subgroup $\Cd{d}$ and the dihedral subgroup $\Dm{d} \equiv \Dmj{d}{0}$ are not identical as abstract subgroups. However, if $\x$ is invariant under the action of either subgroup, then $\x$ must be constant. Consequently, according to Definition~\ref{def:signal_symmetry}, the symmetry group of $\x$ is $\Dm{d}$, since it is the maximal subgroup preserving the signal. If $d$ is even, two additional cases arise:
\begin{itemize}
    \item Invariance under $\Dmj{d/2}{1}$ also forces $\x$ to be constant. Hence, in this case the symmetry group of $\x$ is also $\Dm{d}$.
    \item Invariance under either $\Cd{d/2}$ or $\Dmj{d/2}{0}$ is equivalent to $\x$ having period dividing $2$. If $\x$ is nonconstant, its exact symmetry group is $\Dmj{d/2}{0}$, which we also denote by $\Dm{d/2}^0$; if $\x$ is constant, its exact symmetry group is $\Dm{d}$.
\end{itemize}
\end{proposition}

Thus, the discussion above clarifies the circumstances under which invariance under $\Cd{m}$ and $\Dmj{m}{j}$ leads to the same signal symmetry. In particular, it resolves the case $m=d/2$. Therefore, in the sequel, we restrict attention to the cases $m\leq d/3$. 

From our earlier analysis of the C-MRA setting, we already know that the signal symmetries induced by cyclic subgroups are mutually distinct in terms of the signal sets they preserve. Turning to the dihedral case, note that $\Cd{m}<\Dmj{m}{j}$, which implies that dihedral symmetries with different values of $m$ are also distinct from one another. For dihedral subgroups sharing the same $m$, whether they induce distinct (or ultimately equivalent under D-MRA) signal symmetries is less immediate and will be addressed below.

To study symmetry detection from observations under model~\eqref{eq:MRA_model}, we introduced the notion of \textit{(detectable signal) symmetry classes} (see Definition~\ref{def:detectable_symmetry}). To determine these classes, we need to examine equivalence relations among all possible signal symmetries. As a result, such equivalences can only occur among dihedral subgroups $\Dmj{m}{j}$ with the same $m$. The precise relationships within this family are summarized in the following lemma, whose proof is deferred to Appendix~\ref{sec:proof_equivalence_dihedral_uniform}.

\begin{lemma}[Relationship among dihedral symmetries]\label{lem:equivalence_dihedral_uniform}
Let $m\leq d/3$ be a divisor of $d$, and fix distinct $j,j'\in\{0,1,\ldots,d/m-1\}$. Suppose that $\x\in\R^d$ has exact symmetry group $\Sym(\x)=\Dmj{m}{j}$. There exists $k\in\{0,1,\ldots,d/m-1\}$ such that $\Sym(r^k\x)=\Dmj{m}{j'}$ if and only if $\gcd(2,d/m)$ divides $j'-j$. In particular, every $j'$ is reachable when $d/m$ is odd, whereas only indices of the same parity as $j$ are reachable when $d/m$ is even.
\end{lemma}

Lemma~\ref{lem:equivalence_dihedral_uniform} shows that, for a fixed $m$, the sets of signals induced by different subgroups $\Dmj{m}{j}$ are essentially related by rotations. In other words, there is a one-to-one correspondence between these signal sets up to a cyclic shift. In the context of the D-MRA model, if the underlying group distribution $\rho$ is unknown, not of inferential interest, or assumed to be uniform, the signal can only be identified up to its orbit under the group action. As a consequence, the signal symmetry can only be identified up to the group action: if two subgroups induce signal sets that are related by a single group action, their corresponding symmetries cannot be distinguished from the observations. The following corollary summarizes the implications of Lemma~\ref{lem:equivalence_dihedral_uniform}.

\begin{corollary}[Dihedral equivalence]\label{cor:dihedral_equivalence}
Let $m\leq d/3$ be any divisor of $d$. Under the D-MRA model, we have
\begin{itemize}
    \item If $2m \nmid d$, or equivalently $d/m$ is odd, then all subgroups $ \Dmj{m}{j} $ for $ j \in \{0,1,\dots,d/m-1\} $ are indistinguishable from the perspective of symmetry detection. We denote this symmetry class by $\Dm{m}$-symmetry.
    \item If $2m \mid d$, or equivalently $d/m$ is even, then the subgroups $ \Dmj{m}{j} $ with even $ j $ are indistinguishable from each other, and those with odd $ j $ are likewise indistinguishable. However, the two classes (even vs. odd $ j $) are not equivalent. We denote these two symmetry classes by $\Dm{m}^0$-symmetry and $\Dm{m}^1$-symmetry, respectively.
\end{itemize}
\end{corollary}

\begin{remark}[Regime classification for dihedral symmetry in $(d,m)$]
In view of Corollary~\ref{cor:dihedral_equivalence}, we restrict to $m \le d/3$ and consider some signal $\x$ that is invariant under $\Cd{m}$. There are two regimes:
\begin{itemize}
\item \textbf{Collapsed regime ($d/m$ is odd):} All dihedral subgroups $\Dmj{m}{j}$ are indistinguishable. In this case, the task is to determine whether $\x$ exhibits dihedral symmetry $\Dm{m}$ (beyond cyclic $\Cd{m}$-symmetry).

\item \textbf{Parity-split regime ($d/m$ is even):} The dihedral subgroups $\Dmj{m}{j}$ split into two symmetry classes according to the parity of $j$. In this case, the task is to determine whether $\x$ exhibits dihedral symmetry, and if so, to distinguish between the two classes $\Dm{m}^0$ and $\Dm{m}^1$.
\end{itemize}
\end{remark}

Combining Proposition~\ref{prop:signal_symmetry_equivalence} with 
Corollary~\ref{cor:dihedral_equivalence}, and recalling 
Definition~\ref{def:detectable_symmetry}, the \textit{(detectable signal) symmetry classes} under the D-MRA model 
can be characterized by the parity of $d$ as follows:

\medskip
\noindent
\textbf{Case 1: $d$ is odd.}
\begin{align*}
\bigcup_{m \mid d } \{ \Dm{m} \} \;\cup\; \bigcup_{\substack{m \mid d \\ m < d}} \{ \Cd{m}\}.
\end{align*}
\emph{Example:} If $d = 9$, then $m \in \{1,3\}$ are the divisors less than $9$, so the set of \textit{symmetry classes} is $\{\Dm{9}, \Cd{3}, \Dm{3}, \Cd{1}, \Dm{1}\}$. In particular, example patterns of signals are
\begin{align*}
&\Dm{9}: (x_0, x_0, x_0, x_0, x_0, x_0, x_0, x_0, x_0), \\
&\Cd{3}: (x_0, x_1, x_2, x_0, x_1, x_2, x_0, x_1, x_2), \\
&\Dm{3}: (x_0, x_1, x_1, x_0, x_1, x_1, x_0, x_1, x_1) \\
&\Cd{1}: (x_0, x_1, x_2, x_3, x_4, x_5, x_6, x_7, x_8),\\
&\Dm{1}: (x_0, x_1, x_2, x_3, x_4, x_4, x_3, x_2, x_1), \\
\end{align*}
where all $x_i\in\R$ are distinct. We summarize this case in Table~\ref{tab:detectable-odd}.

\begin{table}[tbh]
\centering
\begin{tabular}{llc}
\toprule
$m$ condition & Detectable symmetry classes & Notes \\
\midrule
$m=d$ & $\Dm{d}$ & Constant signal \\
$1<m<d$, \, $m\mid d$ & $\Dm{m}$ and $\Cd{m}$ & $\Dm{m}$ implies $\Cd{m}$ \\ 
$m=1$ & $\Dm{1}$ and $\Cd{1}$ & $\Cd{1}$ is ``No symmetry'' class \\
\bottomrule
\end{tabular}
\caption{Detectable symmetry classes in Dihedral MRA when $d$ is \textbf{odd}.}
\label{tab:detectable-odd}
\end{table}

\noindent
\textbf{Case 2: $d$ is even.}
\begin{align*}
\{ \Dm{d} \} \;\cup\; \{ \Dm{d/2}^0 \}  
\;\cup\; 
\bigcup_{\substack{m \mid d \\ m < d/2}}
\begin{cases}
\{ \Cd{m}, \Dm{m} \}, & 2m \nmid d, \\[2mm]
\{ \Cd{m}, \Dm{m}^0, \Dm{m}^1 \}, & 2m \mid d .
\end{cases}
\end{align*}
\emph{Example:} If $d = 12$, then $m \in \{1,2,3,4,6\}$ are the divisors less than $12$, so the set of \textit{symmetry classes} is $\{\Dm{12}, \Dm{6}^0, \Cd{4}, \Dm{4}, \Cd{3}, \Dm{3}^0, \Dm{3}^1, \Cd{2}, \Dm{2}^0, \Dm{2}^1, \Cd{1}, \Dm{1}^0, \Dm{1}^1\}$. In particular, example patterns of signals are
\begin{align*}
&\Dm{12}: (x_0, x_0, x_0, x_0, x_0, x_0, x_0, x_0, x_0, x_0, x_0, x_0), \\
&\Dm{6}^0: (x_0, x_1, x_0, x_1, x_0, x_1, x_0, x_1, x_0, x_1, x_0, x_1), \\
&\Cd{4}: (x_0, x_1, x_2, x_0, x_1, x_2, x_0, x_1, x_2, x_0, x_1, x_2), \\
&\Dm{4}: (x_0, x_1, x_1, x_0, x_1, x_1, x_0, x_1, x_1, x_0, x_1, x_1),\\
&\Cd{3}: (x_0, x_1, x_2, x_3, x_0, x_1, x_2, x_3, x_0, x_1, x_2, x_3), \\
&\Dm{3}^0: (x_0, x_1, x_2, x_1, x_0, x_1, x_2, x_1, x_0, x_1, x_2, x_1), \\
&\Dm{3}^1: (x_0, x_0, x_1, x_1, x_0, x_0, x_1, x_1, x_0, x_0, x_1, x_1), \\
&\Cd{2}: (x_0, x_1, x_2, x_3, x_4, x_5, x_0, x_1, x_2, x_3, x_4, x_5), \\
&\Dm{2}^0: (x_0, x_1, x_2, x_3, x_2, x_1, x_0, x_1, x_2, x_3, x_2, x_1), \\
&\Dm{2}^1: (x_0, x_0, x_1, x_2, x_2, x_1, x_0, x_0, x_1, x_2, x_2, x_1), \\
&\Cd{1}: (x_0, x_1, x_2, x_3, x_4, x_5, x_6, x_7, x_8, x_9, x_{10}, x_{11}), \\
&\Dm{1}^0: (x_0, x_1, x_2, x_3, x_4, x_5, x_6, x_5, x_4, x_3, x_2, x_1), \\
&\Dm{1}^1: (x_0, x_0, x_1, x_2, x_3, x_4, x_5, x_5, x_4, x_3, x_2, x_1), 
\end{align*}
where all $x_i\in\R$ are distinct. We summarize this case in Table~\ref{tab:detectable-even}. 

\begin{table}[tbh]
\centering
\begin{tabular}{lll}
\toprule
$m$ condition & Detectable symmetry classes & Notes \\
\midrule
$m=d$ & $\Dm{d}$ & Constant signal \\
$m=d/2$ & $\Dm{d/2}^0$ & Period-$2$ class \\
& & (coincides with $\Cd{d/2}$ at signal level) \\
$1<m<d/2$, $m\mid d$, $2m\nmid d$ & $\Dm{m}$ and $\Cd{m}$ & No $j$-split in $\Dm{m}$ \\
$1<m<d/2$, $m\mid d$, $2m\mid d$ & $\Dm{m}^0,\ \Dm{m}^1$ and $\Cd{m}$ & $\Dm{m}$ splits into two classes \\
& & (where $j$ even/odd) \\
$m=1$ & $\Dm{1}^0,\ \Dm{1}^1$ and $\Cd{1}$ & $\Cd{1}$ is ``No symmetry'' class  \\
\bottomrule
\end{tabular}
\caption{Detectable symmetry classes in Dihedral MRA when $d$ is \textbf{even}.}
\label{tab:detectable-even}
\end{table}

Among the detectable symmetry classes above, the \textit{implication relationships} can further be summarized as follows (whenever the detectable symmetry class is well-defined):
\begin{itemize}
    \item Cyclic symmetry classes: $\Cd{m_2}$-symmetry $\Rightarrow$ $\Cd{m_1}$-symmetry, when $m_1\mid m_2$;
    \item Dihedral symmetry classes: 
    \begin{itemize}
    \item $\Dm{m_2}$-symmetry implies $\Dm{m_1}$-symmetry, 
    whenever $m_1 \mid m_2$.

    \item $\Dm{m_2}^j$-symmetry implies $\Dm{m_1}^j$-symmetry, 
    whenever $m_1 \mid m_2$ and $j \in \{0,1\}$.

    \item $\Dm{m_2}$-symmetry implies $\Dm{m_1}^j$-symmetry, 
    whenever $m_1 \mid m_2$ and $j \in \{0,1\}$.
    \end{itemize}
    
    \item Cyclic vs. dihedral symmetry classes: $\Dm{m_2}$-symmetry or $\Dm{m_2}^j$-symmetry $\Rightarrow$ $\Cd{m_1}$-symmetry, when $m_1\mid m_2$ and $j=0,1$.
\end{itemize}
For example, when $d$ is divisible by $4$, we have subgroups $\Dm{d},\Dmj{d/2}{1},\Cd{d}$ belong to the detectable symmetry class $\Dm{d}$-symmetry. In this case, $\Dm{d}$-symmetry further implies $\Cd{d/4}$-symmetry, $\Dm{d/4}^0$-symmetry, and $\Dm{d/4}^1$-symmetry.

\subsection{Fourier patterns and the limitation of low-order moments under Dihedral MRA}\label{sec:Limitation_DMRA}

Passing from cyclic rotations to the dihedral group introduces equivalence classes of detectable symmetries and phase constraints beyond the cyclic zero pattern. This subsection formalizes the Fourier-domain signature of dihedral invariance and identifies the moment order at which the reflection-consistent phases become visible: on the sparsest $\Cd{m}$-invariant spectra, no moments of order below $d/m$ separate dihedral from cyclic symmetry, so in particular the first three moments are in general insufficient, whereas moments of order $d/m$ separate the classes on every $\Cd{m}$-invariant signal; on generic signals, moreover, third-order moments already separate the classes, and the separating quantities do not depend on the distribution of the group elements. These observations motivate the phase-revealing statistics developed in the subsection that follows this one.

We begin by presenting the Fourier-domain characterization of signals invariant under dihedral subgroups, formulated in Proposition \ref{prop:structure_in_Fourier_DMRA} below. Note that a signal $\x$ is real if and only if it satisfies the conjugate symmetry relation
 \begin{align*}
\F\x[-\ell]=\overline{\F\x[\ell]},\quad\text{for }\ell\in\cI_d.
\end{align*}
Hence, we always have $\F\x[0]\in\R$ and
\begin{align*}
\F\x[d/2]\in\R, \text{ if }d\text{ is even}.
\end{align*}
It is also convenient to consider only one representative from each conjugate pair $\{\ell,d-\ell\}$. To this end, we define the reduced index sets 
\begin{align}\label{eq:cI_plus}
\cI_d^+\triangleq \{0,1,\ldots,\lfloor d/2\rfloor\},\quad \cI_{m,d}^+\triangleq \cI_{m,d}\cap \cI_d^+,\quad \text{and }\cI_{m,d}^{c,+}\triangleq\cI_{m,d}^c \cap \cI_d^+.
\end{align}
For example, consider $d=12$ and $m=3$. Then the set of indices of free coefficients is $\cI_{12}^+=\{0,1,2,3,4,5,6\}$. And we also have
\begin{align*}
\cI_{3,12}^+=\{0,3,6\},\quad \cI_{3,12}^{c,+}=\{1,2,4,5\}.
\end{align*}
Consider $d=9$ and $m=3$. Then the set of indices of free coefficients is $\cI_{9}^+=\{0,1,2,3,4\}$. And we also have 
\begin{align*}
\cI_{3,9}^+=\{0,3\},\quad\cI_{3,9}^{c,+}=\{1,2,4\}.
\end{align*}
The proof of Proposition~\ref{prop:structure_in_Fourier_DMRA} is given in Section~\ref{sec:proof_structure_in_Fourier_DMRA}.

\begin{proposition}\label{prop:structure_in_Fourier_DMRA}
Let $m$ be any divisor of $d$ and $\x\in\R^d$ be a real signal. Then, the following statements hold.
\begin{enumerate}
    \item[(i)] If $\x$ is invariant under the full dihedral group $\Dm{d}$, then 
    \begin{align*}
    &\F\x[\ell]=0\quad\text{for any }\ell\neq 0,\\
    &\F\x[0]\in\R.
    \end{align*}

    \item[(ii)] If $d$ is even and $\x$ is invariant under $\Dmj{d/2}{1}$, then by Proposition \ref{prop:signal_symmetry_equivalence}, it is equivalent to $\Dm{d}$. Hence, the same conclusion as in the full dihedral group $\Dm{d}$ holds.

    \item[(iii)] If $d$ is even and $\x$ is invariant under $\Dmj{d/2}{0}$, then by Proposition \ref{prop:signal_symmetry_equivalence}, it is equivalent to $\Cd{d/2}$. Hence, 
    \begin{align*}
    &\F\x[\ell]=0\quad\text{for any }\ell \not\in \{0,d/2\},\\
    &\F\x[0],\F\x[d/2]\in\R.  
    \end{align*}
    
    \item[(iv)] Exclude the special cases above. Assume $m\le d/3$ and let $\Dmj{m}{j}$ be the dihedral subgroup with $j\in\{0,1,\ldots,d/m-1\}$. Write $\F\x[\ell]=r_\ell e^{\ri\theta_\ell}$ with $r_\ell\ge 0$ and $\theta_\ell\in(-\pi,\pi]$. If $\x$ is invariant under \(\Dmj{m}{j}\), then its Fourier transform exhibits the following structure:
    \begin{itemize}
        \item $\F\x[0]\in\R$ and $\F\x[\ell]=0$ for all $\ell\in \cI_{m,d}^c$.
        
        \item For parity-split regime (i.e., $d/m$ is even), we have $d/2\in\cI_{m,d}$ and
        \begin{align*}
        \F \x[d/2] \in \mathbb{R} \quad \text{if $j$ is even}, 
        \qquad\F \x[d/2] = 0 \quad \text{if $j$ is odd}.
        \end{align*}

        \item For any index $\ell \in \cI_{m,d}^+\setminus\{0,d/2\}$, the Fourier phase satisfies
        \begin{align*}
        \theta_\ell \equiv -\frac{\ell j \pi}{d} \pmod{\pi},
        \end{align*}
        whenever $r_\ell > 0$ (the phase is undefined when $r_\ell = 0$, in which case the condition is vacuous). Equivalently, for each $\ell \in \cI_{m,d}^+$, there exists a real coefficient $c_\ell \in \mathbb{R}$ such that
        \begin{align*}
        \F \x[\ell] = c_\ell \cdot e^{-\ri \ell j \pi / d}.
        \end{align*}

        \item For the remaining indices, namely those $\ell\in\cI_{m,d}\setminus \cI_{m,d}^+$, the phases are determined via conjugate symmetry, $\theta_\ell = -\theta_{d-\ell}$,
        and hence also satisfy the above phase relation. In particular, for any $\ell\in\cI_{m,d}\setminus \cI_{m,d}^+$,
        \begin{align*}
        \F\x[\ell]= (-1)^{j}c_{d-\ell}\cdot e^{-\ri \ell j\pi/d},
        \end{align*}
        where $c_{d-\ell}$ is the real coefficient associated with the index $d-\ell \in \cI_{m,d}^+$ defined in the previous item.
    \end{itemize}
\end{enumerate}
Furthermore, the above conditions are also sufficient. That is, if a real signal satisfies these conditions, its symmetry group must be a supergroup of the symmetry group corresponding to the given conditions.
\end{proposition} 

Next, we aim to illustrate how the above characterization manifests in the moments. Recall the moment expressions from Proposition~\ref{prop:moment_expression}. The explicit forms of the first three (noiseless) population moments are:
\[
\begin{aligned}
\F M_1(\x,\rho)&= \F \x  \odot \F \p +\overline{\F\x} \odot \F\q, \\
\F M_2(\x,\rho)&= \Db_{\F\x} \Cb_{\F \p} \Db_{\overline{\F\x}} +
\Db_{\overline{\F\x}} \Cb_{\F \q} \Db_{\F\x}, \\
\F M_3(\x,\rho)&= \F\mathbf{T}^{(3)}_{\p} \times_1 \Db_{\F\x} \times_{2} \Db_{\F\x} \times_3 \Db_{\overline{\F \x}}
    + \F\mathbf{T}^{(3)}_{\q} \times_1 \Db_{\overline{\F \x}} \times_{2} \Db_{\overline{\F \x}} \times_3 \Db_{\F \x}.
\end{aligned}
\]
An instructive special case is the uniform distribution, for which the Fourier moments are purely real. This property follows directly from~\eqref{eq:population_moment_tensor_Fourier_DMRA}, since the $r$-th moment under the uniform distribution takes the form
\begin{align} \label{eqn:r_moment_uniform_dist}
    \F M_r(\x, \rho)[i_1,\ldots, i_{r-1},i_r] =& \frac{1}{2}\F\x[i_1] \cdots \F\x[i_{r-1}] \, \overline{\F\x[i_1+\cdots + i_{r-1}]} \notag\\
    &+ \frac{1}{2}\overline{\F\x[i_1]} \cdots \overline{\F\x[i_{r-1}]} \, \F\x[i_1+\cdots + i_{r-1}] \notag\\
    =& \Re\big(\F\x[i_1] \cdots \F\x[i_{r-1}] \, \overline{\F\x[i_1+\cdots + i_{r-1}]} \big),
\end{align}
when $i_r = i_1+\cdots+i_{r-1}$, and equals $0$ otherwise. We now provide an example of two distinct real signals with cyclic and dihedral symmetry that share identical first three moments under the uniform distribution. A general construction follows this example.

Consider $d=12$ and $m=3$. We set all Fourier coefficients of our signals to zero, except at $k=3$ and $k=9$ (a conjugate pair), each of unit magnitude. Our first signal, denoted by $\x^{(C)}$, has cyclic symmetry $\Cd{3}$. We define
\begin{equation*}
\F \x^{(C)}[3] = e^{\ri \pi/6}, \qquad
\F \x^{(C)}[9] = e^{-\ri \pi/6}, \qquad
\F \x^{(C)}[k]=0, \quad k\notin{3,9}.
\end{equation*}
Our second signal, denoted by $\x^{(D)}$, has dihedral symmetry $\Dmj{3}{1}$. We define
\begin{equation*}
\F \x^{(D)}[3] = e^{-\ri \pi/4}, \qquad
\F \x^{(D)}[9] = e^{\ri \pi/4}, \qquad
\F \x^{(D)}[k]=0, \quad k\notin{3,9}.
\end{equation*}

Recall Proposition~\ref{prop:structure_in_Fourier_DMRA}(4): for dihedral symmetry $\Dmj{m}{j}$, the nonzero $\F \x[\ell]$, for $\ell \in \cI_{m,d}$ with $\ell \neq d/2$, lies on the line of angle $-\ell j\pi/d$ ($\mathrm{mod}~\pi$).
Here $d=12$, $m=3$, and $j=1$, so the required angles are $-\pi/4$ at $\ell=3$ and $+\pi/4$ at $\ell=9$, consistent with the definition above. In contrast, $\x^{(C)}$ shares the same $\Cd{3}$ zero pattern but is \emph{not} $\Dmj{3}{j}$ for any $j$. To see this, for $\ell=3$, the dihedral condition would require
$\arg(\F \x[3]) \equiv -(3j/12)\pi ~(\mathrm{mod}~\pi)$, i.e., $\tfrac{1}{6} \equiv -\tfrac{j}{4} ~(\mathrm{mod}~12)$,
which rearranges to $3j \equiv 10 ~(\mathrm{mod}~12)$.
Therefore, there is no $j \in {0,1,2,3}$ satisfying this condition. Finally, one can verify directly via the inverse Fourier transform that both signals are real-valued.

While the above signals $\x^{(C)}$ and $\x^{(D)}$ are both real and different, the first three moments of $\x^{(C)}$ and $\x^{(D)}$, under the uniform distribution, coincide. In particular,
\begin{itemize}
  \item $\F M_1$: both have $\F \x[0]=0$ and under the uniform distribution $\F\p[0]$ and $\F\q[0]$ are the only non-zero entries, so $\F M_1$ are identically zero.
  \item $\F M_2$: since all nonzero entries are unit norm, $|\F \x[3]|=|\F \x[9]|=1$, and the diagonal of $\F M_2$  has ones at $k=3,9$ and zero otherwise.
  \item $\F M_3$: the only nonzero triplets (setting the first two indices by nonzeros entries) are $(3,3,6)$, $(9,9,6)$, $(3,9,0)$ (up to permutation). However, since $\F \x[6]=\F \x[0]=0$ for both signals, the relevant triple products vanish. Therefore, the entries are $\Re(\cdot)$ of those same triple products also vanish. Namely, both $\F M_3$ are zero and match.
\end{itemize}

With similar arguments, we can construct a general counterexample.  
Let $m$ be a divisor of $d$ such that $m\leq d/4$, thereby excluding the cases $m=d/2$ and $m=d/3$. Define two spectra supported only at the conjugate pair $\{m, d{-}m\}$:
\begin{equation} \label{eqn:counterexample}
  \F \x^{(D)}[m] = e^{\ri\phi_D}, \quad \F \x^{(D)}[d{-}m] = e^{-\ri\phi_D}, 
  \qquad
  \F \x^{(C)}[m] = e^{\ri\phi_C}, \quad \F \x^{(C)}[d{-}m] = e^{-\ri\phi_C},
\end{equation}
with all other $\F \x[\cdot] = 0$.  
Choose the phases as
\begin{equation*}
  \phi_D \in \Big\{-\frac{m j \pi}{d},\, \pi - \frac{m j \pi}{d}\Big\}
  \quad \text{for some } j \in \{0, \dots, d/m - 1\},
  \qquad \text{and} \qquad
  \phi_C = \frac{\pi}{2d}.
\end{equation*}
Thus, $\phi_C \not\equiv \phi_D \ (\mathrm{mod}\ \pi)$.  
Then, $\x^{(D)}$ is invariant under the dihedral subgroup $\Dmj{m}{j}$, whereas $\x^{(C)}$ has cyclic symmetry $\Cd{m}$ but is \textit{not} dihedral.  
Both time-domain signals are real, and under the uniform distribution of transformations, their $\F M_1$ and $\F M_2$ coincide.  
Furthermore, $\F M_3 = 0$ for both, since any nonzero triple would require an index of $0$ or $2m \notin \{m, d{-}m\}$, which is ensured by the condition $3m \not\equiv 0 \ (\mathrm{mod}\ d)$. The same computation applies at every order below $d/m$: a nonzero entry of order $r$ requires signed multiples of $m$ whose signed sum vanishes modulo $d$, which for $r<d/m$ forces the sum to vanish exactly, so that the entry does not depend on the phase at all. Hence, we conclude with the following theorem, and the proof is given in Section \ref{sec:proof_third_moment_failure}.

\begin{theorem}[Low-order non-separation on sparse spectra] \label{thm:third_moment_failure}
Let $m$ be any divisor of $d$ such that $m\leq d/4$ and let $\rho_0$ denote the uniform distribution on $\Dm{d}$. Then, for every
$j\in\{0,\ldots,d/m-1\}$, there exist real signals
$\x^{(D)},\x^{(C)}\in\mathbb{R}^d$, with $\x^{(C)}$ of exact symmetry $\Cd{m}$ (in particular, $\x^{(C)}$ is not $\Dmj{m}{j'}$-symmetric for any $j'$) and $\x^{(D)}$ $\Dmj{m}{j}$-symmetric, whose population moments of order below $d/m$ under the uniform distribution $\rho_0$ coincide:
    \[
    M_r(\x^{(C)},\rho_0) = M_r(\x^{(D)},\rho_0), \qquad \text{for }r = 1,2,\ldots,d/m-1.
    \]
In particular, the first three population moments do not suffice to distinguish between detectable symmetry classes \(\Dm{m}\) (or $\Dm{m}^0$ and $\Dm{m}^1$) and \(\Cd{m}\), and neither do the moments of any order below $d/m$: since the moments under $\rho_0$ are the Haar-invariant moments of Section~\ref{sec:general_invariant_moment_principle}, Theorem~\ref{thm:general_invariant_moment_criterion} shows that no procedure based only on invariant moments of order at most $d/m-1$ can recover the symmetry class on all $\Cd{m}$-invariant signals.
\end{theorem}

\begin{remark}
The assumptions exclude the cases $2m = d$ and $3m = d$, which require separate consideration.  
When $2m = d$, the dihedral symmetry reduces to a cyclic symmetry; see Proposition~\ref{prop:signal_symmetry_equivalence}.  
In contrast, when $3m = d$, the first three moments already contain enough information to distinguish the cyclic and dihedral cases, consistent with the order $d/m=3$ at which separation first becomes possible. Indeed, consider the third-order moment $\F M_3[m,m,2m]$ under the uniform distribution. Since $2m \equiv -m \ (\mathrm{mod}\ d)$, we obtain
\[
\F M_3[m,m,2m]
= \Re\big( \F\x[m] \, \F\x[m] \, \overline{\F\x[2m]} \big)
= \Re\big( (\F\x[m])^3 \big),
\]
which depends on the phase of $\F\x[m]$, thereby allowing one to distinguish between the two symmetries; by \eqref{eq:tildeM3} below, the same quantity is available for an arbitrary distribution $\rho$ after conjugate symmetrization.
\end{remark}

Theorem~\ref{thm:third_moment_failure} is a lower bound: it exhibits signals on which no order below $d/m$ works. The next proposition proves the matching upper bound: moments of order at most $d/m$ determine the dihedral symmetry label on every $\Cd{m}$-invariant signal. Its proof (Section~\ref{sec:proof_order_dm_sufficient}) rests on an elementary zero-sum argument: every phase relation that a dihedral signal must satisfy is generated by relations involving at most $d/m$ Fourier coefficients.

\begin{proposition}[Order $d/m$ separates the classes on all signals]\label{prop:order_dm_sufficient}
Let $m\le d/3$ be a divisor of $d$, let $\rho_0$ be the uniform distribution on $\Dm{d}$, and let $\x,\x'\in\R^d$ be $\Cd{m}$-invariant signals with
\[
M_r(\x,\rho_0)=M_r(\x',\rho_0),\qquad r=1,\ldots,d/m .
\]
Then $\x$ is invariant under some dihedral subgroup $\Dmj{m}{j}$ if and only if $\x'$ is. If, in addition, $d/m$ is even and the rotational stabilizer of $\x$ is exactly $\Cd{m}$, then the rotational stabilizer of $\x'$ is also $\Cd{m}$, and $\x\in\Dm{m}^{\epsilon}$ if and only if $\x'\in\Dm{m}^{\epsilon}$, for $\epsilon\in\{0,1\}$. Consequently, invariant moments of order at most $d/m$ determine the dihedral label on all $\Cd{m}$-invariant signals (and, in the parity-split regime, the parity class on all signals with exact rotational symmetry $\Cd{m}$), whereas by Theorem~\ref{thm:third_moment_failure} (and the remark above when $3m=d$) no order below $d/m$ does.
\end{proposition}

This proposition concerns population identification of the symmetry label, not recovery of the signal or its orbit. It requires no nonvanishing assumption, but does not imply uniform statistical consistency without a separation condition, such as that in Proposition~\ref{prop:uniform_upper}. Under the nonvanishing conditions below, a lower moment order suffices.

Fix a divisor $m$ of $d$ in the collapsed regime ($d/m$ odd, $d/m\ge 3$, so $m\le d/3$). For a $\Cd{m}$-invariant real signal $\x$, write its nonzero Fourier coefficients as $\F\x[km]=r_{km}e^{\ri\theta_{km}}$. The discriminating information is carried by third-order phases, and it can be extracted without knowledge of $\rho$. Indeed, by Proposition~\ref{prop:moment_expression}(b), for $\ell_1,\ell_2,\ell_3\in\cI_d$ and an arbitrary distribution $\rho=(\p^\T,\q^\T)^\T$ on $\Dm{d}$,
\[
\F M_3(\x,\rho)[\ell_1,\ell_2,\ell_3]
=\F\p[\ell_1+\ell_2-\ell_3]\,\F\x[\ell_1]\F\x[\ell_2]\overline{\F\x[\ell_3]}
+\F\q[\ell_1+\ell_2-\ell_3]\,\overline{\F\x[\ell_1]\F\x[\ell_2]}\,\F\x[\ell_3].
\]
Take $\ell_3\equiv\ell_1+\ell_2 \pmod d$ and set $z\triangleq\F\x[\ell_1]\F\x[\ell_2]\overline{\F\x[\ell_1+\ell_2]}$. Then $\F M_3(\x,\rho)[\ell_1,\ell_2,\ell_1+\ell_2]=\F\p[0]\,z+\F\q[0]\,\bar z$, and, by the conjugate symmetry of $\F\x$, $\F M_3(\x,\rho)[-\ell_1,-\ell_2,-\ell_1-\ell_2]=\F\p[0]\,\bar z+\F\q[0]\,z$. Since $\F\p[0]+\F\q[0]=\sum_{g\in\Dm{d}}\rho(g)=1$, the conjugate-symmetrized entry
\begin{align}
\widetilde{M}_3\big(\x,(\ell_1,\ell_2)\big)&\triangleq
\F M_3(\x,\rho)[\ell_1,\ell_2,\ell_1+\ell_2]+\F M_3(\x,\rho)[-\ell_1,-\ell_2,-\ell_1-\ell_2]\notag\\
&=2\Re(z)
=2\,r_{\ell_1}r_{\ell_2}r_{\ell_1+\ell_2}\cos\!\big(\theta_{\ell_1}+\theta_{\ell_2}-\theta_{\ell_1+\ell_2}\big)\label{eq:tildeM3}
\end{align}
does not depend on $\rho$; under the uniform distribution it equals twice the entry in \eqref{eqn:r_moment_uniform_dist}. In particular,
\[
\widetilde{M}_3\big(\x,(m,m)\big) = 2\Re\!\big((\F\x[m])^2\,\overline{\F\x[2m]}\big) = 2\,r_m^2\, r_{2m}\cos\!\big(2\theta_m-\theta_{2m}\big)
\]
attains its extreme value $\pm 2r_m^2 r_{2m}$ exactly when the phases are reflection-consistent, $2\theta_m-\theta_{2m}\equiv 0\ (\mathrm{mod}\ \pi)$. Thus, the low-order non-separation of Theorem~\ref{thm:third_moment_failure} occurs only where some connecting coefficient vanishes. Throughout the paper, genericity refers to explicit nonvanishing conditions on Fourier coefficients (Definition~\ref{def:generic_signals} below); here, $\x$ is a \emph{generic $\Cd{m}$ signal} if $r_{km}>0$ for all $km\in\cI_{m,d}^{+}$. This condition is sufficient for the chain argument below, but it is not necessary for third-order separation: spectra containing zeros remain identifiable at third order whenever the closed index triples supported on their nonzero coefficients still enforce a common reflection axis, and Proposition~\ref{prop:order_dm_sufficient} determines the dihedral symmetry label from moments of order at most $d/m$, without a nonvanishing assumption.

\begin{proposition}[Generic third-order separation, distribution-free]\label{prop:third_order_generic}
Let $m\le d/3$ be a divisor of $d$ with $d/m$ odd (collapsed regime), let $\rho$ be an arbitrary distribution on $\Dm{d}$, and let $\x$ be a generic $\Cd{m}$ signal, i.e., $\Cd{m}$-invariant with $r_{km}>0$ for all $km\in\cI_{m,d}^+$. Then $\x$ has $\Dm{m}$-symmetry if and only if
\[
\widetilde{M}_3\big(\x,(m,km)\big)=\pm\,2\,r_m\,r_{km}\,r_{(k+1)m} ,
\quad 1\le k\le \tfrac{d/m-1}{2}.
\]
In particular, invariant moments of order at most three separate $\Dm{m}$ from $\Cd{m}$ on generic $\Cd{m}$ signals, and the separating quantities $\widetilde{M}_3(\x,(m,km))$ are independent of $\rho$.
\end{proposition}
\begin{proof}
By \eqref{eq:tildeM3}, for any $\rho$ each third-order chain entry is
\[
\widetilde{M}_3\big(\x,(m,km)\big)=2\,r_m\,r_{km}\,r_{(k+1)m}\,\cos\!\big(\theta_m+\theta_{km}-\theta_{(k+1)m}\big),
\]
so, since $r_\ell=r_{d-\ell}>0$ for all $\ell\in\cI_{m,d}$ for a generic $\Cd{m}$ signal, the entry is extreme ($=\pm 2r_m r_{km} r_{(k+1)m}$) if and only if
\begin{equation}\label{eqn:chain_phase}
\theta_m+\theta_{km}-\theta_{(k+1)m}\equiv 0\ (\mathrm{mod}\ \pi).
\end{equation}

($\Rightarrow$) If $\x$ has $\Dm{m}$-symmetry it is invariant under some $\Dmj{m}{j}$, so by Proposition~\ref{prop:structure_in_Fourier_DMRA}(iv), $\theta_{km}\equiv -kmj\pi/d\ (\mathrm{mod}\ \pi)$ for every $km\in\cI_{m,d}$. Substituting gives $\theta_m+\theta_{km}-\theta_{(k+1)m}\equiv 0\ (\mathrm{mod}\ \pi)$, so \eqref{eqn:chain_phase} holds and every chain entry is extreme.

($\Leftarrow$) Suppose \eqref{eqn:chain_phase} holds for all $1\le k\le (d/m-1)/2$. Set $f(km)=e^{2\ri\theta_{km}}$; then \eqref{eqn:chain_phase} reads $f((k+1)m)=f(m)f(km)$, whence $f(km)=f(m)^k$ by induction. Conjugate symmetry $f(d-km)=f(km)^{-1}$ evaluated at $k=(d/m-1)/2$ forces $f(m)^{d/m}=1$ (as in the proof of Corollary~\ref{cor:DMRA_4th_population_reduced}). Hence $\theta_m\equiv -mj\pi/d\ (\mathrm{mod}\ \pi)$ for some $j\in\{0,\dots,d/m-1\}$, and $\theta_{km}\equiv k\theta_m\equiv -kmj\pi/d\ (\mathrm{mod}\ \pi)$; the conjugate frequencies follow by conjugate symmetry. By the sufficiency clause of Proposition~\ref{prop:structure_in_Fourier_DMRA}(iv), $\x$ is $\Dmj{m}{j}$-invariant, i.e., has $\Dm{m}$-symmetry.
\end{proof}

\begin{remark}[Harder than generic signal recovery]\label{rmk:harder}
Generic (non-symmetric) signals are determined, up to the group action, by their first three moments~\cite{edidin2025orbit}, and Proposition~\ref{prop:third_order_generic} shows that the same order suffices to decide dihedral versus cyclic symmetry for generic signals within the class of $\Cd{m}$-invariant ones. Symmetry detection is nevertheless harder in a uniform sense: the classes to be separated contain arbitrarily sparse spectra, and Theorem~\ref{thm:third_moment_failure} shows that on the sparsest ones no moment of order below $d/m$ carries the relevant phase information. By Proposition~\ref{prop:order_dm_sufficient}, the order needed to separate the classes on all signals is therefore exactly the period $d/m$, whereas generic recovery is settled at order three.
\end{remark}

\begin{remark}[Third order versus the fourth-order statistics used below]\label{rmk:third_vs_fourth}
Proposition~\ref{prop:third_order_generic} shows that the fourth order is \emph{not} necessary for distribution-free detection: the conjugate-symmetrized bispectrum entries $\widetilde M_3$ are independent of $\rho$ for exactly the same reason as the fourth-order quantity $\widetilde M_4$ of Section~\ref{sec:population_analysis_collapsed}. For indices with $\ell_1+\cdots+\ell_{r-1}\equiv\ell_r \pmod d$, both dihedral terms of the $r$-th Fourier moment carry the weights $\F\p[0]$ and $\F\q[0]$, which sum to one, so the conjugate-symmetrized entry equals $2\Re\big(\F\x[\ell_1]\cdots\F\x[\ell_{r-1}]\overline{\F\x[\ell_r]}\big)$ at every order $r\ge2$; see \eqref{eq:tildeM3} and \eqref{eq:tildeM4}. The fourth-order chain statistics on which our tests are built are in fact the third-order ones rescaled by the DC coefficient: for the tuples $(0,m,km)$ of \eqref{eq:chain_set},
\[
\widetilde M_4\big(\x,(0,m,km)\big)=\F\x[0]\cdot\widetilde M_3\big(\x,(m,km)\big),
\]
since $\F\x[0]\in\R$; this is why the genericity condition $\F\x[0]\neq0$ enters Theorem~\ref{thm:DMRA_4th_population_general}. We nevertheless adopt the fourth-order form because it yields a single framework for both dihedral regimes: in the parity-split regime the signed parity witness of Section~\ref{sec:population_analysis_parity_split} is of fourth order, and when $d/m=4$ no third-order parity witness exists (Remark~\ref{rmk:worst_case_d_m}; for $d/m=4$ the signals in $\Dm{m}^1$ have spectra supported on the DC term and a single conjugate pair). A third-order version of the collapsed-regime test is immediate: replace $\widehat D_t$ by the sample analog of $\widetilde M_3(\x,(m,km))$ and drop the factor $\widehat S_0$ in $\widehat H_t$. By the polynomial-feature variance bound in Appendix~\ref{sec:proof_uniform_upper}, one expects its sample complexity to grow like $\sigma^{6}$ rather than $\sigma^{8}$ for fixed, separated, nondegenerate signals, which would make it preferable at very low SNR. We have not established its calibration, its dependence on the separation margin, or its power in a joint high-noise/large-sample limit, and we leave a systematic comparison with the fourth-order test to future work. What no fixed low order provides is uniformity over all $\Cd{m}$-invariant signals: by Theorem~\ref{thm:third_moment_failure} and Proposition~\ref{prop:order_dm_sufficient}, the order required on the whole class is exactly $d/m$.
\end{remark}

The single-pair construction shows that no fixed moment order handles all sparse spectra: in the $d=12$ example above ($d/m=4$) the fourth-order entry $\F M_4[3,3,3,9]$ is nonzero and takes distinct values for the two signals, whereas for $d/m\ge5$ the first phase-sensitive entry appears only at order $d/m$ (Theorem~\ref{thm:third_moment_failure}); order $d/m$, on the other hand, always suffices (Proposition~\ref{prop:order_dm_sufficient}). On generic signals, in contrast, the reflection-consistent phases are visible at third order and can be extracted without knowledge of $\rho$ (Proposition~\ref{prop:third_order_generic}). In the analysis that follows we work with a fourth-order form of these distribution-free phase invariants, which treats the collapsed and the parity-split regimes within a single framework (Remark~\ref{rmk:third_vs_fourth}).
   
\subsection{Distribution-free symmetry detection for Dihedral MRA}\label{sec:Detection_DMRA}

\subsubsection{Generating hypothesis DAG}\label{sec:DAG_construction_dihedral}

As in the C-MRA setting, we construct a DAG with a proper topological ordering over all candidate symmetry hypotheses, that is, the set of (detectable) symmetry classes excluding the trivial symmetry $\Cd{1}$. In this DAG, each vertex represents a symmetry class. We add direct edges to respect the implication relationships among symmetry classes, so that if symmetry class $H$ implies symmetry class $H'$, then $H'$ is a descendant of $H$ in the DAG. Note that $H$ and $H'$ may not be connected by a single edge; the descendant relationship may be realized via a directed path, similar to the case in C-MRA.

The construction proceeds in three main steps: (i) Build the DAG for dihedral symmetry classes, (ii) Augment the DAG with cyclic symmetry classes, together with the corresponding directed edges, and (iii) Complete the DAG by adding directed edges between cyclic symmetry classes. We implement the details separately for the two cases when $d$ is odd and even.

\medskip
\noindent
\textbf{Case 1: $d$ is odd.} The set of all candidate symmetry equivalence classes, which form the vertices of the DAG, is given by 
\begin{align*} 
\bigcup_{m \mid d } \{ \Dm{m} \} \;\cup\; \bigcup_{\substack{m \mid d \\ 1<m < d}} \{ \Cd{m}\}.
\end{align*}

{\bf Step (i):} We first find all the prime factors of $d$, exactly as in Section \ref{sec:DAG_construction_cyclic}, Step (i). We then construct the DAG over all $\Dm{m}$ using the same iterative procedure described in Section \ref{sec:DAG_construction_cyclic}, Step (ii), with two modifications: (1) the vertices now correspond to $\Dm{m}$ rather than $\Cd{m}$, and (2) the construction terminates with the additional leaf vertex $\Dm{1}$, which serves as the common child of all $\Dm{p}$ with $p$ prime. Thus, starting from the root $\Dm{d}$, the DAG is built by repeatedly dividing indices by prime factors until reaching the only leaf $\Dm{1}$.

{\bf Step (ii):} We add all the cyclic symmetry classes $\{\Cd{m} : m \mid d, ~1<m<d\}$ to the DAG. For each divisor $m$ with $1<m<d$, we create a new vertex $\Cd{m}$ and add a directed edge from $\Dm{m}$ to $\Cd{m}$, thereby embedding the cyclic symmetry classes into the existing dihedral structure.

{\bf Step (iii):} We complete the DAG by adding the remaining directed edges among cyclic symmetry classes. Since both the subgroups $\Dm{d}$ and $\Cd{d}$ induce constant signals, and for each prime factor $p$ of $d$, there already exists a directed path $\Dm{d} (\Cd{d}) \to \Dm{d/p} \to \Cd{d/p}$, it is unnecessary to add a new direct edge from $\Dm{d}$ to $\Cd{d/p}$. The directed edges among the remaining cyclic vertices $\Cd{m}$ are then added exactly as in Section~\ref{sec:DAG_construction_cyclic}, Step (ii), that is, without adding the edges starting from $\Cd{d}$.

\vspace{0.2cm}
Consequently, the resulting structure is a DAG whose leaf vertices consist of the cyclic vertices $\Cd{p}$ for all prime $p$, along with the additional leaf vertex $\Dm{1}$.

\medskip
\noindent
\textbf{Case 2: $d$ is even.} The set of all candidate symmetry equivalence classes, which form the vertices of the DAG, is given by 
\begin{align*}
\{ \Dm{d} \} \;\cup\;  \{ \Dm{d/2}^0 \}  
\;\cup\; 
\bigcup_{\substack{m \mid d \\ m < d/2}}
\begin{cases}
\{ \Dm{m} \}, & 2m \nmid d, \\[2mm]
\{ \Dm{m}^0, \Dm{m}^1 \}, & 2m \mid d,
\end{cases}
\;\cup\; \bigcup_{\substack{m \mid d \\ 1<m < d/2}} \{\Cd{m}\}.
\end{align*}

{\bf Step (i):} We first find all the prime factors of $d$, exactly as in Section~\ref{sec:DAG_construction_cyclic}, Step (i). We then construct the DAG over all dihedral symmetry classes as follows. We begin with the root vertex $\Dm{d}$ and create a new vertex $\Dm{d/2}^0$ and add a directed edge from $\Dm{d}$ to $\Dm{d/2}^0$. For every other prime factor $p\not= 2$ of $d$, we create a new vertex $\Dm{d/p}$ and add a directed edge from $\Dm{d}$ to it. If $4\mid d$, create a new vertex $\Dm{d/4}^1$ and add an additional directed edge from $\Dm{d}$ to $\Dm{d/4}^1$ (this corresponds to the implication relationship between $\Dmj{d/2}{1}$-symmetry and $\Dm{d/4}^1$-symmetry).

Next, we proceed to each newly created vertex $\Dm{m}$, $\Dm{m}^0$ or $\Dm{m}^1$ (which has not yet been assigned children) in the DAG. For each prime factor $p\not= d$ of $d$, we split into two cases:
\begin{itemize}
    \item \textbf{Current vertex $\Dm{m}$.} If $p$ divides $m$:
    \begin{itemize}[itemsep=2pt, topsep=2pt]
        \item If $(2m/p) \nmid d$, then create the child vertex $\Dm{m/p}$ if it is not already in the DAG, and add a directed edge from $\Dm{m}$ to $\Dm{m/p}$;
        \item If $(2m/p) \mid d$, then create the child vertices $\Dm{m/p}^0$ and $\Dm{m/p}^1$ if they are not already in the DAG, and add two directed edges, one from $\Dm{m}$ to $\Dm{m/p}^0$ and the other from $\Dm{m}$ to $\Dm{m/p}^1$.
    \end{itemize}

    \item \textbf{Current vertex $\Dm{m}^j$ with $j=0$ or $1$.} If $p$ divides $m$, then create the child vertex $\Dm{m/p}^j$ if it is not already in the DAG, and add a directed edge from $\Dm{m}^j$ to $\Dm{m/p}^j$.
\end{itemize}
The process terminates once all the vertices of dihedral symmetry classes, except for the leaf vertices $\Dm{1}^0$ and $\Dm{1}^1$, have been assigned children. 

{\bf Step (ii):} We add all the candidate cyclic symmetry classes $\{\Cd{m} : m \mid d,~1<m<d/2\}$ to the DAG. 
For each divisor $m$ with $1<m<d/2$, we create a new vertex $\Cd{m}$. If $2m \nmid d$, we add a directed edge from $\Dm{m}$ to $\Cd{m}$; if $2m \mid d$, we instead add two directed edges, one from $\Dmj{m}{0}$ to $\Cd{m}$ and the other from $\Dmj{m}{1}$ to $\Cd{m}$.

{\bf Step (iii):} We complete the DAG by adding the remaining directed edges among cyclic symmetry classes. Since both the subgroups $\Dm{d}$ and $\Cd{d}$ induce constant signals, and for each prime factor $p$ of $d$ there already exists a directed \textit{path} from $\Dm{d}$ to $\Cd{d/p}$, it is unnecessary to add a new direct edge from $\Dm{d}$ to $\Cd{d/p}$. Similarly, since both the subgroups $\Dm{d/2}^0$ and $\Cd{d/2}$ induce signals having period $2$, and for each prime factor $p$ of $d$, there already exists a directed \textit{path} from $\Dm{d/2}^0$ to $\Cd{d/(2p)}$, it is unnecessary to add a new direct edge from $\Dm{d/2}$ to $\Cd{d/(2p)}$. The directed edges among the remaining cyclic vertices $\Cd{m}$ are then added exactly as in Section~\ref{sec:DAG_construction_cyclic}, Step (ii), that is, without adding the edges starting from $\Cd{d}$ and $\Cd{d/2}$. Consequently, the resulting structure is a DAG whose leaf vertices consist of the cyclic vertices $\Cd{p}$ for all prime $p$, along with the two additional leaf vertices $\Dm{1}^0$ and $\Dm{1}^1$. 

Figure~\ref{fig:dihedral_DAG} shows a visual demonstration of the DAG for the two cases of odd and even $d$.
\newlength{\dagheight}

\begin{figure}[ht]
    \centering
    \setlength{\dagheight}{0.28\textwidth}

    \begin{subfigure}[b]{0.44\textwidth}
        \centering
        \includegraphics[height=\dagheight]{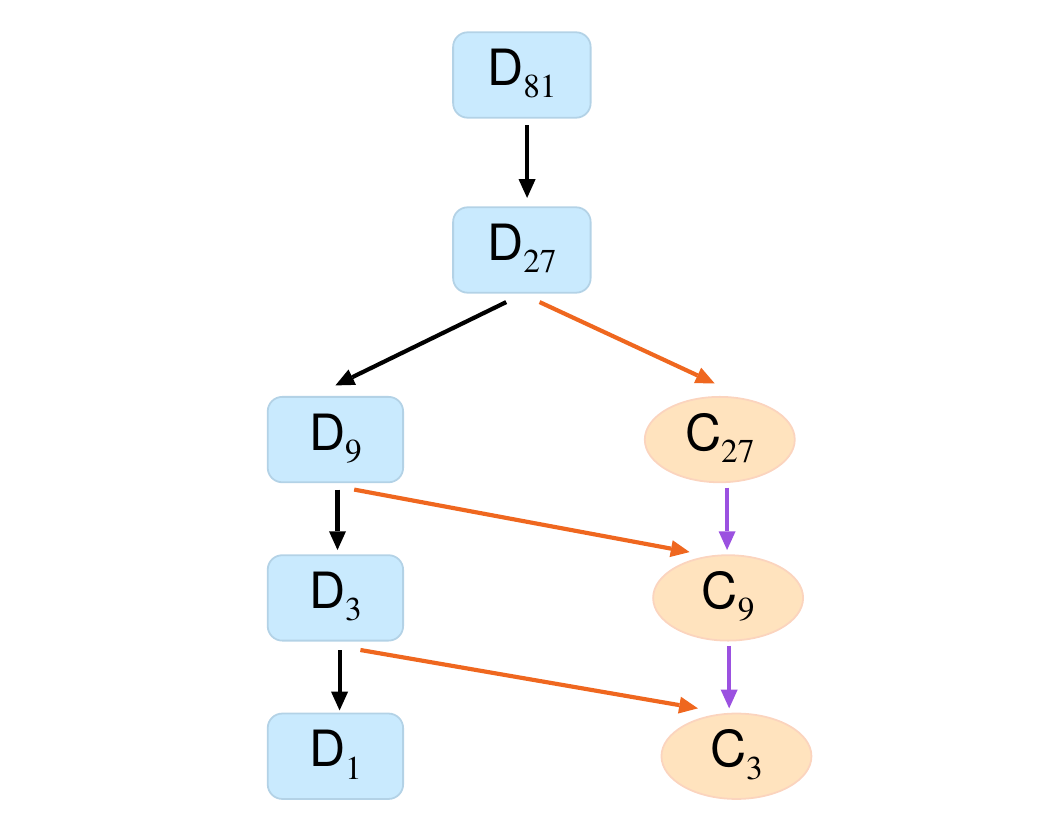}
        \caption{DAG for odd $d=81$}
        \label{fig:dihedral_DAG_odd}
    \end{subfigure}
    \hfill
    \begin{subfigure}[b]{0.52\textwidth}
        \centering
        \includegraphics[height=\dagheight]{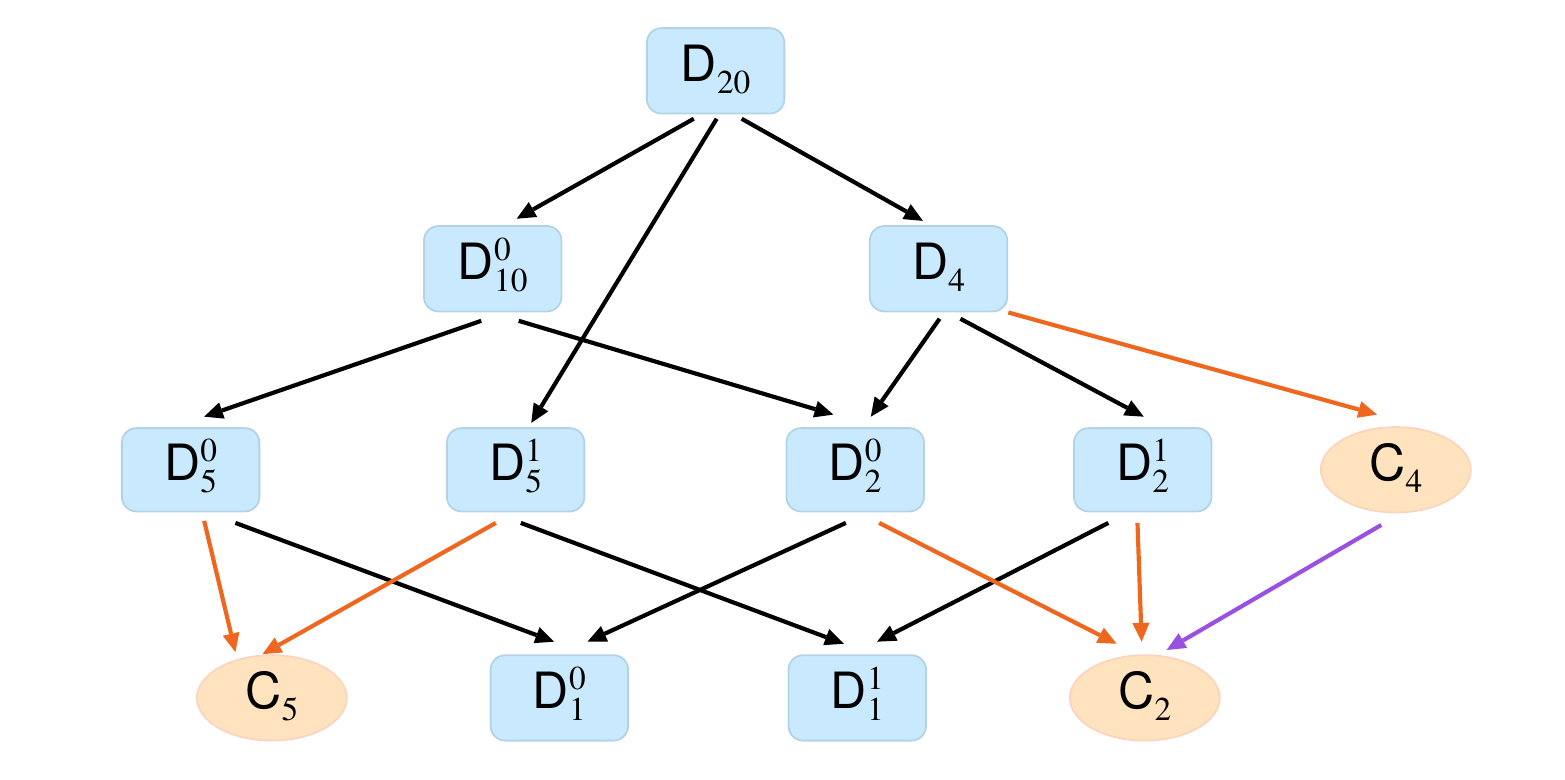}
        \caption{DAG for even $d=20$}
        \label{fig:dihedral_DAG_even}
    \end{subfigure}

    \caption{Examples of the hypothesis DAG for the dihedral group
    in the odd and even cases. Black, orange, and purple arrows
    correspond to Steps~(i), (ii), and (iii), respectively.}
    \label{fig:dihedral_DAG}
\end{figure}

\subsubsection{Generic cyclic and dihedral signals}\label{sec:generic_signals}

When $m = d/2$ is a valid divisor of $d$, the symmetry classes reduce to either constant signals or cyclic symmetry, and thus second-order information is sufficient for testing. Therefore, in the following, we restrict attention to the case $m \le d/3$. 

To make subsequent statements precise, we fix the notion of \textit{generic cyclic} and \textit{dihedral signals} used throughout. It consists of explicit nonvanishing conditions on Fourier coefficients: exactly the conditions under which the population characterizations of Sections~\ref{sec:population_analysis_collapsed} and~\ref{sec:population_analysis_parity_split} are exhaustive and the Wald statistics of Sections~\ref{sec:p-value_calculation_DMRA_collapsed} and~\ref{sec:p-value_calculation_DMRA_parity_split} have nonsingular limits. In the parity-split regime, the Nyquist coefficient $\F\x[d/2]$ is deliberately left free, because odd-parity dihedral symmetry forces it to vanish (Proposition~\ref{prop:structure_in_Fourier_DMRA}(iv)).

\begin{definition}[Generic cyclic and dihedral signals] \label{def:generic_signals}
Let $m \leq d/3$ be any divisor of $d$, and write $\cI_{m,d}^{+,\circ}\triangleq \cI_{m,d}^+\setminus\{d/2\}$ (so $\cI_{m,d}^{+,\circ}=\cI_{m,d}^+$ in the collapsed regime, where $d/2\notin\cI_{m,d}$).

\begin{itemize}
    \item[(a)] A real signal $\x\in\R^d$ is a \textbf{generic $\Cd{m}$ signal} if it is $\Cd{m}$-invariant, i.e., $\F\x[\ell] = 0$ for all $\ell \in \cI_{m,d}^c$ (the cyclic zero pattern~\eqref{eqn:zeros_in_Fourier_MRA}), and
    \begin{align*}
    \F\x[\ell]\neq 0\qquad\text{for all }\ell\in\cI_{m,d}^{+,\circ}.
    \end{align*}
    No condition is imposed on the Nyquist coefficient $\F\x[d/2]\in\R$ when $2m\mid d$; it may vanish or not.

    \item[(b)] For $j\in\{0,1,\ldots,d/m-1\}$, a real signal $\x\in\R^d$ is a \textbf{generic $\Dmj{m}{j}$ signal} if it is invariant under $\Dmj{m}{j}$, i.e., it satisfies the conditions of Proposition~\ref{prop:structure_in_Fourier_DMRA}(iv), and its real coefficients there satisfy
    \begin{align*}
    \F\x[0]\neq 0,\qquad c_\ell\neq 0\quad\text{for all }\ell\in\cI_{m,d}^{+,\circ}\setminus\{0\}.
    \end{align*}
    When $2m\mid d$, the Nyquist coefficient $\F\x[d/2]\in\R$ is unrestricted for even $j$ and equals zero for odd $j$ (as it must).

    In the collapsed regime, $\x$ is a \textbf{generic $\Dm{m}$ signal} if it is a generic $\Dmj{m}{j}$ signal for some $j$. In the parity-split regime, $\x$ is a \textbf{generic $\Dm{m}^0$ signal} if it is a generic $\Dmj{m}{j}$ signal for some even $j$, and a \textbf{generic $\Dm{m}^1$ signal} if it is a generic $\Dmj{m}{j}$ signal for some odd $j$.
\end{itemize}
\end{definition}

\begin{remark}[On the definition]\label{rmk:genericity_definition}
Three comments are in order. First, a generic $\Dmj{m}{j}$ signal is in particular a generic $\Cd{m}$ signal, and a generic $\Cd{m}$ signal has exact rotational symmetry $\Cd{m}$, since $\F\x[m]\neq0$ rules out invariance under any larger cyclic subgroup. Second, within the linear space of $\Cd{m}$-invariant signals (respectively, of $\Dmj{m}{j}$-invariant signals), the non-generic signals form a finite union of coordinate hyperplanes, a proper algebraic subset of Lebesgue measure zero, see, e.g., \cite{mityagin2015zero}; hence every statement made for generic signals holds for almost every signal with the prescribed symmetry, and the generic class is nonempty. Third, and importantly, the definition does \emph{not} ask the free coordinates to avoid every nontrivial polynomial relation: dihedral signals necessarily satisfy the polynomial phase relations of Proposition~\ref{prop:structure_in_Fourier_DMRA}(iv), and the whole point of Theorem~\ref{thm:DMRA_4th_population_general} below is to characterize those relations \emph{within} the class of generic $\Cd{m}$ signals. The nonvanishing conditions are sufficient for the characterizations and tests that follow; they are not necessary for identifiability, since some spectra containing zeros remain identifiable (see the remark following Corollary~\ref{cor:DMRA_4th_population_reduced}), and the dihedral symmetry label is determined by moments of order at most $d/m$ without these conditions (Proposition~\ref{prop:order_dm_sufficient}). Genericity in Remark~\ref{rmk:harder}, in contrast, refers to the full ambient space, as in the orbit-recovery literature.
\end{remark}

\subsubsection{Phase-revealing fourth-order moments and population-level analysis for the collapsed regime}
\label{sec:population_analysis_collapsed}

For generic $\Cd{m}$ signals, third-order moments already provide sufficient, distribution-free information to separate $\Dm{m}$ from $\Cd{m}$ (Proposition~\ref{prop:third_order_generic}). We nevertheless build the tests on fourth-order moments, for the sake of methodological uniformity with the parity-split regime, where the signed parity witness of Section~\ref{sec:population_analysis_parity_split} is of fourth order; this does not incur a significant increase in computational complexity (see Remark~\ref{rmk:third_vs_fourth} for the expected trade-off in noise scaling).

We begin with the \textit{collapsed regime}, where either $d$ is odd or $d$ is even and $2m \nmid d$. In this case, all dihedral subgroups $\Dm{m,j}$ collapse into a single class $\Dm{m}$. As in the C-MRA analysis, second-order moments can also determine that the signal is invariant under the cyclic subgroup $\Cd{m}$. Determining whether the signal further belongs to the dihedral symmetry class $\Dm{m}$ requires phase information, which second-order moments cannot see and which, on non-generic spectra, is invisible to all moments of order below $d/m$ (Theorem~\ref{thm:third_moment_failure}).

We now study a fourth-order moment expression that reveals the Fourier phases while remaining independent of the distribution $\rho$; it is the fourth-order analog of \eqref{eq:tildeM3}. Recall that the fourth-order Fourier moment has the following form:
\begin{equation} \label{eqn:fourth_moment_entry}
\F M_4(\x,\rho)[\ell_1,\ell_2,\ell_3,\ell_4]
= \F\p[\ell']\,\F\x[\ell_1] \F\x [\ell_2] \F\x[\ell_3] \overline{\F\x[\ell_4]}
+ \F\q[\ell']\,\overline{\F\x[\ell_1]\F\x[\ell_2] \F\x[\ell_3] }\,\F\x[\ell_4],
\end{equation}
where $\ell' \equiv \ell_1 + \ell_2 + \ell_3 - \ell_4 \pmod d$ and $\ell',\ell_1,\ell_2,\ell_3 \in\cI_d$. Denote $t = (\ell_1,\ell_2,\ell_3)$, and define
\begin{align}\label{eq:l123}
\ell_1(t)\triangleq \ell_1,\quad
\ell_2(t)\triangleq \ell_2,\quad
\ell_3(t)\triangleq \ell_3,
\end{align}
and let
\begin{align}\label{eq:l4}
\ell_4(t) \equiv \ell_1 + \ell_2 + \ell_3 \pmod d, 
\end{align}
where $\ell_4(t) \in \cI_d$. For notational simplicity, we will omit the dependence on $t$ for $\ell_1(t),\ell_2(t),\ell_3(t)$ when it is clear from the context. Consider the sum of the conjugate pair of entries
\begin{align*}
\widetilde{M}_4(\x,t)\triangleq\F M_4(\x,\rho)[\ell_1,\ell_2,\ell_3,\ell_4(t)]
+ \F M_4(\x,\rho)[-\ell_1,-\ell_2,-\ell_3,-\ell_4(t)].
\end{align*}
Noting that for this choice of indices we have $\ell' \equiv 0 \pmod d$ and $\F\p[0] + \F\q[0]=1$, we obtain the following quantity that is independent of the distribution~$\rho$:
\begin{align}
\widetilde{M}_4(\x,t)&=\F M_4(\x,\rho)[\ell_1,\ell_2,\ell_3,\ell_4(t)]
+ \F M_4(\x,\rho)[-\ell_1,-\ell_2,-\ell_3,-\ell_4(t)] \notag\\
&= (\F\p[0] + \F\q[0])\cdot\,
2\Re\!\big( \F\x[\ell_1]\F\x[\ell_2]\F\x[\ell_3]\overline{\F\x[\ell_4(t)]} \big)\notag\\
&= 2\Re\!\big( \F\x[\ell_1]\F\x[\ell_2]\F\x[\ell_3]\overline{\F\x[\ell_4(t)]} \big)\notag\\
&=2r_{\ell_1}r_{\ell_2}r_{\ell_3}r_{\ell_4(t)}\cos(\theta_{\ell_1}+\theta_{\ell_2}+\theta_{\ell_3}
- \theta_{\ell_4(t)}),\label{eq:tildeM4}
\end{align}
where we write $\F\x[\ell] = r_\ell e^{\ri\theta_\ell}$ with $r_\ell\geq 0$ and $\theta_\ell\in(-\pi,\pi]$. 

Combining the above derivations with the Fourier phase constraints established in Proposition~\ref{prop:structure_in_Fourier_DMRA}, we obtain the following characterization of dihedral symmetry in terms of the fourth-order moment information. The detailed proof is given in Section~\ref{sec:proof_DMRA_4th_population_general}.

\begin{theorem}\label{thm:DMRA_4th_population_general}
Let $m \leq d/3$ be any divisor of $d$. Let $\x \in \R^d$ be a generic cyclic $\Cd{m}$ signal, in the sense that 
\begin{align*}
\F\x[\ell]= 0,&\quad \text{for all }\ell\in\cI_{m,d}^c,\\
\F\x[\ell]\neq 0,&\quad \text{for all }\ell\in\cI_{m,d}^+.
\end{align*}
Then, under the collapsed regime in the D-MRA setting, a signal $\x$ has $\Dm{m}$-symmetry if and only if
\begin{align*}
\widetilde{M}_4(\x,t)
\in \{\pm 2r_{\ell_1}r_{\ell_2}r_{\ell_3}r_{\ell_4(t)}\}
\end{align*}
for all $t=(\ell_1,\ell_2,\ell_3)\in \cI_{m,d}^3\triangleq \cI_{m,d}\times \cI_{m,d}\times \cI_{m,d}$.
\end{theorem}

For subsequent hypothesis testing, we further refine the set of frequency triplets used in Theorem~\ref{thm:DMRA_4th_population_general}. On the one hand, evaluating the condition over all $(\ell_1,\ell_2,\ell_3)\in \cI_{m,d}^3$ incurs a computational cost on the order of $(d/m)^3$, which becomes computationally intensive as $d$ grows. On the other hand, incorporating all such triplets introduces substantial redundancy, which in turn creates degeneracy in the covariance structure of the resulting test statistics and causes singularity issues in the testing procedure.

To address both concerns, we restrict attention to a smaller and carefully chosen set of chain tuples,
\begin{align}\label{eq:chain_set}
\cJ_{m,d} \triangleq \Big\{ (\ell_1,\ell_2,\ell_3) = (0, m, km) : 1 \leq k \leq  \frac{d/m - 1}{2}  \Big\}\subset \cI_{m,d}^3.
\end{align}
Note that the largest possible $\ell_3$ in the tuple is not larger than $d/2$. This reduced set preserves the essential discriminatory information needed for testing while significantly lowering computational complexity and avoiding covariance singularity; see Section~\ref{sec:p-value_calculation_DMRA_collapsed} for further details. The following corollary shows that, in addition to the cyclic structure, only 
$\lfloor (d/m-1)/2 \rfloor$ fourth-order moments are sufficient to characterize the dihedral invariance. We defer the proof to Section~\ref{sec:proof_DMRA_4th_population_reduced}.

\begin{corollary}\label{cor:DMRA_4th_population_reduced}
Let $m\leq d/3$ be any divisor of $d$. Let $\x\in\R^d$ be a generic signal invariant under $\Cd{m}$ as in Theorem~\ref{thm:DMRA_4th_population_general}.
Then, under the collapsed regime in the D-MRA setting, a signal $\x$ has $\Dm{m}$-symmetry if and only if
\begin{align*}
\widetilde{M}_4(\x,t)
\in \{\pm 2r_{\ell_1}r_{\ell_2}r_{\ell_3}r_{\ell_4(t)}\},
\quad \text{for all }t=(\ell_1,\ell_2,\ell_3)\in \cJ_{m,d}.
\end{align*}
\end{corollary}

\begin{remark}[On the generic condition of $\Cd{m}$ signal]
We comment on the generic conditions in Theorem~\ref{thm:DMRA_4th_population_general} and Corollary~\ref{cor:DMRA_4th_population_reduced}. The condition $\F\x[0] \neq 0$ is needed because the chain tuples of $\cJ_{m,d}$ start with the index $0$, so that $\widetilde M_4(\x,(0,m,km))=\F\x[0]\,\widetilde M_3(\x,(m,km))$ (Remark~\ref{rmk:third_vs_fourth}); it is without loss of generality, since $\F\x[0]$ is the signal mean, which does not affect symmetry and can always be made nonzero by adding a constant to the observations (constant vectors are fixed by every element of $\Dm{d}$). The condition $\F\x[\ell] \neq 0$ for all $\ell\in\cI_{m,d}^+\setminus\{0\}$ is what makes the chain-based characterization above exhaustive: the tuples in $\cJ_{m,d}$ pin down the phase $\theta_\ell$ at every admissible coordinate, so if $\F\x[\ell]=0$ then $\theta_\ell$ is undefined and the chain tuples through $\ell$ carry no phase information. It is therefore necessary for that construction (equivalently, for recovering every admissible phase), but not for fourth-order discrimination in general. A single vanishing coordinate $\ell$ only voids the tuples through $\ell$, while every tuple on the remaining nonzero support still tests reflection-consistency. Surviving constraints can detect some departures from reflection symmetry; complete separation, however, requires sufficiently many constraints to enforce a common reflection axis across the whole nonzero support, and fourth-order tuples may fall short of this on sparse spectra. In the extreme case where the nonzero coordinates collapse onto a single conjugate pair, phase-sensitive information first appears at order $d/m$ (Theorem~\ref{thm:third_moment_failure} and Remark~\ref{rmk:worst_case_d_m}); by Proposition~\ref{prop:order_dm_sufficient}, order $d/m$ always restores complete separation.
\end{remark}

For hypothesis testing, it is convenient to reformulate
Corollary~\ref{cor:DMRA_4th_population_reduced} as a collection of polynomial
equalities involving population moments. To this end, we introduce the
following notation. Define
\begin{align*}
s_\ell\triangleq \E\Big[\F\widehat{M}_2[\ell,\ell]\Big]-d\sigma^2, \qquad \text{and }\qquad \s=(s_0,\ldots,s_{d-1})^\T \in\R^d. 
\end{align*}
By Lemma \ref{lem:DMRA_2nd_mean}, $s_\ell=r_{\ell}^2$ and $r_\ell=r_{d-\ell}$ for real $\x$. Recall $\cI_{m,d}=\{0,m,\ldots,d-m\}$, $\cI_{m,d}^c=\cI_d\setminus \cI_{m,d}$, and $\cI_{m,d}^3=\cI_{m,d}\times \cI_{m,d}\times \cI_{m,d}$. For $t=(\ell_1,\ell_2,\ell_3)\in \cI_{m,d}^3$, also recall
\begin{align*}
\cI_{m,d}\ni \ell_4(t) \equiv \ell_1+\ell_2+\ell_3 \pmod{d},
\end{align*}
and define
\begin{align*}
d_t\triangleq \E\big[\F\widehat{M}_4[\ell_1,\ell_2,\ell_3,\ell_4(t)\big]+\F\widehat{M}_4\big[-\ell_1,-\ell_2,-\ell_3,-\ell_4(t)]\big].
\end{align*}
For any $\ell\in\cI_d$, define
\begin{align*}
\ell^{\sharp}\triangleq \min\{\ell,d-\ell\}\in\cI_d^+,
\end{align*}
and write
\begin{align}\label{eq:l4sharp}
\ell_4^{\sharp}(t)\triangleq (\ell_4(t))^{\sharp}
\end{align}
to denote the folded representative. We reformulate Corollary~\ref{cor:DMRA_4th_population_reduced} as conditions under the null hypothesis for the purpose of hypothesis testing. The proof is given in Section~\ref{sec:proof_DMRA_mean_null_collapsed}.

\begin{proposition}[Population null constraints in the collapsed regime]\label{prop:DMRA_mean_null_collapsed}
Let $m\leq d/3$ be any divisor of $d$. Let $\cI_{m,d}^{c,+}$ and $\cJ_{m,d}$ be defined as in \eqref{eq:cI_plus} and \eqref{eq:chain_set}, respectively. Under the collapsed regime in the D-MRA setting, if a signal \( \x \in \R^d \) belongs to the symmetry class $\Dm{m}$, then
\begin{align}\label{eq:DMRA_2nd_population_condition}
s_\ell = 0, \qquad \text{for all }\ell\in\cI_{m,d}^{c,+}, 
\end{align}
and
\begin{align}\label{eq:DMRA_4th_population_condition}
4s_{\ell_1}s_{\ell_2}s_{\ell_3}s_{\ell_4^{\sharp}(t)} - d_t^2 = 0, 
\qquad \text{for all } t=(\ell_1,\ell_2,\ell_3)\in \cJ_{m,d}.
\end{align}
Moreover, the conditions above are sufficient for generic $\Cd{m}$ signals as in Theorem~\ref{thm:DMRA_4th_population_general}.
\end{proposition}

\subsubsection{Raw p-value calculation for single dihedral hypothesis in the collapsed regime}\label{sec:p-value_calculation_DMRA_collapsed}

Motivated by the population-level characterization of fourth-order moments, we perform hypothesis testing at each vertex of the DAG constructed in Section~\ref{sec:DAG_construction_dihedral}, obtaining the corresponding raw p-values.
For cyclic symmetry classes, the testing problem can already be addressed using second-order information as developed in Section~\ref{sec:p-value_calculation_CMRA}. Therefore, in this section, we focus on testing for dihedral symmetry classes, leveraging the fourth-order phase invariants of Section~\ref{sec:population_analysis_collapsed}, which carry the phase information that second-order moments lack. The obtained p-values are then used as input to the sequential rejection algorithm (see Appendix~\ref{sec:SRA}).

Building on Proposition~\ref{prop:DMRA_mean_null_collapsed}, we now consider how to handle the sample version in order to obtain a valid testing procedure. For $j=1,\ldots,N$, and $\ell\in\cI_d$, define
\begin{align*}
s_{\ell,j}\triangleq& |\F\y_j[\ell]|^2 - d\sigma^2,\\
\s_j\triangleq& (s_{0,j},s_{1,j},\ldots,s_{d-1,j})^\T \in\R^d,
\end{align*}
and
\begin{align*}
\widehat{S}_{\ell}\triangleq& \frac{1}{N}\sum_{j=1}^N s_{\ell,j}=\F\widehat{M}_2[\ell,\ell]-d\sigma^2,\\
\widehat{\boldsymbol{S}}\triangleq& (\widehat{S}_{0},\widehat{S}_{1},\ldots,\widehat{S}_{d-1})^\T \in\R^d.
\end{align*}
It is clear that $\E[s_{\ell,j}]=\E[\widehat{S}_\ell]=s_\ell$. For $j=1,\ldots,N$ and $t=(\ell_1,\ell_2,\ell_3)\in\cI_{m,d}^3$, define
\begin{align*}
d_{t,j}\triangleq& \F\y[\ell_1] \F\y[\ell_2] \F\y[\ell_3] \overline{\F\y[\ell_4(t)]}+ \overline{\F\y[\ell_1]} \overline{\F\y[\ell_2]} \overline{\F\y[\ell_3]} \F\y[\ell_4(t)],\\
\widehat{D}_t\triangleq&\frac{1}{N}\sum_{j=1}^N d_{t,j}=\F\widehat{M}_4[\ell_1,\ell_2,\ell_3,\ell_4(t)]+\F\widehat{M}_4[-\ell_1,-\ell_2,-\ell_3,-\ell_4(t)].
\end{align*}
It is also clear that $\E[d_{t,j}]=\E[\widehat{D}_t]=d_t$. Recall the definitions of $\ell_i(t)$ for $t\in\cI_d$ and $i=1,2,3,4$ in \eqref{eq:l123} and \eqref{eq:l4}. For chain tuples $\cJ_{m,d}$, since no bias correction term is needed as in \eqref{eq:DMRA_4th_population_condition}, further define the fourth-order constraint statistics
\begin{align*}
\widehat{H}_t\triangleq 4\widehat{S}_{\ell_1(t)} \widehat{S}_{\ell_2(t)} \widehat{S}_{\ell_3(t)} \widehat{S}_{\ell_4^{\sharp}(t)}-\widehat{D}_t^2,\quad\text{for }t\in\cJ_{m,d}.
\end{align*}
The collapsed-regime test vector is then defined as
\begin{align}
\widehat{\g}_m\triangleq \Big( (\widehat{S}_{\ell})_{\ell\in\cI_{m,d}^{c,+}}, (\widehat{H}_{t})_{t\in\mathcal{J}_{m,d}} \Big)^\T \in\R^{q_m},\qquad q_m\triangleq |\cI_{m,d}^{c,+}|+\frac{d/m-1}{2}.\label{eq:q}
\end{align}
Define the covariance matrix 
\begin{align}\label{eq:Omega_g}
\bOmega_{\g} \triangleq \Var(\bpsi_j),
\end{align}
where $\bpsi_j\in\R^{q_m}$ has coordinates
\begin{align}
\psi_{\ell,j}  \triangleq & s_{\ell,j} - s_\ell,\quad\text{for }\ell\in\cI_{m,d}^{c,+},\label{eq:psi_ell}\\
\psi_{t,j} \triangleq& 4\Big(
s_m s_{km} s_{((k+1)m)^{\sharp}}(s_{0,j}-s_0)+
s_0 s_{km} s_{((k+1)m)^{\sharp}}(s_{m,j}-s_m)\notag\\
&+
s_0 s_m s_{((k+1)m)^{\sharp}}(s_{km,j}-s_{km})+
s_0 s_m s_{km}(s_{((k+1)m)^{\sharp},j}-s_{((k+1)m)^{\sharp}})
\Big)\notag\\
&-2 d_t(d_{t,j}-d_t),\quad\text{for }t=(0,m,km)\in\mathcal{J}_{m,d}.\label{eq:psi_t}
\end{align}
We further define the empirical counterpart $\widehat{\bpsi}_j\in\R^{q_m}$ by replacing the population quantities $(s_\ell, d_t)$ with their estimators $(\widehat{S}_\ell, \widehat{D}_t)$, while keeping the sample-level terms unchanged. That is,
\begin{align}
\widehat{\psi}_{\ell,j}  \triangleq & s_{\ell,j} - \widehat{S}_\ell,\quad\text{for }\ell\in\cI_{m,d}^{c,+},\label{eq:psihat_ell}\\
\widehat{\psi}_{t,j} \triangleq& 4\Big(
\widehat{S}_m \widehat{S}_{km} \widehat{S}_{((k+1)m)^{\sharp}}(s_{0,j}-\widehat{S}_0)+
\widehat{S}_0 \widehat{S}_{km} \widehat{S}_{((k+1)m)^{\sharp}}(s_{m,j}-\widehat{S}_m)\notag\\
&+
\widehat{S}_0 \widehat{S}_m \widehat{S}_{((k+1)m)^{\sharp}}(s_{km,j}-\widehat{S}_{km})+
\widehat{S}_0 \widehat{S}_m \widehat{S}_{km}(s_{((k+1)m)^{\sharp},j}-\widehat{S}_{((k+1)m)^{\sharp}})
\Big)\notag\\
&-2 \widehat{D}_t(d_{t,j}-\widehat{D}_t),\quad\text{for }t=(0,m,km)\in\mathcal{J}_{m,d}.\label{eq:psihat_t}
\end{align}
The second expression simplifies depending on the value of $k$: if $1\leq k< (d/m-1)/2$, one can drop the $\sharp$ notation, whereas for $k=(d/m-1)/2$, one substitutes $((k+1)m)^{\sharp}=km$. In addition, if indices coincide, the corresponding terms add. With the notation above, define the sample covariance matrix
\begin{align}\label{eq:Omegahat_g}
\widehat{\bOmega}_\g\triangleq \frac{1}{N}\sum_{j=1}^N \widehat{\bpsi}_j \widehat{\bpsi}_j^\T.
\end{align}
We now introduce a Wald test for $\Dm{m}$-symmetry, whose validity is established in the following theorem. The proof is deferred to Section \ref{sec:proof_DMRA_Dm_testing}.
\begin{theorem}\label{thm:DMRA_Dm_testing}
Let $m\leq d/3$ be any divisor of $d$ such that $2m \nmid d$. Assume the collapsed regime in the D-MRA setting. Under the null hypothesis that $\x\in\R^d$ is invariant under the class $\Dm{m}$,
\begin{align*}
\sqrt{N}\cdot \widehat{\g}_m\xrightarrow{d} \mathcal{N}(\0,\bOmega_\g).
\end{align*}
If $\x$ is a further generic $\Dm{m}$ signal in the sense of Lemma~\ref{lem:Jacobian_full_rankness}(i), then $\bOmega_\g$ is positive definite, and $\widehat{\bOmega}_\g \xrightarrow{p} \bOmega_\g$. Moreover, the Wald statistic
\begin{align}\label{eq:chi_g}
\widehat{W}_m\triangleq N \cdot \widehat{\g}_m^\T \widehat{\bOmega}_\g^{-1} \widehat{\g}_m \xrightarrow{d} \chi^2_{q_m}.
\end{align}
Consequently, an asymptotically valid (raw) $p$-value for testing the null hypothesis of $\Dm{m}$-invariant is given by
\begin{align}\label{eq:raw_pvalue_dihedral}
p_m^{D}\triangleq 1- F_{\chi_{q_m}^2} (\widehat{W}_m).
\end{align}
Moreover, under any fixed alternative where at least one of the conditions in \eqref{eq:DMRA_2nd_population_condition} and \eqref{eq:DMRA_4th_population_condition} fails\footnote{This condition holds for any signal that is not invariant under $\Cd{m}$, as well as for any generic $\Cd{m}$-invariant signal in the sense of Theorem~\ref{thm:DMRA_4th_population_general} that does not belong to the $\Dm{m}$-symmetry class.}, and the condition in Lemma~\ref{lem:Jacobian_full_rankness}(ii) holds, we have
\begin{align*}
\widehat{W}_m \xrightarrow{p} \infty.
\end{align*}
In particular, the Wald test based on $\widehat{W}_m$ is consistent against any such fixed alternative.
\end{theorem}

\subsubsection{Population-level analysis for the parity-split regime}\label{sec:population_analysis_parity_split}

We now turn to the parity-split regime, i.e., when $d/m \geq 4$ is even. In this regime, the dihedral subgroups split into two distinct detectable symmetry classes, $\Dm{m}^0$ and $\Dm{m}^1$. Our goal is not only to test whether a signal exhibits dihedral symmetry beyond $\Cd{m}$-symmetry, as in the collapsed regime, but also to distinguish between these two dihedral symmetry classes. We first restrict attention to generic $\Cd{m}$ signals in the sense of Theorem~\ref{thm:DMRA_4th_population_general}, except that we treat the nonvanishing condition at coordinate $d/2$ separately (note that $\cI_{m,d}^+$ includes $d/2$ in this regime). 

Equivalently, we consider
\begin{align*}
\F\x[\ell]= 0,&\quad \text{for all }\ell\in\cI_{m,d}^c,\\
\F\x[\ell]\neq 0,&\quad \text{for all }\ell\in\cI_{m,d}^+\setminus\{d/2\},
\end{align*}
where $s_{d/2}=|\F\x[d/2]|^2$ is allowed either to vanish or to be nonzero.

We first consider the case where $s_{d/2}>0$. By Proposition~\ref{prop:structure_in_Fourier_DMRA}(iv), an odd-parity dihedral symmetry $\Dmj{m}{j}$, with $j$ odd, forces $s_{d/2}=0$ whereas an even-parity dihedral symmetry only requires this coordinate to be real. Hence, once the fourth-order phase constraints certify that the signal is dihedral rather than merely $\Cd{m}$-symmetric, the condition $s_{d/2}>0$ immediately rules out $\Dm{m}^1$ and identifies the symmetry class as $\Dm{m}^0$. In this sense, a nonzero $s_{d/2}$ gives a second-order shortcut for the parity split. It should not, however, be viewed as a substitute for the fourth-order dihedral constraints: a generic $\Cd{m}$-symmetric signal may have $s_{d/2}>0$ without satisfying any reflection-consistent phase relation.

The more challenging case is when $s_{d/2}=0$. In this case, second-order moment information alone no longer separates $\Dm{m}^0$ from $\Dm{m}^1$. Nevertheless, under the generic nonvanishing condition on the remaining coordinates, fourth-order information is still sufficient to
distinguish the two parity classes. The reason is that the unsigned fourth-order equations like \eqref{eq:DMRA_4th_population_condition} used
to propagate reflection-consistent phase relations square away the sign that encodes the parity of $j$. To recover this missing sign, we keep one signed fourth-order coordinate. A convenient choice is 
\begin{align*}
t_\star\triangleq (m,m,d/2-m),\qquad \ell_4(t_\star)=d/2+m.
\end{align*}
For a signal invariant under $\Dmj{m}{j}$, Proposition~\ref{prop:structure_in_Fourier_DMRA}(iv) gives
\[
    \F\x[\ell]
    = c_\ell \cdot e^{-\ri \ell j \pi / d},
    \qquad c_\ell\in\mathbb R,
    \qquad \ell\in \cI_{m,d}^+\setminus\{0,d/2\},
\]
together with conjugate symmetry. Therefore by Lemma~\ref{lem:DMRA_4th_mean}, the distribution-free fourth-order Fourier moment quantity
\[
    \widetilde{M}_4(\x,t)
    =2\Re\!(
        \F\x[\ell_1]
        \F\x[\ell_2]
        \F\x[\ell_3]
        \overline{\F\x[\ell_4(t)]}
    )
\]
yields, at $t=t_\star$,
\begin{align}
    d_{t_\star}=2(-1)^j s_m s_{d/2-m}.
\end{align}
Since $s_m s_{d/2-m}>0$ under the generic nonvanishing assumption away from $d/2$, the sign of $d_{t_\star}$ identifies the parity:
\[
    \frac{d_{t_\star}}{2s_m s_{d/2-m}}
    =
    \begin{cases}
        +1, & \x\in \Dm{m}^0,\\
        -1, & \x\in \Dm{m}^1.
    \end{cases}
\]
Equivalently, the two parity classes can be separated by the signed constraints
\[
    \kappa_m^0 \coloneqq 2s_m s_{d/2-m}-d_{t_\star}=0,
    \qquad
    \kappa_m^1 \coloneqq 2s_m s_{d/2-m}+d_{t_\star}=0.
\]
Thus the condition $s_{d/2}>0$ is useful but not necessary for detecting $\Dm{m}^0$: even when $s_{d/2}=0$, the signed fourth-order witness $K_m^0$ still separates $\Dm{m}^0$ from $\Dm{m}^1$, provided the other relevant Fourier coordinates are generic and nonzero.

More generally, if the particular witness $t_\star$ degenerates, one may use any index
\[
    a\in \cI_{m,d}^+\setminus\{0,d/2\}
    \quad\text{with}\quad
    d/2-a\in \cI_{m,d}^+\setminus\{0,d/2\}
\]
and define
\[
    t_a\coloneqq (a,a,d/2-a),
    \qquad
    \ell_4(t_a)=d/2+a.
\]
The same calculation yields
\[
    d_{t_a}=2(-1)^j s_a s_{d/2-a}.
\]
For generic $\Cd{m}$ signals away from $d/2$, such witnesses are nondegenerate. Higher-order moments are only needed in more sparse non-generic cases where all fourth-order parity witnesses vanish. In the extreme case where the only nonzero Fourier coordinates occur at a single conjugate pair, the first parity-sensitive moment information may appear only at order $d/m$; see Remark~\ref{rmk:worst_case_d_m} below.

\begin{remark}[A third-order parity witness when $d/m\ge 6$]\label{rmk:third_order_parity}
The fourth order of the witness $t_\star$ is not intrinsic. If $d/m\ge 6$, then $2m$ and $d/2-m$ both lie in $\cI_{m,d}^+\setminus\{0,d/2\}$, and for a signal invariant under $\Dmj{m}{j}$, Proposition~\ref{prop:structure_in_Fourier_DMRA}(iv) together with the distribution-free third-order quantity \eqref{eq:tildeM3} gives
\begin{align*}
\widetilde M_3\big(\x,(m,m)\big)&=2\Re\big(\F\x[m]^2\,\overline{\F\x[2m]}\big)=2c_m^2c_{2m},\\
\widetilde M_3\big(\x,(\tfrac d2-m,\tfrac d2-m)\big)&=2\Re\big(\F\x[\tfrac d2-m]^2\,\overline{\F\x[d-2m]}\big)=2(-1)^j c_{d/2-m}^2c_{2m},
\end{align*}
where the second identity uses $\overline{\F\x[d-2m]}=\F\x[2m]$, so that the total phase is $e^{-\ri[2(d/2-m)+2m]j\pi/d}=e^{-\ri j\pi}=(-1)^j$. Hence
\[
\widetilde M_3\big(\x,(m,m)\big)\cdot\widetilde M_3\big(\x,(\tfrac d2-m,\tfrac d2-m)\big)=4(-1)^j\,s_m\,s_{2m}\,s_{d/2-m},
\]
whose sign identifies the parity of $j$ whenever $s_ms_{2m}s_{d/2-m}>0$ (for $d/m=6$ one has $d/2-m=2m$, and the right-hand side reads $4(-1)^js_ms_{2m}^2$). Combined with the chain argument of Proposition~\ref{prop:third_order_generic}---closed across $d/2$ by the tuple $(d/2-m,d/2-m)$, which through conjugate symmetry forces $(d/m)\,\theta_m\equiv 0 \pmod{\pi}$, and complemented by the tuple $(m,d/2-m)$ when $s_{d/2}>0$---one checks that third-order invariant moments separate $\Cd{m}$, $\Dm{m}^0$, and $\Dm{m}^1$ when $d/m\ge6$ on every $\Cd{m}$-invariant signal with $r_{km}>0$ for $k=0,1,\ldots,d/(2m)-1$, the Nyquist coefficient $\F\x[d/2]$ being allowed to vanish (as it must for $\Dm{m}^1$), and do so without knowledge of $\rho$; these are the generic $\Cd{m}$ signals of Definition~\ref{def:generic_signals}. For $d/m=4$, the signals in $\Dm{m}^1$ have spectra supported on the DC term and a single conjugate pair, and the computation of Remark~\ref{rmk:worst_case_d_m} shows that no moment of order below four is parity-sensitive (the DC term is real and carries the index $0$, so it does not alter that computation); the fourth-order witness is then genuinely needed. We keep the fourth-order witness $t_\star$ throughout for uniformity.
\end{remark}

\begin{remark}[Worst case: a single conjugate pair] \label{rmk:worst_case_d_m}
Assume the parity-split regime, so that $d/m\ge 4$ is even. For $j\in\{0,1\}$, define a real signal $\x^{(Dj)}\in\R^d$ through its Fourier transform $\F \x^{(Dj)}$ by
\[
    \F \x^{(Dj)}[m]
        = e^{-\ri m j \pi / d},
    \qquad
    \F \x^{(Dj)}[d-m]
        = e^{\ri m j \pi / d},
\]
and
\[
    \F \x^{(Dj)}[\ell]=0,
    \qquad
    \ell\notin\{m,d-m\}.
\]
The two nonzero coefficients form a conjugate pair, so both signals are real-valued.  Moreover, all nonzero Fourier coordinates are multiples of $m$, so both signals are $\Cd{m}$-symmetric. By Proposition~\ref{prop:structure_in_Fourier_DMRA}(iv), $\x^{(D0)}$ satisfies the phase constraint for the even dihedral subgroups, whereas $\x^{(D1)}$ satisfies the phase constraint for the odd dihedral subgroups. Thus $\x^{(D0)}$ further belongs to the detectable class $\Dm{m}^0$, while $\x^{(D1)}$ belongs to the detectable class $\Dm{m}^1$.  Notice that both signals have
\[
    \F \x^{(Dj)}[d/2]=0,
\]
so the second-order moment information does not separate them.

We now show that all population moments of order strictly smaller than $d/m$ coincide under the uniform distribution $\rho_0$; the computation is the same as in the proof of Theorem~\ref{thm:third_moment_failure}, which does not use the parity of $d/m$. 
For a uniform group distribution, an entry in the Fourier moment of order $r\ge 2$ can be nonzero only when the indices satisfy the frequency constraint (see Proposition~\ref{prop:moment_expression})
\[
    i_r \equiv i_1+\cdots+i_{r-1} \pmod d.
\]
Because the spectrum of $\x^{(Dj)}$ is supported only on $\{m,d-m\}$, every nonzero contribution must have
\[
    i_j=\sigma_j m,
    \qquad
    \sigma_j\in\{+1,-1\},
\]
where $d-m$ is identified with $-m$ modulo $d$.  The frequency constraint then becomes
\[
    (\sigma_1+\cdots+\sigma_{r-1}-\sigma_r)m\equiv 0 \pmod d,
\]
or equivalently
\[
    \sigma_1+\cdots+\sigma_{r-1}-\sigma_r\equiv 0 \pmod{d/m}.
\]
If $r<d/m$, the integer
\[
    \sigma_1+\cdots+\sigma_{r-1}-\sigma_r
\]
has absolute value at most $r<d/m$.  Hence the above congruence forces
\[
    \sigma_1+\cdots+\sigma_{r-1}-\sigma_r=0.
\]
The corresponding Fourier product is therefore
\begin{align*}
\F\x^{(Dj)}[i_1]\cdots\F\x^{(Dj)}[i_{r-1}]\overline{\F\x^{(Dj)}[i_r]} =\exp\Big(
        -\frac{\ri mj\pi}{d}(\sigma_1+\cdots+\sigma_{r-1}-\sigma_r)\Big)=1,
\end{align*}
independently of $j$.  Thus every nonzero Fourier moment entry of order
$r<d/m$ is identical for $\x^{(D0)}$ and $\x^{(D1)}$, and the zero entries are also
identical.  The first moment also agrees, since both signals have zero DC
coefficient.  Consequently,
\[
    M_r(\x^{(D0)},\rho_0)=M_r(\x^{(D1)},\rho_0),
    \qquad
    1\le r\le d/m-1.
\]
At order $d/m$, however, the parity becomes visible.  Consider the distribution-free Fourier moment quantity of order $(d/m)$:
\begin{align*}
&\widetilde{M}_{d/m}\big(\x,(\underbrace{m,m,\ldots,m}_{(d/m-1) \ \text{entries}}) \big)\\
&\triangleq\F M_{d/m}(\x,\rho)[\underbrace{m,\ldots,m}_{(d/m-1) \ \text{entries}},d-m]
+ \F M_{d/m}(\x,\rho)[\underbrace{-m,\ldots,-m}_{(d/m-1) \ \text{entries}},m]\\
&=2\Re\big(
        (\F\x[m])^{d/m}\big)
\end{align*}
For the two signals above,
\[
\widetilde{M}_{d/m}\big(\x^{(Dj)},(\underbrace{m,m,\ldots,m}_{(d/m-1) \ \text{entries}}) \big)
    =2(-1)^j.
\]
Therefore, in this very sparse non-generic example, moments of order less than $d/m$ cannot distinguish the two parity classes, while a moment of order $d/m$ does. This does not contradict the fourth-order characterization below, because the present signals violate the generic nonvanishing condition away from $d/2$: all Fourier coefficients in $\cI_{m,d}^+\setminus\{m\}$ vanish.
\end{remark}

The preceding discussion motivates the following population characterization. The proof is deferred to Section~\ref{sec:proof_DMRA_mean_null_parity_split}. Similar to \eqref{eq:chain_set} in the collapsed regime, we introduce the set of chain tuples
\begin{align}\label{eq:chain_set_smaller}
\cJ_{m,d}^{<} \triangleq \Big\{ (\ell_1,\ell_2,\ell_3) = (0, m, km) : 1 \leq k \leq  \frac{d}{2m}-2  \Big\}\subset \cI_{m,d}^3,
\end{align}
with convention that $\cJ_{m,d}^{<}=\varnothing$ when $d/m=4$. Note that, compared with \eqref{eq:chain_set}, we do not include the tuple corresponding to $k=d/(2m)-1$. This is because the remaining phase relation is recovered from the signed fourth-order constraint, so the unsigned and signed constraints together provide a complete set of phase constraints.

\begin{proposition}[Population null constraints in the parity-split regime]\label{prop:DMRA_mean_null_parity_split}
Let $m\leq d/4$ be any divisor of $d$, and assume the parity-split regime, i.e., $d/m$ is even. Let $\x\in\R^d$ be a generic $\Cd{m}$ signal, except that the coordinate $\F\x[d/2]$ is allowed to vanish. More precisely, assume the zero pattern of $\Cd{m}$-invariant signal,
\begin{align}
s_\ell=|\F\x[\ell]|^2=0, \quad &\text{for all } \ell\in \cI_{m,d}^{c,+},\label{eq:cyclic_parity_split}
\end{align}
and the generic condition
\begin{align}
s_{km}=|\F\x[km]|^2\neq 0, \quad &\text{for all } k=0,1,\ldots,d/(2m)-1\label{eq:generic_parity_split}
\end{align}
hold. Then we have
\begin{itemize}
    \item $\x$ has $\Dm{m}^0$-symmetry if and only if
    \begin{equation}\label{eq:D0_condition}
    \begin{aligned}
    &4s_{\ell_1}s_{\ell_2}s_{\ell_3}s_{\ell_4^{\sharp}(t)} - d_t^2 =0,
    \quad\text{for all } t\in \cJ_{m,d}^{<},\\
    &\kappa_m^0 =0.
    \end{aligned}
    \end{equation}
    \item $\x$ has $\Dm{m}^1$-symmetry if and only if
    \begin{equation}\label{eq:D1_condition}
    \begin{aligned}
    &4s_{\ell_1}s_{\ell_2}s_{\ell_3}s_{\ell_4^{\sharp}(t)} - d_t^2=0\quad\text{for all }t\in \cJ_{m,d}^{<},\\
    &\kappa_m^1=0, \qquad s_{d/2}=0.
    \end{aligned}
    \end{equation}
\end{itemize}
\end{proposition}

\subsubsection{Raw p-value calculation for single dihedral hypothesis in the parity-split regime}\label{sec:p-value_calculation_DMRA_parity_split}

Recall that the (unsigned) fourth-order constraint statistics
\begin{align*}
\widehat{H}_t=4\widehat{S}_{\ell_1(t)} \widehat{S}_{\ell_2(t)} \widehat{S}_{\ell_3(t)} \widehat{S}_{\ell_4^{\sharp}(t)}-\widehat{D}_t^2,\quad\text{for }t\in\cJ_{m,d}^{<}.
\end{align*}
Additionally, define the signed fourth-order constraint statistics
\begin{align*}
\widehat{K}_m^0\triangleq2\widehat{S}_m\widehat{S}_{d/2-m}-\widehat{D}_{t_\star}, \qquad \widehat{K}_m^1\triangleq 2\widehat{S}_m\widehat{S}_{d/2-m}+\widehat{D}_{t_\star},
\end{align*}
where $t_\star=(m,m,d/2-m)$. The test vectors for the two parity-split nodes are
\begin{align}
\widehat{\g}_m^0\triangleq\Bigl( (\widehat{S}_\ell)_{\ell\in \cI^{c,+}_{m,d}},(\widehat{H}_t)_{t\in \cJ^{<}_{m,d}},\widehat{K}_m^0 \Bigr)^\T,\qquad q_m^0\triangleq |\cI^{c,+}_{m,d}|+\frac{d}{2m}-1,\label{eq:q0}
\end{align}
and
\begin{align}
\widehat{\g}_m^1\triangleq\Bigl( (\widehat{S}_\ell)_{\ell\in \cI^{c,+}_{m,d}}, \widehat{S}_{d/2},
(\widehat{H}_t)_{t\in \cJ^{<}_{m,d}},\widehat{K}_m^1 \Bigr)^\T, \qquad q_m^1\triangleq |\cI^{c,+}_{m,d}|+\frac{d}{2m}.\label{eq:q1}
\end{align}
For each test vector, the first block enforces the cyclic zero pattern, the $\widehat{H}_t$ block enforces phase propagation away from the $d/2$-th coordinate, and the signed constraint part $\widehat{K}_m^0$ or $\widehat{K}_m^1$ closes the parity. The additional $\widehat{S}_{d/2}$ appears only for the second test vector because $\Dm{m}^1$-symmetry forces $s_{d/2}=0$ whereas $\Dm{m}^0$-symmetry allows this coordinate to be arbitrary.

Define the covariance matrices
\begin{align}\label{eq:Omega_g01}
\bOmega_{\g}^0 \triangleq \Var(\bpsi_j^0), \qquad \text{and}\qquad \bOmega_{\g}^1 \triangleq \Var(\bpsi_j^1).
\end{align}
Here $\bpsi_j^0\in\R^{q_m^0}$ has the same coordinates as \eqref{eq:psi_ell} and \eqref{eq:psi_t}:
\begin{align}
\psi_{\ell,j}^0  \triangleq & s_{\ell,j} - s_\ell,\quad\text{for }\ell\in\cI_{m,d}^{c,+},\label{eq:psi_ell0}\\
\psi_{t,j}^0 \triangleq& 4\Big(
s_m s_{km} s_{((k+1)m)}(s_{0,j}-s_0)+
s_0 s_{km} s_{((k+1)m)}(s_{m,j}-s_m)\notag\\
&+
s_0 s_m s_{((k+1)m)}(s_{km,j}-s_{km})+
s_0 s_m s_{km}(s_{((k+1)m),j}-s_{((k+1)m)})
\Big)\notag\\
&-2 d_t(d_{t,j}-d_t),\quad\text{for }t=(0,m,km)\in\mathcal{J}_{m,d}^{<},\label{eq:psi_t0}
\end{align}
and additionally has the signed parity coordinate
\begin{align}
\psi_{t_{\star},j}^0 \triangleq& 2\Big(s_{d/2-m}(s_{m,j}-s_m)+s_m(s_{d/2-m,j}-s_{d/2-m})\Big) - (d_{t_\star,j}-d_{t_\star}).\label{eq:psi_tstar0}
\end{align}
Likewise, $\bpsi_j^1\in\R^{q_m^1}$ has the same coordinates as \eqref{eq:psi_ell} and \eqref{eq:psi_t}:
\begin{align}
\psi_{\ell,j}^1  \triangleq & s_{\ell,j} - s_\ell,\quad\text{for }\ell\in\cI_{m,d}^{c,+},\label{eq:psi_ell1}\\
\psi_{t,j}^1 \triangleq& 4\Big(
s_m s_{km} s_{((k+1)m)}(s_{0,j}-s_0)+
s_0 s_{km} s_{((k+1)m)}(s_{m,j}-s_m)\notag\\
&+
s_0 s_m s_{((k+1)m)}(s_{km,j}-s_{km})+
s_0 s_m s_{km}(s_{((k+1)m),j}-s_{((k+1)m)})
\Big)\notag\\
&-2 d_t(d_{t,j}-d_t),\quad\text{for }t=(0,m,km)\in\mathcal{J}_{m,d}^{<},\label{eq:psi_t1}
\end{align}
additionally has the $d/2$-th coordinate
\begin{align}\label{eq:psi_d21}
\psi_{d/2,j}^1\triangleq s_{d/2,j}-s_{d/2},
\end{align}
and the signed parity coordinate
\begin{align}\label{eq:psi_tstar1}
\psi_{t_{\star},j}^1 \triangleq& 2\Big(s_{d/2-m}(s_{m,j}-s_m)+s_m(s_{d/2-m,j}-s_{d/2-m})\Big) + (d_{t_\star,j}-d_{t_\star}).
\end{align}
Moreover, we define the empirical counterparts $\widehat{\bpsi}_j^0\in\R^{q_m^0}$and $\widehat{\bpsi}_j^1\in\R^{q_m^1}$ by replacing the population quantities $(s_\ell, d_t)$ with their estimators $(\widehat{S}_\ell, \widehat{D}_t)$, while keeping the sample-level terms above unchanged. Finally, we define the sample covariance matrices
\begin{align}\label{eq:Omegahat_g01}
\widehat{\bOmega}_\g^0\triangleq \frac{1}{N}\sum_{j=1}^N \widehat{\bpsi}_j^0 (\widehat{\bpsi}_j^0)^\T,\qquad\text{and}\qquad  \widehat{\bOmega}_\g^1\triangleq \frac{1}{N}\sum_{j=1}^N \widehat{\bpsi}_j^1 (\widehat{\bpsi}_j^1)^\T.
\end{align}
We can now introduce similar Wald tests for $\Dm{m}^0$-symmetry and $\Dm{m}^1$-symmetry as in Theorem~\ref{thm:DMRA_Dm_testing}, and the proof of the following theorem is given in Section~\ref{sec:proof_DMRA_Dm_testing_parity_split}.

\begin{theorem}\label{thm:DMRA_Dm_testing_parity_split}
Let $d/m\geq 4$ be an even integer. Assume the parity-split regime in the D-MRA setting. Under the null hypothesis that $\x\in\R^d$ belongs to the symmetry class $\Dm{m}^{\epsilon}$,
\begin{align*}
\sqrt{N}\cdot \widehat{\g}_m^{\epsilon}\xrightarrow{d} \mathcal{N}(\0,\bOmega_{\g}^{\epsilon}),\qquad\text{for }\epsilon\in\{0,1\}.
\end{align*}
If $\x$ is a further generic $\Dm{m}^{\epsilon}$ signal in the sense of Lemma~\ref{lem:Jacobian_full_rankness_parity_split}(i), then $\bOmega_{\g}^{\epsilon}$ is positive definite, and $\widehat{\bOmega}_{\g}^{\epsilon} \xrightarrow{p} \bOmega_{\g}^{\epsilon}$. Consequently, the Wald statistics
\begin{align}
\widehat{W}_m^{\epsilon}\triangleq N \cdot (\widehat{\g}_m^{\epsilon})^\T (\widehat{\bOmega}_\g^{\epsilon})^{-1} \widehat{\g}_m^{\epsilon} \xrightarrow{d} \chi^2_{q_m^{\epsilon}}, \qquad\text{for }\epsilon\in\{0,1\}.\label{eq:chi_g01}
\end{align}
Thus, asymptotically valid (raw) $p$-values for testing the null hypotheses of $\Dm{m}^0$-symmetry and $\Dm{m}^1$-symmetry are given by
\begin{align}\label{eq:raw_pvalue_dihedral_parity_split}
p_m^{D0}\triangleq 1- F_{\chi_{q_m^0}^2} (\widehat{W}_m^0),\qquad p_m^{D1}\triangleq 1- F_{\chi_{q_m^1}^2} (\widehat{W}_m^1),
\end{align}
respectively. Moreover, under any fixed alternative where at least one of the conditions in \eqref{eq:cyclic_parity_split} and \eqref{eq:D0_condition} fails, and the condition in Lemma~\ref{lem:Jacobian_full_rankness_parity_split}(ii) holds, we have
\begin{align*}
\widehat{W}_m^0\xrightarrow{p} \infty. 
\end{align*}
Similarly, under any fixed alternative where at least one of the conditions in \eqref{eq:cyclic_parity_split} and \eqref{eq:D1_condition} fails, and the condition in Lemma~\ref{lem:Jacobian_full_rankness_parity_split}(ii) holds, we have
\begin{align*}
\widehat{W}_m^1\xrightarrow{p} \infty. 
\end{align*}
In particular, the Wald tests based on $\widehat{W}_m^0$ and $\widehat{W}_m^1$ are consistent against any such fixed alternative, respectively.
\end{theorem}

\subsubsection{Family-wise error control for the dihedral procedure}\label{sec:FWER_dihedral}

The raw $p$-values of the previous two subsections are fed to the sequential rejection algorithm of Appendix~\ref{sec:SRA}, exactly as in the cyclic case. However, the nature of the guarantee differs. The cyclic $p$-values are exactly valid in finite samples for every $\rho$ (Theorem~\ref{thm:test_size_power}), so the cyclic procedure of Section~\ref{sec:Detection_PMRA} controls the FWER strongly and in finite samples. The dihedral $p$-values are only asymptotically valid, and only on nondegenerate signals: Theorems~\ref{thm:DMRA_Dm_testing} and~\ref{thm:DMRA_Dm_testing_parity_split} require $\x$ to be a generic $\Dm{m}$ (resp.\ $\Dm{m}^\epsilon$) signal in the sense of Definition~\ref{def:generic_signals}. This matters because a signal typically satisfies several nested null hypotheses and is degenerate for the smaller ones. For instance, an exact $\Dm{9}$-symmetric signal with $d=81$ is also $\Dm{3}$- and $\Dm{1}$-invariant, but it has structural zeros at the multiples of $3$ that are not multiples of $9$; at the vertex $\Dm{3}$ every chain tuple $(0,3,3k)$ then has $s_3=0$ and $d_t=0$, all dihedral rows of the Jacobian in Lemma~\ref{lem:Jacobian_full_rankness} vanish, $\sqrt N\,\widehat H_t\to0$ in probability, and the $\chi^2$ calibration of $\widehat W_3$ is not available. The following proposition identifies exactly which true hypotheses must be nondegenerate for the sequential rejection argument to go through.

Fix $\x$ and let $\mathcal T(\x)\subseteq\mathcal H$ denote the set of vertices whose invariance hypothesis is true for $\x$, and $\mathcal F(\x)\triangleq\mathcal H\setminus\mathcal T(\x)$. Since edges encode implications, $\mathcal T(\x)$ is closed under taking descendants and $\mathcal F(\x)$ under taking ancestors. Let $H_1$ denote the root. Define the reachable and maximal true vertices by
\begin{align*}
\mathcal T_1(\x)
&\triangleq
\{H\in\mathcal T(\x):
H=H_1 \text{ or }
\mathrm{parent}(H)\cap\mathcal F(\x)\neq\varnothing\},\\
\mathcal T_{\max}(\x)
&\triangleq
\{H\in\mathcal T(\x):
\mathrm{parent}(H)\subseteq\mathcal F(\x)\}
\subseteq\mathcal T_1(\x).
\end{align*}
A true root belongs to both sets because its parent set is empty. Before the first rejection of a true hypothesis, positive critical values can occur only at vertices in $\mathcal T_1(\x)$ under the any-parent rule, and in $\mathcal T_{\max}(\x)$ under the all-parents rule.

\begin{proposition}[Asymptotic FWER control for the dihedral procedure]\label{prop:FWER_dihedral}
Consider the DAG of Section~\ref{sec:DAG_construction_dihedral}, where the cyclic vertices, the root $\Dm{d}$, and (for even $d$) the vertex $\Dm{d/2}^0$ are tested by second-order statistics whose $p$-values are valid in finite samples for every $\rho$ (Theorem~\ref{thm:test_size_power}), and every other dihedral vertex is tested by the Wald $p$-value of Theorem~\ref{thm:DMRA_Dm_testing} or~\ref{thm:DMRA_Dm_testing_parity_split}. Fix a signal $\x\in\R^d$, a distribution $\rho$ on $\Dm{d}$, and a level $\alpha\in(0,1)$, and run Algorithm~\ref{alg:SR} on $N$ observations.
\begin{enumerate}
\item[(i)] If $\x$ is nondegenerate at every Wald-tested
dihedral vertex in $\mathcal T_1(\x)$, in the sense of
Lemma~\ref{lem:Jacobian_full_rankness}(i) or
Lemma~\ref{lem:Jacobian_full_rankness_parity_split}(i),
as applicable, then
\[
\limsup_{N\to\infty}\ \mathbb P\big(\text{some } H\in\mathcal T(\x)\text{ is rejected}\big)\ \le\ \alpha .
\]
\item[(ii)] If Algorithm~\ref{alg:SR} uses $\alpha_H^{\mathrm{all}}$ from~\eqref{eq:parent_rule_thresholds}, the same conclusion holds whenever $\x$ is nondegenerate, in the same sense, at every Wald-tested dihedral vertex in $\mathcal T_{\max}(\x)$. If $\x$ has exact symmetry group $\Dmj{m'}{j}$ with $m'\le d/3$, then $\mathcal T_{\max}(\x)$ consists of the single vertex $\Dm{m'}$ (resp.\ $\Dm{m'}^{\,j \bmod 2}$), so that (ii) holds whenever $\x$ is a generic member of its own exact symmetry class.
\item[(iii)] If $\mathcal T(\x)$ contains no dihedral vertex tested by a Wald statistic, in particular whenever $\x$ has no dihedral symmetry, then the FWER is at most $\alpha$ in finite samples, for every $N$ and every $\rho$.
\end{enumerate}
\end{proposition}

The proof is given in Appendix~\ref{sec:proof_FWER_dihedral}. It is the sequential rejection argument of \cite{goeman2010sequential}: on the event that no true hypothesis has a $p$-value below its critical value computed at the rejection set $\mathcal F(\x)$, the procedure never rejects a true hypothesis, and the union bound over the complement of that event only involves the true vertices with positive critical values, namely $\mathcal T_1(\x)$ under the any-parent rule and $\mathcal T_{\max}(\x)$ under the all-parents rule.

\begin{remark}[Nested nulls and structural zeros]\label{rmk:nested_nulls}
Proposition~\ref{prop:FWER_dihedral} is weaker than the finite-sample strong control enjoyed by the cyclic procedure in two respects, and we do not claim strong FWER control for the dihedral procedure beyond it. First, the guarantee is asymptotic, as the nodewise dihedral tests are. Second, the nondegeneracy condition in (i) can fail because of nesting. For the $\Dm{9}$ signal with $d=81=3^4$, the dihedral vertices form a chain $\Dm{81}\to\Dm{27}\to\Dm{9}\to\Dm{3}\to\Dm{1}$, so the degenerate true vertices $\Dm{3}$ and $\Dm{1}$ have a true parent only, are not reachable, and (i) reduces to genericity of $\x$ as a $\Dm{9}$ signal; more generally, when $d$ is an odd prime power the only reachable true dihedral vertex is the one of the exact class of $\x$, and (i) coincides with (ii). When $d$ has two distinct prime factors, however, reachable degenerate true vertices do occur: for $d=45$ and an exact $\Dm{9}$ signal, the vertex $\Dm{3}$ has the false parent $\Dm{15}$, and the vertex $\Dm{1}$ has the false parent $\Dm{5}$; both are true, and both are degenerate, since $\F\x[3]=\F\x[1]=0$. The same mechanism operates in the parity-split regime: for $d=36$ and an exact $\Dm{9}$ signal of parity $\epsilon$, the true vertex $\Dm{3}^{\epsilon}$ has the false parent $\Dm{6}^{\epsilon}$ and is degenerate. In such configurations, the any-parent procedure tests a degenerate true dihedral hypothesis whose Wald statistic has no established null distribution, and (i) does not apply. Two observations mitigate this. The cyclic vertices are exactly valid regardless, so false rejections of cyclic hypotheses remain controlled. The all-parents rule of (ii) removes the issue, at the cost of power, since a vertex is then tested only after all of its parents have been rejected. Note that degeneracy need not indicate additional symmetry: a signal supported on $\{\pm m\}$ with a reflection-consistent phase can have exact symmetry $\Dmj{m}{j}$ while violating the nonvanishing conditions of the Wald test. Thus, the all-parents guarantee still requires nondegeneracy within the signal's exact symmetry class.
\end{remark}

\subsection{Numerical Demonstration in the Dihedral Case}
\label{sec:numerical_demonstration}

We evaluate the proposed DAG sequential rejection algorithm for Dihedral MRA through a numerical demonstration in the collapsed regime defined in Section~\ref{sec:p-value_calculation_DMRA_collapsed}. We set the signal length to $d=81$ and test for $D_m$-symmetry with $m=9$. Since $d/m=9$ is odd, the cyclic and dihedral frequency conditions do not overlap. Therefore, in this case the full feature vector can be formed by directly concatenating the two families of constraints. The hypothesis DAG is the odd-$d$ construction of Section~\ref{sec:DAG_construction_dihedral} (Case~1): it has one dihedral vertex $D_{m'}$ for every divisor $m'\mid d$ and one cyclic vertex $C_{m'}$ for every divisor $1<m'<d$, with a directed edge between two vertices of the same type whose orders differ by a single prime factor, together with an edge $D_{m'}\to C_{m'}$ encoding that $D_{m'}$-symmetry implies $C_{m'}$-symmetry. Note that the numerical implementation we present uses a singular-value--truncated approximation to the theoretical Wald statistic; its calibration and the distinction from Theorem~\ref{thm:DMRA_Dm_testing} are described in Remark~\ref{rmk:num_stab}.

We consider a balanced two-class discrimination task for the combined cyclic and dihedral testing pipeline. A ground-truth $D_9$-symmetric signal $x_{\mathrm{sym}} \in \mathbb{R}^{81}$ is generated by drawing a random Gaussian vector and symmetrizing it over the $D_9$ action, namely by summing over the nine cyclic shifts by multiples of $m=9$ together with their reflected copies, and then normalizing to unit average energy per entry ($\|x_{\mathrm{sym}}\|_2^2=d$). To probe the procedure's ability to reject symmetry that is nearly, but not exactly, present, we also construct an asymmetric decoy $x_{\mathrm{asym}} = x_{\mathrm{sym}} + \epsilon\,\xi$, where $\xi$ is a fresh Gaussian perturbation and $\epsilon=0.4$ controls the deviation from $D_9$ symmetry. The decoy is renormalized to the same energy as $x_{\mathrm{sym}}$, so the two classes cannot be distinguished by signal energy and share an identical SNR at every noise level; with this normalization, $\mathrm{SNR}=\|\x\|_2^2/(d\sigma^2)=1/\sigma^2$ in~\eqref{eq:SNR}. The perturbation generally changes both the cyclic zero pattern and the reflection constraints. Equal total energy does not imply equal power spectra, so this comparison does not isolate the contribution of the dihedral phase statistics. A dedicated reflection-discrimination experiment would compare a cyclic signal and a dihedral signal with the same power spectrum and different Fourier phases.

For each signal-to-noise ratio (SNR) we run independent trials; in each trial, the ground-truth signal is drawn uniformly at random from $\{x_{\mathrm{sym}}, x_{\mathrm{asym}}\}$, we generate $N=100{,}000$ noisy observations under a randomly drawn (non-uniform) distribution over the $2d$ group elements, and we apply the full DAG sequential-rejection procedure to all relevant nodes. A trial is scored as a success if the algorithm declares $D_9$ when the input is $x_{\mathrm{sym}}$ (true positive) and declares any label other than $D_9$ when the input is $x_{\mathrm{asym}}$ (true negative); otherwise it is scored as a classification error. This binary classification score is distinct from the family-wise rejection error assessed below. The reported success rate is the fraction of correct classifications, averaged over $500$ trials per SNR level.

For each tested node, we compute the empirical feature vector
\[
    \hat{g}_m
    =
    \left(
    \left(\hat{S}_l\right)_{l \in \mathcal{I}_{m,d}^{c,+}},
    \left(\hat{H}_t\right)_{t \in \mathcal{J}_{m,d}}
    \right),
\]
where $\hat{S}_l$ are the second-order cyclic power-spectrum features and $\hat{H}_t$ are the fourth-order dihedral phase features. The former test the cyclic zero pattern, while the latter test the additional reflection constraints required for dihedral invariance.

\begin{remark}[Numerical stabilization and calibration] \label{rmk:num_stab}
In finite samples, direct evaluation of the Wald statistic may be unstable because the second- and fourth-order features have different scales and may be strongly correlated. We therefore standardize the features and use a singular-value truncated pseudoinverse.

Let $\hat D$ be the diagonal matrix of empirical feature standard deviations, and define
\[
\tilde g_m=\hat D^{-1}\hat g_m,
\qquad
\hat R_g=\hat D^{-1}\hat\Omega_g\hat D^{-1}.
\]
Since $\hat R_g$ is symmetric positive semidefinite, we write
\[
\hat R_g=\sum_{i=1}^{q_m}\hat\sigma_i\hat u_i\hat u_i^T,
\qquad
\hat\sigma_1\geq\cdots\geq\hat\sigma_{q_m}\geq0,
\]
we fix $\tau=10^{-6}$ and set
\[
r_m=\#\{i:\hat\sigma_i>\tau\hat\sigma_1\},
\qquad
\hat R_{g,\tau}^{+}
=
\sum_{\hat\sigma_i>\tau\hat\sigma_1}
\hat\sigma_i^{-1}\hat u_i\hat u_i^T.
\]
The statistic used in the numerical experiments is
\[
\hat W_{m,\tau}
=
N\tilde g_m^T\hat R_{g,\tau}^{+}\tilde g_m.
\]
If no eigenvalues are truncated, standardization is only an invertible reparameterization, and the statistic coincides with the ordinary Wald statistic covered by Theorem~\ref{thm:DMRA_Dm_testing}. When truncation is active, however, the retained subspace and rank $r_m$ are data-dependent, so Theorem~\ref{thm:DMRA_Dm_testing} does not formally establish its null distribution. We use $\chi^2_{r_m}$ as an effective-rank approximation for numerical calibration. Thus, the corresponding p-values and DAG decisions should be interpreted as stabilized numerical approximations rather than as inheriting the formal asymptotic calibration or FWER guarantee of the untruncated statistic.
\end{remark}

We use $\chi^2_{r_m}$ as an approximate reference distribution for the stabilized statistic and pass the resulting $p$-values to the DAG sequential rejection algorithm.

\begin{figure}[htp]
    \centering
    \includegraphics[width=0.5\textwidth]{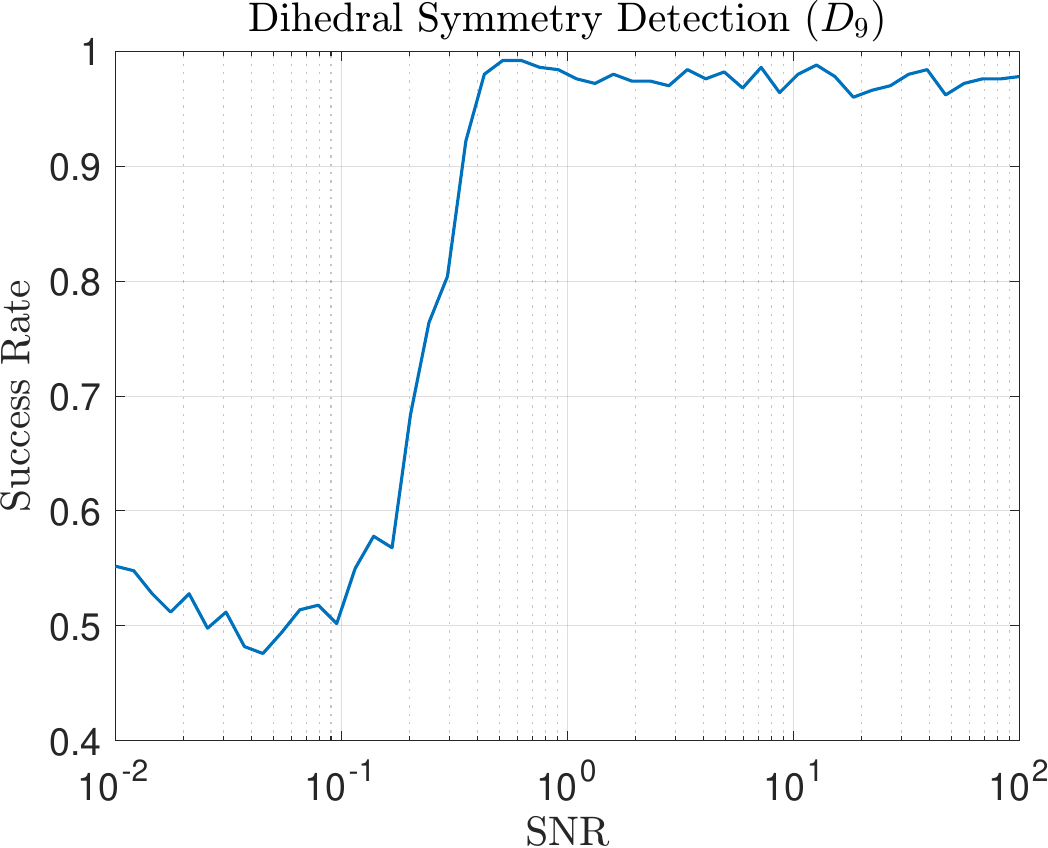}
    \caption{Numerical demonstration of Dihedral MRA in the collapsed regime with $d=81$ and $m=9$. The plot shows the empirical accuracy of the combined cyclic and dihedral DAG procedure in classifying inputs drawn equally from a $D_9$-symmetric signal and an equal-energy asymmetric decoy, over a logarithmic range of SNR values. The accuracy is close to the chance level of $0.5$ at very low SNR and approaches one at higher SNR.}
    \label{fig:success_rate_d9}
\end{figure}

\begin{figure}[htbp]
    \centering
    \includegraphics[width=0.9\textwidth]{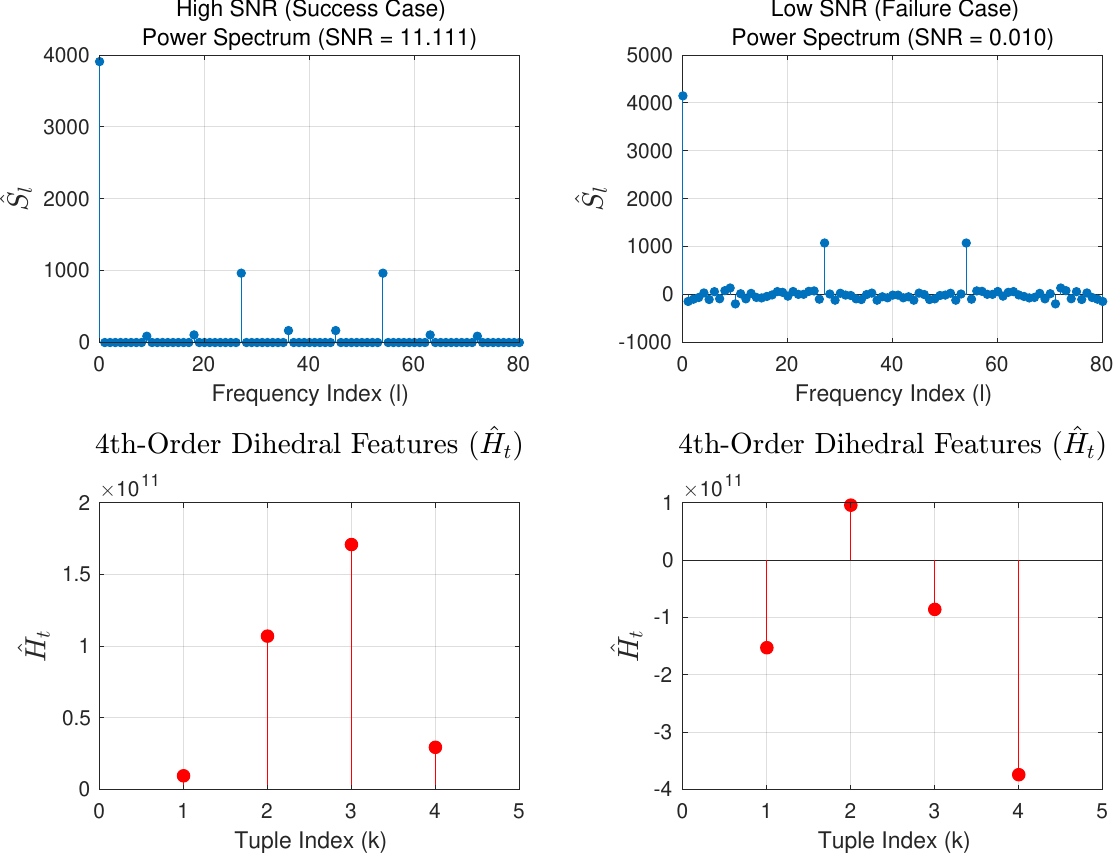}
    \caption{Feature magnitudes for the $D_9$ numerical demonstration at high and low SNR. The left column corresponds to a high-SNR success case with $\mathrm{SNR}=100$, and the right column corresponds to a low-SNR failure case with $\mathrm{SNR}=0.010$. The top row displays the empirical second-order power-spectrum features $\hat{S}_l$, while the bottom row displays the fourth-order dihedral features $\hat{H}_t$. The lower panels show substantial changes in the fourth-order constraint features between these two noise levels.}
    \label{fig:features_d9}
\end{figure}

Figure~\ref{fig:success_rate_d9} shows the empirical classification accuracy across $30$ logarithmically spaced SNR values ranging from $10^{-2}$ to $10^2$. At very low SNR, the accuracy is close to the chance level of $0.5$, indicating that the implemented procedure does not reliably distinguish the two inputs. As the SNR increases, the accuracy undergoes a sharp transition and approaches one for the chosen signal pair.

Figure~\ref{fig:features_d9} displays representative empirical features at high and low SNR. The second-order features preserve the dominant power-spectrum peaks in these snapshots, while the fourth-order constraint features depart substantially from their high-SNR values. These plots illustrate the variability of the estimated feature vectors across noise levels.

\paragraph{Empirical family-wise error rate.} We assess the finite-sample calibration of the implemented DAG procedure, which combines exact cyclic tests with the stabilized dihedral statistics of Remark~\ref{rmk:num_stab}. A \emph{family-wise error} is the event that the procedure rejects at least one hypothesis that is in fact true for the signal, that is, one of its genuine invariances. We consider two ground-truth signals of length $d=81$: a $\Cd{9}$-symmetric signal and a $\Dm{9}$-symmetric signal, constructed as above. For the $\Cd{9}$ signal, the true hypotheses are the nodes $\{\Cd{9},\Cd{3}\}$, and for the $\Dm{9}$ signal they are $\{\Dm{9},\Dm{3},\Dm{1},\Cd{9},\Cd{3}\}$; a family-wise error occurs whenever any of these is rejected. 

To probe the distribution-free property, we run each configuration under two transformation distributions $\rho$ over the $2d$ dihedral group elements: the uniform distribution and a strongly nonuniform, sharply peaked one. For every SNR, we draw $N=2\times 10^4$ observations, run the full procedure at nominal level $\alpha=0.05$, and estimate the empirical FWER over $T=500$ independent repetitions (Monte Carlo standard error $\approx 0.01$).

\begin{figure}[htbp]
    \centering
    \includegraphics[width=0.6\textwidth]{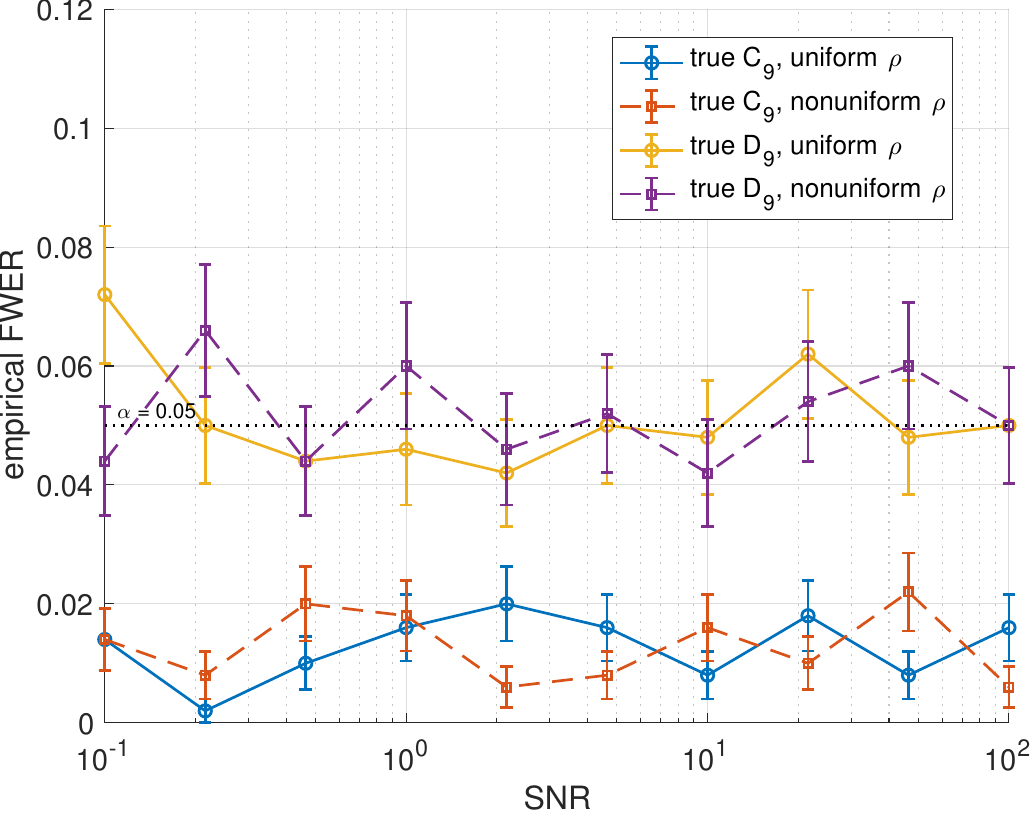}
    \caption{Empirical FWER of the implemented DAG procedure for $\Cd{9}$- and $\Dm{9}$-symmetric signals ($d=81$), under uniform and strongly nonuniform transformation distributions. The procedure uses exact cyclic tests and stabilized dihedral statistics (Remark~\ref{rmk:num_stab}). Each point is based on $500$ repetitions with $N=2\times10^4$ observations; error bars show $\pm1$ Monte Carlo standard error. The dotted line marks the nominal level $\alpha=0.05$.}
    \label{fig:fwer_d9}
\end{figure}

Figure~\ref{fig:fwer_d9} shows that the empirical FWER for the $\Cd{9}$ configuration remains below the nominal level, while the $\Dm{9}$ curves fluctuate around $\alpha=0.05$. The uniform and strongly nonuniform sampling distributions yield similar results. These experiments support the numerical calibration of the stabilized procedure in the settings examined; Proposition~\ref{prop:FWER_dihedral} applies to the untruncated Wald procedure and does not establish FWER control for the truncated implementation.

\section{Outlook: Towards symmetry detection in cryo-EM} \label{sec:conclusion}

Our MRA results motivate a possible symmetry-detection pipeline for cryo-EM based on low-order invariant statistics, tests of candidate symmetry classes, and DAG-structured multiple testing. This section records the three-dimensional imaging model and identifies the additional steps needed to develop such a procedure. The tests and error guarantees established in Sections~\ref{sec:cyclic} and~\ref{sec:dihedral} concern full-observation MRA; no symmetry test for the projected cryo-EM model is constructed or validated here. The reconstructions in Figure~\ref{fig:wrong_prior} illustrate the consequences of symmetry assignment rather than evaluate a cryo-EM implementation of our tests.

\subsection{Imaging model and steerable expansions}
\label{sec:cryoem-model}

Let $V\colon\mathbb{R}^3\to\mathbb{R}$ denote the (unknown) Coulomb potential of a macromolecule. A cryo-EM micrograph yields particle images that, after picking and extraction, we idealize as
\begin{equation}
I_t(\mathbf{r})
= \mathcal{C}_t \circ \mathcal{P} \big(R_{t}V\big) 
\;+\; \varepsilon_t(\mathbf{r}),
\quad \mathbf{r}\in\mathbb{R}^2,
\label{eqn:proj_model}
\end{equation}
where $\mathcal{P}$ is a tomographic projection along the Z-axis, $\mathcal{C}_t$ is the (known per-particle or per-micrograph) CTF operator, $R_{t}$ is an unknown rotation, distributed uniformly in-plane, and $\varepsilon_t$ is additive noise. We assume centering has been performed (e.g., see~\cite{heimowitz2021centering}) and that images are normalized and noise-whitened within defocus groups. Note that in a more advanced setting, one should account for the possibility of small-magnitude 2-D shifts.

An essential question is the relation between the unknown 3-D volume and the 2-D data, where the Fourier Slice Theorem plays a significant role~\cite{singer2018mathematics}. Let \(\F{V}(r,\theta,\phi)\) be the three-dimensional Fourier transform of the function \(V\). This 3-D volume can be expressed using Fourier-Bessel~\cite{kileel2025fast} or 3-D steerable PCA~\cite{fraiman2026so3pca} as,
\begin{equation} \label{eq:2.1}
\F{V}(r,\theta,\phi) = \sum_{s=1}^\infty \sum_{l=0}^\infty \sum_{m=-l}^l A_{lms} \Psi_{lms}(r,\theta,\phi),
\end{equation} 
where The basis functions are called ball harmonics~\cite{kileel2025fast}, 
\begin{equation} \label{eq:2.2}
\Psi_{lms}(r,\theta,\phi) = j_{ls}(r)Y_{lm}(\theta,\phi).
\end{equation}
Here, \(Y_{lm}(\theta,\phi)\) represents the spherical harmonics, which describe the angular dependence of the function, and the radial component is a variant of the spherical Bessel function of order $l$.

A common approach to represent the projection images, after being picked from the microscope's micrographs, is to use a rotationally separable (steerable) basis on the particle support, e.g., Fourier-Bessel or 2-D steerable PCA~\cite{zhao2016fast}. Then, in polar coordinates $(r,\theta)$, we also expand the images in Fourier,
\begin{equation}
\F I_t(r,\theta) = \sum_{\ell\in\mathbb{Z}_+} \sum_{q=1}^{Q(\ell)} a_{t}(\ell,q)\,\psi_{\ell q}(r)\,e^{i\ell\theta},
\quad 
\psi_{\ell q}(r)\in\mathbb{R},
\label{eqn:steerable-expansion}
\end{equation}
with $Q(\ell)$ being the number of radial channels in the bandlimit. We define the coefficient vectors $\mathbf{a}_t(\ell) = \big(a_t(\ell,1),\dots,a_t(\ell,Q(\ell))\big)^\top\in\mathbb{C}^{Q(\ell)}$. Since in-plane rotations act as $a_{t}(\ell,q)\mapsto e^{i\ell\alpha_t}a_{t}(\ell,q)$, the population second moment decomposes into angular-frequency blocks,
\begin{equation}
\Sigma_\ell = \mathbb{E}\!\left[\,\mathbf{a}_t(\ell)\,\mathbf{a}_t(\ell)^{\!H}\,\right] \in \mathbb{C}^{Q(\ell)\times Q(\ell)},
\quad
\mathbb{E}\!\left[\,\mathbf{a}_t(\ell)\,\mathbf{a}_t(\ell')^{\!H}\,\right]=\mathbf{0}\ \text{for }\ell\neq \ell'.
\label{eqn:block-cov}
\end{equation}
Below, expectations are over the latent viewing/rotation distribution and noise; in practice, we use empirical averages over particles within defocus groups, after noise whitening.

We recall the main difference between the cryo-EM model~\eqref{eqn:proj_model} and the MRA model~\eqref{eq:MRA_model}, which is the projection operator. Namely, while we target the symmetries of a 3-D volume, we only observe its 2-D data images. Drawing the connection between essentially the autocorrelation 3-D $C_{l}   = \sum_{m=l}^{l}A_{lms_1}A_{lms_2}^{H}$ and $\Sigma_{\ell}$ of~\eqref{eqn:block-cov} appears in several previous studies, e.g.,~\cite{kam1980reconstruction, sharon2020method}, and remain a key step in formulating the symmetry detection problem as well. 

\subsection{3-D symmetries}
\label{sec:cryoem-cyclic}

In three-dimensional space, the symmetry groups relevant to cryo-electron microscopy are the \emph{finite subgroups} of $\SO{3}$, which is the group of all proper rotations about the origin. These groups, known as the \emph{point groups}, describe the possible rotational symmetries of a rigid object that leave at least one point fixed (typically the origin). They encompass all the ways an object can be rotated and still appear identical to its original orientation.

The classification of finite subgroups of $\SO{3}$ includes both \emph{cyclic groups} $\Cd{n}$, which here correspond to rotations around a single fixed axis by multiples of $2\pi/n$ radians, and \emph{dihedral groups} $\Dm{n}$, which extend the cyclic symmetries by adjoining $n$ additional $\pi$ rotations about axes perpendicular to the principal axis. Note that plane reflections are orientation-reversing, with negative determinant. Therefore, the reflection in our 1-D dihedral model does not exist in $\SO{3}$ and the cryo-EM extension is an analogy of group structure rather than a direct identification of the reflection with a 3-D symmetry.

Beyond these infinite families, there exist the \emph{polyhedral groups} associated with the symmetries of regular (Platonic) solids: the \emph{tetrahedral} ($T_{12}$), \emph{octahedral} ($O_{24}$), and \emph{icosahedral} ($I_{60}$) groups, which represent higher-order rotational symmetries observed in molecular and viral structures. Together, these groups constitute all possible finite rotational symmetries of three-dimensional objects under $\SO{3}$.

The structure of the directed acyclic graph (DAG) for symmetry testing would reflect the hierarchy among the relevant three-dimensional point groups: cyclic $\Cd{n}$, dihedral $\Dm{n}$, and the polyhedral groups. The nodes of the DAG represent these candidate symmetry classes, and the directed edges encode inclusion relations such as \(\Cd{2} < \Dm{2} < T_{12}\) or \(\Cd{3} < \Dm{3} < O_{24}\), depending on the rotational axes that the molecule may exhibit. The precise topology of the DAG therefore depends on the range of subgroups considered in the analysis, whether one restricts attention to axial symmetries $\Cd{n}$ and $\Dm{n}$ or includes higher-order polyhedral groups. In practice, constructing such a DAG provides a structured roadmap for statistical testing: broader hypotheses (e.g., \(O_{24}\)) are tested first, and more specific ones (e.g., \(\Cd{3}\) or \(\Dm{3}\)) are examined only if the broader symmetry is rejected, enabling efficient and interpretable inference of three-dimensional molecular symmetries. Table~\ref{table:DAG_SO3} provides the relations between the polyhedral groups and their cyclic and dihedral subgroups.

\begin{table}[ht]
\centering
\small
\begin{tabularx}{\textwidth}{lXXX}
\toprule
\textbf{Polyhedral groups} & \textbf{Cyclic subgroups} & \textbf{Dihedral subgroups} & \textbf{Inclusion relations} \\
\midrule
Tetrahedral & $\Cd{2},\, \Cd{3}$ & $\Dm{2}$ & $\Cd{2} < \Dm{2} < T_{12},$ \\ & & & $\Cd{3} < T_{12}$ \\
Octahedral & $\Cd{2},\, \Cd{3},\, \Cd{4}$ & $\Dm{2},\, \Dm{3},\, \Dm{4}$ & $\Cd{2} < \Dm{2} < O_{24},$ \\ & & & $ \Cd{3} < \Dm{3} < O_{24}$,
\\ & & & $\Cd{4} < \Dm{4} < O_{24}$ \\
Icosahedral & $\Cd{2},\, \Cd{3},\, \Cd{5}$ & $\Dm{2},\, \Dm{3},\, \Dm{5}$ & $\Cd{2} < \Dm{2} < I_{60},$ \\ & & & $ \Cd{3} < \Dm{3} < I_{60}$, \\ & & & $\Cd{5} < \Dm{5} < I_{60}$ \\
\bottomrule
\end{tabularx}
\caption{Inclusion relations between cyclic and dihedral subgroups of the polyhedral rotation groups.} \label{table:DAG_SO3}
\end{table}

\subsection{Concluding remarks}

As we saw above, in cryo-EM, direct access to underlying three-dimensional molecular structures is inherently limited, as data consist solely of two-dimensional projection images. However, empirical correlations among these projections can offer rich statistical insights into structural symmetries. Namely, by estimating the empirical correlations from projection data and understanding the connection between $\Sigma_l$ and $C_{l}$, we aim to characterize the symmetry pattern, analogous to Proposition~\ref{prop:structure_in_Fourier_DMRA}, and the revealing symmetries invariant of the moments, similar to Proposition~\ref{prop:DMRA_mean_null_collapsed}. Therefore, future methodological research could focus on:
\begin{itemize}
  \item Deriving precise theoretical connections between observable two-dimensional image statistics and their corresponding three-dimensional autocorrelation structures.
  \item Identifying and formulating statistical invariants specifically designed to capture relevant symmetries.
  \item Design a p-value mechanism and apply multiple testing under the DAG structure sequential rejection principle, based on the proper DAG, as discussed above.
\end{itemize}
Two directions remain open within the MRA setting itself. First, the distribution-free third-order phase invariants of Proposition~\ref{prop:third_order_generic} yield a third-order version of the collapsed-regime dihedral test whose noise scaling is expected to be more favorable than that of the fourth-order test we implemented (Remark~\ref{rmk:third_vs_fourth}); its calibration and a systematic comparison with the fourth-order test remain to be carried out. Second, the family-wise error guarantee of the dihedral procedure (Proposition~\ref{prop:FWER_dihedral}) is asymptotic and requires nondegeneracy at the reachable true hypotheses; an adaptive variant that avoids testing degenerate nested hypotheses (Remark~\ref{rmk:nested_nulls}) would extend it to all nested configurations.

\section*{Acknowledgment}
The authors are deeply grateful to Amit Singer for his generous support, encouragement, and inspiring insights throughout this project. His work helped plant the seeds of the central idea developed here. The authors also thank Yoel Shkolnisky for his contribution to Figure 1 and for his valuable insights and discussions.

NS is partially supported by NSF-BSF award 2024791, BSF award 2024266, and DFG award 514588180.

\paragraph{Declaration of generative AI and AI-assisted technologies in the manuscript preparation process.} During the preparation of this work, the authors used ChatGPT (OpenAI) to assist with manuscript organization and language revision. After using this service, the authors reviewed and edited the content as needed and take full responsibility for the publication's content.

\paragraph{Code and data availability.} The code implementing the proposed transforms and reproducing all figures and examples of this paper is openly available at \url{https://github.com/nirsharon/symmetry_detection_MRA.git}. For the MRA examples, we used no external datasets; the accompanying code generates all data analyzed in this work. For the cryo-EM examples, we used EMPIAR-10028 and EMPIAR-10089~\cite{empiar10028,empiar10089}.

\bibliographystyle{plain}
\bibliography{refs}

\begin{thebibliography}{10}

\bibitem{abbe2018multireference}
Emmanuel Abbe, Tamir Bendory, William Leeb, Jo{\~a}o~M Pereira, Nir Sharon, and
  Amit Singer.
\newblock Multireference alignment is easier with an aperiodic translation
  distribution.
\newblock {\em IEEE Transactions on Information Theory}, 65(6):3565--3584,
  2018.

\bibitem{ashburner2000gene}
Michael Ashburner, Catherine~A Ball, Judith~A Blake, David Botstein, Heather
  Butler, J~Michael Cherry, Allan~P Davis, Kara Dolinski, Selina~S Dwight,
  Janan~T Eppig, et~al.
\newblock Gene ontology: tool for the unification of biology.
\newblock {\em Nature genetics}, 25(1):25--29, 2000.

\bibitem{balding2008handbook}
David~J Balding, Martin Bishop, and Chris Cannings.
\newblock {\em Handbook of statistical genetics}.
\newblock John Wiley \& Sons, 2008.

\bibitem{bandeira2023estimation}
Afonso~S Bandeira, Ben Blum-Smith, Joe Kileel, Jonathan Niles-Weed, Amelia
  Perry, and Alexander~S Wein.
\newblock Estimation under group actions: recovering orbits from invariants.
\newblock {\em Applied and Computational Harmonic Analysis}, 66:236--319, 2023.

\bibitem{bendory2017bispectrum}
Tamir Bendory, Nicolas Boumal, Chao Ma, Zhizhen Zhao, and Amit Singer.
\newblock Bispectrum inversion with application to multireference alignment.
\newblock {\em IEEE Transactions on signal processing}, 66(4):1037--1050, 2017.

\bibitem{bendory2022dihedral}
Tamir Bendory, Dan Edidin, William Leeb, and Nir Sharon.
\newblock Dihedral multi-reference alignment.
\newblock {\em IEEE Transactions on Information Theory}, 68(5):3489--3499,
  2022.

\bibitem{berger1999integrated}
James~O Berger, Brunero Liseo, and Robert~L Wolpert.
\newblock Integrated likelihood methods for eliminating nuisance parameters.
\newblock {\em Statistical Science}, 14(1):1--28, 1999.

\bibitem{chen2026archsym}
Hanyu Chen, Ruojin Cai, Steve Marschner, and Noah Snavely.
\newblock Archsym: Detecting 3d-grounded architectural symmetries in the wild.
\newblock In {\em Proceedings of the IEEE/CVF Conference on Computer Vision and
  Pattern Recognition}, pages 36561--36570, 2026.

\bibitem{chang2013}
Xiuyuan Cheng.
\newblock {\em Random Matrices in High-dimensional Data Analysis}.
\newblock diplomathesis, Princeton University, 2013.

\bibitem{conway2003quaternions}
John~H Conway and Derek~A Smith.
\newblock {\em On quaternions and octonions}.
\newblock AK Peters/CRC Press, 2003.

\bibitem{costa2008novel}
Alessandro Costa and Ardan Patwardhan.
\newblock A novel mirror-symmetry analysis approach for the study of
  macromolecular assemblies imaged by electron microscopy.
\newblock {\em Journal of molecular biology}, 378(1):273--283, 2008.

\bibitem{edidin2025orbit}
Dan Edidin and Josh Katz.
\newblock Orbit recovery from invariants of low degree in representations of
  finite groups.
\newblock {\em arXiv preprint arXiv:2503.00009}, 2025.

\bibitem{fan2024maximum}
Zhou Fan, Roy~R Lederman, Yi~Sun, Tianhao Wang, and Sheng Xu.
\newblock Maximum likelihood for high-noise group orbit estimation and
  single-particle cryo-em.
\newblock {\em Annals of statistics}, 52(1):52, 2024.

\bibitem{fraiman2026so3pca}
Michael Fraiman, Paulina Hoyos, Tamir Bendory, Joe Kileel, Oscar Mickelin, Nir
  Sharon, and Amit Singer.
\newblock {SO}$(3)$-invariant {PCA} with application to molecular data.
\newblock In {\em Proceedings of the 2026 IEEE 23rd International Symposium on
  Biomedical Imaging (ISBI 2026)}. IEEE, 2026.

\bibitem{goeman2010sequential}
Jelle~J Goeman and Aldo Solari.
\newblock The sequential rejection principle of familywise error control.
\newblock {\em The Annals of Statistics}, pages 3782--3810, 2010.

\bibitem{gurevich2009representation}
Shamgar Gurevich, Ronny Hadani, and Amit Singer.
\newblock Representation theoretic patterns in three dimensional cryo-electron
  microscopy iii-presence of point symmetries, 2009.

\bibitem{heimowitz2021centering}
Ayelet Heimowitz, Nir Sharon, and Amit Singer.
\newblock Centering noisy images with application to cryo-em.
\newblock {\em SIAM journal on imaging sciences}, 14(2):689--716, 2021.

\bibitem{iudin2016empiar}
Andrii Iudin, Paul~K Korir, Jos{\'e} Salavert-Torres, Gerard~J Kleywegt, and
  Ardan Patwardhan.
\newblock {EMPIAR}: a public archive for raw electron microscopy image data.
\newblock {\em Nature methods}, 13(5):387, 2016.

\bibitem{kam1980reconstruction}
Zvi Kam.
\newblock The reconstruction of structure from electron micrographs of randomly
  oriented particles.
\newblock {\em Journal of Theoretical Biology}, 82(1):15--39, 1980.

\bibitem{kileel2025fast}
Joe Kileel, Nicholas~F Marshall, Oscar Mickelin, and Amit Singer.
\newblock Fast expansion into harmonics on the ball.
\newblock {\em SIAM Journal on Scientific Computing}, 47(2):A1117--A1144, 2025.

\bibitem{kwok1999static}
Yu-Kwong Kwok and Ishfaq Ahmad.
\newblock Static scheduling algorithms for allocating directed task graphs to
  multiprocessors.
\newblock {\em ACM Computing Surveys (CSUR)}, 31(4):406--471, 1999.

\bibitem{liu2010computational}
Yanxi Liu, Hagit Hel-Or, Craig~S Kaplan, and Luc Van~Gool.
\newblock Computational symmetry in computer vision and computer graphics.
\newblock {\em Foundations and Trends in Computer Graphics and Vision},
  5(1-2):1--195, 2010.

\bibitem{loy2006detecting}
Gareth Loy and Jan-Olof Eklundh.
\newblock Detecting symmetry and symmetric constellations of features.
\newblock In {\em European Conference on Computer Vision}, pages 508--521.
  Springer, 2006.

\bibitem{meijer2015multiple}
Rosa~J Meijer and Jelle~J Goeman.
\newblock A multiple testing method for hypotheses structured in a directed
  acyclic graph.
\newblock {\em Biometrical Journal}, 57(1):123--143, 2015.

\bibitem{meinshausen2008hierarchical}
Nicolai Meinshausen.
\newblock Hierarchical testing of variable importance.
\newblock {\em Biometrika}, 95(2):265--278, 2008.

\bibitem{milne2021group}
James~S Milne.
\newblock Group theory (v4. 00).
\newblock {\em group}, 1:139, 2021.

\bibitem{mitra2013symmetry}
Niloy~J. Mitra, Mark Pauly, Michael Wand, and Duygu Ceylan.
\newblock Symmetry in 3{D} geometry: Extraction and applications.
\newblock {\em Computer Graphics Forum}, 32(6):1--23, 2013.

\bibitem{mityagin2015zero}
Boris Mityagin.
\newblock The zero set of a real analytic function.
\newblock {\em arXiv preprint arXiv:1512.07276}, 2015.

\bibitem{empiar10089}
Takanori Moriya, Martin Saur, Markus Stabrin, Felipe Merino, Horia Voicu,
  Zhenning Huang, Pawel~A. Penczek, Stefan Raunser, and Christos Gatsogiannis.
\newblock Empiar-10089: Single particle cryo-{EM} dataset of {TcdA1} in prepore
  state ({SPHIRE Tutorial}).
\newblock \url{https://www.ebi.ac.uk/empiar/EMPIAR-10089/}, 2017.
\newblock Accessed: 2025-05-14.

\bibitem{neyman1948consistent}
Jerzy Neyman and Elizabeth~L Scott.
\newblock Consistent estimates based on partially consistent observations.
\newblock {\em Econometrica: Journal of the Econometric Society}, pages 1--32,
  1948.

\bibitem{perry2019sample}
Amelia Perry, Jonathan Weed, Afonso~S Bandeira, Philippe Rigollet, and Amit
  Singer.
\newblock The sample complexity of multireference alignment.
\newblock {\em SIAM Journal on Mathematics of Data Science}, 1(3):497--517,
  2019.

\bibitem{peters2017elements}
Jonas Peters, Dominik Janzing, and Bernhard Sch{\"o}lkopf.
\newblock {\em Elements of causal inference: foundations and learning
  algorithms}.
\newblock The MIT press, 2017.

\bibitem{ram2013git}
Karthik Ram.
\newblock Git can facilitate greater reproducibility and increased transparency
  in science.
\newblock {\em Source code for biology and medicine}, 8(1):7, 2013.

\bibitem{rcsbPDB}
{RCSB PDB}.
\newblock Rcsb protein data bank.
\newblock \url{https://www.rcsb.org/}, 2025.
\newblock Accessed: 2025-10-10.

\bibitem{reboul2020point}
Cyril~F Reboul, Simon Kiesewetter, Dominika Elmlund, and Hans Elmlund.
\newblock Point-group symmetry detection in three-dimensional charge density of
  biomolecules.
\newblock {\em Bioinformatics}, 36(7):2237--2243, 2020.

\bibitem{scheres2012relion}
Sjors~HW Scheres.
\newblock Relion: implementation of a bayesian approach to cryo-em structure
  determination.
\newblock {\em Journal of structural biology}, 180(3):519--530, 2012.

\bibitem{sharon2020method}
Nir Sharon, Joe Kileel, Yuehaw Khoo, Boris Landa, and Amit Singer.
\newblock Method of moments for 3d single particle ab initio modeling with
  non-uniform distribution of viewing angles.
\newblock {\em Inverse Problems}, 36(4):044003, 2020.

\bibitem{shkolnisky2012viewing}
Yoel Shkolnisky and Amit Singer.
\newblock Viewing direction estimation in cryo-em using synchronization.
\newblock {\em SIAM journal on imaging sciences}, 5(3):1088--1110, 2012.

\bibitem{sigworth1998maximum}
Fred~J Sigworth.
\newblock A maximum-likelihood approach to single-particle image refinement.
\newblock {\em Journal of structural biology}, 122(3):328--339, 1998.

\bibitem{singer2018mathematics}
Amit Singer.
\newblock Mathematics for cryo-electron microscopy.
\newblock In {\em Proceedings of the International Congress of Mathematicians:
  Rio de Janeiro 2018}, pages 3995--4014. World Scientific, 2018.

\bibitem{van2000asymptotic}
Aad~W Van~der Vaart.
\newblock {\em Asymptotic statistics}, volume~3.
\newblock Cambridge university press, 2000.

\bibitem{van1984multivariate}
Marin Van~Heel.
\newblock Multivariate statistical classification of noisy images (randomly
  oriented biological macromolecules).
\newblock {\em Ultramicroscopy}, 13(1-2):165--183, 1984.

\bibitem{empiar10028}
Wilson Wong, Xiao-Chen Bai, Alan Brown, Israel~S Fernandez, Eric Hanssen,
  Melanie Condron, Yan~Hong Tan, Jake Baum, and Sjors~HW Scheres.
\newblock {EMPIAR}-10028: {Cryo-EM} structure of the {P}lasmodium falciparum
  80s ribosome bound to the anti-protozoan drug emetine.
\newblock \url{https://www.ebi.ac.uk/empiar/EMPIAR-10028/}, 2015.
\newblock Accessed: 2025-05-14.

\bibitem{xu2025misspecified}
Sheng Xu, Anderson~Ye Zhang, and Amit Singer.
\newblock Misspecified maximum likelihood estimation for non-uniform group
  orbit recovery.
\newblock {\em arXiv preprint arXiv:2509.22945}, 2025.

\bibitem{zhao2016fast}
Zhizhen Zhao, Yoel Shkolnisky, and Amit Singer.
\newblock Fast steerable principal component analysis.
\newblock {\em IEEE transactions on computational imaging}, 2(1):1--12, 2016.

\bibitem{zivanov2022bayesian}
Jasenko Zivanov, Joaqu{\'\i}n Ot{\'o}n, Zunlong Ke, Andriko von K{\"u}gelgen,
  Euan Pyle, Kun Qu, Dustin Morado, Daniel Castanno-D{\'\i}ez, Giulia Zanetti,
  Tanmay~AM Bharat, et~al.
\newblock A bayesian approach to single-particle electron cryo-tomography in
  relion-4.0.
\newblock {\em Elife}, 11:e83724, 2022.

\end{thebibliography}

\appendix

\section{Proofs for Section~\ref{sec:general_invariant_moment_principle}}

\subsection{Proof of Lemma~\ref{lem:invariant_moments_polynomials}}\label{sec:proof_invariant_moments_polynomials}

\begin{proof}
Every homogeneous polynomial $p$ of degree $r$ can be written as
$p(\x)=\langle\mathbf B,\x^{\otimes r}\rangle$ for a coefficient tensor $\mathbf B$, where the inner product is the sum of products of corresponding entries. Consequently,
\[
(\mathcal R_G p)(\x)=\langle\mathbf B,M_r(\x,\rho_0)\rangle.
\]
If $p$ is invariant, then $\mathcal R_G p=p$, so equality of $M_r(\x,\rho_0)$ and $M_r(\x',\rho_0)$ implies $p(\x)=p(\x')$. Conversely, each coordinate of $M_r(\,\cdot\,,\rho_0)$ is an invariant homogeneous polynomial of degree $r$. Since the group action is linear, the terms of each total degree in an invariant polynomial form an invariant homogeneous polynomial. Applying the argument degree by degree proves the claim.
\end{proof}

\subsection{Proof of Theorem~\ref{thm:general_invariant_moment_criterion}}\label{sec:proof_general_invariant_moment_criterion}

\begin{proof}
Suppose first that a map $\tau_R$ with the stated property exists. If
$M_{\leq R}(\x,\rho_0)=M_{\leq R}(\x',\rho_0)$, applying $\tau_R$ to the common moment vector gives $\kappa(\x)=\kappa(\x')$, proving separation.

Conversely, suppose separation holds. For an attainable moment vector $\boldsymbol u\in M_{\leq R}(\cX,\rho_0)$, choose any $\x\in\cX$ with $M_{\leq R}(\x,\rho_0)=\boldsymbol u$ and set
\[
\tau_R(\boldsymbol u)=\kappa(\x).
\]
Separation makes this definition independent of the chosen signal. Thus
$\tau_R(M_{\leq R}(\x,\rho_0))=\kappa(\x)$ for every $\x\in\cX$.
If separation fails, two signals with different labels have identical moment vectors. Any deterministic rule receiving only this common vector must return the same label for both signals, so it cannot be correct for every signal in $\cX$.

It remains to establish finite-order orbit identification. By the Hilbert--Weyl finite-generation theorem
(see \cite[Theorem~A.31]{bandeira2023estimation}),
the invariant polynomial algebra admits a finite set
of homogeneous generators $q_1,\ldots,q_s$. This means that every invariant polynomial $p$ has the form
\[
p(\z)=F\bigl(q_1(\z),\ldots,q_s(\z)\bigr)
\]
for some polynomial $F$ in $s$ variables. Set
\[
R_0=\max\{1,\deg(q_1),\ldots,\deg(q_s)\}.
\]
If $M_{\leq R_0}(\x,\rho_0)=M_{\leq R_0}(\x',\rho_0)$, Lemma~\ref{lem:invariant_moments_polynomials} implies that every generator takes the same value at $\x$ and $\x'$. The displayed representation then gives equality of every invariant polynomial at these two signals.

Invariant polynomials distinguish distinct compact orbits. To verify this, suppose $G\circ\x$ and $G\circ\x'$ are distinct. They are disjoint compact sets and have positive distance. The function
\[
f(\z)=\mathrm{dist}(\z,G\circ\x)^2
=\inf_{g\in G}\|\z-g\circ\x\|_2^2
\]
is continuous and invariant because the group action is orthogonal. It equals zero on the first orbit and a positive constant $c$ on the second. On the compact set $K=(G\circ\x)\cup(G\circ\x')$, the Stone--Weierstrass theorem allows continuous functions to be uniformly approximated by real polynomials. Choose a polynomial $p$ with
\[
\sup_{\z\in K}|p(\z)-f(\z)|<c/3.
\]
Since $K$ is $G$-invariant, averaging $p$ over the group preserves this error bound. The invariant polynomial $\mathcal R_G p$ is therefore within $c/3$ of zero on the first orbit and within $c/3$ of $c$ on the second, and so takes different values on them.

Equality of all invariant polynomials consequently implies $G\circ\x=G\circ\x'$. Hence $\x'=g\circ\x$ for some $g\in G$, and
\[
\Sym(\x')=g\Sym(\x)g^{-1}.
\]
The moments through order $R_0$ determine the signal orbit and its stabilizer conjugacy class.
\end{proof}

\subsection{Proof of Lemma~\ref{lem:estimation_invariant}}
\begin{proof}
For any polynomial $f$ and $\beps\sim\mathcal N(0,\sigma^2\mathbf I_d)$,
\[
\E_{\beps}[f(\z+\beps)]
=\Big(\exp\!\Big(\frac{\sigma^2}{2}\Delta\Big)f\Big)(\z).
\]
This identity follows by expanding the finite Taylor series of $f$ and taking Gaussian moments: odd-order terms vanish, and the even-order terms give the displayed exponential. Both exponentials are finite series on a space of polynomials of bounded degree, and their compositions are the identity. Orthogonality implies
$\Delta(f\circ g)=(\Delta f)\circ g$, where $(f\circ g)(\z)=f(g\circ\z)$; hence $\mathcal D_\sigma p$ is invariant whenever $p$ is. Therefore, for every fixed $g\in G$,
\[
\E_{\beps}[q(g\circ\x+\beps)]
=p(g\circ\x)=p(\x).
\]
Averaging over $g\sim\rho$ gives
\[
\E_{g\sim\rho,\beps}[q(g\circ\x+\beps)]
=\E_{g\sim\rho}\!\left[\E_{\beps}[q(g\circ\x+\beps)]\right]
=p(\x),
\]
proving unbiasedness for arbitrary $\rho$. Orthogonality gives $\|g\circ\x\|_2=\|\x\|_2$, and Gaussian noise has moments of all orders. The corrected polynomial therefore has finite moments of every order. Convergence follows from the law of large numbers.

For a fixed $g\in G$, invariance of each coordinate of $\boldsymbol q$ gives
\[
\begin{aligned}
\boldsymbol q(g\circ\x+\beps)
&=\boldsymbol q(\x+g^{-1}\circ\beps)\\
&\stackrel{d}{=}\boldsymbol q(\x+\beps).
\end{aligned}
\]
The last equality uses the fact that an orthogonal transformation preserves the isotropic Gaussian noise distribution. The conditional feature distribution is therefore the same for every $g$, proving the final assertion.
\end{proof}

\subsection{Proof of Proposition~\ref{prop:uniform_upper}}\label{sec:proof_uniform_upper}
\begin{proof}
Let $B$ bound $\|\x\|$ on $\cX$. Use the notation of Lemma~\ref{lem:estimation_invariant} with
$\boldsymbol p$ consisting of all invariant-moment coordinates of
orders $1,\ldots,R$. Thus, with tensor entries concatenated as in
Section~\ref{sec:recovery},
\[
\boldsymbol p(\z)=M_{\leq R}(\z,\rho_0),\qquad
\boldsymbol q(\z)
=\bigl(\mathcal D_\sigma[M_{\leq R}(\,\cdot\,,\rho_0)]\bigr)(\z).
\]
Here $\boldsymbol q$ is the corrected feature vector, with $\mathcal D_\sigma$ acting coordinatewise. Lemma~\ref{lem:estimation_invariant} gives
\[
\E_\rho[\boldsymbol q(\y_j)]=M_{\leq R}(\x,\rho_0),\qquad j=1,\ldots,N,
\]
for every probability measure $\rho$ on $G$, where we abbreviate $\E_{g\sim\rho,\beps}$ as $\E_\rho$ whenever no ambiguity arises.

We first bound the second moment of $\boldsymbol q(\y_j)$.
For a coordinate polynomial $p$ of an invariant moment of order
$r\leq R$, Gaussian correction gives
\[
(\mathcal D_\sigma p)(\y)
=\sum_{k=0}^{\lfloor r/2\rfloor}
\frac{(-\sigma^2/2)^k}{k!}(\Delta^k p)(\y).
\]
After substituting $\y=g\circ\x+\sigma\widetilde\beps$, with
$\widetilde\beps\sim\mathcal N(0,\mathbf I_d)$, each summand is a
polynomial in $\sigma$ of degree at most $r$, with coefficients
polynomial in $g\circ\x$ and $\widetilde\beps$. Orthogonality gives
$\|g\circ\x\|\leq B$, so Gaussian moment bounds imply
\[
\sup_{\x\in\cX,\,\rho}
\E_\rho\bigl[|(\mathcal D_\sigma p)(\y_j)|^2\bigr]
\leq K_p(1+\sigma^{2r}),\qquad \sigma\geq0,
\]
where $K_p$ depends only on $p$ and $B$. Since a vector's squared
norm is the sum of its squared coordinates and
$1+\sigma^{2r}\leq2(1+\sigma^{2R})$, summing over the finitely many
coordinates yields a constant $V_R>0$ such that
\[
\sup_{\x\in\cX,\,\rho}
\E_\rho\|\boldsymbol q(\y_j)\|^2
\leq V_R(1+\sigma^{2R}).
\]
Here $V_R$ depends only on $G$, $R$, and $B$.

Recall that the population moment image of label class $\cX_a$ is
\[
M_{\leq R}(\cX_a,\rho_0)
=\{M_{\leq R}(\x,\rho_0):\x\in\cX_a\}.
\]
Since
\[
\|M_r(\x,\rho_0)\|_F
\leq\E_{g\sim\rho_0}\|(g\circ\x)^{\otimes r}\|_F
=\|\x\|^r\leq B^r,
\]
the definition of $\delta_R$ gives, for any differently labelled
signals $\x,\x'\in\cX$,
\[
\begin{aligned}
\delta_R^2
&\leq\|M_{\leq R}(\x,\rho_0)-M_{\leq R}(\x',\rho_0)\|^2\\
&=\sum_{r=1}^R\|M_r(\x,\rho_0)-M_r(\x',\rho_0)\|_F^2
\leq4\sum_{r=1}^R B^{2r}.
\end{aligned}
\]
Enlarging $V_R$ if necessary so that $V_R\geq\sum_{r=1}^R B^{2r}$
preserves the preceding second-moment bound and ensures
$\delta_R^2\leq4V_R$.

Choose the number of observations per block and the number of full
blocks, respectively, as
\[
m=\left\lceil\frac{16V_R(1+\sigma^{2R})}{\delta_R^2}\right\rceil,
\qquad k=\lfloor N/m\rfloor.
\]
The inequality $\delta_R^2\leq4V_R$ implies
\[
m\leq\frac{16V_R(1+\sigma^{2R})}{\delta_R^2}+1
\leq\frac{20V_R(1+\sigma^{2R})}{\delta_R^2}.
\]
If $k=0$, define $\widehat\kappa_N$ to be any fixed label. For
$k\geq1$, use the disjoint blocks
\[
I_l=\{(l-1)m+1,\ldots,lm\},\qquad l=1,\ldots,k,
\]
and define
\[
\begin{aligned}
\widehat M_{\leq R}^{\mathrm{corr},(l)}(\rho_0)
&\triangleq\frac1m\sum_{j\in I_l}\boldsymbol q(\y_j)\\
&=\frac1m\sum_{j\in I_l}
\bigl(\mathcal D_\sigma[M_{\leq R}(\,\cdot\,,\rho_0)]\bigr)(\y_j).
\end{aligned}
\]
This is the corrected invariant-moment estimator, computed
using only observations in $I_l$. Unbiasedness and independence of the observations give
\begin{align*}
\E_\rho\|\widehat M_{\leq R}^{\mathrm{corr},(l)}(\rho_0)-M_{\leq R}(\x,\rho_0)\|^2
&=\frac1{m^2}\sum_{j\in I_l}
\E_\rho\|\boldsymbol q(\y_j)-M_{\leq R}(\x,\rho_0)\|^2\\
&\leq\frac1{m^2}\sum_{j\in I_l}
\E_\rho\|\boldsymbol q(\y_j)\|^2
\leq\frac{V_R(1+\sigma^{2R})}{m}.
\end{align*}

Assign block $l$ the label
\[
\widetilde\kappa^{(l)}\in
\underset{a\in\cC}{\arg\min}\;\mathrm{dist}\Big(\widehat M_{\leq R}^{\mathrm{corr},(l)}(\rho_0),M_{\leq R}(\cX_a,\rho_0)\Big),
\]
with ties resolved in a fixed order. Put $e_l=\|\widehat M_{\leq R}^{\mathrm{corr},(l)}(\rho_0)-M_{\leq R}(\x,\rho_0)\|$.
Since distinct label images are separated by
at least $\delta_R$, we have
\[
\begin{aligned}
\mathrm{dist}\Big(\widehat M_{\leq R}^{\mathrm{corr},(l)}(\rho_0),
M_{\leq R}(\cX_{\kappa(\x)},\rho_0)\Big)&\leq e_l,\\
\mathrm{dist}\Big(\widehat M_{\leq R}^{\mathrm{corr},(l)}(\rho_0),
M_{\leq R}(\cX_a,\rho_0)\Big)&\geq\delta_R-e_l
\quad(a\ne \kappa(\x)).
\end{aligned}
\]
Hence $e_l<\delta_R/2$ forces the correct label. Applying Markov's
inequality to $e_l^2$ gives
\[
\begin{aligned}
\P_\rho\{\widetilde\kappa^{(l)}\ne\kappa(\x)\}
&\leq\P_\rho\{e_l^2\geq\delta_R^2/4\}\\
&\leq\frac{4\E_\rho[e_l^2]}{\delta_R^2}
\leq\frac{4V_R(1+\sigma^{2R})}{m\delta_R^2}
\leq\frac14.
\end{aligned}
\]
All these bounds are uniform in $\x$ and $\rho$.

For $k\geq1$, let $\widehat\kappa_N$ be the most frequent among the block labels
$\widetilde\kappa^{(1)},\ldots,\widetilde\kappa^{(k)}$, again resolving
ties in a fixed order. For fixed $\x,\rho$, the indicators
\[
Z_l=\1\{\widetilde\kappa^{(l)}\ne\kappa(\x)\},
\qquad l=1,\ldots,k,
\]
are independent and satisfy $\P_\rho\{Z_l=1\}\leq1/4$.
If more than half the blocks are correct, the true label has the
most votes. Thus an incorrect final label requires at least $k/2$
block errors. Also,
$\E_\rho[3^{Z_l}]=1+2\P_\rho\{Z_l=1\}\leq3/2$.
Markov's inequality and independence therefore give
\[
\begin{aligned}
\P_\rho\{\widehat\kappa_N\ne\kappa(\x)\}
&\leq\P_\rho\left\{\sum_{l=1}^k Z_l\geq k/2\right\}\\
&\leq3^{-k/2}\prod_{l=1}^k\E_\rho[3^{Z_l}]
\leq\left(\frac{\sqrt3}{2}\right)^k
\leq e^{-k/8},
\end{aligned}
\]
where the last inequality uses $\log(4/3)\geq1/4$. For $k=0$,
the same bound holds trivially since $e^{-k/8}=1$. Finally,
$k\geq N/m-1$ and the upper bound on $m$ imply
\[
e^{-k/8}\leq e^{1/8}\exp\!\left(-\frac{N}{8m}\right)
\leq e^{1/8}\exp\!\left(
-\frac{N\delta_R^2}{160V_R(1+\sigma^{2R})}\right).
\]
Combining this with the bound by one, taking $K_R=160V_R$, and
using $e^{1/8}\leq2$ proves \eqref{eq:uniform_upper}.

For completeness, distance to any nonempty set is continuous. The
finite-label selection rule with fixed tie-breaking is therefore
measurable even if the moment images are not closed, and the final finite
voting rule is measurable as well. The final estimator uses only corrected
invariant-moment estimates of orders up to $R$, the specified signal class, and the known
noise level, and does not use $\rho$.
\end{proof}

\subsection{Proof of Proposition~\ref{prop:matching_lower}}\label{sec:proof_matching_lower}

\begin{proof}
Write $\x_0=\x$ and $\x_1=\x'$. Let $P_i$ denote the law of
one observation at $\x_i$ under Haar sampling $\rho=\rho_0$,
and let $p_i$ be its density, for $i=0,1$.
We first bound the relative entropy between these two laws.

Choose $B>0$ such that
$\max\{\|\x_0\|,\|\x_1\|\}\leq B$, and set
\[
\boldsymbol\mu_1=M_1(\x_1,\rho_0),\qquad
\boldsymbol V_i=g\circ\x_i-\boldsymbol\mu_1,
\quad g\sim\rho_0.
\]
Let $\varphi_\sigma$ denote the density of
$\mathcal N(\boldsymbol\mu_1,\sigma^2\mathbf I_d)$.
Orthogonality of the group action gives
\[
\|\boldsymbol\mu_1\|\leq B,\qquad
\|\boldsymbol V_i\|\leq2B,
\]
and
\[
\E\boldsymbol V_1=0,\qquad
\E\|\boldsymbol V_1\|^2
=\|\x_1\|^2-\|\boldsymbol\mu_1\|^2
\leq B^2,
\]
where the expectations are taken over $g\sim \rho_0$.
The Gaussian density ratio and Jensen's inequality therefore yield
the pointwise bound
\[
\begin{aligned}
\frac{p_1(\z)}{\varphi_\sigma(\z)}
&=\E\exp\!\left(
\frac{\langle\z-\boldsymbol\mu_1,\boldsymbol V_1\rangle}{\sigma^2}
-\frac{\|\boldsymbol V_1\|^2}{2\sigma^2}
\right)\\
&\geq
\exp\!\left(-\frac{\E\|\boldsymbol V_1\|^2}{2\sigma^2}\right)
\geq \exp\!\left(-\frac{B^2}{2\sigma^2}\right).
\end{aligned}
\]

For $k\geq0$, define the centered noiseless moment tensors
\[
\mathring{M}_k^{(i)}
\triangleq\E[\boldsymbol V_i^{\otimes k}],
\qquad i=0,1,
\]
with $\mathring{M}_0^{(i)}=1$.
For $i,j\in\{0,1\}$, let $\boldsymbol V_j'$ be an independent
copy of $\boldsymbol V_j$, independent of $\boldsymbol V_i$.
Gaussian integration gives
\[
\int_{\R^d}
\frac{p_i(\z)p_j(\z)}{\varphi_\sigma(\z)}\,\mathrm{d}\z
=
\E\exp\!\Big(
\frac{\langle\boldsymbol V_i,\boldsymbol V_j'\rangle}{\sigma^2}
\Big).
\]
Since $\|\boldsymbol V_i\|,\|\boldsymbol V_j'\|\leq2B$,
the exponential series is absolutely integrable. Consequently,
\[
\int_{\R^d}
\frac{p_i(\z)p_j(\z)}{\varphi_\sigma(\z)}\,\mathrm{d}\z
=
\sum_{k=0}^{\infty}
\frac{
\langle\mathring{M}_k^{(i)},\mathring{M}_k^{(j)}\rangle_F
}{k!\,\sigma^{2k}}.
\]
Taking the corresponding difference yields
\[
\int_{\R^d}
\frac{(p_0(\z)-p_1(\z))^2}{\varphi_\sigma(\z)}\,\mathrm{d}\z
=
\sum_{k=0}^{\infty}
\frac{
\|\mathring{M}_k^{(0)}-\mathring{M}_k^{(1)}\|_F^2
}{k!\,\sigma^{2k}}.
\]

Under the assumed moment matching, a common translation preserves
equality of moments below order $R$ and preserves their difference
at order $R$. Since both signals are translated by the same vector
$\boldsymbol\mu_1$, we have
\[
\mathring{M}_k^{(0)}=\mathring{M}_k^{(1)},
\qquad 0\leq k<R,
\]
and
\[
\mathring{M}_R^{(0)}-\mathring{M}_R^{(1)}
=M_R(\x_0,\rho_0)-M_R(\x_1,\rho_0).
\]
These identities also hold for $R=1$, since the order-zero
moments are both one and the common translation cancels from
the first-moment difference. Therefore
\[
\int_{\R^d}
\frac{(p_0-p_1)^2}{\varphi_\sigma}
=
\frac{\Delta_R^2}{R!\,\sigma^{2R}}
+O_{B,R}(\sigma^{-2R-2})
\qquad\text{as }\sigma\to\infty.
\]
Here $O_{B,R}(\sigma^{-2R-2})$ denotes a remainder whose absolute
value is bounded by $C_{B,R}\sigma^{-2R-2}$ for all sufficiently
large $\sigma$, where $C_{B,R}>0$ depends only on $B$ and $R$. Indeed,
\[
\|\mathring{M}_k^{(0)}-\mathring{M}_k^{(1)}\|_F
\leq2(2B)^k,
\]
so the tail starting at $k=R+1$ has the stated order.

Combining this expansion with the pointwise lower bound on
$p_1/\varphi_\sigma$ gives
\[
\begin{aligned}
D_{\mathrm{KL}}(P_0\|P_1)
&\leq\chi^2(P_0\|P_1)\\
&=\int_{\R^d}\frac{(p_0-p_1)^2}{p_1}\\
&\leq e^{B^2/(2\sigma^2)}
\int_{\R^d}\frac{(p_0-p_1)^2}{\varphi_\sigma}\\
&=\frac{\Delta_R^2}{R!\,\sigma^{2R}}
+O_{B,R}(\sigma^{-2R-2}).
\end{aligned}
\]
The last equality uses
$e^{B^2/(2\sigma^2)}=1+O_B(\sigma^{-2})$
and $\Delta_R\leq2B^R$, so the correction introduced by the
exponential factor is absorbed into the remainder. Since $\Delta_R>0$, there exists $\sigma_0>0$, depending on
$\x_0,\x_1$, and $R$, such that
\[
D_{\mathrm{KL}}(P_0\|P_1)
\leq\frac{2\Delta_R^2}{R!\,\sigma^{2R}},
\qquad \sigma\geq\sigma_0.
\]

It remains to translate this bound into a lower bound on label
recovery. Since $\x_0$ and $\x_1$ have different labels and Haar
sampling is included among the admissible distributions,
the two-point testing bound gives
\[
\mathcal E_N(\sigma;\cX)
\geq
\frac{1-\operatorname{TV}
(P_0^{\otimes N},P_1^{\otimes N})}{2}.
\]
For completeness, write
$A(P,Q)=\int\sqrt{pq}$ for the Hellinger affinity.
Jensen's inequality and the Cauchy--Schwarz inequality imply
\[
A(P,Q)\geq e^{-D_{\mathrm{KL}}(P\|Q)/2},
\qquad
\operatorname{TV}(P,Q)\leq\sqrt{1-A(P,Q)^2}.
\]
Hence
\[
1-\operatorname{TV}(P,Q)
\geq1-\sqrt{1-A(P,Q)^2}
\geq\frac12 A(P,Q)^2
\geq\frac12 e^{-D_{\mathrm{KL}}(P\|Q)}.
\]
Applying this inequality to the product laws and using additivity
of relative entropy, we obtain
\[
\begin{aligned}
\mathcal E_N(\sigma;\cX)
&\geq\frac14
\exp\!\left(-N D_{\mathrm{KL}}(P_0\|P_1)\right)\\
&\geq\frac14
\exp\!\left(-\frac{2N\Delta_R^2}{R!\,\sigma^{2R}}\right),
\qquad \sigma\geq\sigma_0.
\end{aligned}
\]
This proves the claimed lower bound for every $N\geq1$.
\end{proof}

\subsection{Proof of Theorem~\ref{thm:sample_complexity}}\label{sec:proof_sample_complexity}

\begin{proof}
Suppose first that $R>1$. Since $R$ is the smallest order
that separates the symmetry labels on $\cX$, moments through
order $R-1$ fail to separate them. Hence there exist
signals $\x,\x'\in\cX$ with different labels such that
\[
M_r(\x,\rho_0)=M_r(\x',\rho_0),
\qquad 1\leq r<R.
\]
Separation through order $R$ then implies
\[
\Delta_R
=\|M_R(\x,\rho_0)-M_R(\x',\rho_0)\|_F>0.
\]
If $R=1$, choose any two signals with different labels.
Separation by the first moment gives $\Delta_1>0$,
and the lower-order moment-matching condition is vacuous.

Set
\[
C=\frac{2\Delta_R^2}{R!}>0.
\]
Proposition~\ref{prop:matching_lower} provides a threshold $\sigma_0>0$ such that
\[
\mathcal E_N(\sigma;\cX)
\geq\frac14\exp\!\left(-\frac{CN}{\sigma^{2R}}\right)
\]
for every $\sigma\geq\sigma_0$ and $N\geq1$.
Increase $\sigma_0$ if necessary so that $\sigma_0\geq1$.
If $\mathcal E_N(\sigma;\cX)\leq\eta<1/4$, the preceding
bound forces
\[
N\geq\frac{\sigma^{2R}}{C}\log\frac1{4\eta}.
\]

For the upper bound, Proposition~\ref{prop:uniform_upper} and \eqref{eq:uniform_upper}
give
\[
\mathcal E_N(\sigma;\cX)
\leq
2\exp\!\left(
-\frac{N\delta_R^2}{K_R(1+\sigma^{2R})}
\right).
\]
Since $\sigma\geq1$, we have
$1+\sigma^{2R}\leq2\sigma^{2R}$. Therefore, with
\[
C'=\frac{\delta_R^2}{2K_R}>0,
\]
we obtain
\[
\mathcal E_N(\sigma;\cX)
\leq2\exp\!\left(-\frac{C' N}{\sigma^{2R}}\right).
\]
In particular, the minimax sample complexity is finite and satisfies
\[
\frac{\sigma^{2R}}{C}\log\frac1{4\eta}
\leq N_\eta(\sigma;\cX)
\leq
\left\lceil
\frac{\sigma^{2R}}{C'}\log\frac2\eta
\right\rceil.
\]

For $0<\eta\leq1/16$, the logarithms satisfy
\[
\log\frac1{4\eta}
\geq\frac12\log\frac1\eta,
\qquad
\log\frac2\eta
\leq\frac54\log\frac1\eta.
\]
Also, since $\sigma\geq1$,
\[
1\leq\frac{\sigma^{2R}\log(1/\eta)}{\log16}.
\]
Using $\lceil u\rceil\leq u+1$, we conclude that
\[
c_1\sigma^{2R}\log\frac1\eta
\leq N_\eta(\sigma;\cX)
\leq c_2\sigma^{2R}\log\frac1\eta,
\]
where one may take
\[
c_1=\frac1{2C},
\qquad
c_2=\frac5{4C'}+\frac1{\log16}.
\]
\end{proof}

\section{Proofs for Section~\ref{sec:invariants}}

\subsection{Proof of Proposition \ref{prop:moment_expression}}\label{sec:proof_moment_expression}
\begin{proof}
We only prove for the C-MRA case; The D-MRA case follows similarly by additionally noting that $\F s\x=\overline{\F\x}$. By the definition of the population moment in \eqref{eq:population_moment}, we have 
\begin{align*}
M_r(\x,\rho)=&\E_{g\sim \rho} \Big[(g\circ \x)\otimes \cdots \otimes (g\circ \x)\otimes (\overline{g\circ \x})\Big]\\
=&\sum_{\ell=0}^{d-1} \rho[\ell] \Big[(\Cb^{\ell}\x) \otimes \ldots \otimes (\Cb^{\ell}\x) \otimes (\overline{\Cb^{\ell}\x}) \Big],
\end{align*}
which implies elementwise,
\begin{align*}
M_r(\x,\rho)[i_1,\ldots,i_r]=\sum_{\ell=0}^{d-1} \rho[\ell] \x[i_1-\ell]\ldots\x[i_{r-1}-\ell] \overline{\x[i_r-\ell]}.
\end{align*}
Note that $\Cb_\x = (\x,\Cb\x,\ldots,\Cb^{d-1}\x)$ is the circulant matrix generated by $\x$. A direct calculation then shows that 
\begin{align*}
(\mathbf{T}^{(r)}_{\rho} \times_1 \Cb_{\x}  \cdots \times_{r-1} \Cb_{\x} \times_r \Cb_{\overline{\x}})[i_1,\ldots,i_r]=\sum_{\ell=0}^{d-1} \rho[\ell] \x[i_1-\ell]\ldots\x[i_{r-1}-\ell] \overline{\x[i_r-\ell]}.
\end{align*}
This verifies that the expression in the proposition indeed recovers the $r$-th population moment as defined. For the Fourier transform, by Lemma \ref{lem:n-mode_product_associative},
\begin{align*}
M_r(\x,\rho)\times_1 \Fb\ldots\times_{r-1}\Fb\times_r \overline{\Fb}=&\mathbf{T}_\rho^{(r)} \times_1 (\Fb\Cb_\x) \ldots\times_{r-1}(\Fb\Cb_\x)\times_r (\overline{\Fb\Cb_\x})\\
=&\big(\mathbf{T}_\rho^{(r)} \times_1 \Fb \ldots\times_{r-1}\Fb\times_r \overline{\Fb}\big) \\
&\times_1 (\Fb\Cb_\x\overline{\Fb}/d) \ldots\times_{r-1}(\Fb\Cb_\x\overline{\Fb}/d)\times_r (\overline{\Fb\Cb_\x\overline{\Fb}/d}) \\
=& \big(\mathbf{T}_\rho^{(r)} \times_1 \Fb \ldots\times_{r-1}\Fb\times_r \overline{\Fb}\big) \\
&\times_1 \Db_{\F\x} \ldots\times_{r-1}\Db_{\F\x} \times_r \Db_{\overline{\F\x}},\\
=&\F \mathbf{T}_\rho^{(r)}\times_1 \Db_{\F\x} \ldots\times_{r-1}\Db_{\F\x} \times_r \Db_{\overline{\F\x}},
\end{align*}
where we apply Lemma \ref{lem:Cv} in the penultimate equality. By direct calculation, elementwise,
\begin{align*}
\F \mathbf{T}_\rho^{(r)}[i_1,\ldots,i_{r-1},i_r]=&\sum_{k=0}^{d-1} \rho[k]\Fb[i_1,k]\ldots\Fb[i_{r-1},k]\overline{\Fb}[i_r,k]\\
=&\sum_{k=0}^{d-1} \rho[k] w_d^{(i_1+\ldots+i_{r-1}-i_r)k}\\
=&(\F\rho) [i_1+\ldots+i_{r-1}-i_r].
\end{align*}
\end{proof}

\subsection{Proof of Corollary \ref{cor:population_moment_PMRA_uniform}}\label{sec:proof_population_moment_PMRA_uniform}
\begin{proof}
Under the uniform distribution over $\Cd{d}$, we have $\rho[\ell]=1/d$ for all $\ell=0,1,\ldots,d-1$. By direct calculation, for $r\geq 1$,
\begin{align*}
M_r(\x,\rho_0)[i_1,\ldots,i_{r-1},i_r] =& \frac{1}{d}\sum_{\ell=0}^{d-1} \Cb_\x[i_1,\ell] \ldots  \Cb_\x[i_{r-1},\ell] \Cb_{\overline{\x}}[i_r,\ell]\\
=& \frac{1}{d}\sum_{\ell=0}^{d-1} \x[i_1-\ell]\ldots\x[i_{r-1}-\ell]\overline{\x[i_r-\ell]},
\end{align*}
which yields the desired expressions for the first three population moments. For the Fourier transforms, by direct calculation, for $r\geq 2$,
\begin{align*}
(\F M_r(\x,\rho_0))[i_1,\ldots,i_{r-1},i_r]=& (\F \mathbf{T}_{\rho_0}^{(r)})[i_1,\ldots,i_{r-1},i_r]\cdot (\F\x)[i_1]\ldots(\F\x)[i_{r-1}]\cdot \overline{(\F\x)[i_r]}\\
=& (\F\rho_0) [i_1+\ldots+i_{r-1}-i_r]\cdot (\F\x)[i_1]\ldots(\F\x)[i_{r-1}]\cdot \overline{(\F\x)[i_r]}.
\end{align*}
Noting that $(\F\rho_0)[0]=1$ and $(\F\rho_0)[\ell]=0$ for any $\ell=1,2,\ldots,d-1$, we obtain
\begin{align*}
(\F M_r(\x,\rho_0))[i_1,\ldots,i_{r-1},i_1+\ldots+i_{r-1}]=(\F\x)[i_1]\ldots(\F\x)[i_{r-1}]\cdot \overline{(\F\x)[i_1+\ldots+i_{r-1}]},
\end{align*}
and zeros elsewhere. In particular, we have
\begin{align*}
(\F M_1(\x,\rho_0))[0]=& (\F\x)[0]=\sum_{\ell=0}^{d-1}\x[\ell], \\
(\F M_2(\x,\rho_0))[k,k]=&|(\F\x)[k]|^2, \quad \text{for } k=0,1,\ldots,d-1,\\
(\F M_3(\x,\rho_0))[k_1,k_2,k_1+k_2]=& (\F\x)[k_1]\cdot (\F\x)[k_2] \cdot \overline{(\F\x)[k_1+k_2]},\quad \text{for } k_1,k_2=0,1,\ldots,d-1.
\end{align*}
\end{proof}

\section{Proofs for Section~\ref{sec:cyclic}}

\subsection{Proof of Proposition \ref{prop:zeros_in_Fourier_MRA}}\label{sec:proof_zeros_in_Fourier_MRA}
\begin{proof}
Since $\Cd{m}$ is generated by $r^{d/m}$, the signal $\x$
is $\Cd{m}$-invariant if and only if $r^{d/m}\circ\x=\x$.
Using the Fourier shift identity,
\[
\F(r^{d/m}\circ\x)[\ell]
=
\exp(-2\pi i\ell/m)\,\F\x[\ell],
\]
and the invertibility of $\F$, this is equivalent to
\[
\bigl(\exp(-2\pi i\ell/m)-1\bigr)\F\x[\ell]=0
\quad\text{for every }\ell\in\cI_d.
\]
The exponential factor equals one precisely when $m\mid\ell$.
Thus, invariance is equivalent to
\eqref{eqn:zeros_in_Fourier_MRA}, and its negation gives
the stated nonzero-coefficient characterization.

Finally, if $\Sym(\x)=\Cd{m'}$ in the C-MRA setting, then
$\x$ is $\Cd{m}$-invariant if and only if
$\Cd{m}\leq\Cd{m'}$, which holds precisely when $m\mid m'$.
The final conclusion follows from the Fourier characterization.
\end{proof}

\subsection{Proof of Lemma \ref{lem:population_moments_MRA}}
\begin{proof}
The first and second population moments expressions follow immediately from Proposition \ref{prop:moment_expression} and Remark \ref{rmk:tensor_multiplication_order2}. Their Fourier transforms can be similarly derived. Specifically,
\begin{align*}
\F M_2(\x,\rho)=\Db_{\F \x}  (\F\mathbf{T}^{(2)}_\rho) \Db_{\overline{\F \x}}=\Db_{\F \x}  (\F\Db_\rho) \Db_{\overline{\F \x}}.
\end{align*}
Applying Lemma \ref{lem:Cv} to $\vv=\F\rho$ gives
\begin{align*}
\F M_2(\x,\rho)=&\Db_{\F \x}  (\F (\Fb^{\sH} \Cb_{\F\rho} \Fb)) \Db_{\overline{\F \x}}/d^2\\
=&\Db_{\F \x}  (\Fb \Fb^{\sH} \Cb_{\F\rho} \Fb \Fb^{\sH})) \Db_{\overline{\F \x}}/d^2\\
=&\Db_{\F \x}  \Cb_{\F\rho} \Db_{\overline{\F \x}}.
\end{align*}
\end{proof}

\subsection{Proof of Theorem \ref{thm:test_size_power}}\label{sec:proof_test_size_power}
\begin{proof}
Note that the second-order sample moment has the form
\begin{align*}
\widehat{M}_2=&\frac{1}{N}\sum_{j=1}^N \y_j\y_j^{\sH}\\
=&\frac{1}{N}\sum_{j=1}^N (g_j\circ\x)(g_j\circ\x)^{\sH}+ \frac{1}{N}\sum_{j=1}^N (g_j\circ\x)\varepsilon_j^{\sH}+ \frac{1}{N}\sum_{j=1}^N \varepsilon_j (g_j\circ\x)^{\sH}+\frac{1}{N}\sum_{j=1}^N \varepsilon_j\varepsilon_j^{\sH}\\
\triangleq&\frac{1}{N}\sum_{j=1}^N A_{1j} +\frac{1}{N}\sum_{j=1}^N A_{2j} +\frac{1}{N}\sum_{j=1}^N A_{2j}^{\sH} +\frac{1}{N}\sum_{j=1}^N A_{3j}. 
\end{align*}
Its Fourier transform equals
\begin{align}\label{eq:second_moment_decomposition}
\F\widehat{M}_2=\frac{1}{N}\sum_{j=1}^N \F A_{1j} +\frac{1}{N}\sum_{j=1}^N \F A_{2j} +\frac{1}{N}\sum_{j=1}^N \F A_{2j}^{\sH} +\frac{1}{N}\sum_{j=1}^N \F A_{3j},
\end{align}
where
\begin{align*}
\F A_{1j}=& \Fb(g_j\circ\x)(g_j\circ\x)^{\sH} \Fb^{\sH}\\
=& (\Fb \Cb^{S_j}\x) (\Fb \Cb^{S_j}\x)^{\sH}\\
=& (\Wb^{S_j}\Fb\x) (\Wb^{S_j}\Fb\x)^{\sH}\\
=& (\Db_{\Fb\x} \w^{S_j}) (\Db_{\Fb\x} \w^{S_j})^{\sH}\\
=& \Db_{\Fb\x} \Big(\w^{S_j} (\w^{S_j})^{\sH}\Big) \Db_{\overline{\Fb\x}},
\end{align*}
\begin{align*}
\F A_{2j}=& \Fb(g_j\circ\x)\varepsilon_j^{\sH} \Fb^{\sH}\\
=& \Db_{\Fb\x} \w^{S_j} (\Fb\varepsilon_j)^{\sH},
\end{align*}
and
\begin{align*}
\F A_{3j}=  \Fb\varepsilon_j\varepsilon_j^{\sH}\Fb^{\sH}.
\end{align*}
Note that if $\x$ has symmetry $\Cd{m}$, then $\Fb\x[\ell]=0$ for any $\ell\in \cI_{m,d}^c$. Hence, under the hypothesis that $\x$ has symmetry $\Cd{m}$ and any density $\rho$, we obtain
\begin{align}\label{eq:moment1}
\F\widehat{M}_2[\ell,\ell]=\frac{1}{N}\sum_{j=1}^N\F A_{3j}[\ell,\ell]=\frac{1}{N}\sum_{j=1}^N \beps_j^\T \Ab_\ell \beps_j, ~~~\text{for}~~~\ell\in \cI_{m,d}^c.
\end{align}
Here $\Ab_{\ell}\in\C^{d\times d}$ with $\Ab_\ell[k_1,k_2]=w^{\ell(k_2-k_1)}$ for $k_1,k_2=0,1,\ldots,d-1$. Note that
\begin{align*}
\sum_{\ell\in \cI_{m,d}^c} \Ab_\ell [k_1,k_2] = \sum_{\ell=0}^{d-1} w^{\ell(k_2-k_1)}- \sum_{t=0}^{d/m-1} w^{tm(k_2-k_1)} =
\begin{cases}
    d-d/m & \text{if }k_1=k_2,\\
    -d/m & \text{else if } k_1-k_2\equiv 0~(\text{mod } d/m),\\
    0 & \text{else}.
\end{cases}
\end{align*}
Therefore, under hypothesis $\Cd{m}$-symmetry,
\begin{align*}
\widehat{B}_m=\sum_{\ell\in \cI_{m,d}^c}\F\widehat{M}_2[\ell,\ell]=&\frac{1}{N}\sum_{j=1}^N \beps_j^\T \big(\sum_{\ell\in \cI_{m,d}^c}\Ab_\ell\big) \beps_j\\
=&\frac{1}{N}\sum_{j=1}^N \beps_j^\T \big(d\cdot\Ib_d - \frac{d}{m}\cdot\mathbf{J}_{d,m}\big) \beps_j
\end{align*}
where $\mathbf{J}_{d,m}\in\R^{d\times d}$ has $(i,j)$-th entry one if $i-j\equiv 0~(\text{mod } d/m)$, and zero elsewhere. In particular, $\mathbf{J}_{d,m}$ can be diagonalized as
\[
\mathbf{J}_{d,m} = \mathbf{Q} \, \mathrm{diag}(0, \ldots, 0, \underbrace{m, \ldots, m}_{d/m \ \text{entries}}) \, \mathbf{Q}^{\T},
\]
for some orthogonal matrix $\mathbf{Q}$. It follows
\begin{align*}
\widehat{B}_m=\frac{1}{N}\sum_{j=1}^N \beps_j^\T \mathbf{Q} \, \mathrm{diag}(d, \ldots, d, \underbrace{0, \ldots, 0}_{d/m \ \text{entries}}) \, \mathbf{Q}^{\T} \beps_j.
\end{align*}
Noting that $\tilde\beps_j\triangleq \mathbf{Q}^{\T} \beps_j\stackrel{i.i.d.}{\sim} \mathcal{N}(0,\sigma^2 \Ib_d)$, we get
\begin{align}\label{eq:A3}
\widehat{B}_m=\frac{d}{N}\sum_{j=1}^N \sum_{k=1}^{d-d/m}\tilde\varepsilon_{jk}^2,
\end{align}
where  $\tilde\varepsilon_{jk}$ denotes $k$-th entry of $\tilde\beps_j$. Hence, the size of the proposed test satisfies
\begin{align*}
\E_{\Cd{m}}[\phi_m]= \alpha.
\end{align*}

For the power, it follows from \eqref{eq:second_moment_decomposition} that
\begin{align*}
\F\widehat{M}_2[\ell,\ell]=\frac{1}{N}\sum_{j=1}^N \F A_{1j}[\ell,\ell] +\frac{1}{N}\sum_{j=1}^N \big(\F A_{2j}[\ell,\ell]+(\F A_{2j}^{\sH})[\ell,\ell]\big)  +\frac{1}{N}\sum_{j=1}^N \F A_{3j}[\ell,\ell].
\end{align*}
Note that for any $j=1,\ldots,N$,
\begin{align*}
\F A_{1j}[\ell,\ell]=|\F \x[\ell]|^2.
\end{align*}
According to Proposition \ref{prop:zeros_in_Fourier_MRA}, under the hypothesis that $\x$ has symmetry $\Cd{m'}$ where $m' \mid d$ and $m  \nmid m'$,
\begin{align}\label{eq:A4}
\widehat{B}_{m1}\triangleq\sum_{\ell\in \cI_{m,d}^c} \frac{1}{N}\sum_{j=1}^N\F A_{1j}[\ell,\ell]=\sum_{\ell\in \cI_{m,d}^c}|\F \x[\ell]|^2>0.
\end{align}
Similar to \eqref{eq:moment1}, we have
\begin{align*}
\widehat{B}_{m2}\triangleq&\sum_{\ell\in \cI_{m,d}^c} \frac{1}{N}\sum_{j=1}^N \Big(\F A_{2j}[\ell,\ell]+(\F A_{2j}^{\sH})[\ell,\ell]\Big)\\
=&\frac{1}{N}\sum_{j=1}^N \Big(\sum_{\ell\in \cI_{m,d}^c} (\x_j^{\T} \Ab_{\ell}\varepsilon_j + \x_j^{\T} \overline{\Ab_{\ell}}\beps_j)\Big)\\
=&\frac{2}{N}\sum_{j=1}^N \x_j^\T \big(d\cdot\Ib_d - \frac{d}{m}\cdot\mathbf{J}_{d,m}\big) \beps_j\\
=&\frac{2}{N}\sum_{j=1}^N \x_j^\T \mathbf{Q} \, \mathrm{diag}(d, \ldots, d, \underbrace{0, \ldots, 0}_{d/m \ \text{entries}}) \, \mathbf{Q}^{\T} \beps_j\\
=&\frac{2}{N}\sum_{j=1}^N \x_j^\T \mathbf{Q} \, \mathrm{diag}(d, \ldots, d, \underbrace{0, \ldots, 0}_{d/m \ \text{entries}}) \, \tilde\beps_j,
\end{align*}
where $\x_j\triangleq g_j\circ \x$ and $\tilde\beps_j\stackrel{i.i.d.}{\sim}\mathcal{N}(0,\sigma^2\Ib_d)$. The matrix $d\Ib_d-\tfrac{d}{m}\mathbf{J}_{d,m}$ acts as $d$ on the Fourier modes indexed by $\cI_{m,d}^c$ and as $0$ on those indexed by $\cI_{m,d}$; since each rotation $g_j$ preserves the magnitude of every Fourier coefficient, the relevant projection satisfies $\big\|\mathrm{diag}(d,\ldots,d,0,\ldots,0)\,\mathbf{Q}^\T\x_j\big\|_2^2 = d\,E_m(\x)$. Therefore, for any $t>0$,
\begin{align}\label{eq:A5}
\P(|\widehat{B}_{m2}|>t) \leq 2 \exp\Big(-\frac{Nt^2}{8\,\sigma^2\, d\, E_m(\x)}\Big).
\end{align}
Additionally, previous derivations give
\begin{align*}
\widehat{B}_{m3}\triangleq \sum_{\ell\in \cI_{m,d}^c} \frac{1}{N}\sum_{j=1}^N\F A_{3j}[\ell,\ell]\sim \frac{\sigma^2 d}{N}\chi_{Nd(1-1/m)}^2.
\end{align*}
Therefore,
\begin{align*}
\E_{\Cd{m'}} [\phi_m] =& \P_{\Cd{m'}}\Big(\widehat{B}_{m1}+\widehat{B}_{m2}+ \frac{\sigma^2 d}{N}\chi_{Nd(1-1/m)}^2 > \frac{\sigma^2 d}{N} t_{Nd(1-1/m)}(\alpha)\Big)\\
=&\P_{\Cd{m'}}\Big(\chi_{Nd(1-1/m)}^2> t_{Nd(1-1/m)}(\alpha) -\frac{N}{\sigma^2 d} (\widehat{B}_{m1}+\widehat{B}_{m2}) \Big)\\
=&\P_{\Cd{m'}}\Big(\chi_{Nd(1-1/m)}^2- Nd(1-1/m)> t_{Nd(1-1/m)}(\alpha)-Nd(1-1/m) -\frac{N}{\sigma^2 d} (\widehat{B}_{m1}+\widehat{B}_{m2}) \Big)
\end{align*}
Using the chi-squared quantile bound, we know
\begin{align*}
t_{Nd(1-1/m)}(\alpha)-Nd(1-1/m) =  O(\sqrt{N}).
\end{align*}
On the other hand, combining \eqref{eq:A4} and \eqref{eq:A5} yields
\begin{align*}
\frac{N}{\sigma^2 d} (\widehat{B}_{m1}+\widehat{B}_{m2})=\Theta(N)-O_p(\sqrt{N}).
\end{align*}
Applying a tail bound for the chi-squared distribution, we get
\begin{align}
\E_{\Cd{m'}} [\phi_m]\geq 1- \exp(-CN),
\end{align}
where the constant $C$ may depend on $d,m,m',\sigma$ and $E_m(\x)$.
\end{proof}

\section{Proofs for Section~\ref{sec:dihedral}}

\subsection{Proof of Lemma \ref{lem:equivalence_dihedral_uniform}}\label{sec:proof_equivalence_dihedral_uniform}

\begin{proof}
Write $n=d/m$. The stabilizer transforms by conjugation, so
\begin{align}\label{eq:invariant_j}
\Sym(r^k\x)
&=r^k\Dmj{m}{j}r^{-k}
=\langle r^n,r^{j+2k}s\rangle.
\end{align}
Here $r^ksr^{-k}=r^{2k}s$ follows from the dihedral group relations. Since the reflections in $\Dmj{m}{j'}$ are exactly $r^{j'+tn}s$, the stabilizer in \eqref{eq:invariant_j} equals $\Dmj{m}{j'}$ if and only if
\begin{align}\label{eq:reflected_invariant}
2k\equiv j'-j\pmod n.
\end{align}
This congruence has a solution precisely when $\gcd(2,n)$ divides $j'-j$, and a solution can be chosen in $\{0,1,\ldots,n-1\}$. The parity conclusions follow.
\end{proof}

\subsection{Proof of Proposition \ref{prop:structure_in_Fourier_DMRA}}\label{sec:proof_structure_in_Fourier_DMRA}
\begin{proof}
The proof for the zero patterns in Fourier space induced by the action of rotations, applicable to both the cyclic and dihedral symmetries, follows directly from the proof of Proposition \ref{prop:zeros_in_Fourier_MRA}. The conjugate symmetry condition forces $\F\x[0]\in\R$ and $\F\x[d/2]\in\R$ if $d$ is even. 

For the dihedral symmetry $\Dmj{m}{j}$, we focus on the structure of these nonzero Fourier coefficients, i.e., the indices $\ell\in\cI_{m,d}^+\setminus\{0,d/2\}$ together with the additional index $\ell = d/2$ when $d$ is even. First, for $k\in\cI_d$, note that the action of a reflection $r^k s$ on a signal $\x$ translates to the following in Fourier space 
\begin{align*}
(\F (r^k s\cdot \x))[\ell]= w_d^{\ell k}\cdot (\F (s\cdot \x))[\ell]= w_d^{\ell k}\cdot \overline{\F \x [\ell]}~~~\text{for}~\ell\in\cI_d.
\end{align*}
Thus, if a signal $\x$ satisfies $r^k s\cdot \x=\x$, it follows that 
\begin{align}\label{eq:Fourier_reflection}
\F \x [\ell]=w_d^{\ell k}\cdot \overline{\F \x [\ell]},
\end{align}
for all $\ell\in\cI_d$. For a signal invariant under $\Dmj{m}{j}$, consider those indices $\ell=t m$ for some integer $t$. Since $r^{j+\tilde{t} d/m}s\in \Dmj{m}{j}$ for any integer $\tilde{t}$, the above relation implies 
\begin{align*}
\F \x [\ell]=w_d^{\ell (j+\tilde{t}d/m)}\cdot \overline{\F \x [\ell]}=w_d^{\ell j+t\tilde{t}d}\cdot \overline{\F \x [\ell]}=w_d^{\ell j} \cdot \overline{\F \x [\ell]},
\end{align*}
where the last equality holds because $w_d^d=1$ by definition. Writing $\F \x [\ell]= r_\ell e^{i\theta_\ell}$, we obtain
\begin{align*}
\theta_\ell = -\frac{\ell j}{d}\pi ~\text{or}~\big(1-\frac{\ell j}{d}\big)\pi, \quad\text{for}~\ell\in\cI_{m,d}, ~~\text{whenever } r_\ell>0.
\end{align*}
In other word,
\begin{align}\label{eq:dihedral_phases}
\theta_\ell\equiv -\frac{\ell j}{d}\pi\pmod{\pi},\quad\text{for }\ell\in\cI_{m,d},~~\text{whenever } r_\ell>0.
\end{align}
Lastly, we need to ensure that the above phase characterization is consistent with the conjugate symmetry condition:
\begin{itemize}
\item For $\ell=0$, \eqref{eq:dihedral_phases} corresponds to $\F\x[0]\in\R$.

\item If $d$ is even and $\ell=d/2$, \eqref{eq:dihedral_phases} yields
\begin{align*}
\theta_{d/2}\equiv -\frac{j}{2}\pi \pmod{\pi}.
\end{align*}
If $j$ is even, this is equivalent to $\F\x[d/2]\in\R$. If $j$ is odd, this condition contradicts the fact that $\F\x[d/2]\in\R$, and hence we must have $\F\x[d/2]=0$.

\item For $\ell \in \cI_{m,d}^+\setminus\{0,d/2\}$, \eqref{eq:dihedral_phases} is exactly the condition stated in the proposition. Translating this phase constraint into the expression for Fourier coefficients, we obtain
\begin{align*}
\F \x[\ell] = c_\ell \cdot  e^{-\ri \ell j \pi / d},
\end{align*}
for some $c_\ell \in \mathbb{R}$.

\item For the remaining indices, namely those $\ell\in\cI_{m,d}\setminus\cI_{m,d}^+$, the conjugate symmetry relation $\theta_\ell = -\theta_{d-\ell}$ ensures that \eqref{eq:dihedral_phases} is automatically satisfied. In particular, it implies the corresponding expression for the Fourier coefficients stated in the proposition.
\end{itemize}
This proves all the conditions for the dihedral cases.

The sufficiency follows from the sufficiency part of Proposition \ref{prop:zeros_in_Fourier_MRA}, along with the fact that \eqref{eq:Fourier_reflection} is equivalent to the symmetry condition $r^k s\cdot \x=\x$. Thus, we complete the proof.
\end{proof}

\subsection{Proof of Theorem \ref{thm:third_moment_failure}}\label{sec:proof_third_moment_failure}
\begin{proof}
Fix \(j\in\{0,\ldots,d/m-1\}\), and let
\(\x^{(D)}\) and \(\x^{(C)}\) be the signals defined in \eqref{eqn:counterexample}, with
\[
    \phi_D=-\frac{mj\pi}{d},
    \qquad
    \phi_C=\frac{\pi}{2d}.
\]
Their Fourier transforms satisfy conjugate symmetry, so both signals
are real.

Since the Fourier coefficient at frequency \(m\) is nonzero for both
signals, a rotation \(r^k\) fixes either signal if and only if
\[
    e^{-2\pi\ri mk/d}=1,
\]
or equivalently \(k\equiv0\pmod{d/m}\). Thus, the rotational stabilizer
of each signal is exactly \(\Cd{m}\).

By Proposition~\ref{prop:structure_in_Fourier_DMRA}(iv), a reflection
\(r^k s\) fixes a signal supported on \(\{\pm m\}\) with phase \(\phi\)
if and only if
\[
    \phi\equiv-\frac{mk\pi}{d}\pmod{\pi}.
\]
For \(\phi_D=-mj\pi/d\), this condition holds precisely when
\(k\equiv j\pmod{d/m}\). Therefore, $\x^{(D)}$ has $\Dmj{m}{j}$ symmetry. In contrast, if some reflection \(r^k s\) fixed \(\x^{(C)}\), then
\[
    \frac{\pi}{2d}
    \equiv-\frac{mk\pi}{d}\pmod{\pi}.
\]
This would imply
\[
    1+2mk=2d\ell ,
\]
for some \(\ell\in\mathbb{Z}\), which is impossible because the
left-hand side is odd and the right-hand side is even. Therefore, $\x^{(C)}$ has no dihedral symmetry and it is exact $\Cd{m}$ symmetry.

It remains to compare their moments under the uniform distribution
\(\rho_0\). Both signals have zero DC coefficient, so their first
population moments vanish. For \(r\ge2\) we use the representation
\eqref{eqn:r_moment_uniform_dist}: the entry of \(\F M_r(\cdot,\rho_0)\) at
\((i_1,\ldots,i_r)\) vanishes unless \(i_r\equiv i_1+\cdots+i_{r-1}\pmod d\),
in which case it equals
\(\Re\bigl(\F\x[i_1]\cdots\F\x[i_{r-1}]\,\overline{\F\x[i_r]}\bigr)\).
Both spectra are supported on \(\{m,d-m\}=\{\pm m\}\), so a nonzero entry
requires \(i_a\equiv\sigma_a m\pmod d\) with signs
\(\sigma_a\in\{\pm1\}\), \(a=1,\ldots,r\). For a signal of the form
\eqref{eqn:counterexample} with phase \(\phi\) we have
\(\F\x[\sigma m]=e^{\ri\sigma\phi}\), hence
\[
    \F\x[i_1]\cdots\F\x[i_{r-1}]\,\overline{\F\x[i_r]}
    =
    e^{\ri\phi S},
    \qquad
    S\coloneqq\sigma_1+\cdots+\sigma_{r-1}-\sigma_r\in\mathbb Z,
    \qquad |S|\le r .
\]
The index constraint \(i_r\equiv i_1+\cdots+i_{r-1}\pmod d\) reads
\(Sm\equiv0\pmod d\), that is, \(S\equiv0\pmod{d/m}\). If
\(r\le d/m-1\), then \(|S|\le r<d/m\) forces \(S=0\), so every
nonzero entry equals \(\Re(e^{0})=1\) and does not depend on the phase
\(\phi\) at all. (Since \(S\equiv r\pmod 2\), the odd-order moments in
fact vanish identically, while the even-order ones have the entry \(1\) at
every balanced sign pattern; for \(r=2\) this is the power spectrum, and
for \(r=3\) it recovers the direct observation that
\(\{0,\pm2m\}\) is disjoint from \(\{\pm m\}\) modulo \(d\) when
\(d/m\ge4\).) Applying this to \(\phi=\phi_C\) and
\(\phi=\phi_D\) gives
\[
    M_r\bigl(\x^{(C)},\rho_0\bigr)
    =
    M_r\bigl(\x^{(D)},\rho_0\bigr),
    \qquad r=1,2,\ldots,d/m-1 .
\]
The bound is sharp: at \(r=d/m\) the sign pattern
\(\sigma_1=\cdots=\sigma_{r-1}=1\), \(\sigma_r=-1\) has \(S=d/m\) and
satisfies the index constraint, and the corresponding entry
\(\Re\bigl(e^{\ri\phi d/m}\bigr)=\cos(\phi d/m)\) takes different values at
\(\phi_C\) and \(\phi_D\): \(\cos(\phi_D d/m)=\cos(j\pi)=(-1)^j\), whereas
\(\cos(\phi_C d/m)=\cos\bigl(\pi/(2m)\bigr)\in[0,1)\).

Since \(j\) was arbitrary, the construction applies to every detectable
dihedral class represented by \(\Dmj{m}{j}\). Finally, \(\rho_0\) is the
Haar probability measure on \(\Dm{d}\), so
\(M_r(\x,\rho_0)=\mathcal M_r(\x)\) is the invariant moment tensor of
Section~\ref{sec:general_invariant_moment_principle}. The pair
\(\x^{(C)},\x^{(D)}\) therefore violates the separation condition of
Definition~\ref{def:moment_separation} for \(R=d/m-1\) on the class of
\(\Cd{m}\)-invariant signals, and the impossibility clause of
Theorem~\ref{thm:general_invariant_moment_criterion} yields the last
assertion.
\end{proof}

\subsection{Proof of Proposition~\ref{prop:order_dm_sufficient}}\label{sec:proof_order_dm_sufficient}
\begin{proof}
Write $n\triangleq d/m$ and $\omega\triangleq e^{2\pi\ri/n}$. Since $\x$ is $\Cd{m}$-invariant, its Fourier transform is supported on $\cI_{m,d}=\{am:\ a\in\Z_n\}$, and we index its coefficients by $a\in\Z_n$: $\F\x[am]=r_ae^{\ri\theta_a}$ with $r_a\ge0$; we write $r'_a,\theta'_a$ for $\x'$. Put
\[
B\triangleq\{a\in\Z_n:\ r_a>0\},\qquad B^+\triangleq\{a\in B:\ 1\le a\le n/2\}.
\]
Conjugate symmetry gives $B=-B$, $r_{-a}=r_a$ and $\theta_{-a}\equiv-\theta_a$.

\emph{Step 1 (orders one and two).} By \eqref{eqn:r_moment_uniform_dist}, $\F M_2(\x,\rho_0)[i,i]=|\F\x[i]|^2$ for every $i$, and $\F M_1(\x,\rho_0)[0]=\Re\F\x[0]=\F\x[0]$. Hence $r_a=r'_a$ for all $a$, so $\x$ and $\x'$ have the same support $B$ and the same DC coefficient. In particular, they have the same rotational stabilizer, since $r^k$ fixes a signal with support $B$ if and only if $ak\equiv0\pmod n$ for all $a\in B$.

\emph{Step 2 (closed monomials).} Let
\[
L\triangleq\Big\{c=(c_a)_{a\in B^+}\in\Z^{B^+}:\ \sum_{a\in B^+}c_a\,a\equiv0\pmod n\Big\},
\qquad |c|\triangleq\sum_{a\in B^+}|c_a| ,
\]
and for $c\in L$ define the \emph{closed monomial}
\[
P_c(\x)\triangleq\prod_{a\in B^+:\ c_a>0}\F\x[am]^{c_a}\ \prod_{a\in B^+:\ c_a<0}\F\x[-am]^{|c_a|}
=\Big(\prod_{a\in B^+}r_a^{|c_a|}\Big)\exp\Big(\ri\sum_{a\in B^+}c_a\theta_a\Big).
\]
Listing $c_a$ copies of $am$ for $c_a>0$ and $|c_a|$ copies of $-am$ for $c_a<0$ gives indices $u_1,\ldots,u_{|c|}\in\cI_{m,d}$ with $\sum_iu_i\equiv0\pmod d$ and $P_c(\x)=\prod_{i=1}^{|c|}\F\x[u_i]$. Since $\F\x[u_{|c|}]=\overline{\F\x[-u_{|c|}]}$ and $-u_{|c|}\equiv u_1+\cdots+u_{|c|-1}\pmod d$, formula \eqref{eqn:r_moment_uniform_dist} shows that
\begin{equation}\label{eq:closed_monomial_entry}
\Re P_c(\x)=\F M_{|c|}(\x,\rho_0)\big[u_1,\ldots,u_{|c|-1},-u_{|c|}\big],
\end{equation}
an entry of the moment of order $|c|$. Note that $|c|\ge2$ for every nonzero $c\in L$, because $a\not\equiv0\pmod n$ for $a\in B^+$.

\emph{Step 3 (reflections as characters).} By Proposition~\ref{prop:structure_in_Fourier_DMRA}(iv) and its sufficiency clause, a $\Cd{m}$-invariant real signal with support $B$ is fixed by the reflection $r^ks$ if and only if $\theta_a\equiv-ak\pi/n\pmod\pi$ for all $a\in B$, that is, if and only if
\begin{equation}\label{eq:char_condition}
f(a)\triangleq e^{2\ri\theta_a}=\omega^{-ak}\qquad\text{for all }a\in B^+ .
\end{equation}
(When $n$ is even and $n/2\in B^+$, the coefficient $\F\x[d/2]$ is real, so $f(n/2)=1$, and \eqref{eq:char_condition} forces $k$ to be even, in agreement with the vanishing of $\F\x[d/2]$ under odd reflections.) If \eqref{eq:char_condition} holds, write $\theta_a=-ak\pi/n+\pi\epsilon_a$ with $\epsilon_a\in\Z$; then for every $c\in L$, with $\sum_ac_aa=tn$,
\[
\sum_{a\in B^+}c_a\theta_a=-k\pi t+\pi\sum_{a\in B^+}c_a\epsilon_a\in\pi\Z,
\]
so that $P_c(\x)$ is real and $|\Re P_c(\x)|=\prod_ar_a^{|c_a|}$.

\emph{Step 4 (short relations generate $L$).} We claim that every $c\in L$ is a sum $c=c^{(1)}+\cdots+c^{(s)}$ of elements of $L$ with $|c^{(i)}|\le n$ and $\sum_i|c^{(i)}|=|c|$, each $c^{(i)}_a$ having the sign of $c_a$ or vanishing. It suffices to split off one such summand when $|c|>n$. Consider the indices $u_1,\ldots,u_{|c|}$ of Step~2 read in $\Z_n$, i.e., $v_i\triangleq u_i/m\in\Z_n$, so that $\sum_iv_i\equiv0$, and take any $n$ of them, say $v_1,\ldots,v_n$. Among the $n+1$ partial sums $S_0\triangleq0,\ S_p\triangleq v_1+\cdots+v_p\in\Z_n$, $p=1,\ldots,n$, two coincide, $S_p=S_q$ with $p<q$; hence $v_{p+1}+\cdots+v_q\equiv0\pmod n$. Let $c^{(1)}\in\Z^{B^+}$ record these $q-p$ indices with signs (a copy of $am$ contributes $+1$ to $c^{(1)}_a$, a copy of $-am$ contributes $-1$). Then $c^{(1)}\in L$ and $|c^{(1)}|=q-p\le n$, because for each $a$ only one sign occurs among the indices of $c$; for the same reason $c-c^{(1)}\in L$ has the same sign pattern as $c$ and $|c-c^{(1)}|=|c|-|c^{(1)}|$. Iterating proves the claim; in particular, $L$ is generated as a group by its elements of $\ell^1$-norm at most $n$.

\emph{Step 5 (the dihedral label).} Suppose $\x$ is fixed by $r^ks$, and let $c\in L$ with $|c|\le n$. By \eqref{eq:closed_monomial_entry} and the hypothesis, $\Re P_c(\x)=\Re P_c(\x')$, both being entries of the order-$|c|$ moments; by Step~1, $|P_c(\x')|=\prod_ar_a^{|c_a|}$; and by Step~3, $|\Re P_c(\x)|=\prod_ar_a^{|c_a|}$. Hence $|\Re P_c(\x')|=|P_c(\x')|$, so $P_c(\x')$ is real, that is,
\[
\prod_{a\in B^+}f'(a)^{c_a}=1,\qquad f'(a)\triangleq e^{2\ri\theta'_a}.
\]
The map $c\mapsto\prod_af'(a)^{c_a}$ is a group homomorphism from $\Z^{B^+}$ to the unit circle, so by Step~4 it is trivial on all of $L$. Consequently
\[
\chi\Big(\sum_{a\in B^+}c_aa\bmod n\Big)\triangleq\prod_{a\in B^+}f'(a)^{c_a}
\]
is a well-defined character of the subgroup $H\triangleq\langle B^+\rangle\le\Z_n$. Every character of a subgroup of the cyclic group $\Z_n$ is the restriction of a character of $\Z_n$,\footnote{If $H=g\Z_n$ with $g\mid n$, then $H\cong\Z_{n/g}$, and the $n$ characters $a\mapsto\omega^{-ka}$ of $\Z_n$ restrict to $a\mapsto e^{-2\pi\ri k(a/g)/(n/g)}$ on $H$, which as $k$ varies gives all $n/g$ characters of $H$.} so there is $k'\in\Z_n$ with $f'(a)=\omega^{-ak'}$ for all $a\in B^+$, and $\x'$ is fixed by $r^{k'}s$ by Step~3. Exchanging the roles of $\x$ and $\x'$ proves the first assertion.

\emph{Step 6 (parity).} Let $n$ be even and suppose the rotational stabilizer of $\x$ is $\Cd{m}$, so that, by Step~1, $B^+$ generates $\Z_n$ and the rotational stabilizer of $\x'$ is $\Cd{m}$ as well. Then $B^+$ contains an odd element $a$, for otherwise $\langle B^+\rangle\subseteq2\Z_n\neq\Z_n$. Let $g\triangleq\gcd(a,n)$, which is odd; then $g\mid n/2$, so $n/2\in g\Z_n=\langle a\rangle$, and there is $q\in\Z$ with $qa\equiv n/2\pmod n$; since $\langle a\rangle$ has order $n/g$, we may take $|q|\le n/(2g)\le n/2$. The vector $c\triangleq2q\,e_a$ lies in $L$ and satisfies $|c|=2|q|\le n$. Suppose $\x$ is fixed by $r^ks$ and $\x'$ by $r^{k'}s$ (both exist by Step~5 whenever one of them does). Writing $qa=n/2+sn$ and $\theta_a=-ak\pi/n+\pi\epsilon_a$,
\[
\sum_bc_b\theta_b=2q\theta_a\equiv-\frac{2qak\pi}{n}=-(1+2s)k\pi\pmod{2\pi},
\qquad\text{hence}\qquad
P_c(\x)=(-1)^k\,r_a^{2|q|},
\]
and likewise $P_c(\x')=(-1)^{k'}r_a^{2|q|}$. Since $\Re P_c(\x)=\Re P_c(\x')$ by \eqref{eq:closed_monomial_entry} and $r_a>0$, we conclude $(-1)^k=(-1)^{k'}$. For a signal with rotational stabilizer $\Cd{m}$, the reflections fixing it form the single coset $\{r^{k+tn}s:\ t\in\Z\}$, so the parity of $k$ is well defined and determines the class $\Dm{m}^\epsilon$ (Corollary~\ref{cor:dihedral_equivalence}); hence $\x$ and $\x'$ lie in the same parity class.

Finally, $M_r(\cdot,\rho_0)=\mathcal M_r(\cdot)$ is the invariant moment tensor of Section~\ref{sec:general_invariant_moment_principle}, so the statements just proved are the separation condition of Definition~\ref{def:moment_separation} with $R=d/m$, and the last assertion follows from Theorem~\ref{thm:general_invariant_moment_criterion} together with Theorem~\ref{thm:third_moment_failure}.
\end{proof}

\subsection{Proof of Theorem \ref{thm:DMRA_4th_population_general}}\label{sec:proof_DMRA_4th_population_general}

\begin{proof}
We first prove the ``only if'' direction. Suppose that $\x$ has $\Dm{m}$-symmetry, and hence $\x$ is invariant under $\Dmj{m}{j}$ for some $j$. Since we consider the case where either $d$ is odd, or $d$ is even and $2m \nmid d$, it follows that $d/2$ is either not an integer or does not belong to $\cI_{m,d}$.
Then by Proposition~\ref{prop:structure_in_Fourier_DMRA}(4),  
the Fourier phases satisfy
\[
\theta_\ell \equiv -\frac{\ell j \pi}{d} \pmod{\pi},\qquad\text{for }\ell\in\cI_{m,d}.
\]
Substituting these into the \eqref{eq:tildeM4} gives
\[
\theta_{\ell_1}+\theta_{\ell_2}+\theta_{\ell_3}
- \theta_{\ell_4(t)}
\equiv -\frac{j\pi}{d}(\ell_1+\ell_2+\ell_3 - (\ell_1+\ell_2+\ell_3))
\equiv 0 \pmod{\pi},
\]
so that for any $t\in\cI_{m,d}^3$
\begin{align*}
\widetilde{M}_4(\x,t)=\pm 2r_{\ell_1}r_{\ell_2}r_{\ell_3}r_{\ell_4(t)}.
\end{align*}

We now prove the ``if'' direction.  
Suppose that 
$\widetilde{M}_4(\x,t)=\pm 2r_{\ell_1}r_{\ell_2}r_{\ell_3}r_{\ell_4(t)}$ 
for all $t\in\cI_{m,d}^3$. By definition and the generic condition of $\x$, this implies that
\[
\theta_{\ell_1}+\theta_{\ell_2}+\theta_{\ell_3}
- \theta_{\ell_4(t)} \equiv 0 \pmod{\pi}.
\]
Extending the definition of $\theta_\ell$ to all integers $\ell\in\Z_{\geq 0}$ by setting
\[
\theta_\ell \triangleq \theta_{\ell \bmod d},
\]
we equivalently have
\[
e^{\,\ri 2(\theta_{\ell_1}+\theta_{\ell_2}+\theta_{\ell_3}
- \theta_{\ell_1+\ell_2+\ell_3})} = 1,
\quad\text{so that}\quad
e^{\,\ri 2\theta_{\ell_1+\ell_2+\ell_3}}
= e^{\,\ri 2\theta_{\ell_1}}\, e^{\,\ri 2\theta_{\ell_2}}\, e^{\,\ri 2\theta_{\ell_3}}.
\]
Define $f(\ell)= e^{\,\ri 2\theta_\ell}$ for $\ell\in\{0,m,2m,\dots\}$.  
Then the above relation shows that 
\[
f(\ell_1+\ell_2+\ell_3) = f(\ell_1) f(\ell_2) f(\ell_3)
\quad\text{for all such triples.}
\]
Since $\x$ is real, $\theta_0$ equals either $0$ or $\pi$, so $f(0)=1$.  
Setting $\ell_3=0$ yields
$f(\ell_1+\ell_2)=f(\ell_1)f(\ell_2)$,  
which means that $f$ satisfies the Cauchy multiplicative equation.  
Because $|f(\ell)|=1$, it defines a group character on the cyclic group
generated by $m$, which has order $d/m$.  
Hence, there exists an integer $j\in\{0,\dots,d/m-1\}$ such that
\[
e^{\,\ri 2\theta_k} = e^{-\,\ri\frac{2\pi j k}{d}},
\qquad\text{for } k\in\cI_{m,d}.
\]
This in turn implies that 
$\theta_k \equiv -\tfrac{j k\pi}{d} \pmod{\pi}$,
which is exactly the phase structure corresponding to
dihedral symmetry $\Dmj{m}{j}$, by Proposition~\ref{prop:structure_in_Fourier_DMRA}(4). This further implies $\x$ has $\Dm{m}$-symmetry by Corollary~\ref{cor:dihedral_equivalence}.
\end{proof}

\subsection{Proof of Corollary \ref{cor:DMRA_4th_population_reduced}}\label{sec:proof_DMRA_4th_population_reduced}
\begin{proof}
The ``only if'' direction follows immediately from Theorem~\ref{thm:DMRA_4th_population_general}. For the ``if'' direction, recall that
\begin{align*}
\cJ_{m,d} = \Big\{ (\ell_1,\ell_2,\ell_3) = (0, m, km) : 1 \leq k \leq  \frac{d/m - 1}{2} \Big\}.
\end{align*}
Similarly, extend the definition of $\theta_\ell$ to all integers $\ell\in\Z_{\geq 0}$ by setting $\theta_\ell=\theta_{\ell \bmod d}$. Define $f(\ell)=e^{\,\ri 2\theta_\ell}$ for $\ell\in\{0,m,2m,\ldots\}$. Since $\widetilde{M}_4(\x,t)=\pm 2 r_{\ell_1}r_{\ell_2}r_{\ell_3}r_{\ell_4(t)}$ 
for all $t\in\cJ_{m,d}$, 
\begin{align*}
\theta_0+\theta_m+\theta_{km}-\theta_{(k+1)m}\equiv 0\pmod{\pi},
\end{align*}
which implies
\begin{align*}
f\big((k+1)m\big)=f(m)f(km),\quad\text{for }k=1,\ldots,\frac{d/m - 1}{2}.
\end{align*}
Hence, by induction,
\begin{align}\label{eq:induction_f}
f(km) = (f(m))^k,\quad \text{for }k=1,\ldots,\frac{d/m +1}{2}.
\end{align}
Now use the conjugate symmetry condition, which implies
\begin{align*}
f(d-km) = \frac{1}{f(km)},\quad\text{for }k=1,\ldots,\frac{d/m - 1}{2}.
\end{align*}
Therefore, taking $k=(d/m-1)/2$ gives
\begin{align*}
f\big(\frac{d+m}{2}\big) f\big(\frac{d-m}{2}\big)=1
\end{align*}
Combining with \eqref{eq:induction_f} implies
\begin{align*}
(f(m))^{d/m}=1.
\end{align*}
Hence,
\begin{align*}
\theta_m\equiv -\frac{j\pi}{d/m} \pmod{\pi}, 
\end{align*}
for some $j\in\{0,\ldots,d/m-1\}$. Applying \eqref{eq:induction_f} and Proposition~\ref{prop:structure_in_Fourier_DMRA}(iv) completes the proof.

\end{proof}

\subsection{Proof of Proposition \ref{prop:DMRA_mean_null_collapsed}}\label{sec:proof_DMRA_mean_null_collapsed}
Write $\F\x[\ell] = r_\ell e^{\ri\theta_\ell}$ with $r_\ell\geq 0$ and $\theta_\ell\in(-\pi,\pi]$ for $\ell\in\cI_d$. We first establish two lemmas concerning the expectations of the second- and fourth-order sample moments. The proposition then follows immediately from these results.

\begin{lemma}\label{lem:DMRA_2nd_mean}
Let $m\leq d/2$ be any divisor of $d$. Then, under the D-MRA model, we have
\[
\E\Big[\F \widehat{M}_2[\ell,\ell]\Big] -d \sigma^2 =
r_\ell^2,\qquad\text{for }\ell\in\cI_d.
\]
\end{lemma}

\begin{proof}
Similar to the proof in Section \ref{sec:proof_test_size_power}, we have 
\begin{align*}
\F\widehat{M}_2=\frac{1}{N}\sum_{j=1}^N \F A_{1j} +\frac{1}{N}\sum_{j=1}^N \F A_{2j} +\frac{1}{N}\sum_{j=1}^N \F A_{2j}^{\sH} +\frac{1}{N}\sum_{j=1}^N \F A_{3j}.
\end{align*}
Here
\begin{align*}
\F A_{1j}[\ell,\ell]
&= \big(\Fb(g_j\circ x)(g_j\circ x)^{\sH}\Fb^{\sH}\big)[\ell,\ell] \\
&=
\begin{cases}
\e_{\ell}^{\T}\,\Db_{\Fb x}\,\big(\w^{S_j}(\w^{S_j})^{\sH}\big)\,\Db_{\overline{\Fb \x}}\,\e_{\ell}
= \F\x[\ell] \, w^{\ell S_j}_d w^{-\ell S_j}_d \,\overline{\F\x[\ell]}
= \big|\F \x[\ell]\big|^2, & \text{if } g_j \text{ is a c. shift},\\[6pt]
\e_{\ell}^{\T}\,\Db_{\overline{\Fb \x}}\,\big(\w^{S_j}(\w^{S_j})^{\sH}\big)\,\Db_{\Fb \x}\,\e_{\ell}
= \overline{\F\x[\ell]}\, w^{\ell S_j}_d w^{-\ell S_j}_d  \,\F\x[\ell]
= \big|\F\x[\ell]\big|^2, & \text{if } g_j \text{ is a reflection},
\end{cases}
\end{align*}
\begin{align*}
\F A_{2j}[\ell,\ell] = \begin{cases}
\e_\ell^\T \Db_{\Fb\x} \w^{S_j} (\Fb\beps_j)^{\sH}\e_\ell=\F\x[\ell] w_d^{\ell S_j} \overline{\F\beps_j[\ell]}, & \text{if } g_j \text{ is a cyclic shift},\\[6pt]
\e_\ell^\T \Db_{\overline{\Fb\x}} \w^{S_j} (\Fb\beps_j)^{\sH}\e_\ell=\overline{F\x[\ell]} w_d^{\ell S_j} \overline{\F\beps_j[\ell]}, & \text{if } g_j \text{ is a reflection},
\end{cases}
\end{align*}
and by \eqref{eq:moment1},
\begin{align*}
\F A_{3j}[\ell,\ell] = \beps_j^\T \Ab_\ell \beps_j,
\end{align*}
where $\Ab_{\ell}\in\C^{d\times d}$ with $\Ab_\ell[k_1,k_2]=w^{\ell(k_2-k_1)}$ for $k_1,k_2=0,1,\ldots,d-1$. By leveraging the independence of $\beps_j$ and $S_j$, we arrive at
\begin{align}\label{eq:second_moment_mean_DMRA}
\E\big[\F\widehat{M}_2[\ell,\ell]\big] = 
|\F\x[\ell]|^2 +d\sigma^2=r_\ell^2+d\sigma^2,
\end{align}
which proves the claim.

\end{proof}

\begin{lemma}\label{lem:DMRA_4th_mean}
Let $m\leq d/3$ be any divisor of $d$ and let $\ell_4(t)$ be defined as $\eqref{eq:l4}$ for $t=(\ell_1,\ell_2,\ell_3)$.  Then, under the D-MRA model, we have
\begin{align*}
d_t&=\E\Big[ \F\widehat{M}_4[\ell_1,\ell_2,\ell_3,\ell_4(t)] + \F\widehat{M}_4[-\ell_1,-\ell_2,-\ell_3,-\ell_4(t)] \Big]\\
&= 2r_{\ell_1}r_{\ell_2}r_{\ell_3}r_{\ell_4(t)} \cos\!\big(
\theta_{\ell_1}+\theta_{\ell_2}+\theta_{\ell_3}
- \theta_{\ell_4(t)}
\big) \\
&\quad + \1\{\ell_1+\ell_2\equiv 0 \ (\mathrm{mod}\ d) \} \cdot\big( 2d\sigma^2(r_{\ell_1}^2/2+r_{\ell_2}^2/2+ r_{\ell_3}^2)+ 2d^2\sigma^4 \big)\\
&\quad + \1\{\ell_1+\ell_3\equiv 0 \ (\mathrm{mod}\ d) \} \cdot\big( 2d\sigma^2(r_{\ell_1}^2/2+r_{\ell_3}^2/2+ r_{\ell_2}^2)+ 2d^2\sigma^4 \big)\\
&\quad + \1\{\ell_2+\ell_3\equiv 0 \ (\mathrm{mod}\ d) \} \cdot\big( 2d\sigma^2(r_{\ell_2}^2/2+r_{\ell_3}^2/2+ r_{\ell_1}^2)+ 2d^2\sigma^4 \big),
\end{align*}
for $\ell_1,\ell_2,\ell_3\in\cI_d$.
\end{lemma}

\begin{proof}
By definition,
\begin{align*}
\F\widehat{M}_4=&\frac{1}{N}\sum_{j=1}^N (\y_j^{\otimes 4})\times_1 \Fb\times_2 \Fb\times_3\Fb\times_4 \overline{\Fb}\\
=& \frac{1}{N}\sum_{j=1}^N (\F\y_j) \otimes (\F\y_j)\otimes (\F\y_j)\otimes (\overline{\F\y_j}).
\end{align*}
Since $\y_j=g_j\circ \x+\beps_j$, each factor contains a signal component and a noise component. Expanding the four-fold tensor, we can group all resulting terms according to the number of noise factors they contain. In particular, for each $j$ we write 
\begin{align*}
(\F\y_j) \otimes (\F\y_j)\otimes (\F\y_j)\otimes (\overline{\F\y_j})= B_{0j}+B_{1j}+B_{2j}+B_{3j}+B_{4j},
\end{align*}
where $B_{ij}$ collects all terms that contain exactly $i$ factors of the noise $\F\beps_j$ and $(4-i)$ factors of the signal $\F(g_j\circ\x)$. Since any term involving an odd number of noise factors has zero mean, we immediately obtain 
\begin{align*}
    \E [B_{1j}]=0,\qquad \E [B_{3j}]=0.
\end{align*}
Thus, we turn to study the expectation of the rest terms. For $B_{0j}$ we have
\begin{align*}
\E \big[B_{0j}[\ell_1,\ell_2,\ell_3,\ell_4(t)]\big]=&\E\Big[
\F(g_j\circ \x)[\ell_1] \F(g_j\circ \x)[\ell_2] \F(g_j\circ \x)[\ell_3] \overline{\F(g_j\circ \x)[\ell_4(t)]}\Big]\\
=& \F M_4(\x,\rho)[\ell_1,\ell_2,\ell_3,\ell_4(t)].
\end{align*}
This implies that
\begin{align}
&\E \big[B_{0j}[\ell_1,\ell_2,\ell_3,\ell_4(t)]+ B_{0j}[-\ell_1,-\ell_2,-\ell_3,-\ell_4(t)]\big] \notag  \\
&= 2\Re\!\big( \F\x[\ell_1]\F\x[\ell_2]\F\x[\ell_3]\overline{\F\x[\ell_4(t)]} \big)\notag\\
&=2r_{\ell_1}r_{\ell_2}r_{\ell_3}r_{\ell_4(t)} \cos\!\big(
\theta_{\ell_1}+\theta_{\ell_2}+\theta_{\ell_3}
- \theta_{\ell_4(t)}
\big). \label{eq:B0j_mean}
\end{align}
For $B_{4j}$, according to Lemma \ref{lem:Gaussian_Fourier_moment}, we have
\begin{align}
&\E\big[B_{4j}[\ell_1,\ell_2,\ell_3,\ell_4(t)]\big] \notag\\
&=\E\Big[\F\beps_j[\ell_1]\F\beps_j[\ell_2]\F\beps_j[\ell_3]\overline{\F\beps_j[\ell_4(t)]}\Big] \notag\\ 
&= d^2\sigma^4 \cdot \big(\1\{ \ell_1+\ell_2\equiv 0 \ (\mathrm{mod}\ d) \}+ \1\{ \ell_2+\ell_3\equiv 0 \ (\mathrm{mod}\ d) \}+\1\{ \ell_1+\ell_3\equiv 0 \ (\mathrm{mod}\ d) \}\big)\label{eq:B4j_mean}
\end{align}
For $B_{2j}$, we have
\begin{align*}
&\E\big[B_{2j}[\ell_1,\ell_2,\ell_3,\ell_4(t)]\big] \\
&=\E\Big[ \F (g_j\circ \x)[\ell_1]\F(g_j\circ \x)[\ell_2]\F\beps_j[\ell_3]\overline{\F\beps_j[\ell_4(t)]} \Big]\\
&\quad +\E\Big[ \F (g_j\circ \x)[\ell_1]\F\beps_j[\ell_2]\F(g_j\circ \x)[\ell_3]\overline{\F\beps_j[\ell_4(t)]} \Big]\\
&\quad +\E\Big[ \F\beps_j[\ell_1]\F (g_j\circ \x)[\ell_2]\F(g_j\circ \x)[\ell_3]\overline{\F\beps_j[\ell_4(t)]} \Big]\\
&\quad +\E\Big[ \F (g_j\circ \x)[\ell_1]\F\beps_j[\ell_2]\F\beps_j[\ell_3]\overline{\F(g_j\circ \x)[\ell_4(t)]} \Big]\\
&\quad +\E\Big[ \F\beps_j[\ell_1]\F (g_j\circ \x)[\ell_2]\F\beps_j[\ell_3]\overline{\F(g_j\circ \x)[\ell_4(t)]} \Big]\\
&\quad +\E\Big[ \F\beps_j[\ell_1]\F\beps_j[\ell_2]\F (g_j\circ \x)[\ell_3]\overline{\F(g_j\circ \x)[\ell_4(t)]} \Big].
\end{align*}
According to Lemma \ref{lem:Gaussian_Fourier_moment}, we further have
\begin{align*}
&\E\big[B_{2j}[\ell_1,\ell_2,\ell_3,\ell_4(t)]\big] \\
&=d\sigma^2\cdot \1\{\ell_1+\ell_2\equiv 0 \ (\mathrm{mod}\ d) \} \cdot\E\Big[ \F (g_j\circ \x)[\ell_1]\F (g_j\circ \x)[-\ell_1] \Big]\\
&\quad + d\sigma^2\cdot \1\{\ell_1+\ell_3\equiv 0 \ (\mathrm{mod}\ d) \} \cdot\E\Big[ \F (g_j\circ \x)[\ell_1]\F (g_j\circ \x)[-\ell_1] \Big]\\
&\quad + d\sigma^2\cdot \1\{\ell_2+\ell_3\equiv 0 \ (\mathrm{mod}\ d) \} \cdot\E\Big[ \F (g_j\circ \x)[\ell_2]\F (g_j\circ \x)[-\ell_2] \Big]\\
&\quad + d\sigma^2\cdot \1\{\ell_2+\ell_3\equiv 0 \ (\mathrm{mod}\ d) \} \cdot\E\Big[ \F (g_j\circ \x)[\ell_1] \overline{\F (g_j\circ \x)[\ell_1]} \Big]\\
&\quad + d\sigma^2\cdot \1\{\ell_1+\ell_3\equiv 0 \ (\mathrm{mod}\ d) \} \cdot\E\Big[ \F (g_j\circ \x)[\ell_2] \overline{\F (g_j\circ \x)[\ell_2]} \Big]\\
&\quad + d\sigma^2\cdot \1\{\ell_1+\ell_2\equiv 0 \ (\mathrm{mod}\ d) \} \cdot\E\Big[ \F (g_j\circ \x)[\ell_3] \overline{\F (g_j\circ \x)[\ell_3]} \Big]
\end{align*}
Noting that under the D-MRA setting, 
\begin{align*}
\E\Big[ \F (g_j\circ \x)[\ell] \overline{\F (g_j\circ \x)[\ell]} \Big]= \F M_2(\x,\rho)[\ell,\ell]=|\F\x[\ell]|^2=r_{\ell}^2,~~~\text{for }0\leq \ell\leq d-1,
\end{align*}
we further obtain
\begin{align}
&\E\big[B_{2j}[\ell_1,\ell_2,\ell_3,\ell_4(t)]\big]\notag\\
&=d\sigma^2\cdot \1\{\ell_1+\ell_2\equiv 0 \ (\mathrm{mod}\ d) \} \cdot \big( r_{\ell_1}^2/2+r_{\ell_2}^2/2+ r_{\ell_3}^2\big)\notag\\
&\quad + d\sigma^2\cdot \1\{\ell_1+\ell_3\equiv 0 \ (\mathrm{mod}\ d) \} \cdot \big( r_{\ell_1}^2/2+r_{\ell_3}^2/2+ r_{\ell_2}^2\big)\notag\\
&\quad + d\sigma^2\cdot \1\{\ell_2+\ell_3\equiv 0 \ (\mathrm{mod}\ d) \} \cdot \big( r_{\ell_2}^2/2+r_{\ell_3}^2/2+ r_{\ell_1}^2\big).\label{eq:B2j_mean}
\end{align}
Combining \eqref{eq:B0j_mean}, \eqref{eq:B2j_mean}, and \eqref{eq:B4j_mean} gives the result.
\end{proof}

With these two lemmas, we return to the proof of Proposition~\ref{prop:DMRA_mean_null_collapsed}.
\begin{proof}
Due to Lemma \ref{lem:DMRA_2nd_mean}, we have $s_\ell=r_\ell^2$ for $\ell\in\cI_d$. By Proposition~\ref{prop:structure_in_Fourier_DMRA}(4), we get $s_\ell=0$ for $\ell\in\cI_{m,d}^c$, and hence for all $\ell \in \cI_{m,d}^{c,+}$.

For $t=(\ell_1,\ell_2,\ell_3)\in \cI_{m,d}^3$ and any vector $\u=(u_\ell)\in\R^d$, further define $b_t:\R^d\to \R$ by
\begin{align}
b_t(\u)\triangleq &\1\{\ell_1 + \ell_2 \equiv 0 \pmod d\} \cdot\bigl[d\sigma^2 (u_{\ell_1} + u_{\ell_2} + 2u_{\ell_3}) + 2d^2\sigma^4\bigr] \notag\\
  &+ \1\{\ell_1 + \ell_3 \equiv 0 \pmod d\} \cdot\bigl[d\sigma^2 (u_{\ell_1} + u_{\ell_3} + 2u_{\ell_2}) + 2d^2\sigma^4\bigr] \notag\\
  &+ \1\{\ell_2 + \ell_3 \equiv 0 \pmod d\} \cdot\bigl[d\sigma^2 (u_{\ell_2} + u_{\ell_3} + 2u_{\ell_1}) + 2d^2\sigma^4\bigr].\label{eq:bt}
\end{align}
Note that for every $t=(\ell_1,\ell_2,\ell_3)\in\cJ_{m,d}$, 
\begin{align*}
0+m\not\equiv 0, \quad 0+km\not\equiv 0,\quad m+km\not\equiv 0 \pmod{d},
\end{align*}
for $r=1,\ldots,(d/m-1)/2$
we get $b_t(\u)=0$ for $t\in\cJ_{m,d}$ and $\u\in\R^d$. Due to Lemma \ref{lem:DMRA_4th_mean}, we have
\begin{align*}
d_t=d_t-b_t(\s)=2r_{\ell_1}r_{\ell_2}r_{\ell_3}r_{\ell_4(t)} \cos\!\big(
\theta_{\ell_1}+\theta_{\ell_2}+\theta_{\ell_3}
- \theta_{\ell_4(t)}
\big).
\end{align*}
By Proposition~\ref{prop:structure_in_Fourier_DMRA}(4), we get, for $t\in\cJ_{m,d}$,
\begin{align}\label{eq:dt}
d_t^2=4 r_{\ell_1}^2 r_{\ell_2}^2 r_{\ell_3}^2 r_{\ell_4(t)}^2=4s_{\ell_1}s_{\ell_2}s_{\ell_3}s_{\ell_4(t)}=4s_{\ell_1}s_{\ell_2}s_{\ell_3}s_{\ell_4^{\sharp}(t)},
\end{align}
where the last equality follows from the conjugate symmetry.

In view of Corollary~\ref{cor:DMRA_4th_population_reduced} and the conjugate symmetry, the two conditions \eqref{eq:DMRA_2nd_population_condition} and \eqref{eq:DMRA_4th_population_condition} are also sufficient for $\Dm{m}$-symmetry when $\x$ is a generic $\Cd{m}$ signal in the sense of Theorem~\ref{thm:DMRA_4th_population_general}.
\end{proof}

\subsection{Proof of Theorem~\ref{thm:DMRA_Dm_testing}}\label{sec:proof_DMRA_Dm_testing}
We first introduce some additional notation. For $j=1,\ldots,N$, define the auxiliary vectors
\begin{align*}
\bmu_j\triangleq  \Big( (s_{\ell,j})_{\ell\in\cI_d^+}, (d_{t,j})_{t\in\mathcal{J}_{m,d}} \Big)^\T \in\R^{|\cI_d^+|+(d/m-1)/2},\quad\text{and } \widehat{\bmu}\triangleq \frac{1}{N}\sum_{j=1}^N \bmu_j.
\end{align*}
Denote the corresponding expectation by
\begin{align}\label{eq:mu}
\bmu\triangleq\E[\bmu_j]=\Big( (s_{\ell})_{\ell\in\cI_d^+}, (d_{t})_{t\in\mathcal{J}_{m,d}} \Big)^\T.
\end{align}
Denote the corresponding covariance matrix by
\begin{align}\label{eq:Omega_mu}
\bOmega_{\bmu}\triangleq \Var(\bmu_j).
\end{align}
We have the following lemma, which establishes that $\bOmega_{\bmu}$ is positive definite.
\begin{lemma}\label{lem:positive_Sigma_mu}
Assume the collapsed regime in the D-MRA setting. The covariance matrix $\bOmega_{\bmu}$ defined in \eqref{eq:Omega_mu} is positive definite.
\end{lemma}
\begin{proof}
Let 
\begin{align*}
L_j=\sum_{\ell\in\cI_d^+} \alpha_\ell s_{\ell,j}+\sum_{t\in\cJ_{m,d}} \beta_t d_{t,j}
\end{align*}
be a real linear combination of the coordinates of $\bmu_j$, and suppose $\Var(L_j)=0$. We aim to show all coefficients $\alpha_\ell$ and $\beta_t$ vanish.

Condition on the latent group element $g_j=g$. In the Fourier domain,
\begin{align*}
\F\y_j[\ell] = \F(g\circ \x)[\ell]+\F\beps_j[\ell],
\end{align*}
where the first quantity is deterministic and the second is centered Gaussian Fourier noise. For real observations, the independent noise coordinates are exactly the folded representatives: $\F\beps_j[0]=a_0$ and, when present, $\F\beps_j[d/2]=a_{d/2}$, are real Gaussians, while $\F\beps_j[\ell]=a_{\ell}+\ri b_{\ell}$ for $1\leq \ell< d/2$ are independent complex Gaussian coordinates with $\F\beps_j[d-\ell]=\overline{\F\beps_j[\ell]}$. All the random variables $\{a_\ell,b_{\ell}:1\leq \ell< d/2\}$ and $a_0, a_{d/2}$ are real-valued and independent Gaussian distributed.

Since $L_j$ has zero variance, its conditional version is almost surely constant for every $g$ with respect to the underlying distribution $\rho$. As a polynomial in the independent Gaussian coordinates, it must therefore be identically constant. Inspect the highest-degree part. The second-order coordinates $s_{\ell,j}$ contribute only quadratic leading terms, whereas the fourth-order coordinates $d_{t,j}$ contribute quartic leading terms. Thus the quartic leading terms must cancel among themselves.

For $t=(0,m,km)$ with $k=1,\ldots,(d/m-3)/2$, each $d_{t,j}$ gives
\begin{align*}
2\Re\big(a_0 (a_m+\ri b_m) (a_{km}+\ri b_{km}) (a_{(k+1)m}-\ri b_{(k+1)m})\big),
\end{align*}
and the last $d_{t,j}$ with $t=(0,m, (d/m-1)m/2)$ gives
\begin{align*}
&2\Re\Big(a_0 (a_m+\ri b_m) \big(a_{\frac{(d/m-1)m}{2}}+\ri b_{\frac{(d/m-1)m}{2}}\big) \big(a_{\frac{(d/m+1)m}{2}}-\ri b_{\frac{(d/m+1)m}{2}}\big)\Big)\\
&=2\Re\Big(a_0 (a_m+\ri b_m) \big(a_{\frac{(d/m-1)m}{2}}+\ri b_{\frac{(d/m-1)m}{2}}\big) \big(a_{\frac{(d/m-1)m}{2}}+\ri b_{\frac{(d/m-1)m}{2}}\big)\Big),
\end{align*}
by the conjugate symmetry condition. The pair $(km,(k+1)m)$ for $k=1,\ldots,(d/m-3)/2$ and the last one $((d/m-1)m/2,(d/m-1)m/2)$ uniquely identify the leading terms. Hence these leading polynomials are linearly independent across $t\in\cJ_{m,d}$. Therefore, all $\beta_t=0$.

After the quartic part is removed, the remaining leading part is quadratic
\begin{align*}
\sum_{\ell\in\cI_{d}^+} \alpha_\ell |\F\beps_j[\ell]|^2.
\end{align*}
Clearly, these quadratic polynomials are also linearly independent, so $\alpha_{\ell}=0$.

In conclusion, no nontrivial linear combination has zero variance, and $\bOmega_{\bmu}$ is positive definite.
\end{proof}

We then link the auxiliary vector $\widehat{\bmu}$ with the test vector $\widehat{\g}$. Recall the definitions of $\ell_i(t)$ for $t\in\cJ_{m,d}$, $i=1,2,3,4$ in \eqref{eq:l123} and \eqref{eq:l4}. For $\u=(u_\ell)_{\ell\in\cI_d^+}$ and $\vv=(v_t)_{t\in\cJ_{m,d}}$, define the polynomial map $\varphi:\R^{|\cI_d^+|+(d/m-1)/2}\to\R^{|\cI_{m,d}^{c,+}|+(d/m-1)/2}$ by
\begin{align}\label{eq:varphi}
\varphi(\u,\vv)\triangleq \Big( (u_\ell)_{\ell\in\cI_{m,d}^{c,+}}, (4u_{\ell_1(t)} u_{\ell_2(t)} u_{\ell_3(t)} u_{\ell_4^{\sharp}(t)}- v_t^2)_{t\in\mathcal{J}_{m,d}} \Big)^\T,
\end{align}
where we refer to the first block as the cyclic part and the second block as the dihedral part. Then by construction
\begin{align*}
\widehat{\g}_m=\varphi(\widehat{\bmu}).
\end{align*}
To facilitate subsequent hypothesis testing, we next derive the explicit form of the differential of $\varphi$. In particular, we will show that $\nabla\varphi$ has full rank at the population parameter. These properties are summarized in the following lemma.

\begin{lemma}[Full row rank of the Jacobian under the null and nondegenerate alternatives for the collapsed regime]
\label{lem:Jacobian_full_rankness}
Assume the collapsed regime in the D-MRA setting. Let $\bmu$
be the population auxiliary vector defined in \eqref{eq:mu} at which the Jacobian
\(\nabla\varphi(\bmu)\) is evaluated, and let \(q_m\) be as defined in
\eqref{eq:q}. Then \(\nabla\varphi(\bmu)\) has full row rank \(q_m\) under either of the following two conditions.
\begin{enumerate}
\item[(i)] The null hypothesis holds that the signal $\x\in\R^d$ belongs to the symmetry class $\Dm{m}$. Moreover, $x\in\R^d$ is a generic $\Dm{m}$
signal in the sense that
\[
r_{km}>0,\qquad k=0,1,\ldots,\frac{d/m-1}{2}.
\]

\item[(ii)] Under a fixed alternative, the selected fourth-order population
coordinates are nonzero:
\[
d_t\neq 0,\qquad t\in \cJ_{m,d}.
\]
\end{enumerate}
\end{lemma}

\begin{proof}
For the cyclic rows, we have $\varphi_\ell(\u,\vv)=u_{\ell}$ for $\ell\in\cI_{m,d}^{c,+}$, and thus
\begin{align*}
\frac{\partial \varphi_\ell}{\partial u_a} (\u,\vv)=& \1\{a=\ell\},\quad\text{for }a\in\cI_d^+,\\
\frac{\partial \varphi_\ell}{\partial v_b} (\u,\vv)=&0,\quad\text{for }b\in\cJ_{m,d}.
\end{align*}
For the dihedral rows, we have, for $t=(0,m,km)\in\cJ_{m,d}$ with $k=1,\ldots,(d/m-1)/2$
\begin{align*}
\varphi_t(\u,\vv)&=4u_{\ell_1(t)} u_{\ell_2(t)} u_{\ell_3(t)} u_{\ell_4^{\sharp}(t)}- v_t^2,\\
&= 4u_0 u_m u_{km} u_{((k+1)m)^{\sharp}} - v_t^2, \\
&=
\begin{cases}
4u_0 u_m u_{km} u_{(k+1)m} - v_t^2, & \text{for } 1\leq k< (d/m-1)/2,\\
4u_0 u_m u_{km}^2 - v_t^2, & \text{for } k = (d/m-1)/2,
\end{cases}
\end{align*}
where we notice $((k+1)m)^{\sharp}=km$ when $k=(d/m-1)/2$ for the second case. Thus, for $t=(0,m,km)\in\cJ_{m,d}$, we have
\begin{align*}
\frac{\partial \varphi_t}{\partial u_a}(\u,\vv) =& 4\Big(
\1\{a=0\}u_m u_{km} u_{((k+1)m)^{\sharp}}+
\1\{a=m\}u_0 u_{km} u_{((k+1)m)^{\sharp}}\\
&+
\1\{a=km\}u_0 u_m u_{((k+1)m)^{\sharp}}+
\1\{a=((k+1)m)^{\sharp}\}u_0u_m u_{km}
\Big),\quad\text{for }a\in\cI_d^+,\\
\frac{\partial \varphi_t}{\partial v_b} (\u,\vv)=&-2v_t\1\{b=t\},\quad\text{for }b\in\cJ_{m,d}.
\end{align*}
As before, the expression of $\partial \varphi_t/\partial u_a$ simplifies depending on the value of $k$: if $1\leq k< (d/m-1)/2$, one can drop the $\sharp$ notation, whereas for $k=(d/m-1)/2$, one substitutes $((k+1)m)^{\sharp}=km$.

Equivalently, define matrices
\begin{align*}
\Eb_{\ell,a}^{(11)}=\1\{a=\ell\},\quad \Eb_{t,a}^{(21)}(\u)=\frac{\partial \varphi_t}{\partial u_a}(\u,\vv),\quad \Eb_{t,b}^{(22)}(\vv)=-2v_t\1\{b=t\}.
\end{align*}
Here $\Eb^{(11)}$ has rows $\ell\in\cI_{m,d}^{c,+}$ and columns $a\in\cI_d^+$, $\Eb^{(21)}$ has rows $t\in \cJ_{m,d}$ and columns $a\in\cI_d^+$, and $\Eb^{(22)}$ is the $|\cJ_{m,d}|\times |\cJ_{m,d}|$ diagonal block indexed by $\cJ_{m,d}$. Then the full $\nabla\varphi$ is 
\begin{align*}
\nabla\varphi(\u,\vv) = \begin{pmatrix}
\Eb^{(11)} & \0 \\
\Eb^{(21)}(\u) & \Eb^{(22)}(\vv)
\end{pmatrix}.
\end{align*}
Plugging in $(\u,\vv)=\bmu$ gives the Jacobian $\nabla\varphi(\bmu)$. Under either of the two conditions in the lemma statement and due to Lemma~\ref{lem:DMRA_4th_mean}, all  $d_t\neq 0$ for $t\in\cJ_{m,d}$, and hence the Jacobian $\nabla\varphi(\bmu)$ is of full rank.
\end{proof}

We are now ready to prove Theorem \ref{thm:DMRA_Dm_testing}.
\begin{proof}
We first consider the case where the null hypothesis holds that $\x$ belongs to $\Dm{m}$-symmetry. Since $\E\|\F\y_j\|^8< \infty$, we have $\E\|\bmu_j\|^2<\infty$. Hence, by the central limit theorem,
\begin{align*}
\sqrt{N}\, (\widehat{\bmu}-\bmu)\xrightarrow{d} \mathcal{N}(\0,\bOmega_{\bmu}),
\end{align*}
where $\bOmega_{\bmu}$ is the covariance matrix defined in \eqref{eq:Omega_mu}. Now applying the polynomial map $\varphi$ defined in \eqref{eq:varphi}, we have by construction
\begin{align*}
\widehat{\g}_m=\varphi(\widehat{\bmu}).
\end{align*}
Under the null hypothesis, Proposition~\ref{prop:DMRA_mean_null_collapsed} implies that $\varphi(\bmu)=\0$.
Since $\varphi$ is continuously differentiable in a neighborhood of $\bmu$, the delta method (see, e.g., \cite[Theorem 3.1]{van2000asymptotic}) yields
\begin{align*}
\sqrt{N}\, \widehat{\g}_m= \sqrt{N}\, (\varphi(\widehat{\bmu})-\varphi(\bmu))
&= \sqrt{N}\, \nabla\varphi(\bmu) (\widehat{\bmu}-\bmu) + o_p(1)\\
&= \frac{1}{\sqrt{N}}\sum_{j=1}^N \nabla\varphi(\bmu)(\bmu_j-\bmu) + o_p(1)
\xrightarrow{d} \mathcal{N}(\0, \bOmega_\g),
\end{align*}
where
\begin{align*}
\bOmega_\g =& \nabla\varphi(\bmu)\, \bOmega_{\bmu}\, \nabla\varphi(\bmu)^\T\\
=&\Var(\nabla\varphi(\bmu)(\bmu_j-\bmu)).
\end{align*}
Note that, from the explicit Jacobian calculation in the proof of Lemma~\ref{lem:Jacobian_full_rankness}, we have
\begin{align*}
\bpsi_j = \nabla\varphi(\bmu)(\bmu_j-\bmu),
\end{align*}
which corresponds exactly to the coordinate-wise expressions given in \eqref{eq:psi_ell} and \eqref{eq:psi_t}. Therefore,
\begin{align*}
\bOmega_\g = \Var(\bpsi_j).
\end{align*}
Moreover, by Lemmas~\ref{lem:positive_Sigma_mu} and~\ref{lem:Jacobian_full_rankness}, under the generic condition in Lemma~\ref{lem:Jacobian_full_rankness}(i), $\bOmega_\g$ is positive definite.

Next we aim to show $\widehat{\bOmega}_{\g}$ defined in \eqref{eq:Omegahat_g} is a consistent estimator of $\bOmega_\g$. By definition of $\widehat{\bpsi}_j$ in \eqref{eq:psihat_ell} and \eqref{eq:psihat_t}, we obtain it from $\bpsi_j$ by replacing the population quantities $(s_\ell, d_t)$ with their estimators $(\widehat{S}_\ell, \widehat{D}_t)$. Hence,
\begin{align*}
\widehat{\bpsi}_j = \nabla\varphi(\widehat{\bmu})(\bmu_j-\widehat{\bmu}).
\end{align*}
It follows
\begin{align*}
\widehat{\bOmega}_{\g}=&\frac{1}{N}\sum_{j=1}^N \widehat{\bpsi}_j \widehat{\bpsi}_j^\T\\
=&\nabla\varphi(\widehat{\bmu})\big(\frac{1}{N}\sum_{j=1}^N (\bmu_j-\widehat{\bmu})(\bmu_j-\widehat{\bmu})^{\T} \big) (\nabla\varphi(\widehat{\bmu}))^{\T}
\end{align*}
First, the law of large numbers gives
\begin{align*}
\widehat{\bmu}\xrightarrow{p} \bmu.
\end{align*}
Since the entries of the Jacobian $\nabla\varphi(\u,\vv)$ are polynomials in the $\u$-coordinates and linear in the $\vv$-coordinates, the continuous mapping theorem gives
\begin{align*}
\nabla\varphi(\widehat{\bmu})\xrightarrow{p}    \nabla\varphi(\bmu).
\end{align*}
Second, the sample covariance matrix
\begin{align*}
\frac{1}{N}\sum_{j=1}^N (\bmu_j-\widehat{\bmu})(\bmu_j-\widehat{\bmu})^{\T}=\frac{1}{N}\sum_{j=1}^N \bmu_j\bmu_j^{\T} -\widehat{\bmu} \widehat{\bmu}^\T\xrightarrow{p} \E[\bmu_j\bmu_j^{\T}] - \bmu \bmu^\T=\E\big[(\bmu_j-\bmu)(\bmu_j-\bmu)^{\T}\big] 
\end{align*}
Therefore, using the continuous mapping theorem again,
\begin{align*}
\widehat{\bOmega}_{\g}\xrightarrow{p} \nabla\varphi(\bmu) \E\big[(\bmu_j-\bmu)(\bmu_j-\bmu)^{\T}\big] (\nabla\varphi(\bmu))^{\T}=\bOmega_\g.
\end{align*}
Slutsky's theorem then implies the chi-squared limit for the statistic $\widehat{W}_m=N \cdot \widehat{\g}_m^\T \widehat{\bOmega}_\g^{-1} \widehat{\g}_m$ in \eqref{eq:chi_g}, and the p-value formula \eqref{eq:raw_pvalue_dihedral} follows.

Now consider any alternative hypothesis where at least one of the conditions in \eqref{eq:DMRA_2nd_population_condition} and \eqref{eq:DMRA_4th_population_condition} fails. Then the law of large numbers implies $\widehat{\g}_m$ converges to a non-zero limit $\g_m^{\mathrm{alt}}$ under the alternative. By Lemma~\ref{lem:Jacobian_full_rankness}, $\widehat{\bOmega}_\g$ converges to some positive definite matrix $\bOmega_\g^{\mathrm{alt}}$ given condition (ii). Applying the continuous mapping theorem,
\begin{align*}
\frac{1}{N}\widehat{W}_m =\widehat{\g}_m^\T \widehat{\bOmega}_\g^{-1} \widehat{\g}_m \xrightarrow{p} (\g_m^{\mathrm{alt}})^{\T} (\bOmega_\g^{\mathrm{alt}})^{-1}  \g_m^{\mathrm{alt}}>0,
\end{align*}
and hence $\widehat{W}_m\to \infty$ in probability.

\end{proof}

\subsection{Proof of Proposition \ref{prop:DMRA_mean_null_parity_split}}\label{sec:proof_DMRA_mean_null_parity_split}

\begin{proof}
We first prove the ``only if'' direction. Suppose $\x$ is invariant under $\Dmj{m}{j}$. By Proposition~\ref{prop:structure_in_Fourier_DMRA}(iv), $\F\x[0]\in\R$ and $\F\x[d/2]\in\R$. When $j$ is odd, $s_{d/2}=|\F\x[d/2]|^2=0$. For every nonzero frequency $\ell\in\cI_{m,d}^+\setminus\{0,d/2\}$,
\[
    \theta_\ell \equiv -\frac{\ell j\pi}{d}\pmod\pi.
\]
For $t=(0,m,km)\in \cJ^{<}_{m,d}$, this gives
\[
    \theta_m+\theta_{km}-\theta_{(k+1)m}
    \equiv 0 \pmod\pi,
\]
and therefore $4s_{\ell_1}s_{\ell_2}s_{\ell_3}s_{\ell_4^{\sharp}(t)} - d_t^2=0$.

For the signed parity tuple, note that $\ell_4(t_\star)=d/2+m=d-(d/2-m)$. Conjugate symmetry condition gives
\[
    \F\x[d/2+m]=\overline{\F\x[d/2-m]}.
\]
Hence
\[
d_{t_\star}=2\Re\big((\F\x[m])^2 \F\x[d/2-m]\overline{\F\x[d/2+m]}\big)=2\Re\big((\F\x[m])^2 (\F\x[d/2-m])^2 \big).
\]
Using the $\Dmj{m}{j}$ phase relation in the last display yields
\[
    d_{t_\star}=2(-1)^j s_m s_{d/2-m}.
\]
Thus $\kappa_m^0=0$ when $j$ is even and $\kappa_m^1=0$ when $j$ is odd. This proves necessity for both parity classes.

We now prove the ``if'' direction. Assume first that
\[
4s_{\ell_1}s_{\ell_2}s_{\ell_3}s_{\ell_4^{\sharp}(t)} - d_t^2=0\qquad \text{for all }t\in \cJ^{<}_{m,d}.
\]
Given the generic condition of $\x$, this implies for $1\leq k\leq d/(2m)-2$,
\[
    \theta_0+\theta_m+\theta_{km}-\theta_{(k+1)m}
    \equiv 0 \pmod\pi.
\]
Define $f(\ell)=e^{\ri 2\theta_{\ell}}$ for $\ell\in\{0,m,2m,\ldots\}$ for the nonzero coordinates $\ell$. Since $\F\x[0]$ is real and nonzero, $f(0)=1$. Therefore, the preceding congruence is equivalent to
\[
    f((k+1)m)=f(m)f(km),
    \qquad \text{for }k=1,\ldots,d/(2m)-2.
\]
By induction,
\begin{align}\label{eq:induction_f_parity}
f(km)=f(m)^k,
    \qquad \text{for }k=1,\ldots,d/(2m)-1.
\end{align}
Thus the selected chain determines the phase progression up to the last nonvanishing frequency $d/2-m$ before $d/2$, but it does not by itself quantize the base $f(m)$. It remains to close this base. From conjugate symmetry and the definition of $t_\star$,
\[
d_{t_\star}=2\Re\big((\F\x[m])^2 (\F\x[d/2-m])^2 \big) =2s_ms_{d/2-m}\cos\!\big(2\theta_m+2\theta_{d/2-m}\big).
\]
Since $s_ms_{d/2-m}>0$, we can divide by $2s_ms_{d/2-m}$. By \eqref{eq:induction_f_parity},
\[
    e^{2\ri\theta_{d/2-m}}=f(d/2-m)=f(m)^{d/(2m)-1}=e^{2\ri(d/(2m)-1)\theta_m}.
\]
Consequently,
\[
    2\theta_m+2\theta_{d/2-m}
    \equiv \frac{d}{m}\cdot\theta_m \pmod{2\pi}. 
\]

For the even parity class, assume in addition that $\kappa_m^0=0$. Then
\[
    d_{t_\star}=2s_ms_{d/2-m},
\]
so this implies
\[
    \cos\Big(\frac{d}{m}\cdot\theta_m\Big)=1.
\]
Equivalently, $d\theta_m/m\equiv 0\pmod{2\pi}$. Hence there exists an even integer $j$ such that
\[
    \theta_m\equiv -\frac{mj\pi}{d}\pmod\pi.
\]
Combining this with \eqref{eq:induction_f_parity} gives, for $k=1,\ldots,d/(2m)-1$,
\[
    \theta_{km}\equiv k\theta_m
    \equiv -\frac{kmj\pi}{d}
    \pmod\pi.
\]
The coordinate $\F\x[d/2]$, if nonzero, is real because $\x$ is real. The remaining half of the spectrum follows from conjugate symmetry. Therefore $\x$ is invariant under $\Dmj{m}{j}$ for some even $j$, so $\x$ belongs to $\Dm{m}^0$.

For the odd class, assume instead that $\kappa_m^1=0$ and $s_{d/2}=0$. The equation $\kappa_m^1=0$ gives
\[
    d_{t_\star}=-2s_ms_{d/2-m},
\]
and hence
\[
    \cos\Big(\frac{d}{m}\cdot\theta_m\Big)=-1.
\]
Equivalently, $d\theta_m/m\equiv \pi\pmod{2\pi}$. Thus there exists an odd integer $j$ such that
\[
    \theta_m\equiv -\frac{mj\pi}{d}\pmod\pi.
\]
The same induction relation \eqref{eq:induction_f_parity} gives
\[
    \theta_{km}\equiv -\frac{kmj\pi}{d}\pmod\pi,
    \qquad k=1,\ldots,d/(2m)-1.
\]
The additional condition $s_{d/2}=0$ is exactly the condition required by Proposition~\ref{prop:structure_in_Fourier_DMRA}(iv) for odd $j$. Again, conjugate symmetry gives the remaining frequencies. Therefore $\x$ is invariant under $\Dmj{m}{j}$ for some odd $j$, so $\x$ belongs to $\Dm{m}^1$.

The assumed nonvanishing of $\F \x[m]$ rules out any strictly larger cyclic symmetry than $\Cd{m}$; hence under the stated generic assumptions, the corresponding detectable class is exactly $\Dm{m}^0$ or $\Dm{m}^1$ rather than a finer dihedral class. This completes the proof.
\end{proof}

\subsection{Proof of Theorem~\ref{thm:DMRA_Dm_testing_parity_split}}\label{sec:proof_DMRA_Dm_testing_parity_split}

The proof follows the same strategy as that of
Theorem~\ref{thm:DMRA_Dm_testing}, with only a few minor modifications. For $j=1,\ldots,N$, define the auxiliary vectors
\begin{align*}
\bmu_j'\triangleq  \Big( (s_{\ell,j})_{\ell\in\cI_d^+}, (d_{t,j})_{t\in\mathcal{J}_{m,d}^{<}\cup\{t_\star\} } \Big)^\T \in\R^{|\cI_d^+|+d/(2m)-1},\quad\text{and } \widehat{\bmu}'\triangleq \frac{1}{N}\sum_{j=1}^N \bmu_j'.
\end{align*}
Denote the corresponding expectation by
\begin{align}\label{eq:mu_parity_split}
\bmu'\triangleq\E[\bmu_j']=\Big( (s_{\ell})_{\ell\in\cI_d^+}, (d_{t})_{t\in\mathcal{J}_{m,d}^{<}\cup \{t_\star\}} \Big)^\T.
\end{align}
Denote the corresponding covariance matrix by
\begin{align}\label{eq:Omega_mu_parity_split}
\bOmega_{\bmu}'\triangleq \Var(\bmu_j').
\end{align}
We have the following lemma, which establishes that $\bOmega_{\bmu}'$ is positive definite.

\begin{lemma}\label{lem:positive_Sigma_mu_parity_split}
Assume the parity-split regime in the D-MRA setting. The covariance matrix $\bOmega_{\bmu}'$ defined in \eqref{eq:Omega_mu_parity_split} is positive definite.
\end{lemma}
\begin{proof}
Let 
\begin{align*}
L_j'=\sum_{\ell\in\cI_d^+} \alpha_\ell s_{\ell,j}+\sum_{t\in\cJ_{m,d}^{<}} \beta_t d_{t,j}+\beta_\star d_{t_\star,j}
\end{align*}
be a real linear combination of the coordinates of $\bmu_j$, and suppose $\Var(L_j')=0$. As in the proof of Lemma~\ref{lem:positive_Sigma_mu}, we aim to show that all coefficients $\alpha_\ell$ and $\beta_t$ vanish.

Condition on the latent group element $g_j=g$. For real observations, note that the independent noise coordinates are exactly the folded representatives: $\F\beps_j[0]=a_0$ and $\F\beps_j[d/2]=a_{d/2}$, are real Gaussians, while $\F\beps_j[\ell]=a_{\ell}+\ri b_{\ell}$ for $1\leq \ell< d/2$ are independent complex Gaussian coordinates with $\F\beps_j[d-\ell]=\overline{\F\beps_j[\ell]}$. All the random variables $\{a_\ell,b_{\ell}:1\leq \ell< d/2\}$ and $a_0, a_{d/2}$ are real-valued and independent Gaussian distributed.

Since $L_j$ has zero variance, its conditional version is almost surely constant for every $g$ with respect to the underlying distribution $\rho$. As a polynomial in the independent Gaussian coordinates, it must therefore be identically constant. Inspect the highest-degree part. The second-order coordinates $s_{\ell,j}$ contribute only quadratic leading terms, whereas the fourth-order coordinates $d_{t,j}$ contribute quartic leading terms. Thus the quartic leading terms must cancel among themselves.

For $t=(0,m,km)$ with $k=1,\ldots,d/(2m)-2$, each $d_{t,j}$ gives
\begin{align*}
2\Re\big(a_0 (a_m+\ri b_m) (a_{km}+\ri b_{km}) (a_{(k+1)m}-\ri b_{(k+1)m})\big),
\end{align*}
whereas the leading quartic term for $t_\star$ is 
\begin{align*}
2\Re\big( (a_m+\ri b_m)^2(a_{d/2-m}+\ri b_{d/2-m})^2 \big).
\end{align*}
When $d/m=4$, this gives $2\Re\big( (a_m+\ri b_m)^4\big)$. Hence, these leading polynomials are linearly independent, which implies all $\beta_t=0$ and $\beta_{\star}=0$.

After the quartic part is removed, the remaining leading part is quadratic
\begin{align*}
\sum_{\ell\in\cI_{d}^+} \alpha_\ell |\F\beps_j[\ell]|^2.
\end{align*}
Clearly, these quadratic polynomials are also linearly independent, so $\alpha_{\ell}=0$.

In conclusion, no nontrivial linear combination has zero variance, and $\bOmega_{\bmu}'$ is positive definite.
\end{proof}

We then link the auxiliary vectors with the test vectors $\widehat{\g}_m^0$ and $\widehat{\g}_m^1$. For $\u=(u_\ell)_{\ell\in\cI_d^+}$ and $\vv=(v_t)_{t\in\cJ_{m,d}^{<}\cup\{t_\star\}}$, define the polynomial maps $\varphi^0$ and $\varphi^1$ by
\begin{align}\label{eq:varphi_01}
\varphi^0(\u,\vv)\triangleq& \Big( (u_\ell)_{\ell\in\cI_{m,d}^{c,+}}, (4u_{\ell_1(t)} u_{\ell_2(t)} u_{\ell_3(t)} u_{\ell_4^{\sharp}(t)}- v_t^2)_{t\in\mathcal{J}_{m,d}^{<}}, 2u_m u_{d/2-m}-v_{t_\star} \Big)^\T,\\
\varphi^1(\u,\vv)\triangleq& \Big( (u_\ell)_{\ell\in\cI_{m,d}^{c,+}},u_{d/2}, (4u_{\ell_1(t)} u_{\ell_2(t)} u_{\ell_3(t)} u_{\ell_4^{\sharp}(t)}- v_t^2)_{t\in\mathcal{J}_{m,d}^{<}},2u_m u_{d/2-m}+v_{t_\star} \Big)^\T.
\end{align}
Then by construction
\begin{align*}
\widehat{\g}_m^{\epsilon}=\varphi^{\epsilon}(\widehat{\bmu}'), \qquad\text{for }\epsilon\in\{0,1\}.
\end{align*}
Similarly, the corresponding Jacobian $\nabla\varphi^{\epsilon}$ has full rank

\begin{lemma}[Full row rank of the Jacobian under the null and nondegenerate alternatives for the parity-split regime]
\label{lem:Jacobian_full_rankness_parity_split}
Assume the parity-split regime in the D-MRA setting. Let $\bmu'$ be the population auxiliary vector defined in \eqref{eq:mu_parity_split} at which the Jacobian $\nabla\varphi^0(\bmu')$ and $\nabla\varphi^1(\bmu')$ are evaluated, and let $q_m^0$ and $q_m^1$ be as defined in \eqref{eq:q0} and \eqref{eq:q1}. Then $\nabla\varphi^0(\bmu')$ and $\nabla\varphi^1(\bmu')$ have full row rank $q_m^0$ and $q_m^1$, respectively, under either of the following two conditions.
\begin{enumerate}
\item[(i)] The null hypothesis holds that the signal
$\x\in\R^d$ belongs to either symmetry class
$\Dm{m}^0$ or $\Dm{m}^1$.
Moreover, $\x$ is generic within the corresponding dihedral class in the sense that
\[
r_{km}>0,
\qquad
k=0,1,\ldots,\frac{d}{2m}-1.
\]
We do not impose any condition on $r_{d/2}$ when
$\x\in\Dm{m}^0$.
On the other hand, if $\x\in\Dm{m}^1$, then
$r_{d/2}=0$ automatically by the characterization of
$\Dm{m}^1$-symmetry.
\item[(ii)] Under a fixed alternative, the selected fourth-order population
coordinates are nonzero:
\[
d_t\neq 0,\qquad t\in \cJ_{m,d}^{<}.
\]
\end{enumerate}

\end{lemma}

\begin{proof}
We first consider $\varphi^0$. For the cyclic rows, we have $\varphi_\ell^0(\u,\vv)=u_{\ell}$ for $\ell\in\cI_{m,d}^{c,+}$, and thus
\begin{align*}
\frac{\partial \varphi_\ell^0}{\partial u_a} (\u,\vv)=& \1\{a=\ell\},\quad\text{for }a\in\cI_d^+,\\
\frac{\partial \varphi_\ell^0}{\partial v_b} (\u,\vv)=&0,\quad\text{for }b\in\cJ_{m,d}^{<}\cup\{t_\star\}.
\end{align*}
For the dihedral rows, we have, for $t=(0,m,km)\in\cJ_{m,d}^{<}$ with $k=1,\ldots,d/(2m)-2$,
\begin{align*}
\varphi_t^0(\u,\vv)&=4u_{\ell_1(t)} u_{\ell_2(t)} u_{\ell_3(t)} u_{\ell_4^{\sharp}(t)}- v_t^2,\\
&= 4u_0 u_m u_{km} u_{((k+1)m)} - v_t^2.
\end{align*}
Thus, for $t=(0,m,km)\in\cJ_{m,d}^{<}$, we have
\begin{align*}
\frac{\partial \varphi_t^0}{\partial u_a}(\u,\vv) =& 4\Big(
\1\{a=0\}u_m u_{km} u_{((k+1)m)}+
\1\{a=m\}u_0 u_{km} u_{((k+1)m)}\\
&+
\1\{a=km\}u_0 u_m u_{((k+1)m)}+
\1\{a=((k+1)m)\}u_0u_m u_{km}
\Big),\quad\text{for }a\in\cI_d^+,\\
\frac{\partial \varphi_t^0}{\partial v_b} (\u,\vv)=&-2v_t\1\{b=t\},\quad\text{for }b\in\cJ_{m,d}^{<}\cup\{t_\star\}.
\end{align*}
Finally, for the signed parity row, we have $\varphi_{t_\star}^0(\u,\vv)=2u_m u_{d/2-m} -v_{t_\star}$, and thus
\begin{align*}
\frac{\partial \varphi_{t_\star}^0}{\partial u_a} (\u,\vv)=& 2\Big(\1\{a=m\}u_{d/2-m} +\1\{a=d/2-m\}u_m \Big),\quad\text{for }a\in\cI_d^+,\\
\frac{\partial \varphi_{t_\star}^0}{\partial v_b} (\u,\vv)=& -\1\{b=t_\star\},\quad\text{for }b\in\cJ_{m,d}^{<}\cup\{t_\star\}.
\end{align*}
Similar to the proof of Lemma~\ref{lem:Jacobian_full_rankness}, if $d_t\neq 0$ for $t\in\cJ_{m,d}^{<}$, then the Jacobian $\nabla\varphi^0(\bmu')$ has full rank.

Now consider $\varphi^1$. For cyclic rows where $\ell\in\cI_{m,d}^{c,+}$, we additionally have 
\begin{align*}
\frac{\partial \varphi_\ell^1}{\partial u_{d/2}} (\u,\vv)= 0.
\end{align*}
For the row corresponding to $\ell=d/2$,
\begin{align*}\frac{\partial \varphi_{d/2}^1}{\partial u_a} (\u,\vv)=& 0,\quad\text{for }a\in\cI_d^+\setminus \{d/2\},\\
\frac{\partial \varphi_{d/2}^1}{\partial u_{d/2}} (\u,\vv)=& 1,\\
\frac{\partial \varphi_{d/2}^1}{\partial v_b} (\u,\vv)=&0,\quad\text{for }b\in\cJ_{m,d}^{<}\cup\{t_\star\}.
\end{align*}
For the dihedral rows where $t\in\cJ_{m,d}^{<}$, we additionally have
\begin{align*}
\frac{\partial \varphi_t^1}{\partial u_{d/2}}(\u,\vv) =0.
\end{align*}
Finally, for the signed parity row, we have $\varphi_{t_\star}^0=2u_m u_{d/2-m}+v_{t_\star}$, and thus,
\begin{align*}
\frac{\partial \varphi_{t_\star}^1}{\partial u_a} (\u,\vv)=& 2\Big(\1\{a=m\}u_{d/2-m} +\1\{a=d/2-m\}u_m \Big),\quad\text{for }a\in\cI_d^+\setminus \{d/2\},\\
\frac{\partial \varphi_{t_\star}^1}{\partial u_{d/2}} (\u,\vv)=&0,\\
\frac{\partial \varphi_{t_\star}^1}{\partial v_b} (\u,\vv)=& \1\{b=t_\star\},\quad\text{for }b\in\cJ_{m,d}^{<}\cup\{t_\star\}.
\end{align*}
Similarly, if $d_t\neq 0$ for $t\in\cJ_{m,d}^{<}$, then the Jacobian $\nabla\varphi^1(\bmu')$ has full rank.

\end{proof}

We are now ready to prove Theorem~\ref{thm:DMRA_Dm_testing_parity_split}.

\begin{proof}
The proof proceeds along the same lines as that of Theorem~\ref{thm:DMRA_Dm_testing}. We establish the result for both cases $\epsilon\in\{0,1\}$ simultaneously. Suppose the null hypothesis holds that $\x$ belongs to $\Dm{m}^{\epsilon}$-symmetry. Since the observations are Gaussian perturbations of a fixed signal, the central limit theorem gives 
\begin{align*}
\sqrt{N}\, (\widehat{\bmu}'-\bmu')\xrightarrow{d} \mathcal{N}(\0,\bOmega_{\bmu}'),
\end{align*}
where $\bOmega_{\bmu}'$ is the covariance matrix defined in \eqref{eq:Omega_mu_parity_split}. Applying the polynomial map $\varphi^{\epsilon}$, we have by construction $\widehat{\g}_m^{\epsilon}=\varphi^{\epsilon}(\widehat{\bmu}')$. Under the null hypothesis, Proposition\ref{prop:DMRA_mean_null_parity_split} implies that $\varphi^{\epsilon}(\bmu')=0$. Then the delta method yields
\begin{align*}
\sqrt{N}\,\widehat{\g}_m^{\epsilon}\xrightarrow{d} \mathcal{N}(\0, \bOmega_\g^{\epsilon})
\end{align*}
Note that
\begin{align*}
\bpsi_j^{\epsilon} = \nabla\varphi^{\epsilon}(\bmu')(\bmu_j'-\bmu'),
\end{align*}
which corresponds exactly to the coordinate-wise expressions given in \eqref{eq:psi_ell0}-\eqref{eq:psi_tstar1}. Therefore,
\begin{align*}
\bOmega_\g^{\epsilon} = \Var(\bpsi_j^{\epsilon}).
\end{align*}
By Lemma~\ref{lem:positive_Sigma_mu_parity_split} and~\ref{lem:Jacobian_full_rankness_parity_split}, under the generic condition in Lemma~\ref{lem:Jacobian_full_rankness_parity_split}(i), $\bOmega_\g^{\epsilon}$ is positive definite. Similar to the proof of Theorem~\ref{thm:DMRA_Dm_testing},
\begin{align*}
\widehat{\bOmega}_{\g}^{\epsilon}\xrightarrow{p} \bOmega_\g^{\epsilon}.
\end{align*}
Slutsky's theorem then implies the chi-squared limit for the statistic $\widehat{W}_m^{\epsilon}=N \cdot (\widehat{\g}_m^{\epsilon})^\T (\widehat{\bOmega}_\g^{\epsilon})^{-1} \widehat{\g}_m^{\epsilon}$ in \eqref{eq:chi_g01}, and the p-value formula \eqref{eq:raw_pvalue_dihedral_parity_split} follows. The consistency result under the alternative hypothesis, as well as the condition in Lemma~\ref{lem:Jacobian_full_rankness_parity_split}(ii), follow from analogous arguments and are therefore omitted.
\end{proof}

\section{Auxiliary Lemmas}
\begin{lemma}[Associativity of $n$-mode product]\label{lem:n-mode_product_associative}
Let $\mathbf{T}$ be a tensor and let $\mathbf{M}_1$, $\mathbf{M}_2$ be matrices such that all dimensions are compatible for $n$-mode multiplication as defined in Definition~\ref{def:n-mode_tensor_multiplication}. Then, the $n$-mode product is associative in the following sense:
\begin{align*}
(\mathbf{T} \times_n \mathbf{M}_1) \times_n \mathbf{M}_2 = \mathbf{T} \times_n (\mathbf{M}_2 \mathbf{M}_1).
\end{align*}
\end{lemma}

\begin{lemma}[Relationship between circulant and Fourier transform matrices]\label{lem:Cv}
Let $\vv\in\C^d$. Let $\Cb_\vv$ be the circulant matrix generated by $\vv$, and $\Db_{\F\vv},\Db_{\F^{-1}\vv}$ be the diagonal matrix generated by $\F\vv,\F^{-1}\vv$, respectively. Then we have
\begin{align*}
\Db_{\F\vv}=&\Ub_d \Cb_\vv \Ub_d^{\sH} = \Fb_d \Cb_\vv \Fb_d^{\sH}/d,\\
\Db_{\F^{-1}\vv}=&\Ub_d^{\sH} \Cb_\vv \Ub_d/d = \Fb_d^{\sH} \Cb_\vv \Fb_d/d^2.
\end{align*}
\end{lemma}

\begin{proof}
Let $\Cb_d\in\R^{d\times d}$ be the cyclic permutation matrix, i.e.,
\begin{align*}
\Cb_d=
\begin{pmatrix}
0 & 0 & \ldots & 0 & 1 \\
1 & 0 & \ldots & 0 & 0 \\
0 & 1 & \ldots & 0 & 0 \\
\vdots & \vdots & \ddots & \vdots & \vdots \\
0 & 0 & \ldots & 1 & 0
\end{pmatrix}.
\end{align*}
This is also the companion matrix of the polynomial $p(t)=t^d-1$. Thanks to the relationship between a Vandermonde matrix and the corresponding companion matrix,
\begin{align}\label{eq:VC_relationship}
\Cb_d=\Ub_d \Wb_d^{-1} \Ub_d^{\sH} = \Ub_d^{\sH} \Wb_d \Ub_d,
\end{align}
where $\Ub_d$ is the normalized Fourier transform matrix and $\Wb_d$ is the diagonal matrix with diagonal entries $\mathbf{W}_d[\ell,\ell]=w_d^{\ell}$ as defined in Section \ref{sec:Fourier_convention}. Recall that the circulant matrix $\Cb_\x$ has the following decomposition
\begin{align}\label{eq:circulant_decomposition}
\Cb_\vv = \sum_{\ell=0}^{d-1} \vv[\ell] \Cb_d^{\ell},
\end{align}
which implies
\begin{align*}
\Cb_\vv=& \sum_{\ell=0}^{d-1} \vv[\ell]  \Ub_d^{\sH} \Wb_d^{\ell} \Ub_d= \Ub_d^{\sH} \big(\sum_{\ell=0}^{d-1}\vv[\ell]\Wb_d^{\ell}\big) \Ub_d=\Ub_d^{\sH} \Db_{\F\vv} \Ub_d,\\
\Cb_\vv=& \sum_{\ell=0}^{d-1} \vv[\ell]  \Ub_d \Wb_d^{-\ell} \Ub_d^{\sH}=\Ub_d \big(\sum_{\ell=0}^{d-1}\vv[\ell]\Wb_d^{-\ell}\big) \Ub_d^{\sH}=d \Ub_d \Db_{\F^{-1}\vv}\Ub_d^{\sH}.
\end{align*}
This completes the proof.
\end{proof}

\begin{lemma}[Gaussian Fourier moment identity]\label{lem:Gaussian_Fourier_moment}
Let $\bbeta\sim \mathcal{N}(0,\sigma^2 \Ib_d)$. Then we have the following second-order identity
\begin{align*}
\E\big[\F \bbeta[i_1]\,
\overline{\F\bbeta[i_2]}\big]
&= d\sigma^2 \cdot \1\{ i_1-i_2\equiv 0 \ (\mathrm{mod}\ d) \},\\
\E\big[\F \bbeta[i_1]\,
\F\bbeta[i_2]\big]
&= d\sigma^2 \cdot \1\{ i_1 +i_2\equiv 0 \ (\mathrm{mod}\ d) \},
\end{align*}
for any $i_1,i_2\in\cI_d$. Moreover, we have the fourth-order identity
\begin{align*}
&\E\big[\F \bbeta[i_1]\,\F \bbeta[i_2]\,\F \bbeta[i_3]\,
\overline{\F\bbeta[i_1+i_2+i_3 \ (\mathrm{mod}\ d)]}\big]\\
&=d^2\sigma^4 \cdot \big(\1\{ i_1+i_2\equiv 0 \ (\mathrm{mod}\ d) \}+ \1\{ i_2+i_3\equiv 0 \ (\mathrm{mod}\ d) \}+\1\{ i_1+i_3\equiv 0 \ (\mathrm{mod}\ d) \}\big),
\end{align*}
for any $i_1,i_2,i_3\in\cI_d$.
\end{lemma}

\begin{proof}
Note that for any $0\leq i\leq d-1$,
\begin{align*}
\F \bbeta[i] = \sum_{k=0}^{d-1} \bbeta[k] e^{-\frac{2\pi\ri ki}{d}}.
\end{align*}
Hence, we obtain for any $0\leq i_1,i_2\leq d-1$,
\begin{align*}
\E\big[\F \bbeta[i_1]\,
\overline{\F\bbeta[i_2]}\big] =& \sum_{k=0}^{d-1} \sigma^2 \cdot e^{-\frac{2\pi\ri k(i_1-i_2)}{d}}=d\sigma^2 \cdot \1\{ i_1-i_2\equiv 0 \ (\mathrm{mod}\ d) \},\\
\E\big[\F \bbeta[i_1]\,
\F\bbeta[i_2]\big] = &  \sum_{k=0}^{d-1} \sigma^2 \cdot e^{-\frac{2\pi\ri k(i_1+i_2)}{d}}=d\sigma^2 \cdot \1\{ i_1+i_2\equiv 0 \ (\mathrm{mod}\ d) \}.
\end{align*}
For the second identity, we first have for any $0\leq i_1,i_2,i_3\leq d-1$,
\begin{align*}
&\E\big[\F \bbeta[i_1]\,\F \bbeta[i_2]\,\F \bbeta[i_3]\,
\overline{\F\bbeta[i_1+i_2+i_3]}\big]\\
&=\sum_{0\leq k_1,k_2,k_3,k_4\leq d-1} \E\big[\bbeta[k_1] \bbeta[k_2]\bbeta[k_3]\bbeta[k_4] \big]\cdot e^{-\frac{2\pi\ri (k_1i_1+k_2i_2+k_3i_3-k_4(i_1+i_2+i_3))}{d}}\\
&=\sum_{0\leq k_1,k_2,k_3,k_4\leq d-1} \sigma^4\cdot \Big(\1\{k_1\equiv k_2 \ (\mathrm{mod}\ d) \}\cdot\1\{k_3\equiv k_4 \ (\mathrm{mod}\ d) \} \\
&\quad + \1\{k_1\equiv k_3 \ (\mathrm{mod}\ d) \}\cdot\1\{k_2\equiv k_4 \ (\mathrm{mod}\ d) \}+ \1\{k_1\equiv k_4 \ (\mathrm{mod}\ d) \}\cdot\1\{k_2\equiv k_3 \ (\mathrm{mod}\ d) \}\Big)\\
&\quad \cdot e^{-\frac{2\pi\ri (k_1i_1+k_2i_2+k_3i_3-k_4(i_1+i_2+i_3))}{d}},
\end{align*}
where we use the Gaussian moment factoring identity in the last equation. Now we evaluate the contribution for each pairing. For the pairing that $k_1\equiv k_2$ and $k_3\equiv k_4$, the contribution equals
\begin{align*}
\sigma^4\cdot \sum_{k_1=0}^{d-1}\sum_{k_3=0}^{d-1} e^{-\frac{2\pi\ri (k_1(i_1+i_2)-k_3(i_1+i_2))}{d}}=d^2\sigma^4\cdot \1\{ i_1+i_2\equiv 0 \ (\mathrm{mod}\ d) \}.
\end{align*}
For the pairing that $k_1\equiv k_3$ and $k_2\equiv k_4$, the contribution equals
\begin{align*}
\sigma^4\cdot \sum_{k_1=0}^{d-1}\sum_{k_2=0}^{d-1} 
e^{-\frac{2\pi\ri (k_1(i_1+i_3)-k_2(i_1+i_3))}{d}}=d^2\sigma^4\cdot \1\{ i_1+i_3\equiv 0 \ (\mathrm{mod}\ d) \}.
\end{align*}
For the pairing that $k_1\equiv k_4$ and $k_2\equiv k_3$, the contribution equals
\begin{align*}
\sigma^4\cdot \sum_{k_1=0}^{d-1}\sum_{k_2=0}^{d-1} 
e^{-\frac{2\pi\ri (k_2(i_2+i_3)-k_1(i_2+i_3))}{d}}=d^2\sigma^4\cdot \1\{ i_2+i_3\equiv 0 \ (\mathrm{mod}\ d) \}.
\end{align*}
Combining all gives
\begin{align*}
&\E\big[\F \bbeta[i_1]\,\F \bbeta[i_2]\,\F \bbeta[i_3]\,
\overline{\F\bbeta[i_1+i_2+i_3]}\big]\\
&=d^2\sigma^4\cdot\Big( \1\{ i_1+i_2\equiv 0 \ (\mathrm{mod}\ d) \}+\1\{ i_1+i_3\equiv 0 \ (\mathrm{mod}\ d) \}+\1\{ i_2+i_3\equiv 0 \ (\mathrm{mod}\ d) \} \Big).
\end{align*}

\end{proof}

\section{FWER control by sequential rejection principle}

\subsection{Directed acyclic graph}

A directed graph is a graph $G=(V,E)$ consisting of vertices $V$ connected by directed edges $E$. A directed acyclic graph (DAG) is a directed graph with no directed cycles; that is, following that direction according to these directed edges never forms a closed loop. For each directed edge $(u,v)\in E$ from vertex $u$ to $v$ in a DAG, we call $u$ is a \textit{parent vertex} of $v$, and $v$ is a \textit{child vertex} of $u$. For each vertex $u$, we denote the sets consisting of all its parents, ancestors, children, and offspring by $\text{parent}(u)$, $\text{ancestor}(u)$, $\text{child}(u)$, and $\text{offspring}(u)$, respectively. Notably, a direct graph is a DAG if and only if it admits a \textit{topological ordering} of vertices, which is a linear ordering such that for every directed edge, the parent vertex comes before the child vertex in the ordering (a topological ordering may not be unique, but in our paper, any valid ordering suffices). Several algorithms can be used to obtain a valid topological ordering for a given DAG, such as Kahn's algorithm or depth-first search (DFS). DAGs have many scientific and computational applications, including modeling causal relationships~\cite[Chapters 2--4]{peters2017elements}, task scheduling~\cite{kwok1999static}, version control systems~\cite{ram2013git}, genealogy~\cite[Chapters 2--4]{balding2008handbook}, and beyond.

\subsection{Multiple testing under DAG structure and sequential rejection principle}\label{sec:SRA}

Multiple testing procedures based on directed acyclic graph (DAG) structures provide a flexible and powerful framework for testing general hypotheses while accounting for complex logical relationships among them. The hierarchical testing framework introduced by \cite{meinshausen2008hierarchical} was originally designed for tree-structured hypotheses. This foundational approach was later extended in \cite{meijer2015multiple}, guided by the sequential rejection principle \cite{goeman2010sequential}, to handle the more general setting of DAG structures, offering a broader and more versatile framework for hypothesis testing. By incorporating logical dependencies within the DAG, these methods enable a structured ``zooming-in'' strategy: larger, broader hypotheses are tested first, with smaller, more specific hypotheses tested only if their parent hypotheses are rejected. This framework has found applications in various domains, such as analyzing associations among groups of variables or testing gene sets derived from the Gene Ontology \cite{ashburner2000gene}, where hierarchical relationships naturally arise. In such cases, DAG-based methods have been employed to detect associations or to identify significant subsets within larger groups, demonstrating their adaptability and effectiveness in structured testing scenarios.

Building on the framework of \cite{meijer2015multiple}, we apply the method to the symmetry detection setting, where logical relationships naturally arise from the inclusion relationships among different symmetry groups. Specifically, we construct a DAG in which each vertex represents a symmetry group of the signal, and directed edges are defined by the inclusion relationships between these groups. In other words, for any directed edge, the symmetry represented by the parent vertex implies the symmetry represented by the child vertex. Our goal is to detect the symmetry of the signal while providing rigorous statistical guarantees~---~specifically, control of the familywise error rate (FWER), which is strong and exact in finite samples for the cyclic procedure and asymptotic, under the nondegeneracy conditions of Proposition~\ref{prop:FWER_dihedral}, for the dihedral procedure. This approach ensures that the hierarchical structure of symmetry groups is appropriately leveraged in the testing process, maintaining validity across the entire family of tests.

Algorithm~\ref{alg:weight_update_SR} supplies the weights for
the any-parent procedure. For an unrejected hypothesis $H$ and
a rejection set $\mathcal D$, define
\begin{align}
\alpha_H^{\mathrm{any}}(\mathcal D)
&=\frac{\alpha\,w_H(\mathcal D)}
{|\mathcal L\setminus\mathcal D|}, \notag\\
\alpha_H^{\mathrm{all}}(\mathcal D)
&=\alpha_H^{\mathrm{any}}(\mathcal D)\,
\mathbf 1\{\mathrm{parent}(H)\subseteq\mathcal D\}.
\label{eq:parent_rule_thresholds}
\end{align}
The all-parents variant therefore withholds a hypothesis's
testing level until all its parents are rejected, without
redistributing that level. The root is eligible under both
rules because its parent set is empty. The procedure terminates
when no unrejected leaves remain. Unless stated otherwise,
we use the any-parent rule, including in the numerical experiments.

To describe the detailed procedure, denote the full vertex set of the DAG by $\mathcal{H}=\{H_i: i=1,\ldots,K\}$, where $H_i$ represents the hypothesis for each candidate symmetry group of our signal. The vertices are arranged in a topological ordering, with $H_1$ as the root vertex, also ensuring that for any vertex $H \in \mathcal{H}$, all its parent vertices precede it in the ordering. Assume that raw p-values $p_H$ are available for all $H\in\mathcal{H}$, and a significance level $\alpha$ has been specified for controlling the FWER. The sequential rejection method then operates as an iterative multiple testing procedure (see Algorithm \ref{alg:SR} for more details), beginning with an empty rejection set $\mathcal{D}_0$. At each iteration $t$, critical values for all hypotheses, given the current rejection set $\mathcal{D}_{t-1}$, are calculated (see Step 1). Hypotheses with p-values smaller than their assigned $\alpha$-level are added to the current rejection set. Due to the DAG structure~---~where the falsehood of a child vertex implies the falsehood of all its ancestors~---~all ancestors of the newly rejected vertices are also added to the rejection set, a process often referred to as free rejection, forming the updated rejection set $\mathcal{D}_t$ (see Step 2). In the next iteration, the critical values are then updated based on this expanded rejection set, and the process is repeated until no further rejections can be made. Based on \cite{goeman2010sequential}, as long as the critical values satisfy the monotonicity condition and the single-step condition, this method can strongly control the FWER at the given level $\alpha$. The single-step condition requires the raw $p$-values of the true hypotheses that receive positive critical values to be valid, exactly or, for an asymptotic statement, in the limit; which hypotheses these are, and what their validity requires in the dihedral case, is made explicit in Proposition~\ref{prop:FWER_dihedral}, proved in Section~\ref{sec:proof_FWER_dihedral} below. For further details, we refer readers to \cite{goeman2010sequential,meijer2015multiple}.

\begin{algorithm}[htb]
\caption{Sequential Rejection for DAG}\label{alg:SR}
\begin{algorithmic}[1]
\REQUIRE A DAG with vertices
$\mathcal H=\{H_1,\ldots,H_K\}$ in topological order;
raw $p$-values $(p_H)_{H\in\mathcal H}$;
significance level $\alpha$;
rule $b\in\{\mathrm{any},\mathrm{all}\}$.

\STATE Let $\mathcal L$ be the set of leaf vertices.
\STATE Initialize $\mathcal D_0=\varnothing$ and $t=0$.

\REPEAT
    \STATE Set $t=t+1$.
    \IF{$\mathcal L\setminus\mathcal D_{t-1}=\varnothing$}
        \STATE \textbf{return} $\mathcal D_{t-1}$.
    \ENDIF

    \STATE \textbf{Step 1: Update the critical values.}
    \STATE Compute $w_H(\mathcal D_{t-1})$ using
    Algorithm~\ref{alg:weight_update_SR}.
    \FOR{each $H\in\mathcal H\setminus\mathcal D_{t-1}$}
        \STATE Set
        $\displaystyle
        \alpha_H(\mathcal D_{t-1})
        =\frac{\alpha\,w_H(\mathcal D_{t-1})}
        {|\mathcal L\setminus\mathcal D_{t-1}|}$.
        \IF{$b=\mathrm{all}$ and
        $\mathrm{parent}(H)\not\subseteq\mathcal D_{t-1}$}
            \STATE Set $\alpha_H(\mathcal D_{t-1})=0$.
        \ENDIF
    \ENDFOR

    \STATE \textbf{Step 2: Update the rejection set.}
    \STATE Set
    \[
    \mathcal R_t=
    \left\{
    H\in\mathcal H\setminus\mathcal D_{t-1}:
    \alpha_H(\mathcal D_{t-1})>0,\quad
    p_H\leq\alpha_H(\mathcal D_{t-1})
    \right\}.
    \]
    \STATE Add the new rejections and their ancestors:
    \[
    \mathcal D_t
    =\mathcal D_{t-1}\cup\mathcal R_t
    \cup\bigcup_{H\in\mathcal R_t}\mathrm{ancestor}(H).
    \]
\UNTIL{$\mathcal D_t=\mathcal D_{t-1}$}

\STATE \textbf{return} $\mathcal D_t$.
\end{algorithmic}
\end{algorithm}

Algorithm~\ref{alg:weight_update_SR} computes the weights $w_H(\mathcal D)$ used in \eqref{eq:parent_rule_thresholds}. It assigns unit weight to each unrejected leaf and propagates weight toward unrejected parents in reverse topological order. The total weight is $|\mathcal L\setminus\mathcal D|$. The same weights are used for both parent rules.

\begin{algorithm}[htb]
    \caption{Weight Update for Sequential Rejection Algorithm}\label{alg:weight_update_SR}
\begin{algorithmic}[1]
\REQUIRE A directed acyclic graph (DAG) with the full node set $\mathcal{H}=\{H_i: i=1,\ldots,K\}$ satisfying the topological ordering; current rejection set $\mathcal{D}$.

\STATE Initialize weights: \\
\quad \quad $w_H=1$ for all $H\in\mathcal{L}\setminus\mathcal{D}$;\\
\quad \quad $w_H=0$ otherwise.
\FOR{$i = K$ to $1$}
    \IF{$H_i\not\in \mathcal{D}$}
        \IF{$\text{parent}(H_i)\setminus \mathcal{D}\not=\emptyset$}
            \FOR{each $P\in \text{parent}(H_i)\setminus \mathcal{D}$}
                \STATE $w_P = w_P + w_{H_i}/|\text{parent}(H_i)|$.
            \ENDFOR
            \STATE $w_{H_i} = w_{H_i} - |\text{parent}(H_i)\setminus \mathcal{D}|\cdot w_{H_i}/|\text{parent}(H_i)|$.
        \ENDIF
    \ENDIF
\ENDFOR
\STATE \textbf{return} \texttt{$w_H(\mathcal{D})\triangleq w_H$ for all $H\in\mathcal{H}$}.
\end{algorithmic}
\end{algorithm}

\begin{figure}[htb]
    \centering
    \begin{subfigure}[b]{0.48\textwidth}
        \centering
        \includegraphics[width=\textwidth]{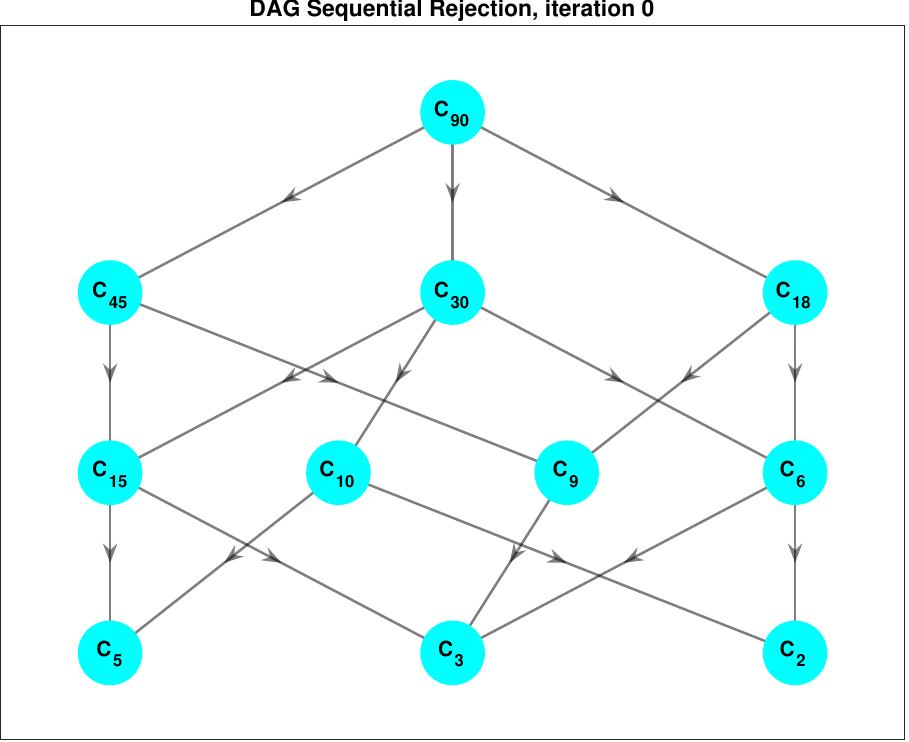}
    \end{subfigure}
    \hfill
    \begin{subfigure}[b]{0.48\textwidth}
        \centering
        \includegraphics[width=\textwidth]{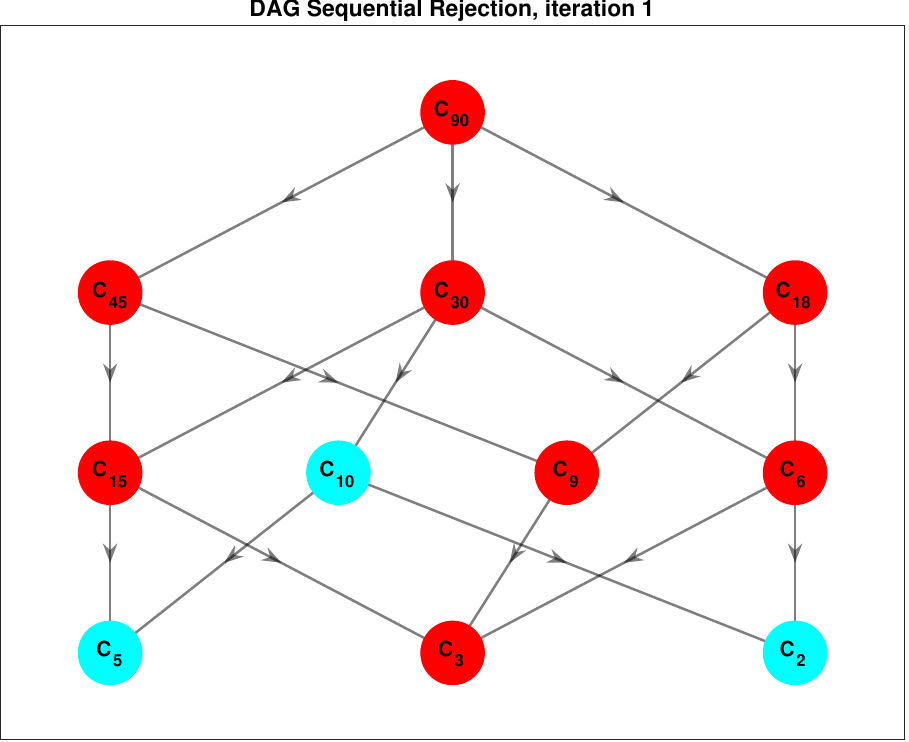}
    \end{subfigure}
    \caption{A two-step sequential rejection for DAG, Algorithm~\ref{alg:SR}, for $d=90$ and symmetry group $\Cd{10}$. The left shows the initialization step. On the right, all red nodes are rejected. The remaining nodes are associated with $\Cd{10}$ and its subgroups.}
    \label{fig:refection_illust}
\end{figure}

Figure~\ref{fig:refection_illust} provides a graphical illustration of sequential rejection for DAG.

\subsection{Proof of Proposition~\ref{prop:FWER_dihedral}}\label{sec:proof_FWER_dihedral}

\begin{proof}
Fix $\x,\rho,\alpha$, and abbreviate
$\mathcal T=\mathcal T(\x)$ and
$\mathcal F=\mathcal F(\x)$.
For ancestor-closed rejection sets
$\mathcal D\subseteq\mathcal D'$ and $H\notin\mathcal D'$,
the any-parent critical values are nondecreasing
\cite{meijer2015multiple}.
The eligibility indicator in
\eqref{eq:parent_rule_thresholds} is also nondecreasing,
so the all-parents critical values inherit this property.
Moreover, for either rule $b$,
\[
\sum_{H\notin\mathcal D}\alpha_H^b(\mathcal D)\leq\alpha:
\]
the any-parent weights exhaust the available level, while
the all-parents restriction can only decrease individual
critical values. Apart from the root, positive critical
values require at least one rejected parent under the
any-parent rule and all parents rejected under the
all-parents rule.

\emph{Step 1 (reduction to the single-step event).} Consider the event
\[
E\triangleq\bigcap_{H\in\mathcal T}\{p_H>\alpha_H(\mathcal F)\}.
\]
We claim that on $E$ no true hypothesis is ever rejected, i.e., $\mathcal D_t\subseteq\mathcal F$ for all $t$. Indeed, $\mathcal D_0=\varnothing\subseteq\mathcal F$. If $\mathcal D_{t-1}\subseteq\mathcal F$, then for every $H\in\mathcal T$, monotonicity gives $\alpha_H(\mathcal D_{t-1})\le\alpha_H(\mathcal F)<p_H$ on $E$, so no true hypothesis is added in Step~2 of Algorithm~\ref{alg:SR} by its $p$-value; the free-rejection step adds ancestors of rejected, hence false, hypotheses, and ancestors of false hypotheses are false because $\mathcal F$ is closed under taking ancestors. Thus $\mathcal D_t\subseteq\mathcal F$, and by induction the claim holds. Consequently,
\begin{equation}\label{eq:fwer_union}
\mathbb P\big(\text{some }H\in\mathcal T\text{ is rejected}\big)\le\mathbb P(E^c)\le\sum_{H\in\mathcal T}\mathbb P\big(p_H\le\alpha_H(\mathcal F)\big)
=\sum_{H\in\mathcal T_+}\mathbb P\big(p_H\le\alpha_H(\mathcal F)\big),
\end{equation}
where $\mathcal T_+\triangleq\{H\in\mathcal T:\ \alpha_H(\mathcal F)>0\}$; the last equality holds because every raw $p$-value is almost surely positive (the cyclic $p$-values have a continuous distribution, and a Wald $p$-value vanishes only if the statistic is infinite). By the third property above, $\mathcal T_+\subseteq\mathcal T_1(\x)$ under the any-parent rule and $\mathcal T_+\subseteq\mathcal T_{\max}(\x)$ under the all-parents rule.

\emph{Step 2 (validity of the relevant $p$-values).} Let $H\in\mathcal T_+$. If $H$ is a cyclic vertex, the root, or $\Dm{d/2}^0$, its $p$-value is exactly valid for every $N$ and $\rho$ (Theorem~\ref{thm:test_size_power}), so $\mathbb P(p_H\le u)\le u$ for all $u\in(0,1)$. If $H=\Dm{m}$ (resp.\ $\Dm{m}^\epsilon$) is a dihedral vertex tested by a Wald statistic and $\x$ is nondegenerate at $H$, then $\x$ satisfies the null hypothesis of Theorem~\ref{thm:DMRA_Dm_testing} (resp.\ Theorem~\ref{thm:DMRA_Dm_testing_parity_split}) together with the genericity condition of Lemma~\ref{lem:Jacobian_full_rankness}(i) (resp.\ Lemma~\ref{lem:Jacobian_full_rankness_parity_split}(i)), so $\widehat W\xrightarrow{d}\chi^2_q$ for the appropriate $q$, and therefore $p_H=1-F_{\chi^2_q}(\widehat W)$ converges in distribution to the uniform law on $(0,1)$: $\mathbb P(p_H\le u)\to u$ for every $u\in(0,1)$.

\emph{Step 3 (conclusion).} Under the hypothesis of (i) (any-parent rule) or (ii) (all-parents rule), every $H\in\mathcal T_+$ falls under Step~2. Since $\mathcal H$ is finite and the numbers $\alpha_H(\mathcal F)$ do not depend on $N$, \eqref{eq:fwer_union} gives
\[
\limsup_{N\to\infty}\mathbb P\big(\text{some }H\in\mathcal T\text{ is rejected}\big)\le\sum_{H\in\mathcal T_+}\alpha_H(\mathcal F)\le\sum_{H\notin\mathcal F}\alpha_H(\mathcal F)\le\alpha,
\]
by the budget bound. If no $H\in\mathcal T_+$ is a Wald-tested dihedral vertex, in particular under the hypothesis of (iii), all $p$-values in \eqref{eq:fwer_union} are exactly valid and the same bound holds for every $N$, without taking limits.

It remains to identify $\mathcal T_{\max}(\x)$ in (ii) when $\x$ has exact symmetry group $\Dmj{m'}{j}$ with $m'\le d/3$. Let $E_\x$ denote the vertex of the corresponding detectable class, $\Dm{m'}$ in the collapsed regime and $\Dm{m'}^{\,j\bmod2}$ in the parity-split regime. A vertex is true for $\x$ if and only if its class is implied by $\Dmj{m'}{j}$-invariance, and by construction of the DAG (Section~\ref{sec:DAG_construction_dihedral}) every class implied by $E_\x$ is a descendant of $E_\x$; hence $\mathcal T=\{E_\x\}\cup\mathrm{offspring}(E_\x)$. Every $H\in\mathrm{offspring}(E_\x)$ has a parent on a directed path from $E_\x$ to $H$, and that parent lies in $\mathcal T$, so $H\notin\mathcal T_{\max}(\x)$. The parents of $E_\x$ correspond to strictly larger symmetry classes, which $\x$ does not possess since $\Dmj{m'}{j}$ is its full symmetry group, so they are false and $E_\x\in\mathcal T_{\max}(\x)$. Thus $\mathcal T_{\max}(\x)=\{E_\x\}$, and nondegeneracy at $E_\x$ means precisely that $\x$ is a generic $\Dm{m'}$ (resp.\ $\Dm{m'}^{\,j\bmod2}$) signal.
\end{proof}

\end{document}